\documentclass[twocolumn,tighten]{aastex63}

\pdfoutput=1 %for arXiv submission
\usepackage{amsmath,amstext,longtable}
\usepackage[T1]{fontenc}
\usepackage{apjfonts} 
\usepackage[figure,figure*]{hypcap}
\received{2020 Oct 11}
\revised{2020 Nov 12}
\accepted{2020 Nov 22}
\submitjournal{ApJS}
\shortauthors{Kirkpatrick et al.}
\shorttitle{20-pc Mass Function}

\usepackage{color,amsmath,longtable}
\begin{document}

\title{The Field Substellar Mass Function Based on the Full-sky 20-pc Census of 525 L, T, and Y Dwarfs}

\correspondingauthor{J.\ Davy Kirkpatrick}
\email{davy@ipac.caltech.edu}

\author[0000-0003-4269-260X]{J.\ Davy Kirkpatrick}
\affiliation{IPAC, Mail Code 100-22, Caltech, 1200 E. California Blvd., Pasadena, CA 91125, USA; davy@ipac.caltech.edu}

% Major co-authors who have contributed a lot

\author{Christopher R.\ Gelino}
\affiliation{IPAC, Mail Code 100-22, Caltech, 1200 E. California Blvd., Pasadena, CA 91125, USA}

\author[0000-0001-6251-0573]{Jacqueline K.\ Faherty}
\affiliation{Department of Astrophysics, American Museum of Natural History, Central Park West at 79th Street, New York, NY 10034, USA}

\author[0000-0002-1125-7384]{Aaron M. Meisner}
\affiliation{NSF's National Optical-Infrared Astronomy Research Laboratory, 950 N. Cherry Ave., Tucson, AZ 85719, USA}

\author[0000-0001-7896-5791]{Dan Caselden}
\affiliation{Gigamon Applied Threat Research, 619 Western Avenue, Suite 200, Seattle, WA 98104, USA}
%Dan replied with above

\author[0000-0002-6294-5937]{Adam C.\ Schneider}
\affiliation{US Naval Observatory, Flagstaff Station, P.O.Box 1149, Flagstaff, AZ 86002, USA}
\affiliation{Department of Physics and Astronomy, George Mason University, MS3F3, 4400 University Drive, Fairfax, VA 22030, USA}

\author[0000-0001-7519-1700]{Federico Marocco}
\affiliation{IPAC, Mail Code 100-22, Caltech, 1200 E. California Blvd., Pasadena, CA 91125, USA}

\author{Alfred J.\ Cayago}
\affiliation{Department of Statistics, University of California Riverside, 900 University Avenue, Riverside, CA, 92521, USA}

\author[0000-0002-4424-4766]{R.\ L.\ Smart}
\affiliation{Istituto Nazionale di Astrofisica, Osservatorio Astrofisico di Torino, Strada Osservatorio 20, 10025 Pino Torinese, Italy}

\author{Peter R.\ Eisenhardt}
\affiliation{Jet Propulsion Laboratory, California Institute of Technology, MS 169-237, 4800 Oak Grove Drive, Pasadena, CA 91109, USA}

\author[0000-0002-2387-5489]{Marc J. Kuchner}
\affiliation{NASA Goddard Space Flight Center, Exoplanets and Stellar Astrophysics Laboratory, Code 667, Greenbelt, MD 20771, USA}

\author[0000-0001-5058-1593]{Edward L.\ Wright}
\affiliation{Department of Physics and Astronomy, University of California Los Angeles, 430 Portola Plaza, Box 951547, Los Angeles, CA, 90095-1547, USA}

\author[0000-0001-7780-3352]{Michael C.\ Cushing}
\affiliation{The University of Toledo, 2801 West Bancroft Street, Mailstop 111, Toledo, OH 43606, USA}

% Others on the CatWISE, BYW, or extended teams

\author[0000-0003-0580-7244]{Katelyn N.\ Allers}
\affiliation{Department of Physics and Astronomy, Bucknell University, Lewisburg, PA 17837, USA}

\author[0000-0001-8170-7072]{Daniella C.\ Bardalez Gagliuffi}
\affiliation{Department of Astrophysics, American Museum of Natural History, Central Park West at 79th Street, New York, NY 10034, USA}

\author[0000-0002-6523-9536]{Adam J.\ Burgasser}
\affiliation{Center for Astrophysics and Space Science, University of California San Diego, La Jolla, CA 92093, USA}

\author[0000-0002-2592-9612]{Jonathan Gagn\'e}
\affiliation{Institute for Research on Exoplanets, Universit\'e de Montr\'eal, Montr\'eal, Canada}

\author[0000-0002-9632-9382]{Sarah E. Logsdon}
\affiliation{NSF's National Optical-Infrared Astronomy Research Laboratory, 950 N. Cherry Ave., Tucson, AZ 85719, USA}

\author[0000-0002-0618-5128]{Emily C.\ Martin}
\affiliation{Department of Astronomy \& Astrophysics, University of California Santa Cruz, 1156 High Street, Santa Cruz, CA 95064, USA}

% Co-authors who helped with Spitzer logistics

\author[0000-0003-4714-1364]{James G.\ Ingalls}
\affiliation{IPAC, Mail Code 314-6, Caltech, 1200 E. California Blvd., Pasadena, CA 91125, USA}

\author[0000-0001-8014-0270]{Patrick J.\ Lowrance}
\affiliation{IPAC, Mail Code 314-6, Caltech, 1200 E. California Blvd., Pasadena, CA 91125, USA}

% Co-authors who helped with spectroscopic follow-up only

\author[0000-0002-9879-1183]{Ellianna S. Abrahams}
\affiliation{Department of Astronomy, University of California, Berkeley, CA 94720-3411, USA}
\affiliation{Department of Statistics, University of California, Berkeley, CA 94720-3411, USA}

\author{Christian Aganze}
\affiliation{Center for Astrophysics and Space Science, University of California San Diego, La Jolla, CA 92093, USA}

\author{Roman Gerasimov}
\affiliation{Center for Astrophysics and Space Science, University of California San Diego, La Jolla, CA 92093, USA}

\author[0000-0003-4636-6676]{Eileen C.\ Gonzales}
%\affiliation{Department of Astrophysics, American Museum of Natural History, New York, NY 10024, USA}
%\affiliation{The Graduate Center, City University of New York, New York, NY 10016, USA}
%\affiliation{Department of Physics and Astronomy, Hunter College, City University of New York, New York, NY 10065, USA}
\affiliation{Department of Astronomy and Carl Sagan Institute, Cornell University, 122 Sciences Drive, Ithaca, NY 14853, USA}
\affiliation{51 Pegasi b Fellow}

\author[0000-0002-5370-7494]{Chih-Chun Hsu}
\affiliation{Center for Astrophysics and Space Science, University of California San Diego, La Jolla, CA 92093, USA}

\author[0000-0002-3233-2451]{Nikita Kamraj}
\affiliation{Cahill Center for Astronomy and Astrophysics, California Institute of Technology, Pasadena, CA 91125, USA}

\author[0000-0003-2102-3159]{Rocio Kiman}
\affiliation{Department of Astrophysics, American Museum of Natural History, Central Park West at 79th St., New York, NY 10024, USA}
\affiliation{Department of Physics, Graduate Center, City University of New York, 365 5th Ave., New York, NY 10016, USA}

\author{Jon Rees}
\affiliation{Center for Astrophysics and Space Science, University of California San Diego, La Jolla, CA 92093, USA}

\author{Christopher Theissen}
\affiliation{Center for Astrophysics and Space Science, University of California San Diego, La Jolla, CA 92093, USA}
\affiliation{NASA Sagan Fellow}

% Co-authors who helped with candidate selection

\author{Kareem Ammar}
\affiliation{Polytechnic School, 1030 E.\ California Blvd., Pasadena, CA 91106, USA}

\author[0000-0003-4714-3829]{Nikolaj Stevnbak Andersen}
\affiliation{Sygehus Lillebalt, Department of Cardiology, Kolding, Denmark}
%Nikolaj responded with the above

\author{Paul Beaulieu}
\affiliation{Backyard Worlds: Planet 9}
%Paul replied with the above info

\author[0000-0002-7630-1243]{Guillaume Colin}
\affiliation{Backyard Worlds: Planet 9}
%Guillaume responded with the above info

\author{Charles A.\ Elachi}
\affiliation{St. Francis High School, 200 Foothill Blvd., La Ca\~nada Flintridge, CA 91011, USA}

\author[0000-0003-2236-2320]{Samuel J. Goodman}
\affiliation{Backyard Worlds: Planet 9}
%Sam replied with the info above

\author[0000-0002-8960-4964]{L\'eopold Gramaize}
\affiliation{Backyard Worlds: Planet 9}

\author[0000-0002-7389-2092]{Leslie K.\ Hamlet}
\affiliation{Backyard Worlds: Planet 9}
%Leslie sent the info above

\author{Justin Hong}
\affiliation{Pasadena High School, 2925 E.\ Sierra Madre Blvd., Pasadena, CA 91107, USA}

\author{Alexander Jonkeren}
\affiliation{Backyard Worlds: Planet 9}

\author{Mohammed Khalil}
%\affiliation{IPAC, Mail Code 100-22, Caltech, 1200 E. California Blvd., Pasadena, CA 91125, USA}
\affiliation{International College, P.O. Box 113-5373 Hamra, Bliss Street, Beirut, Lebanon}
\affiliation{Stanford University, 450 Serra Mall, Stanford, CA 94305, USA}

\author{David W.\ Martin}
\affiliation{Backyard Worlds: Planet 9}
%David replied with the above info

\author{William Pendrill}
\affiliation{Backyard Worlds: Planet 9}
%Billy replied with the above info

\author[0000-0001-9692-7908]{Benjamin Pumphrey}
\affiliation{Augusta Psychological Associates, Suite A1, Fishersville, VA 22939, USA}
%Ben replied with above

\author[0000-0003-4083-9962]{Austin Rothermich}
\affiliation{Physics Department, University of Central Florida, 4000 Central Florida Boulevard, Orlando, FL 32816, USA}
%Austin replied with the info above

\author[0000-0003-4864-5484]{Arttu Sainio}
\affiliation{Backyard Worlds: Planet 9}

\author{Andres Stenner}
\affiliation{Backyard Worlds: Planet 9}

\author{Christopher Tanner}
\affiliation{Backyard Worlds: Planet 9}
%Christopher replied with above

\author[0000-0001-5284-9231]{Melina Th\'evenot}
\affiliation{Backyard Worlds: Planet 9}
%Melina replied with above

\author{Nikita V.\ Voloshin}
\affiliation{Backyard Worlds: Planet 9}
%Nikita replied with above

\author{Jim Walla}
\affiliation{Backyard Worlds: Planet 9}
%Jim replied with above

\author{Zbigniew W{\k e}dracki}
\affiliation{Backyard Worlds: Planet 9}
%Zigi replied with the above info

\author{The Backyard Worlds: Planet 9 Collaboration}

\begin{abstract}

We present final {\it Spitzer} trigonometric parallaxes for 361 L, T, and Y dwarfs. We combine these with prior studies to build a list of 525 known L, T, and Y dwarfs within 20 pc of the Sun, 38 of which are presented here for the first time. Using published photometry and spectroscopy as well as our own follow-up, we present an array of color-magnitude and color-color diagrams to further characterize census members, and we provide polynomial fits to the bulk trends. Using these characterizations, we assign each object a $T_{\rm eff}$ value and judge sample completeness over bins of $T_{\rm eff}$ and spectral type. Except for types $\ge$ T8 and $T_{\rm eff} <$ 600K, our census is statistically complete to the 20-pc limit. We compare our measured space densities to simulated density distributions and find that the best fit is a power law ($dN/dM \propto M^{-\alpha}$) with $\alpha = 0.6{\pm}0.1$. We find that the evolutionary models of Saumon \& Marley correctly predict the observed magnitude of the space density spike seen at 1200K $< T_{\rm eff} <$ 1350K, believed to be caused by an increase in the cooling timescale across the L/T transition. Defining the low-mass terminus using this sample requires a more statistically robust and complete sample of dwarfs $\ge$Y0.5 and with $T_{\rm eff} <$ 400K. We conclude that such frigid objects must exist in substantial numbers, despite the fact that few have so far been identified, and we discuss possible reasons why they have largely eluded detection.

\end{abstract}

%\keywords{stars: low-mass, brown dwarfs -- (stars:) subdwarfs -- stars: fundamental parameters -- (Galaxy:) solar neighborhood -- catalogs}

\section{Introduction}

We now find ourselves at a moment in history where selecting parallax-based censuses of nearby objects from the hottest O stars to the coldest Y dwarfs is almost a reality. With the release of {\it Gaia} Data Release 2 (DR2; \citealt{gaia2018}) and Data Release 3 (scheduled for the first half of 2022), the astronomical community can begin extracting complete, volume-limited samples out to distances which provide exquisite statistics on the distribution of stellar types. As a result of operating at wavelengths $<$1 $\mu$m and selecting a conservative detection threshold, {\it Gaia} provides complete astrometry only for L5 dwarfs out to $\sim$24 pc (\citealt{smart2017}). Extending this census to colder types, though, is more easily accomplished by ground-based or space-based astrometric monitoring at longer wavelengths, where late-L, T, and Y dwarfs are brightest. A complete, volume-limited census across all stellar and substellar types is extremely useful in a variety of investigations, including: (1) analysis of the mass function, (2) determining the frequency of binaries across all types, (3) providing a catalog of host stars around which the nearest habitable planets to our own Solar System can be searched, and (4) establishing correlations among colors, absolute magnitudes, spectral types, effective temperatures, etc.\ that can be applied to other samples whose parallaxes are unknown or not so easily measured.

In this paper we provide the cold dwarf complement to the complete, nearby samples being extracted from {\it Gaia}. Our contribution is twofold. One, we present analysis on a flurry of new discovereies by the Backyard Worlds: Planet 9 (hereafter, "Backyard Worlds") and CatWISE teams  that in the last several months have helped to identify even more previously hidden members of the 20-pc census. Two, we present a set of 361 parallaxes measured by the {\it Spitzer Space Telescope} (hereafter, {\it Spitzer}) that, when combined with astrometric monitoring of other objects by the astronomical community, establishes a complete, full-sky, volume-limited census of L, T and Y dwarfs out to 20 pc. We use this census to establish the shape and functional form of the mass function in the substellar regime.

This paper is organized as follows. In section~\ref{why_study_the_MF} we provide motivation for studying the mass function and describe what can be learned from the results. In section~\ref{building_list} we build the seed list of targets for the 20-pc L, T, and Y census and describe how this parallels historical efforts to catalog nearby stars of types M and earlier. In section~\ref{section_astrometry} we discuss our {\it Spitzer} data acquisition and the subsequent astrometric reductions, and we compare our results to other published parallaxes for objects with independent measurements. In section~\ref{supporting_data} we discuss photometric and spectroscopic follow-up in support of the 20-pc seed list. In section~\ref{section:data_analysis} we construct the final 20-pc census, and in section~\ref{section:characterization} we examine outliers on various color-color and color-magnitude diagrams in order to more carefully characterize objects in the census. In section~\ref{section:temps_densities} we assign values of $T_{\rm eff}$ to each object, then calculate space densities as a function of $T_{\rm eff}$, once we have determined completeness limits and completeness corrections. In section~\ref{determining_the_MF} we provide the best fits of these measured space densities to predictions. These predictions simulate space densities for various forms of the mass function passed through two different sets of evolutionary models. We also discuss the value of the low-mass cutoff and ponder why so few brown dwarfs with $T_{\rm eff} <$ 400K have been uncovered to date. We conclude with future avenues of exploration in section~\ref{conclusions}.

\section{Why Explore the Mass Function?\label{why_study_the_MF}}

What does an analysis of the mass function tell us? The astronomical literature is replete with arguments about the functional form of the overall mass function, but what knowledge do we gain from its determination?

The two main, competing forms for the stellar mass function are the power law and the log-normal. At a fundamental level, a power law would inform us that the physical process is scale-free, meaning that the mass of the natal cloud has no bearing on the final stellar mass distribution, only on the total number of objects formed. That is, the relative distribution of masses formed from a small cloud will be the same as that from a much more massive cloud. A power law functional form would therefore imply a single physical process reigning over all of star production. If a universal power law is the correct form, then averaging results over many different star formation sites -- as we do when looking at an older, well mixed, volume-limited sample near the Sun -- should still result in a mass distribution with a power law form.

Even if a power law form describes the observed data, it is common in Nature to find that it applies only above some minimum value. For example, in investigations such as the peak intensity of solar flares or the magnitudes of earthquakes, a power law fits the data well only if a minimum value is imposed (\citealt{clauset2009}). To employ a {\it reductio ad absurdum} of our own, there must be a minimum value for the cut-off mass of star formation because Nature cannot create a star containing only one atom.

The log-normal form, on the other hand, is the result expected when there are many processes that contribute multiplicatively to the result. (Contrast this to a normal distribution, which is the result of processes that contribute additively.) As \cite{kapteyn1903} elegantly argued, even if some physical processes, like the swelling in diameter of a growing blueberry (or a stellar embryo), appear to be normally distributed -- i.e., a symmetric distribution centered on some mean value -- other quantities, such as the growing {\it volumes} of those blueberries (or stars), would necessarily have skewed distributions. He argued that skewed forms are, in fact, favored over symmetrical ones. Many of Nature's skewed distributions are well characterized by a log-normal form (\citealt{limpert2001}), again implying that several independent processes are working together to produce the final outcome (\citealt{miller1979}).

If a single functional form fails to describe the observed distribution over the entire mass range from O stars to Y dwarfs -- and it is well known that there is a break in the shape of the mass function below 1 M$_{\odot}$ (see Figure 2 of \citealt{bastian2010}, who give an overview of the stellar initial mass function) -- then the inflection in the shape of the mass function roughly corresponds to the mass at which a new set of physical processes is becoming dominant. In fact, the mass function may have several inflection points, indicating that separate sets dominate in different mass regimes.

Even with solid knowledge of the mass function's shape across the entire mass spectrum of interest -- in our case, over the entirety of the brown dwarf masses -- divining the physical causes responsible for that shape will be difficult. Nonetheless, knowing the shape enables a semi-empirical determination of the low-mass cutoff and allows us to build simulations that better reflect true space densities across all spectral types. 

\section{Building the Target List\label{building_list}}

Since the 1988 discovery of GD 165B (\citealt{becklin1988}), large swaths of the astronomical community have contributed to uncovering hidden L, T and Y dwarfs in the immediate solar vicinity. New members of the 20-pc census have been announced not only by brown dwarf researchers specifically looking for examples (e.g., \citealt{kendall2004}), but also by researchers in unassociated fields who have serendipitously found others (e.g., \citealt{hall2002}, \citealt{thorstensen2003}). New additions to the sample have been published as single-object papers (e.g., \citealt{ruiz1997}, \citealt{folkes2007}); as part of large photometric (e.g., \citealt{delfosse1997}, \citealt{lucas2010}), spectroscopic (e.g., \citealt{schmidt2010}), proper motion (e.g., \citealt{smith2014}, \citealt{meisner2020a,meisner2020b}), or parallax surveys (e.g., from {\it Gaia}: \citealt{reyle2018}, \citealt{scholz2020}); or as the result of dedicated searches for companions around higher mass stars (e.g., \citealt{thalmann2009}, \citealt{freed2003}) or around other brown dwarfs (e.g., \citealt{volk2003}, \citealt{gelino2011}). Construction of the census of the closest L, T, and Y dwarfs has been the effort of many dozens of lead authors presenting results in hundreds of publications.

\subsection{A Nearby Census in its Historical Context}

Compiling these results into a volume-limited data set is a difficult task. To place this in historical context, consider that the first parallax -- that of the 3.5-pc distant 61 Cygni AB -- was obtained in 1838 by \cite{bessel1838}. Few stars were bright enough and near enough to the Sun to have accurate astrometry measured, but there was enough information seven decades later for \cite{hertzsprung1907} to compile what may have been the first list of nearby stars (see \citealt{batten1998}). It was not until 1913-1914 that the first M dwarfs with both a parallax and a measured spectral type were published -- Groombridge 34 (\citealt{adams1913}) and Lalande 21185 (\citealt{adams1914}). This prompted \cite{hertzsprung1922} to update his previous paper, the new list having just under thirty stars confirmed to lie within 5 pc of the Sun. Just four years later, nearly a hundred nearby M dwarfs had been identified (\citealt{adams1926}). Occasional updates on the 5.2-pc sample were made for years thereafter by \cite{vandekamp1930, vandekamp1940, vandekamp1945, vandekamp1953, vandekamp1955, vandekamp1969, vandekamp1971}, the last update containing a total of sixty stars, including the Sun. \cite{kuiper1942} published a larger list, pushing out to 10.5 pc, that contained 254 individual objects. In more recent times, similar lists have appeared, such as the online list\footnote{See \url{http://www.recons.org/TOP100.posted.htm}.} of the top one hundred closest systems -- which as of the last update in 2012 extends to a radius of 6.95 pc from the Sun -- by the Research Consortium On Nearby Stars (RECONS) team, or the 8-pc census presented by \cite{kirkpatrick2012} that contained 243 individual objects.

The above lists, however, have inadequate statistics with which to perform any meaningful analysis of the mass function. Other lists, covering a more substantial volume, are clearly needed for this work, and such compilations were amassed in the latter half of the twentieth century. The 20-pc catalog of \cite{gliese1957} contained 1,097 individual objects, and a second catalog was produced over a decade later (\citealt{gliese1969}) to update that number to 1,890. A supplement to the second catalog was published by \cite{gliese1979} and listed an additional 462 objects. A third catalog, produced on CD-ROM (\citealt{gliese1991}) but never published in a refereed journal, contained over 3,800 entries within 25 pc. A fourth catalog, promised around 1999\footnote{See \url{https://wwwadd.zah.uni-heidelberg.de/datenbanken/aricns/cnsprint.htm}.}, never materialized. These catalogs have now been superseded by {\it Gaia}.

The list of nearby L, T, and Y dwarfs, on the other hand, has not been superseded, because {\it Gaia} can acquire accurate astrometry for L5 dwarfs out to only $\sim$24 pc, T0 dwarfs to only $\sim$12 pc, T5 dwarfs to only $\sim$10 pc, and T9 dwarfs to only $\sim$2 pc (\citealt{smart2017}). As argued in \cite{kirkpatrick2019}, a 20-pc census provides adequate statistics for determining the mass function in the L, T, and Y dwarf regime, and 20 pc is also the maximum distance\footnote{\cite{kuiper1942} also advocated for a 20-pc census, albeit to provide adequate statistics at earlier types at a time when the sheer number of nearby M dwarfs was just becoming evident.} at which a census of Y0 dwarfs can be constructed, given the sensitivity limits of {\it Wide-field Infrared Survey Explorer} ({\it WISE}; \citealt{wright2010}) data. \cite{best2020} have argued for a partial-sky 25-pc census for low-mass mass function studies; however, their desire to perform astrometric follow-up from the United Kingdom Infrared Telescope (UKIRT) restricts them to $-30^\circ < \delta < +60^\circ$, so their increase in volume over a full-sky 20-pc census is only $\sim$33\%.

In order to construct a census of nearby, low-mass dwarfs, we began constructing an archive in 2003 (\citealt{kirkpatrick2003}) to amass published discoveries of all L and T dwarfs along with their near-infrared photometry and spectral types. At the time the catalog was begun, the list of L and T dwarfs contained 277 objects. Shortly thereafter, the list had grown into a publicly available online database\footnote{See \url{http://dwarfarchives.org}.} listing 470 L and T dwarfs (\citealt{gelino2004}). By 2009 this number had grown to over 650 L and T dwarfs (\citealt{gelino2009}), and by late 2012, which was the last online update, the list had grown to 1,281 L, T, and Y dwarfs. Other researchers provided their own post-2012 updates; the \cite{mace2014} list had 1,565 entries and the List of UltraCool Dwarfs\footnote{See \url{https://jgagneastro.com/list-of-ultracool-dwarfs/}.} had 1,773, although neither of those has been updated in the last 5+ years. One of us (CRG) maintains an in-house spreadsheet that captures new discoveries from the literature, and at its last update in Oct 2019, it contained 2,513 L, T and Y dwarfs.

\subsection{Building a List of Probable 20-pc L, T, and Y Dwarfs\label{building_list_subsection}}

The efforts above provided the cornerstones for the building of a volume-limited census needed for this paper. For each of the known L, T, and Y dwarfs, the object's spectral type and magnitudes in the {\it WISE} W2 band and in $H$ band, the latter of which is invariant between the 2MASS and MKO filter systems (see \citealt{kirkpatrick2019}), were tabulated. Using the color/spectral type to absolute magnitude relations presented in \cite{kirkpatrick2012} and \cite{looper2008}, we calculated spectrophotometric distance estimates and retained all objects having d $<$ 23 pc. Separately, we combed the literature in search of published trigonometric parallaxes for each of the known L, T, and Y dwarfs, many of which were already compiled in the CRG spreadsheet noted above. Objects with trigonometric parallaxes measured to better than 10\% accuracy and falling within 20 pc were kept in our official nearby census, and those lacking a parallax with 10\% accuracy or lacking astrometric follow-up entirely but having distance estimates within 23 pc were retained for further astrometric monitoring with {\it Spitzer}. This limit was chosen to account for margin of error in the distance estimates, the expectation being that most objects truly within 20 pc would have estimates placing them within 23 pc.

In \cite{kirkpatrick2019}, we used the Infrared Array Camera (IRAC; \citealt{fazio2004}) to measure preliminary trigonometric parallaxes for those objects having spectral types of T6 and later. These results were based on data from {\it Spitzer} programs 70062, 80109, 90007, 11059, and the first year's data from 13012 (all with Kirkpatrick as PI). This left a gap in the L and T dwarf sequence between T6 and the latest type for which {\it Gaia} has complete coverage ($\sim$L5). The aim of {\it Spitzer} program 14000 (Kirkpatrick, PI) was to astrometrically monitor those objects in the gap that lacked published parallaxes of high quality but were believed to fall within 23 pc. An extension to provide additional data points for these objects at the end of the {\it Spitzer} mission was further approved as program 14326 (Kirkpatrick, PI).

Meanwhile, old {\it WISE} data and newer {\it Near Earth Object WISE} ({\it NEOWISE}; \citealt{mainzer2014}) data were being continually processed, searched, re-processed, and re-searched in hopes of uncovering new objects at the coldest types, since \cite{kirkpatrick2019} found that the targets in that paper were not complete to 20 pc for any of the late-T or Y dwarf types. Specifically, their measured completeness limits ranged from 19 pc at T6 to only 8 pc at Y0. Both the Backyard Worlds (\citealt{kuchner2017}) and CatWISE (\citealt{eisenhardt2020}) teams were continuing to identify new candidate late-T and Y dwarfs from {\it WISE} data as {\it Spitzer} hurled toward its assigned decommissioning date in late-Jan 2020. As chronicled in \cite{meisner2020a, meisner2020b}, candidates lacking extant {\it Spitzer} photometry were added to {\it Spitzer} photometric programs 14034 (Meisner, PI), 14076 (Faherty, PI), and 14299 (Faherty, PI). As these new IRAC data became available, we used the new {\it Spitzer} photometry to predict a distance to each candidate using the $M_{\rm ch2}$ vs.\ ch1$-$ch2 color\footnote{For brevity, we refer to IRAC's two short wavelength bands as ch1 for the 3.6 $\mu$m band and as ch2 for the 4.5 $\mu$m band.} relation of \cite{kirkpatrick2019}. Such objects with spectrophotometric distance estimates $<$23 pc were the subject of yet another {\it Spitzer} astrometric follow-up program (14224; Kirkpatrick, PI). 

Not all of the late-type candidates were included in programs 14034, 14076, or 14299, however, either because ch1$-$ch2 data already existed in the {\it Spitzer} Heritage Archive, mainly from our own, earlier programs (70062, 80109, or 11059), or because their discoveries occurred after the end of the {\it Spitzer} mission. These objects, which were selected by the community scientists of Backyard Worlds, team members of CatWISE, or both were uncovered via the same selection criteria discussed in \cite{meisner2020a,meisner2020b} and are listed in Table~\ref{new_discoveries}. Also included in this table are additional late-T and Y dwarf candidates, observed as part of {\it Spitzer} photometric program 14329 (Marocco, PI), that were discovered as part of the CatWISE2020 effort (\citealt{marocco2020b}) and have not previously been published. 

\startlongtable
\begin{deluxetable}{lcl}
\tabletypesize{\scriptsize}
%\tablenum{1}
\tablecaption{New L, T, and Y Dwarf Candidates\label{new_discoveries}}
\tablehead{
\colhead{Object} &
\colhead{Note\tablenotemark{a}} &                          
\colhead{Discoverer Code}  \\
\colhead{(1)} &                          
\colhead{(2)} &
\colhead{(3)}
}
\startdata
CWISE J002727.44$-$012101.7&  astrom&  C, F, J, N, Q, R, S, W         \\
CWISE J004143.77$-$401929.9&  astrom&  C, D, F, G, H, J, K, Q, R\\
CWISE J004311.24$-$382225.0&  astrom&  F, G, J, K, R, V   \\
%CWISE J010650.59+225158.7  &     ---&  F, J, P \\ -- Meisner20a
CWISE J011558.74$-$461620.8&     ---&  A, F, G, J, K, Q \\
CWISE J011931.78$-$493750.4&     ---&  F, G, J, K, W \\
CWISE J011952.82$-$450231.2&     ---&  A, D, K, N            \\
CWISE J014308.73$-$703359.1&     ---&  B, F, G, J, K, Q      \\
CWISE J014837.51$-$104805.6&  astrom&  F, G, K, N               \\
CWISE J015042.24$-$462155.3&     ---&  F, G, I, J, K, N, Q \\
CWISE J015349.89+613746.3  &     new&  V \\
CWISE J021705.51+075849.9  &     new&  A, D \\
%CWISE J025805.30$-$321918.0&     ---&  G \\ -- Meisner20b
%WISE J025934.00$-$034645.7 &     ---&  F, J \\ -- not new, T5 from Mace13
CWISE J031021.61$-$573355.6&     ---&  C, G, J, K, M, Q \\
%WISE J032120.91$-$734758.8 &     ---&  J \\ -- not new, T8 from Kirkpatrick12
%CWISE J032600.46+421058.5  &     ---&  A, C, D, F, J, S \\ -- Meisner20a
CWISE J034146.12+471530.5  &     new&  G, V \\
CWISE J041102.41+471422.6  &     new&  A, D, N, R, W \\
CWISE J042335.38$-$401929.5\tablenotemark{b}&  astrom&  J          \\
%WISEA J042506.66$-$425509.6&     new&  A, G \\ -- not new, L8 from Schneider17
%CWISE J043309.36+100902.3  &     ---&  D, J \\ -- Meisner20a
CWISE J044214.20$-$385515.7\tablenotemark{b}&  astrom&  J \\
CWISE J051427.35+200447.7  &     new&  D, G, S \\
%WISEPA J052536.33+673952.3&     ---&  J \\ -- not new, T6pec from Kirkpatrick11
CWISE J054025.89$-$180240.3&  astrom&  C \\
CWISE J060149.45+141955.2  &     new&  G \\
CWISE J060251.35$-$403534.4&     ---&  C, J, K \\
%CWISE J060515.64+134637.4  &     ---&  J \\ -- not moving
CWISE J061348.70+480820.5  &  astrom&  A, G \\
CWISE J061741.79+194512.8  &     new&  G, Z \\
CWISE J062050.79$-$300620.8&     new&  C, G, V \\
CWISE J062725.28$-$373033.1&     ---&  A, C, G \\
CWISE J063018.23$-$371734.3&     ---&  A, G, J, N, Q \\
CWISE J063031.50$-$600221.0&     ---&  A, C, G, J, K \\
CWISE J063558.52$-$322549.4&   color&  D, F, S, V\\
CWISE J063649.77$-$542429.2&     new&  G, V \\
CWISE J064128.15$-$312359.3&     ---&  J, K, Q \\
CWISE J064223.54+042342.2  &  astrom&  D, Z \\
CWISE J064749.87$-$160022.7&     ---&  D, G, P, N \\
%CWISE J065113.27$-$835501.5&     ---&  J, S, Z \\ -- Meisner20b
%WISE J073347.94+754439.2   &     ---&  D \\ -- not new, T6 from Kirkpatrick12
CWISE J074956.20$-$682722.4&     ---&  B, F, G, J, K \\
CWISE J075648.34$-$600130.9&     ---&  A, G, J, K \\
CWISE J075831.11+571153.9  &     ---&  F, G, J, K, N, Q, S, X, Z \\
CWISE J080436.67$-$000028.6&     ---&  A, D \\
CWISE J080556.14+515330.4  &     ---&  D, G, L, S, V \\
CWISE J081606.70+482822.9  &     ---&  B, D, S \\
CWISE J084506.51$-$330532.7&     new&  G, D, S \\  
CWISE J085401.22$-$502028.1&     ---&  A, E, F, G, J, K \\
%CWISE J085908.26+152526.8  &     ---&  F, G, J \\  -- Meisner20a
%WISEA J090258.99+670833.1  &     new&  A \\ -- not new, L7 from Schneider17
CWISE J091105.02+214645.1  &  astrom&  C, D, F, J, K, S, T \\
CWISE J091735.38$-$634451.2&     new&  A \\
CWISE J092503.20$-$472013.8&     new&  L, S \\
CWISE J093823.15$-$841114.4&   color&  D, F, L, S\\
CWISE J094925.88$-$102601.9&     ---&  A, D, F, J, N, Q \\
CWISE J095316.32$-$094318.9&     ---&  A, F, J, K, Q \\
%CWISE J100628.98+105408.5  &     ---&  C, D, Z \\ -- Meisner20a
%CWISE J105349.12$-$460239.1&     ---&  A, D, J \\ -- Meisner20b
CWISE J105512.11+544328.3  &  astrom&  D, G, J \\
CWISE J110201.76+350334.7  &     new&  A, J, N, S, V \\
%CWISE J110238.85$-$775039.7&     new&  G \\ - likely not moving, in Chamaeleon
%CWISE J111055.13$-$174738.8&     ---&  A, F, J \\ -- Meisner20a
CWISE J112106.36$-$623221.5&     new&  L, S \\
%CWISE J112440.19+663051.1  &     ---&  C, D, S \\ -- Meisner20b
CWISE J113019.19$-$115811.3&  astrom&  B, D, F, J, K \\
CWISE J113717.27$-$532007.9&  astrom&  A, F, G, J, K \\
CWISE J113833.47+721207.8  &  astrom&  F, G, J, K, Q \\
CWISE J114120.42$-$211024.5&  astrom&  A, C, F, G, J, S, V \\
%CWISE J114350.62+401332.0  &     ---&  C, F, J, N, S \\ -- Meisner20b
%CWISE J114601.01+342459.5  &     ---&  F, J, V \\ -- Meisner20b
CWISE J115229.37$-$374157.8&     ---&  A, D, G, J \\
%Gaia 1159$-$3634           &     new&  G, V \\
%CWISE J120444.32$-$235927.3&     ---&  A, F, J \\ -- Meisner20a
CWISE J120502.74$-$180215.5&  astrom&  D, G, J, K, Y \\
CWISE J121557.87+270154.2  &     ---&  F, G, J, K, Q, S \\
CWISE J123228.86+225714.5  &     ---&  C, D, N, Z \\
CWISE J130841.31$-$032157.7&     new&  G, L, V \\
CWISE J131548.23$-$493645.4&     new&  C, S \\
%CWISER J133139.34$-$651305.3&    new&  V \\ -- not new, from Reyle18
%CWISE J133259.92$-$160755.1&     new&  F, G, I, J, K, L, V \\  -- not new
CWISE J141127.70$-$481153.4&  astrom&  A, J \\
%CWISE J144901.78+114709.4  &     ---&  J \\ -- not new, T5 from Scholz10
CWISE J153143.38-330657.3  &     new&  G, S, V \\
CWISE J153347.50+175306.7  &  astrom&  G, J, K, N \\
%CWISE J160835.09+161449.7  &     new&  D, Z \\ -- no motion
%CWISE J161822.80$-$062310.3&     ---&  D \\ -- Meisner20a
%CWISE J161940.54+134751.9\tablenotemark{b}  &  astrom&  C \\ -- not new; in Meisner et al.
CWISE J163041.79$-$064338.3&     new&  A, D, G, U \\
CWISE J165013.37+565257.0  &     new&  A, G, S, V \\
CWISE J170127.12+415805.3  &  astrom&  C, D, F, G, J, N, P, Q, V, Z \\
%CWISE J172104.41+595047.8  &     ---&  F \\ -- no motion?
CWISE J172617.09$-$484424.9&     new&  A, E \\
CWISE J174907.16+554050.3  &   color&  A, F, Z \\
CWISE J175517.35+250147.3  &     ---&  F, G, J, K, L, N, Q \\
%CWISE J175541.44+705010.3  &     new&  Varuna82 \\
CWISE J175628.97+505328.5  &   color&  F \\
CWISE J175800.46+555322.7  &   color&  F, S \\
%Gaia 1807$-$0625           &     new&  V \\
CWISE J182755.05+564507.8  &     new&  G, Q \\
CWISE J183207.94$-$540943.3&  astrom&  C, Y \\
CWISE J185104.34$-$245232.1&     new&  G, S \\
%CWISE J190405.09$-$372616.9&     new&  G \\ -- likely not moving
CWISE J192537.88+290159.0  &   color&  E, F, S \\
CWISE J192636.29$-$342955.7&  astrom&  A, B, K, J, M, Q \\
CWISE J193823.28+663602.7  &     ---&  J, S, Z \\
CWISE J193824.10+350025.0  &   color&  F, L, S \\
CWISE J194201.42+534830.5  &   color&  F, L, S \\
CWISE J195228.45$-$730049.4&     new&  B, D, G, Q \\
%WISE J195436.15+691541.3   &     ---&  D \\ -- not new, T5.5 from Mace13
%WISEP J195550.11$-$441514.2&     ---&  A, C, D, F, J, K \\ -- VVV candidate from Lodieu13
CWISE J200121.21$-$413606.8&     ---&  A, B, C, F, J, Q, T \\
CWISE J201221.32+701740.2  &  astrom&  D, J, L \\
CWISE J201342.27$-$032643.7&     new&  B, F, G, J, K \\
%CWISE J201510.63$-$675005.6&     ---&  A, J \\ -- Meisner20a
CWISE J203859.15$-$570110.3&   color&  F \\
%CWISE J205338.54$-$353922.5&     new&  C \\ -- no motion
CWISE J205701.64$-$170407.3&  astrom&  J, N, S, V \\
%CWISE J205834.45$-$513459.4&     new&  G, Q \\ -- motion star is not the bright one --> junk
%CWISE J205921.42+662724.8  &     ---&  C, F, J, S, V \\ -- Meisner20b
%CWISE J215841.48+732842.8  &     ---&  D \\ -- Meisner20a
%CWISE J221841.26+143002.2  &     ---&  F, H, J, X \\ -- Meisner20b
%CWISE J221859.41+114642.7  &     ---&  C, F, G, J, K, V \\ -- Meisner20b
%CWISE J225109.59$-$074037.2&     ---&  B, C, F, H, J \\ Meisner20a
%CWISE J230158.28$-$645858.8&     ---&  I, Z \\ -- Meisner20a
CWISE J234426.81$-$475502.6&     ---&  G \\
%CWISE J235130.48$-$185800.5&     ---&  F, J \\ -- Meisner20a
CWISE J235448.04$-$814044.6&     ---&  G, J, K, N \\
%CWISE J013525.53+171502.3                    & F, J \\
%CWISE J030533.66+395434.6                    & F, J \\
%CWISE J032553.05+042539.6                    & F \\
%CWISE J154459.21+584202.3                    & J \\
%CWISE J174640.73$-$033818.5                  & J \\
%CWISE J184124.66+700040.7                    & F, J \\
%CWISE J201403.89+042410.4                    & F, J \\
\enddata
\tablecomments{Reference code for discoverer:
A = Andersen,
B = Beaulieu,
C = Colin,
D = Caselden,
E = Stenner,
F = Marocco,
G = Goodman,
H = Hamlet,
I = Voloshin,
J = Kirkpatrick,
K = Khalil,
L = Gramaize,
M = D.\ Martin,
N = Ammar,
P = Pendrill,
Q = Hong,    
R = Rothermich,
S = Sainio,
T = Tanner,
U = Hinckley,       
V = Th\'evenot,
W = Walla,
X = Jonkeren,
Y = Pumphrey,
Z = W{\k e}dracki.
}
\tablecomments{Discoveries in this table were scrutinized using the online WiseView tool (\citealt{caselden2018}).}
\tablenotetext{a}{Codes for Note: ``astrom'' = Object was observed as part of our {\it Spitzer} astrometric monitoring program; ``color'' = Object was observed as part of {\it Spitzer} photometry program 14329 (Marocco, PI); ``---'' = Object was ultimately dropped from {\it Spitzer} follow-up after the time awarded for program 14224 was cut in half; ``new'' = Object was discovered after final {\it Spitzer} target lists were selected. }
\tablenotetext{b}{Astrometric follow-up of this object by {\it Spitzer} shows it to be a background source. See section~\ref{section:data_analysis}.}
\end{deluxetable}

Sometime after {\it Spitzer} program 14224 was selected for 246.5 hours of data collection, we were informed that, for unforeseen logistical reasons at the {\it Spitzer} Science Center, the originally planned 2019 Apr 15 start date of our observations would have to be moved to 2019 Jun 16 and that our allotted time would be halved. This had two ramifications for the intended science: (1) In order to get enough astrometric data points for a meaningful parallactic solution, we had to remove many of the original targets in the program, and (2) the later start date meant that we would only be able to obtain observations at one additional epoch for those targets with a visibility window that closed between Apr 15 and Jun 16, which was roughly one-third of the targets. As a result, we dropped most of the objects in our program with spectrophotometric distance estimates between 20 and 23 pc, along with some of those with the earliest types (around T6). We were also forced to rely more heavily on outside astrometry since our {\it Spitzer} data would now cover an insufficient time baseline to disentangle the effects of parallax and proper motion. More discussion of this can be found in section~\ref{section_astrometry}.

Table~\ref{spitzer_targets} lists all 361 targets that were eventually observed in one of our {\it Spitzer} parallax programs. In total, 98.7\% of the Astronomical Observation Requests\footnote{An AOR is the fundamental scheduling unit for {\it Spitzer} and consists of a fully defined set of observing parameters.} (AORs) in the table were from programs proposed by various {\it WISE}, CatWISE, and Backyard Worlds team members. We used the {\it Spitzer} Heritage Archive to supplement our 5,041 AORs with another 66 from other researchers, which primarily enabled us to extend the time baseline of the {\it Spitzer} data set. Table~\ref{spitzer_programs} lists the individual {\it Spitzer} programs whose data were used. Of these 66 supplementary observations, fifteen were taken during the original {\it Spitzer} cryogenic mission and were reduced using software applicable to that mission phase, as described in more detail in section~\ref{section_astrometry}.

%We have excluded two objects, 0430+2556 and 1721+5950, which have serendipitous {\it Spitzer} observations but for which the data are insufficient for accurate astrometry.

%To measure delta days, one can use https://www.thecalculatorsite.com/time/days-between-dates.php . Then just divide by 365.2425 to get delta years.

%\setcounter{LTchunksize}{81}
%\begin{center}
\startlongtable
\begin{deluxetable*}{lcccl}
\tabletypesize{\scriptsize}
%\tablenum{2}
\tablecaption{Objects on the IRAC ch2 {\it Spitzer} Parallax Programs\label{spitzer_targets}}
\tablehead{
\colhead{Object} &                          
\colhead{First Obs.\ Date} &
\colhead{Last Obs.\ Date} &     
\colhead{Baseline\tablenotemark{b}}  &
\colhead{Program \# (and \# of Epochs)} \\
\colhead{Name\tablenotemark{a}} &                          
\colhead{(UT)} &
\colhead{(UT)} &     
\colhead{(yr)}  &
\colhead{with ch2 Coverage} \\ 
\colhead{(1)} &                          
\colhead{(2)} &  
\colhead{(3)} &     
\colhead{(4)} &
\colhead{(5)}    
}
\startdata
WISE  0005+3737    & 2012 Sep 06 & 2018 May 02 &  5.7    & 80109(1), 90007(12), 13012(12)   \\
WISE  0015$-$4615  & 2010 Dec 17 & 2018 Oct 04 &  7.8    & 70062(2), 90007(12), 13012(12)   \\
CWISE 0027$-$0121  & 2015 Feb 25 & 2019 Nov 26 &  4.8    & 10135(1), 11059(1), 14224(6) \\
WISE  0031+5749    & 2018 Nov 18 & 2020 Jan 15 &  1.2    & 14000(9), 14326(2) \\
PSO   0031+3335    & 2018 Nov 05 & 2019 Nov 25 &  1.1    & 14000(9) \\
WISE  0032$-$4946  & 2012 Jul 28 & 2018 Sep 23 &  6.2    & 80109(1), 90007(12), 13012(12)   \\
2MASS 0034+0523    & 2012 Feb 15 & 2018 Apr 23 &  6.2    & 80109(2), 90007(12), 13012(12)   \\
WISE  0038+2758    & 2012 Mar 22 & 2018 May 05 &  6.1    & 80109(2), 90007(14), 13012(12)   \\
\enddata
\tablenotetext{a}{Full object designations can be found in Table~\ref{table:monster_table}.}
\tablenotetext{b}{The units are Earth-based years. To translate into the number of Spitzer orbits of the Sun, multiply these values by $\sim$0.98.}
\tablecomments{(This table is available in its entirety in a machine-readable form in the online journal. A portion is shown here for guidance regarding its form and content.)}
\end{deluxetable*}
%\twocolumngrid

\begin{center}
\begin{deluxetable}{lccc}
\tabletypesize{\small}
%\tablenum{3}
\tablecaption{{\it Spitzer} Programs with ch2 Data Used in the Astrometric Analysis\label{spitzer_programs}}
\tablehead{
\colhead{Program} & 
\colhead{Type} &
\colhead{\# of ch2} &
\colhead{Principal} \\
\colhead{\#} & 
\colhead{} &
\colhead{AORs} &
\colhead{Investigator} \\
\colhead{(1)} &                          
\colhead{(2)} &  
\colhead{(3)} &
\colhead{(4)} 
}
\startdata
35*   & GTO&    1& Fazio       \\
244*  & DDT&    1& Metchev     \\  
3136* &  GO&    2& Cruz        \\  
20514*&  GO&    3& Golimowski  \\  
30298*&  GO&    3& Luhman      \\
40198*& GTO&    4& Fazio       \\
50059*&  GO&    1& Burgasser   \\
60046 &  GO&   10& Luhman      \\
60093 &  GO&    1& Leggett     \\
551   & DDT&    1& Mainzer     \\
70021 & SNAP&   4& Luhman      \\
70058 &  GO&    1& Leggett     \\
70062 &  GO&  175& Kirkpatrick \\
80077 &  GO&    2& Leggett     \\
80109 &  GO&  212& Kirkpatrick \\
90007 &  GO&  870& Kirkpatrick \\
90095 &  GO&    4& Luhman      \\
10135 &  GO&    3& Pinfield    \\
10168 & DDT&    4& Luhman      \\
11059 &  GO&    9& Kirkpatrick \\
13012 &  GO& 1704& Kirkpatrick \\
14000 &  GO& 1404& Kirkpatrick \\
14034 &  GO&   33& Meisner     \\
14076 &  GO&   18& Faherty     \\
14224 & DDT&  485& Kirkpatrick \\
14299 & DDT&    2& Faherty     \\
14326 & DDT&  131& Kirkpatrick \\
\enddata
\tablecomments{An asterisk indicates a program from the {\it Spitzer} cryogenic mission (ending 2009 May).}
\end{deluxetable}
\end{center}

%However, it was obvious fro K19 that we still lacked adequate statistics for the Y dwarfs. On-going efforts with CatWISE (Prelim and 2020) and BYW have been the main new avenues for Y dwarf discovery. New Y dwarf candidates have continued to be studied in the post-Spitzer era. 

\section{Astrometric Data Acquisition and Reduction\label{section_astrometry}}

The reduction of the {\it Spitzer} astrometry used the same methodology as that outlined in section 5.2 of \cite{kirkpatrick2019}, with the following exceptions. First, the list of possible re-registration stars was paired not against {\it Gaia} Data Release 1 (DR1) but with the newer {\it Gaia} DR2 instead, as the latter contains five-parameter ($\alpha_0$, $\delta_0$, $\varpi_{abs}$, $\mu_{\alpha}$, and $\mu_{\delta}$) solutions for $\sim$70\% of cataloged objects. Second, we used these full astrometric solutions to predict the per-epoch positions of each re-registration star at the observation date of each AOR, thereby enabling us to measure absolute parallaxes and proper motions of the {\it Spitzer} targets directly\footnote{For the twenty-five targets having full five-parameter solutions themselves in {\it Gaia} DR2, special care was taken to remove the target from the list of re-registration stars.}. %(We also measured relative astrometry using just the static 2015.0-epoch {\it Gaia} DR2 positions of the registration stars so that we could compare results to both the previous relative astrometry from \citealt{kirkpatrick2019} and the new absolute astrometry measured here.) 
Third, to assure that we had a sufficient number of five-parameter {\it Gaia} DR2 re-registration stars per frame, we set the signal-to-noise (S/N) requirement to S/N $\ge$ 30 per frame\footnote{Source crowding in a few Galactic Plane fields, such as that for WISE 2000+3629, forced us to impose higher S/N cuts.}; in \cite{kirkpatrick2019}, we used S/N $\ge$ 30 only when the field for that target was starved of S/N $\ge$ 100 background stars. As stated in that paper, however, the inclusion of re-registration stars with 30 $<$ S/N $<$ 100 does not generally degrade the $\chi^2$ values of the final parallax and proper motion solution compared to solutions using S/N $\ge$ 100 stars only. Fourth, one small modification to the astrometric solution was included for these new reductions. In \cite{kirkpatrick2019}, the mean epoch for all solutions was set to 2014.0 because the time span for each of the objects was similar. The time coverage of the new data set, however, varies greatly from object to object (see Table~\ref{spitzer_targets}), so we have chosen to compute and report the mean epoch of each object separately.

For those AORs in Table~\ref{spitzer_programs} that came from the cryogenic portion\footnote{Cryogenic data, which are those prior to mid-2009, currently have a CREATOR software processing tag with prefix of "S18" in their FITS headers, whereas data from the warm mission have "S19." Also, the Astronomical Observation Template type (AOT\_TYPE) in the header will be tagged with a suffix of "PC" (post-cryogenic) for warm data but will lack this tag for cryogenic data.} of the {\it Spitzer} mission, we modified our reductions slightly. During the single-frame reduction step detailed in section 5.2.2 of \cite{kirkpatrick2019}, we ran the MOPEX/APEX software so that the Point Response Function (PRF) fitting made use of the PRF maps measured for cryogenic data. All data from the warm mission were, as before, reduced using PRF maps applicable to the warm phase.

As stated in section~\ref{building_list_subsection}, some of our {\it Spitzer} astrometry from Cycle 14 lacked a sufficient time baseline with which to disentangle proper motion and parallax, so we supplemented the {\it Spitzer} data with positions derived from the unWISE (\citealt{lang2014}) "time-resolved" coadds of \cite{meisner2018a, meisner2018b}. The methodology is the same as that described in section 8.3 of \cite{meisner2020a}, which measures the positions of our sources on the time-resolved unWISE coadds whose astrometry has been re-registered to the {\it Gaia} DR2 reference frame. The unWISE measurements used here are the NEO5 version of the time-resolved coadds, covering early 2010 through late 2018. For this current work, however, the coadds were produced on an epochal basis; that is, because we needed a clearly defined time stamp, positions were not combined across differing time-resolved sets (usually spaced by six months), as was done in \cite{meisner2020a} to increase the S/N of the final detection. 

Because our planned observations between 2019 Apr 15 and 2019 Jun 16 never materialized (see section~\ref{building_list_subsection}), thirteen of our 361 sources had {\it Spitzer} observations sampling only one side of the parallactic ellipse and thus, only a proper motion measurement was possible. For these cases, the same fitting procedure outlined above was used except that the parallax term was set to zero.

For each target, a listing of all of the measured positions from our {\it Spitzer} reductions-- and from the unWISE reductions, if applicable -- is given in Table~\ref{all_registered_positions}. Per the above discussion, all positions are re-registered to the {\it Gaia} DR2 reference frame and have uncertainties and time stamps attached. Additional information regarding registration of the unWISE astrometry can be found in section 8.3 of \cite{meisner2020a}. 

\begin{deluxetable*}{lrrrrccrrr}
\tabletypesize{\scriptsize}
%\tablenum{4}
\tablecaption{Astrometry on the {\it Gaia} DR2 Reference Frame\label{all_registered_positions}}
\tablehead{
\colhead{Object} &                          
\colhead{RA} &
\colhead{Dec} &     
\colhead{$\sigma_{\rm RA}$}  &
\colhead{$\sigma_{\rm Dec}$} &
\colhead{Source} &
\colhead{MJD} &
\colhead{$X$} &
\colhead{$Y$} &
\colhead{$Z$} \\
\colhead{} &                          
\colhead{(deg)} &
\colhead{(deg)} &     
\colhead{(asec)}  &
\colhead{(asec)} &
\colhead{} &
\colhead{(day)} &
\colhead{(km)} &
\colhead{(km)} &
\colhead{(km)} \\
\colhead{(1)} &                          
\colhead{(2)} &  
\colhead{(3)} &     
\colhead{(4)} &
\colhead{(5)} &  
\colhead{(6)} &                          
\colhead{(7)} &  
\colhead{(8)} &     
\colhead{(9)} &
\colhead{(10)}    
}
\startdata
 0005+3737&   1.323697&  37.622181&  0.010&  0.010&   ch2&    56176.54&     28532439.95&   -137033306.05&    -61329018.68\\
 0005+3737&   1.324420&  37.622029&  0.010&  0.010&   ch2&    56923.88&     30795904.89&   -136653138.81&    -61123398.32\\
 0005+3737&   1.324171&  37.622031&  0.010&  0.010&   ch2&    56750.45&    -57605421.64&    127020923.18&     56379531.75\\
 0005+3737&   1.324158&  37.622036&  0.010&  0.010&   ch2&    56736.94&    -24304166.88&    135745713.92&     60811831.23\\
 0005+3737&   1.324143&  37.622043&  0.010&  0.010&   ch2&    56724.04&      8722950.84&    137418753.36&     62061969.36\\
 0005+3737&   1.324141&  37.622048&  0.010&  0.010&   ch2&    56714.29&     33502880.86&    134324880.85&     61039865.73\\
\enddata
\tablecomments{The column Object includes only the first four digits of the sexagesimal RA and the first four digits (plus the sign) of the sexagesimal Dec.}
\tablecomments{References for Source: 
ch2 = {\it Spitzer} ch2 astrometry,
W1 = {\it WISE} astrometry from the {\it Gaia}-registered unWISE time-resolved coadds in W1,
W2 = {\it WISE} astrometry from the {\it Gaia}-registered unWISE time-resolved coadds in W2.}
\tablecomments{(This table is available in its entirety in a machine-readable form in the online journal. A portion is shown here for guidance regarding its form and content.)}
\end{deluxetable*}

Because the two sets of astrometry are taken from different positions within our Solar System -- one from the Earth-orbiting {\it WISE} spacecraft and the other from the Sun-orbiting {\it Spitzer} spacecraft -- all observations were tagged with the $XYZ$ positions within the Solar System corresponding to the Modified Julian Date (MJD) of the data. For {\it Spitzer} observations, these $XYZ$ positions are tabulated by the {\it Spitzer} Science Center in the FITS image headers; for the unWISE epochs, we used the mean MJD of each epochal coadd and assigned to them the $XYZ$ of the Earth at that time, using data available through the JPL Horizons website\footnote{See \url{https://ssd.jpl.nasa.gov/horizons.cgi}.}. Note that the use of the Earth's position is sufficient since the unWISE epochal data themselves are an average over a few days of {\it WISE} observations near that mean epoch. Even with the inclusion of non-{\it Spitzer} astrometry into the astrometric solutions, no special modifications to the fitting routine employed in section 5.2.3 of \cite{kirkpatrick2019} were needed. It should be noted that, with the exception of a very small number of confused observations noted in Table~\ref{spitzer_targets}, {\it all} astrometric data points were used in the fits since no sigma clipping and refitting were performed.

In principle, the unWISE epochal astrometry was needed only for those {\it Spitzer} data sets that had observations covering fewer than three {\it Spitzer} visibility windows. In practice, however, we included unWISE data into the astrometric solutions for all objects in programs 14000, 14224, and 14326; the only exceptions were objects in common to program 13012, as these already had {\it Spitzer} observations spanning multiple years.

\begin{figure*}
\figurenum{1a}
\includegraphics[scale=0.85,angle=0]{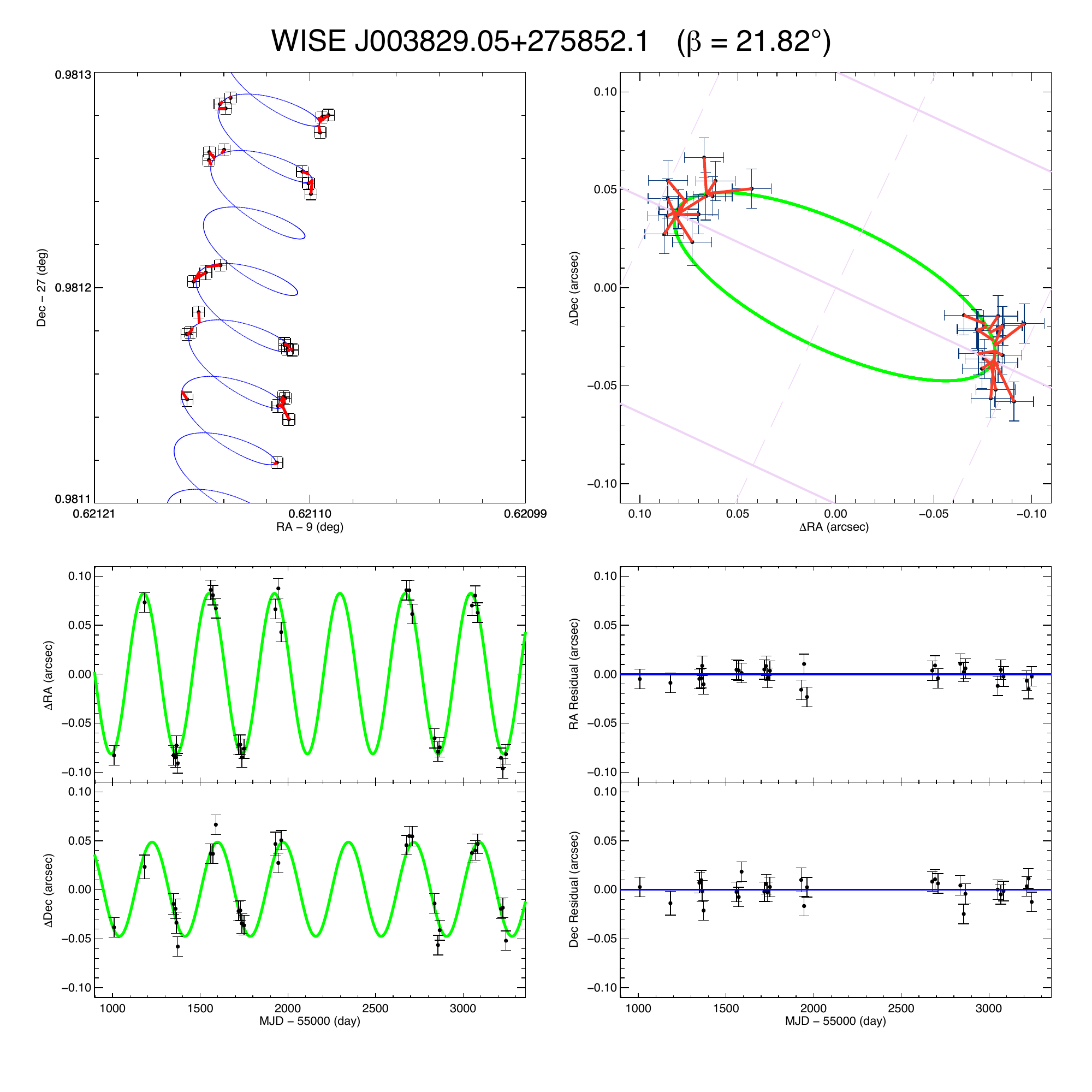}
\caption{Example of a target whose astrometric fit uses only {\it Spitzer} data. (Upper left) A square patch of sky showing the measured astrometry and its uncertainty at each epoch (black points with error bars) plotted 
in RA vs.\ Dec. The blue curve shows the best fit. Red lines connect each observation to its corresponding time point 
along the best-fit curve. (Upper right) A square patch of sky 
centered at the mean equatorial position of the target. The green 
curve is the parallactic fit, which is just the blue curve in the 
previous panel with the proper motion vector removed. Again, red lines connect the time of the observation with its prediction. In the background is the ecliptic coordinate grid, with lines of constant 
$\beta$ shown in solid pale purple and lines of constant $\lambda$ 
shown in dashed pale purple. Grid lines are shown at 0$\farcs$1 
spacing. (Lower left) The change in RA and Dec as a function of time
with the proper motion component removed. The parallactic fit is again 
shown in green. (Lower right) The RA and Dec residuals from the fit as a function of 
time. 
\label{0038p2758_plot}}
\end{figure*}

\begin{figure*}
\figurenum{1b}
\includegraphics[scale=0.85,angle=0]{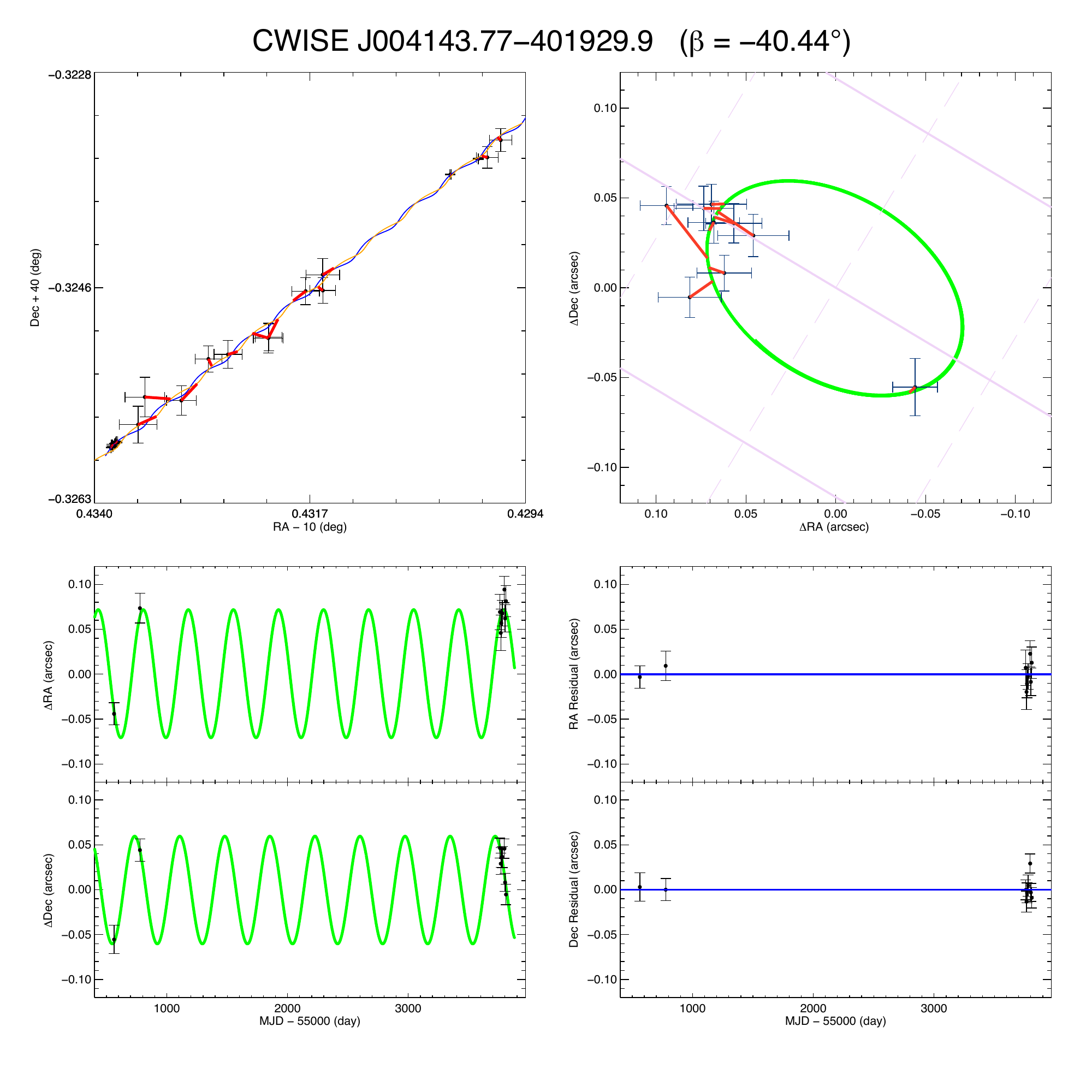}
\caption{Example of a target whose astrometric fit uses both {\it Spitzer} and unWISE data. (Upper left) A square patch of sky showing the measured astrometry and its uncertainty at each epoch (black points with error bars) plotted 
in RA vs.\ Dec. Points with small error bars are the {\it Spitzer} ch2 measurements; those with larger error bars are the {\it WISE} W1 and W2 measurements. The blue curve shows the best fit from the vantage point of {\it Spitzer}, and the orange curve shows this same fit as seen from the vantage point of {\it WISE}. Red lines connect each observation to its corresponding time point 
along the best-fit curve. (Upper right) A square patch of sky 
centered at the mean equatorial position of the target. The green 
curve is the parallactic fit, which is just the blue curve in the 
previous panel with the proper motion vector removed. For clarity, only the {\it Spitzer} astrometric points are shown, again with red lines connecting the time of the observation with its prediction. In the background is the ecliptic coordinate grid, with lines of constant 
$\beta$ shown in solid pale purple and lines of constant $\lambda$ 
shown in dashed pale purple. Grid lines are shown at 0$\farcs$1 
spacing. (Lower left) The change in RA and Dec as a function of time
with the proper motion component removed. The parallactic fit is again 
shown in green and only the {\it Spitzer} astrometry is shown. (Lower right) The RA and Dec residuals from the fit as a function of 
time. As with the lower left panel, only the {\it Spitzer} data are shown.
\label{0041m4019_plot}}
\end{figure*}

\begin{figure*}
\figurenum{1c}
\includegraphics[scale=0.85,angle=0]{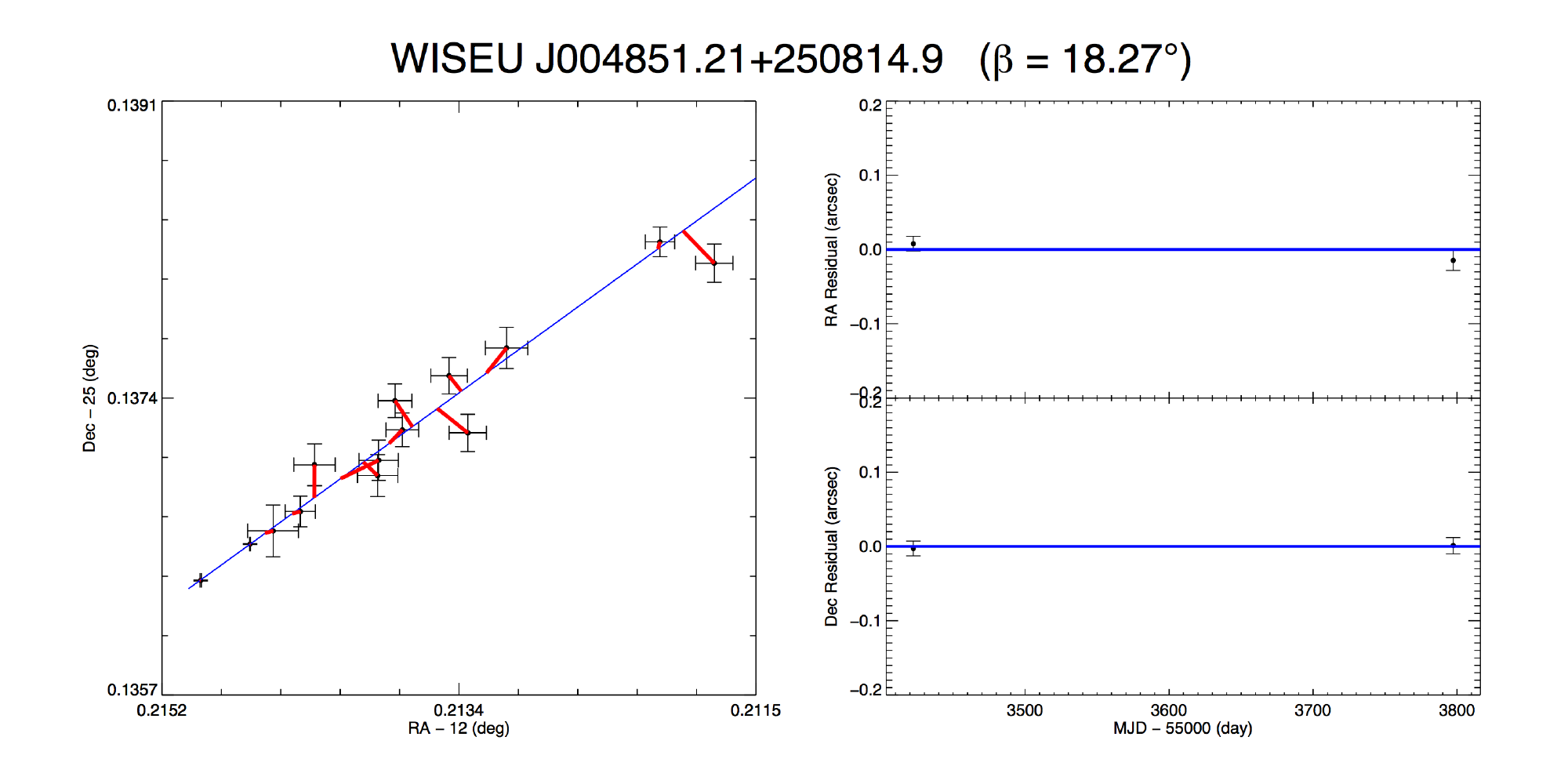}
\caption{Example of a target whose astrometric fit uses both {\it Spitzer} and unWISE data but for which a parallactic fit could not be attempted. (Left) A square patch of sky showing the measured astrometry and its uncertainty at each epoch (black points with error bars) plotted 
in RA vs.\ Dec. Points with small error bars are the {\it Spitzer} ch2 measurements; those with larger error bars are the {\it WISE} W1 and W2 measurements. The blue curve shows the best proper motion fit. Red lines connect each observation to its corresponding time point 
along the best-fit curve. (Right) The RA and Dec residuals from the fit as a function of 
time. Only the {\it Spitzer} data are shown since the error bars of the {\it WISE} points would otherwise dominate the plot.
\label{0048p2508_plot}}
\end{figure*}

Plots of our astrometric measurements and their best fits are shown in Figure Set 1 for each of our 361 targets. Figures~\ref{0038p2758_plot}, \ref{0041m4019_plot}, and \ref{0048p2508_plot} show examples of the three types of plots found within the figure set.

%The 361 PDF files for this figure set can be found at Papers/Spitzer_Final_Parallaxes/figure_set.
\figsetstart
\figsetnum{1}
\figsettitle{Astrometric fits to the 361 objects in the {\it Spitzer} parallax program}

\figsetend

Our astrometric results are summarized in Tables~\ref{spitzer_results_highq}, \ref{spitzer_results_lowq}, and \ref{spitzer_results_poorq}. For each object, the RA and Dec position (in deg) with their uncertainties (in mas) are quoted at the mean epoch, $t_0$, along with the absolute parallax ($\varpi_{abs}$) and absolute proper motions ($\mu_{RA}$ and $\mu_{Dec}$) and their uncertainties. Also listed are the chi-squared value of the best fit ($\chi^2$), the number of degrees of freedom in the fit ($\nu$), and the reduced chi-squared value ($\chi^2_\nu$), along with the number of {\it Spitzer} (\#$_{Spitzer}$) and {\it WISE} (\#$_{WISE}$) astrometric epochs  and the number of {\it Gaia} DR2 five-parameter re-registration stars used (\#$_{Gaia}$). The two values listed in the \#$_{WISE}$ column refer to the number of astrometric epochs in bands W1 (3.4 $\mu$m) and W2 (4.6 $\mu$m), respectively. We find that the median $\chi^2_\nu$ value across all of our solutions in Tables~\ref{spitzer_results_highq}, \ref{spitzer_results_lowq}, and \ref{spitzer_results_poorq} is 1.03, indicating that our uncertainties are properly measured.

Given the wide range of parallax uncertainties found in our final astrometry, we should determine at what point the uncertainty is too large to give a credible result. \cite{lutz1973} looked at populations of objects with differing parallax uncertainties to see at which values these uncertainties become so large that characterizing the true absolute magnitude of the population becomes impossible. For parallax uncertainties of 5\%, the distribution of the ratio of the true parallax to the measured one resembles a Gaussian with a tight variance, but the central value is slightly less than one. This effect is predictable and thus correctable. When the astrometric uncertainty of the population reaches 15\%, the effect is still correctable, but the distribution of true-to-measured parallaxes is broader and centered considerably further from unity than for the case of 5\% uncertainties. \cite{francis2014} improves (and corrects) the formalism of \cite{lutz1973}, showing that the predicted absolute magnitude error is 0.1 mag for an astrometric uncertainty of $\sim$12.5\%. (\citealt{lutz1973} state that for a magnitude error this small, an astrometric uncertainty of $<$10\% is required.) \cite{francis2014} further demonstrates that the effect becomes uncorrectable at astrometric uncertainties between 17.5\% and 20.0\%. With these values in mind, we have chosen "high quality" parallaxes to be those with uncertainties $\le$12.5\%, "low quality" to be those with 12.5-17.5\% uncertainties, and "poor quality (suspect)" to be those with $\ge$17.5\% uncertainties.

Table~\ref{spitzer_results_highq} lists 296 targets for which the uncertainty in the parallax is $\le$12.5\%. Results in this table can be considered robust. Table~\ref{spitzer_results_lowq} lists 18 targets for which the parallax uncertainty falls between 12.5\% and 17.5\%. Results from this table should be used with caution, as additional monitoring is needed to drive these uncertainties lower. Finally, Table~\ref{spitzer_results_poorq} lists 47 targets for which the parallax uncertainties are $\ge$17.5\%. For most of these objects, the $>3\sigma$ detection of a parallax and/or proper motion proves that they are nearby, but derived distances and absolute magnitudes should be regarded as suspect. For these, additional astrometric observations from post-{\it Spitzer} resources are needed to establish credible values. 

%Finally, there were several objects for which we had intended to extract {\it Spitzer} astrometry but for which our efforts were unsuccessful:
%WISE0130-4445: The M dwarf and L dwarf are too badly blended to separate in the Spitzer images.
%WISE0213+3648: There is a lot of bright star chaff near the target, which might destroy any hope of accurate astrometry here. Yeah, I eventually gave up with this target.
%WISE0430+2556: Data in the SHA were insufficient for an astrometric solution.
%WISE0745+2332: is contaminated by a background star.
%WISE0837-0041: Our master table incorrectly listed the W1 and W2 mags of this source, believing it was the much brighter object nearby. Our spectrophotometric distance estimate, which suggested a distance less than 20 pc, was thus incorrect. The Pinfield et al. (2008) distance estimate of 54-76 pc is much more credible.
%WISE1721+5950: Data in the SHA were insufficient for an astrometric solution.
%WISE2349+3459: Both the Spitzer and Gaia astrometry are for the M dwarf, not the T dwarf companion.

%\setcounter{LTchunksize}{81}
%\begin{center}
%\startlongtable
\begin{longrotatetable}
\begin{deluxetable*}{lllcrrrrrrrrrrr}
\tabletypesize{\footnotesize}
%\tablenum{3}
\tablecaption{Parallax and Motion Fits for Objects with High-quality Parallaxes\label{spitzer_results_highq}}
\tablehead{
\colhead{Object} &                          
\colhead{RA} &
\colhead{Dec} &
\colhead{t$_0$} &
\colhead{$\varpi_{abs}$} &
\colhead{$\mu_{RA}$} &
\colhead{$\mu_{Dec}$} &
\colhead{$\chi^2$} &                          
\colhead{$\nu$} &
\colhead{$\chi_{\nu}^2$} &
\colhead{\#$_{Spitzer}$} &
\colhead{\#$_{WISE}$} &
\colhead{\#$_{Gaia}$} \\
\colhead{Name} &
\colhead{at t$_0$} &
\colhead{at t$_0$} &
\colhead{MJD} &
\colhead{(mas)} &
\colhead{(mas/yr)} &
\colhead{(mas/yr)} &
\colhead{} &                          
\colhead{} &
\colhead{} &
\colhead{} &
\colhead{} &
\colhead{} \\
\colhead{} &
\colhead{(deg(mas))} &
\colhead{(deg(mas))} &
\colhead{(day)} &
\colhead{} &
\colhead{} &
\colhead{} &
\colhead{} &                          
\colhead{} &
\colhead{} &
\colhead{} &
\colhead{} &
\colhead{} \\
\colhead{(1)} &                          
\colhead{(2)} &  
\colhead{(3)} &  
\colhead{(4)} &
\colhead{(5)} &
\colhead{(6)} &
\colhead{(7)} &
\colhead{(8)} &
\colhead{(9)} &                          
\colhead{(10)} &
\colhead{(11)} &
\colhead{(12)} &
\colhead{(13)} \\
}
\startdata
WISE J000517.48+373720.5    &   1.324667(2.0) &   37.621951(2.0) &  57226.89& 126.9$\pm$2.1 &    997.3$\pm$1.0 &    -271.6$\pm$1.0 &      24.63&  45&   0.54&       25&      0,0&      45\\
WISE J001505.87$-$461517.6  &   3.775538(2.4) &  -46.255813(2.2) &  57210.91&  75.2$\pm$2.4 &    413.4$\pm$1.1 &    -687.8$\pm$1.0 &      42.76&  47&   0.91&       26&      0,0&      18\\
PSO J007.9194+33.5961       &   7.919347(8.0) &   33.596018(8.9) &  57416.75&  44.4$\pm$3.9 &     -9.2$\pm$2.5 &     -31.8$\pm$2.7 &     131.81&  65&   2.02&        9&    13,13&      19\\
WISE J003110.04+574936.3    &   7.793459(4.7) &   57.826711(5.1) &  57500.34&  71.0$\pm$3.2 &    521.8$\pm$1.5 &     -18.3$\pm$1.6 &      44.16&  69&   0.64&       11&    13,13&     139\\
WISE J003231.09$-$494651.4  &   8.128639(2.4) &  -49.782345(2.3) &  57263.63&  60.8$\pm$2.5 &   -368.6$\pm$1.2 &    -861.7$\pm$1.1 &      60.03&  45&   1.33&       25&      0,0&      16\\
2MASS J00345157+0523050     &   8.717679(2.6) &    5.385476(2.9) &  57179.27& 118.8$\pm$2.7 &    673.6$\pm$1.3 &     178.2$\pm$1.5 &      24.62&  47&   0.52&       26&      0,0&      19\\
WISE J003829.05+275852.1    &   9.621131(1.9) &   27.981211(1.9) &  57166.13&  88.2$\pm$2.0 &    -12.0$\pm$0.9 &      92.4$\pm$1.0 &      47.60&  51&   0.93&       28&      0,0&      27\\
CWISE J004143.77$-$401929.9 &  10.432378(7.5) &  -40.324986(5.0) &  57621.97&  76.7$\pm$9.6 &   1196.6$\pm$1.7 &    -958.1$\pm$1.3 &      26.84&  37&   0.72&        9&     0,12&      14\\
\enddata
\tablecomments{The RA and Dec values are listed on the ICRS coordinate system, and the number in parentheses after each value is the uncertainty in milliarcseconds. The last three columns represent the number of {\it Spitzer} epochs (\#$_{Spitzer}$) and the number of unWISE epochs (\#$_{WISE}$; W1 is the first value and W2 is the second) used in the fits, along with the number of five-parameter Gaia DR2 stars used for the astrometric re-registration (\#$_{Gaia}$).}
\end{deluxetable*}
\end{longrotatetable}
%\end{center}

%\movetabledown=30mm
%\setcounter{LTchunksize}{81}
%\begin{center}
%\startlongtable
\begin{longrotatetable}
\begin{deluxetable*}{lllcrrrrrrrrrrr}
\tabletypesize{\footnotesize}
%\tablenum{4}
\tablecaption{Parallax and Motion Fits for Objects with Low-quality Parallaxes\label{spitzer_results_lowq}}
\tablehead{
\colhead{Object} &                          
\colhead{RA} &
\colhead{Dec} &
\colhead{t$_0$} &
\colhead{$\varpi_{abs}$} &
\colhead{$\mu_{RA}$} &
\colhead{$\mu_{Dec}$} &
\colhead{$\chi^2$} &                          
\colhead{$\nu$} &
\colhead{$\chi_{\nu}^2$} &
\colhead{\#$_{Spitzer}$} &
\colhead{\#$_{WISE}$} &
\colhead{\#$_{Gaia}$} \\
\colhead{Name} &
\colhead{at t$_0$} &
\colhead{at t$_0$} &
\colhead{MJD} &
\colhead{(mas)} &
\colhead{(mas/yr)} &
\colhead{(mas/yr)} &
\colhead{} &                          
\colhead{} &
\colhead{} &
\colhead{} &
\colhead{} &
\colhead{} \\
\colhead{} &
\colhead{(deg(mas))} &
\colhead{(deg(mas))} &
\colhead{(day)} &
\colhead{} &
\colhead{} &
\colhead{} &
\colhead{} &                          
\colhead{} &
\colhead{} &
\colhead{} &
\colhead{} &
\colhead{} \\
\colhead{(1)} &                          
\colhead{(2)} &  
\colhead{(3)} &  
\colhead{(4)} &
\colhead{(5)} &
\colhead{(6)} &
\colhead{(7)} &
\colhead{(8)} &
\colhead{(9)} &                          
\colhead{(10)} &
\colhead{(11)} &
\colhead{(12)} &
\colhead{(13)} \\
}
\startdata
CWISE J002727.44$-$012101.7 &   6.864471(5.6)  &  -1.350482(5.2) &  57678.42&  54.2$\pm$7.9 &    449.3$\pm$2.6  &     57.1$\pm$2.1   &    31.37&  37&   0.84&        8&     0,13&     21\\
2MASSI J0103320+193536      &  15.885101(9.0)  &  19.593484(9.1) &  57418.12&  35.9$\pm$5.7 &    304.3$\pm$2.9  &     18.7$\pm$2.8   &    45.05&  65&   0.69&        9&    13,13&     18\\
WISE J013525.64+171503.4    &  23.856286(5.4)  &  17.250637(5.4) &  57370.55&  65.3$\pm$10.0&    -366.9$\pm$2.3 &    -244.3$\pm$1.4  &    31.58&  35&   0.90&        7&     0,13&     20\\
WISEPC J013836.59$-$032221.2&  24.652649(8.9)  &  -3.373114(5.9) &  57367.54&  38.5$\pm$6.4 &    116.5$\pm$2.6  &   -311.8$\pm$1.7   &    53.67&  67&   0.80&       10&    13,13&     10\\
WISEPA J030533.54+395434.4  &  46.390408(3.5)  &  39.909622(3.5) &  57459.66&  28.9$\pm$4.3 &    277.0$\pm$1.1  &      6.5$\pm$0.9   &    30.98&  37&   0.83&        9&     0,12&     71\\
SDSS J032553.17+042540.1    &  51.470985(4.5)  &   4.427632(4.4) &  57544.73&  28.3$\pm$4.8 &   -178.5$\pm$1.2  &    -66.2$\pm$1.2   &    33.15&  35&   0.94&        8&     0,12&     13\\
WISEA J083011.95+283716.0   & 127.549578(114.9)&  28.617508(74.3)&  57762.10&  90.6$\pm$13.7&    -233.3$\pm$48.6&    -2040.8$\pm$29.9&    20.37&  29&   0.70&        6&     0,11&     18\\
SDSS J104409.43+042937.6    & 161.039137(9.8)  &   4.494191(9.4) &  57617.81&  39.7$\pm$6.7 &    -41.8$\pm$3.6  &     94.5$\pm$3.7   &    32.68&  61&   0.53&        9&    12,12&     17\\
\enddata
\tablecomments{See the comments to Table~\ref{spitzer_results_highq} for additional information.}
\end{deluxetable*}
\end{longrotatetable}
%\end{center}

%\setcounter{LTchunksize}{81}
%\begin{center}
%\startlongtable
\begin{longrotatetable}
\begin{deluxetable*}{lllcrrrrrrrrrrr}
\tabletypesize{\footnotesize}
%\tablenum{5}
\tablecaption{Astrometry for Objects with Poor-quality (or Unmeasurable) Parallaxes that Need Improvement\label{spitzer_results_poorq}}
\tablehead{
\colhead{Object} &                          
\colhead{RA} &
\colhead{Dec} &
\colhead{t$_0$} &
\colhead{$\varpi_{abs}$} &
\colhead{$\mu_{RA}$} &
\colhead{$\mu_{Dec}$} &
\colhead{$\chi^2$} &                          
\colhead{$\nu$} &
\colhead{$\chi_{\nu}^2$} &
\colhead{\#$_{Spitzer}$} &
\colhead{\#$_{WISE}$} &
\colhead{\#$_{Gaia}$} \\
\colhead{Name} &
\colhead{at t$_0$} &
\colhead{at t$_0$} &
\colhead{MJD} &
\colhead{(mas)} &
\colhead{(mas/yr)} &
\colhead{(mas/yr)} &
\colhead{} &                          
\colhead{} &
\colhead{} &
\colhead{} &
\colhead{} &
\colhead{} \\
\colhead{} &
\colhead{(deg(mas))} &
\colhead{(deg(mas))} &
\colhead{(day)} &
\colhead{} &
\colhead{} &
\colhead{} &
\colhead{} &                          
\colhead{} &
\colhead{} &
\colhead{} &
\colhead{} &
\colhead{} \\
\colhead{(1)} &                          
\colhead{(2)} &  
\colhead{(3)} &  
\colhead{(4)} &
\colhead{(5)} &
\colhead{(6)} &
\colhead{(7)} &
\colhead{(8)} &
\colhead{(9)} &                          
\colhead{(10)} &
\colhead{(11)} &
\colhead{(12)} &
\colhead{(13)} \\
}
\startdata
CWISE J004311.24$-$382225.0 &  10.796964(13.5)&   -38.373645(7.0)&    57624.29&  38.1$\pm$15.7&     388.0$\pm$3.5 &      59.1$\pm$2.0 &       20.30&  37&   0.54&        9&     0,12&     22\\
WISEU J004851.21+250814.9   &  12.213738(44.3)&    25.137174(41.5)&   57334.28&  \nodata      &     997.2$\pm$13.2&    -721.7$\pm$11.9&       28.65&  24&   1.19&        2&     0,12&     22\\
2MASSW J0051107$-$154417    &  12.795312(9.0) &   -15.738187(9.6) &   57588.15&  34.0$\pm$6.6 &      65.2$\pm$3.4 &     -29.2$\pm$3.6 &       28.17&  65&   0.43&        9&    13,13&      7\\
WISEA J013217.78$-$581825.9 &  23.075584(23.1)&   -58.307239(26.6)&   57906.21&  27.2$\pm$7.3 &     437.7$\pm$9.2 &     -62.7$\pm$11.1&       25.90&  37&   0.70&        9&     0,12&     25\\
CWISE J014837.51$-$104805.6 &  27.156400(9.6)&    -10.801706(5.6)&    57406.74&   5.4$\pm$49.3&     -22.5$\pm$10.6&    -251.9$\pm$4.3 &       41.69&  33&   1.26&        6&     0,13&      9\\
CWISEP J021243.55+053147.2  &  33.181364(135.1)&    5.529890(87.0)&   57610.86&  24.7$\pm$16.3&     -59.8$\pm$45.0&      57.0$\pm$27.4&       32.28&  33&   0.97&        7&     0,12&     17\\
CWISEP J023842.60$-$133210.7&  39.677455(168.2)&  -13.536971(140.2)&  57653.49&  85.8$\pm$21.4&     -62.0$\pm$58.2&    -768.5$\pm$44.6&       59.55&  29&   2.05&        8&      0,9&     15\\
WISEA J025756.40$-$265528.8 &  44.486076(74.8)&   -26.924653(54.3)&   57226.22&  \nodata      &     520.7$\pm$18.7&      98.9$\pm$14.0&       29.28&  24&   1.22&        2&     0,12&     22\\
\enddata
\tablecomments{See the comments to Table~\ref{spitzer_results_highq} for additional information.}
\end{deluxetable*}
\end{longrotatetable}
%\end{center}

In previous papers -- \cite{kirkpatrick2019} and \cite{martin2018} -- we compared our parallax results to those of other surveys and found excellent agreement with all of those except the {\it Spitzer}/IRAC ch1 results of \cite{dupuy2013}. Below we perform additional checks to assure that our newly measured {\it Spitzer} astrometry is robust.

\subsection{Comparison to the Results of \cite{kirkpatrick2019}}

\begin{figure}
\figurenum{2}
%The script for this plot is pi_comparison_plotter_K19_to_K20.pro
\includegraphics[scale=0.85,angle=0]{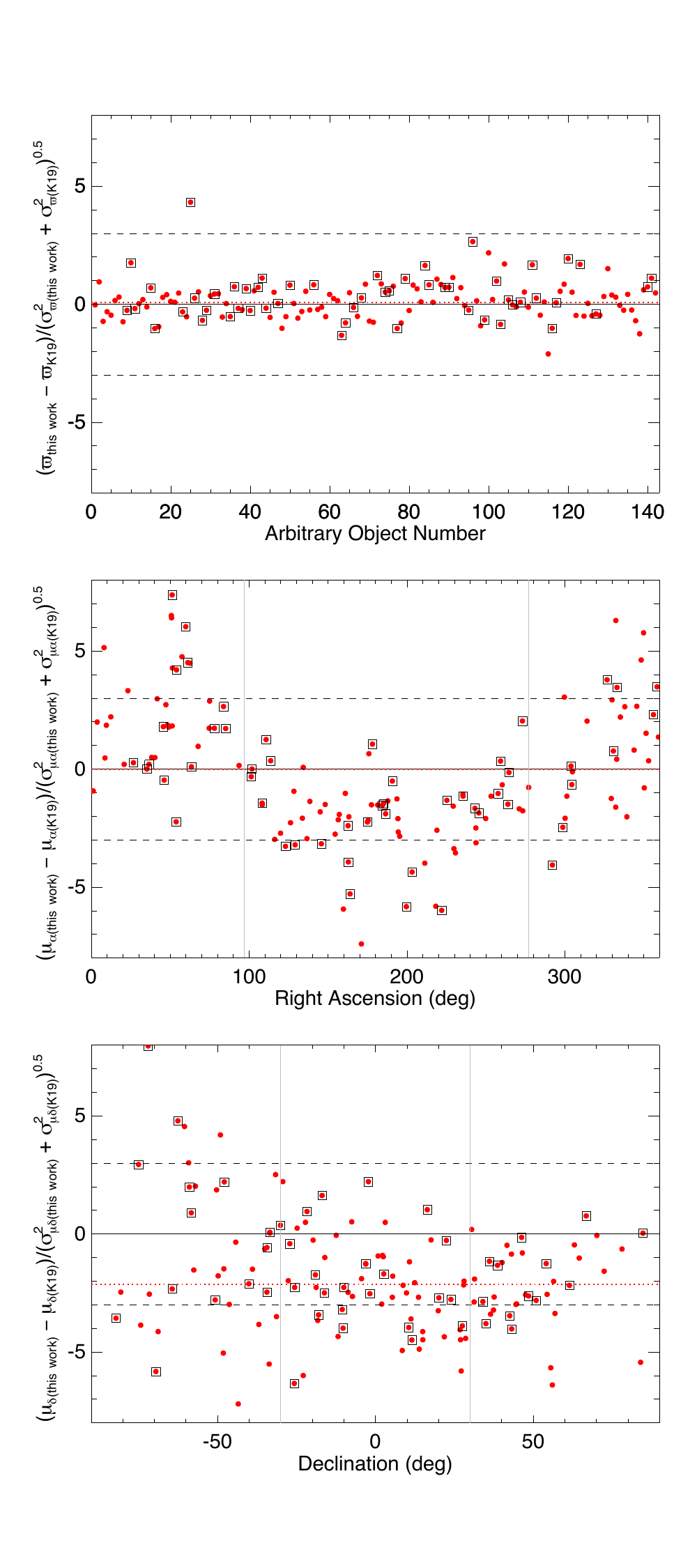}
\caption{Comparison of astrometric results from this paper to those presented in \cite{kirkpatrick2019} ("K19" in the labels) for the 142 objects (red points) in common. The $y$-axis, which shows the difference between the measurements divided by the root-sum-square of the uncertainties in those measurements, indicates the discrepancy between the two values in units of the combined $\sigma$. Mean offsets along the $y$-axis are shown by the dotted red line; the dashed black lines show 3$\sigma$ excursions. Vertical lines in the bottom two panels mark the RA and Dec values of the solar apex and antapex. Objects with $\chi^2_\nu$ values of 1.5 or greater are marked by squares and are not included in the computation of the mean. For ease of comparison across figures, the vertical scale is kept constant for Figures~\ref{K19_to_K20_astrometric_comparison} through \ref{B20_to_K20_astrometric comparison}.
\label{K19_to_K20_astrometric_comparison}}
\end{figure}

All 142 {\it Spitzer} targets from \cite{kirkpatrick2019} have new measurements in this paper. A comparison between the measured astrometry for these objects is shown in Figure~\ref{K19_to_K20_astrometric_comparison}. No bias in the measured parallaxes is seen between the two sets of results, as shown in the top panel of the figure.

Biases are evident in the measured proper motions, however, in both Right Ascension (middle panel) and Declination (bottom panel). These differences are small; the offset (dotted red line) in the lower panel of Figure~\ref{K19_to_K20_astrometric_comparison}, for example, corresponds to a motion difference in Declination of $-$4.6 mas yr$^{-1}$. Other than the longer time baseline, the only difference between our new results and those of \cite{kirkpatrick2019} is the methodology for calculating absolute parallaxes. In \cite{kirkpatrick2019}, a correction from relative to absolute was applied after the fact, whereas in this paper the {\it Gaia} DR2 parallax and motion values of the re-registration stars were used to measure the absolute astrometry of target objects directly. In \cite{kirkpatrick2019}, the {\it post facto} corrections were applied only to the parallaxes. Therefore, the differences in motion values between the two papers are just a reflection of the fact that the \cite{kirkpatrick2019}  motions were deliberately reported as relative whereas the ones in this paper are absolute. 

We can illustrate this as follows. By not correcting the proper motions to absolute, the solar motion is imprinted on the values reported in \cite{kirkpatrick2019}, and this is reflected in the way the differences between the \cite{kirkpatrick2019} relative motions and this paper's absolute motions behave around the celestial sphere. If we were to invent a coordinate system having the solar apex and antapex as its poles, then the difference between relative and absolute motions would be smallest toward the poles and largest at locations on the sphere 90$^\circ$ away from the poles -- i.e., along this coordinate system's equator, where the solar motion is reflected in an apparent "streaming" motion of the background stars. The solar apex is located toward (RA, Dec) = (18$^h$28$^m$, +30$^\circ$), meaning that this invented coordinate system is within 30$^\circ$ of orthogonal to the equatorial system.

This means that the differences between relative and absolute motions will be near zero at the apex (RA $\approx$ 277$^\circ$) and antapex (RA $\approx$ 97$^\circ$). Likewise, the relative proper motions will be maximally too high relative to the absolute ones near RA = 7$^\circ$ (where the true motion and reflex solar motion add constructively) and maximally too low near RA = 187$^\circ$ (where they add destructively). This is the same qualitative behavior exhibited in the middle panel of Figure~\ref{K19_to_K20_astrometric_comparison}. The uncorrected solar reflex motion itself will be a more constant offset along Declination, and the difference between relative and absolute motions in Declination will be negative since the solar apex lies north of the celestial equator. The bottom panel in  Figure~\ref{K19_to_K20_astrometric_comparison} qualitatively shows this behavior, too. 
%As described below, our newly measured motions compare well with those of {\it Gaia} for objects in common to both surveys.

%We can quantitatively check that the magnitudes match as follows. The maximum offset between the the K19 and K20 motions in RA is $\sim$6$\sigma$, which corresponds to $\sim$18 mas/yr. In Dec, the maximum offset between the K19 and K20 motions is $\sim$6$\sigma$, which corresponds to $\sim$18 mas/yr. In quadrature, this is $\sim$25 mas/yr total. The median correction from relative to absolute for these same low-$\chi^2_\nu$ objects was 1.4 mas (Table 4 of \citealt{kirkpatrick2019}), indicating that on average the registration stars were at a distance of $\sim$715 pc. An uncorrected value of 25 mas/yr at this distance corresponds to a velocity of 85 km/s. The canonical values of the Sun's motion through the Milky Way is 229 km/s (Hayes, Law, and Majewski 2018).
 
\subsection{Comparison to {\it Gaia} Results}
 
\begin{figure}
\figurenum{3}
%The script for this plot is pi_comparison_plotter_Gaia_to_K20.pro
\includegraphics[scale=0.85,angle=0]{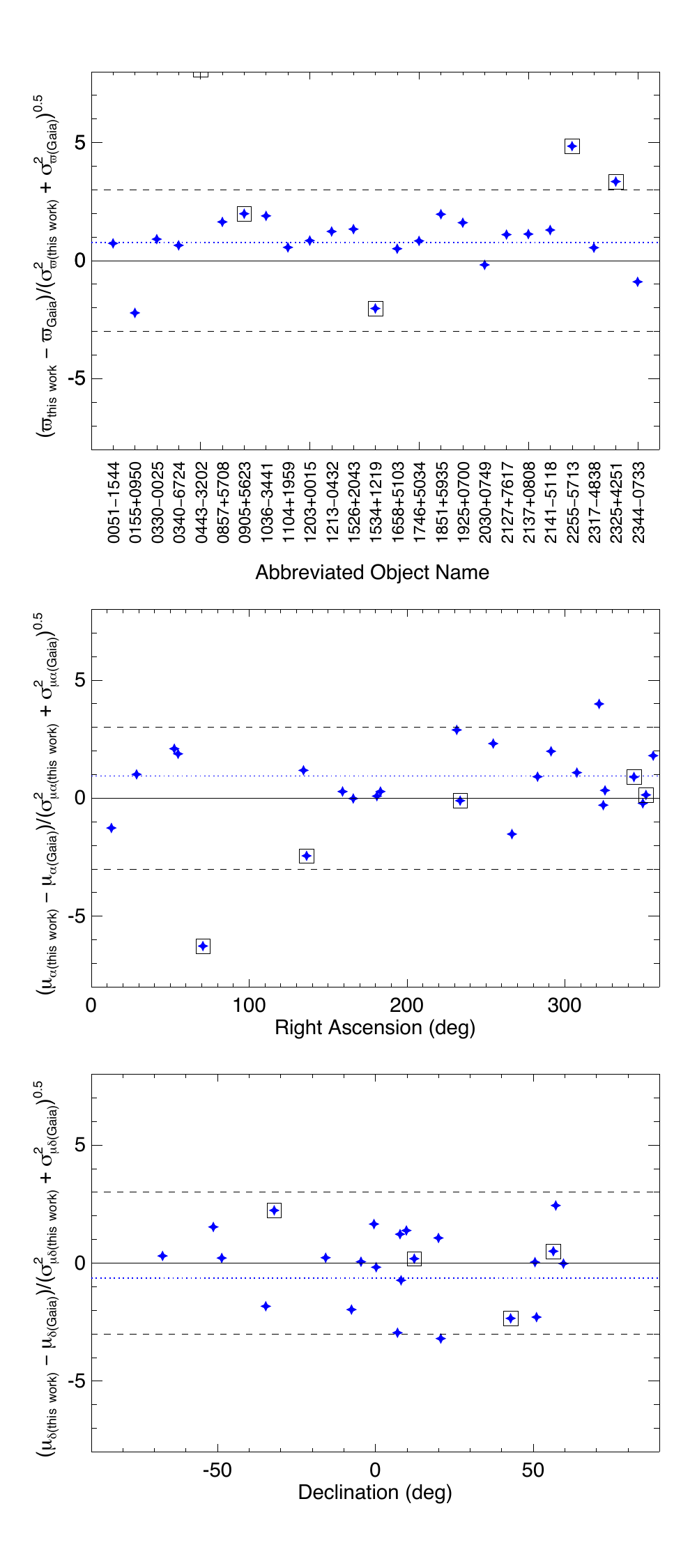}
\caption{Comparison of the astrometric results from this paper to those presented in {\it Gaia} DR2 for the twenty-five objects (blue stars) in common. Mean offsets along the $y$-axis are shown by the dotted blue line; the dashed black lines show 3$\sigma$ excursions. Objects with $\chi^2_\nu$ values of 1.5 or greater are marked by squares and are not included in the computation of the mean. 
\label{Gaia_to_K20_astrometric comparison}}
\end{figure}

At the time objects were chosen for {\it Spitzer} program 14000, {\it Gaia} DR2 had not yet been released and the magnitude limit at which {\it Gaia} astrometry could be reliably measured was still unclear. Making a conservative guess resulted in an overlap of twenty-five objects that, fortunately, now enables a direct comparison to {\it Gaia} (Figure~\ref{Gaia_to_K20_astrometric comparison}). As all three panels of the figure illustrate, the differences between our measured absolute astrometry and that of {\it Gaia} are only marginally significant, those differences falling at the 0.8$\sigma$ (where $\sigma$ refers to the combined value; $\Delta{\varpi_{abs}}=2.8$ mas), 0.9$\sigma$ ($\Delta{\mu_{\alpha}}=2.7$ mas yr$^{-1}$), and 0.6$\sigma$ ($\Delta{\mu_{\delta}}=-1.9$ mas yr$^{-1}$) levels for the top, middle, and bottom panels, respectively. These values of the significance would shrink even further if, for example, it were found that the {\it Gaia} astrometric uncertainties for objects this faint were underestimated. For reference, these twenty-five targets have {\it Gaia} $G$-band values between 19.1 and 20.9 mag and quoted parallax uncertainties between 0.4 and 2.1 mas, the latter of which are typically only 3-4$\times$ smaller than those we measure with {\it Spitzer}.

\begin{figure}
\figurenum{4}
%The script for this plot is pi_comparison_plotter_publishedMs_to_Gaia_new2.pro
\includegraphics[scale=0.33,angle=0]{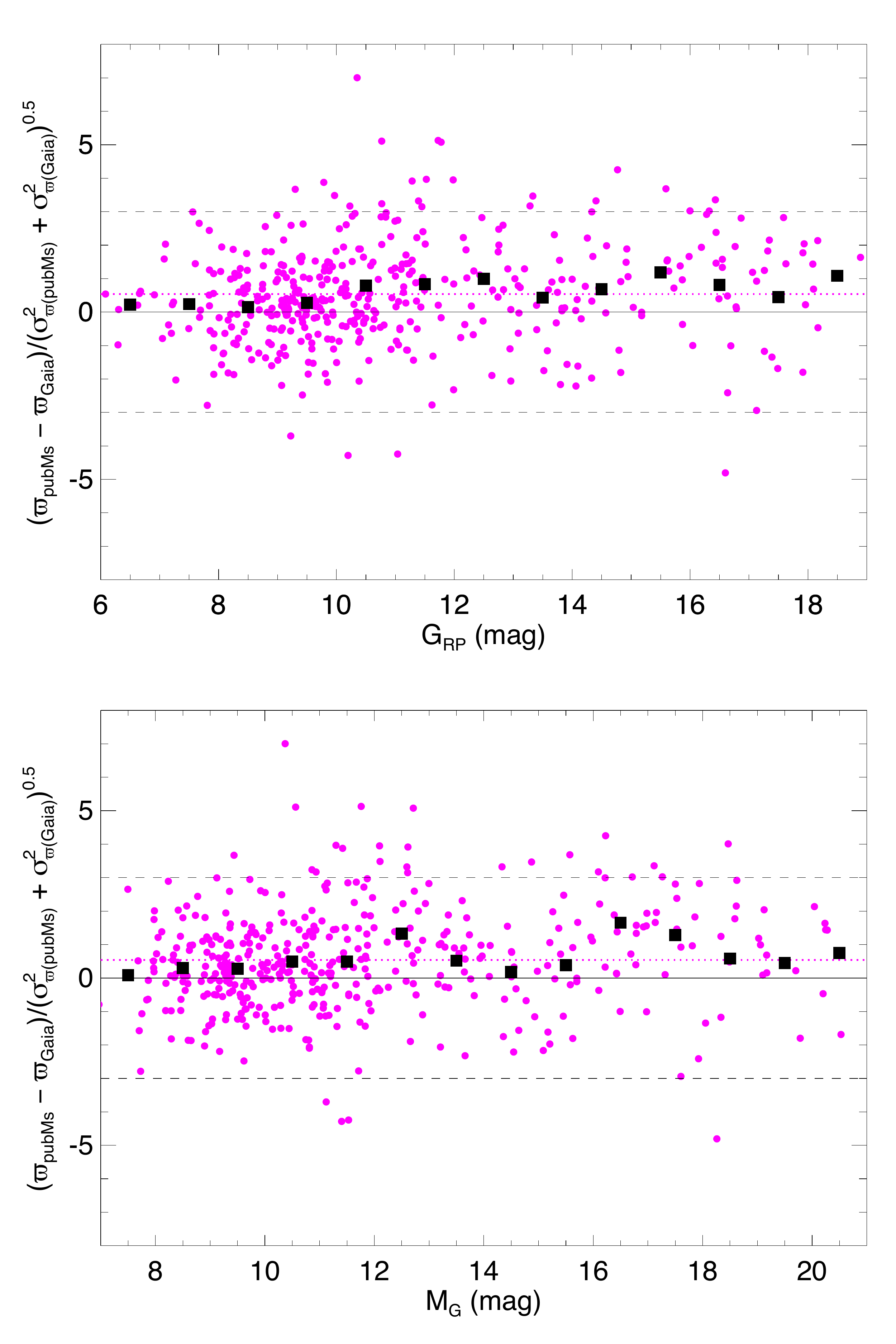}
\caption{Comparison of the {\it Gaia} DR2 astrometric results to other published astrometry for a wider range of spectral types (M0 to L8)\footnote{See also \url{https://www.pas.rochester.edu/~emamajek/EEM_dwarf_UBVIJHK_colors_Teff.txt}.} than that shown in Figure~\ref{Gaia_to_K20_astrometric comparison}. Mean offsets along the $y$-axis are shown by the dotted magenta line; the dashed black lines show 3$\sigma$ excursions. Black squares show the median values along integral magnitude intervals in apparent $G_{RP}$ magnitude (top panel) and absolute $G$ magnitude (bottom panel). Trends suggest that the median offset increases with fainter apparent  magnitude and with fainter absolute magnitude (which is used here as a proxy for color).
\label{pubMs_to_Gaia_astrometric_comparison}}
\end{figure}

The objects whose {\it Gaia} parallaxes we are using for comparison in Figure~\ref{Gaia_to_K20_astrometric comparison} are among the reddest and faintest objects that {\it Gaia} can detect. We can test whether the offsets seen between {\it Gaia} and our {\it Spitzer} results are pointing to an issue with the {\it Gaia} parallaxes themselves by comparing other {\it Gaia} parallaxes to independent literature values. Figure~\ref{pubMs_to_Gaia_astrometric_comparison} illustrates this using parallaxes from \cite{dahn2002}, \cite{dieterich2014}, \cite{winters2015}, and \cite{bartlett2017}. Most of these parallaxes were measured by ground-based CCD programs, with the exception of those from \cite{winters2015}, who presented weighted parallax results using ground-based astrometry measured from photographic plates, CCDs, and infrared arrays as well as astrometry from {\it Hipparcos}\footnote{We retained only those \cite{winters2015} parallaxes built on {\it absolute} parallax values so that no additional relative-to-absolute bias would be introduced}. In our figure, care was taken not to double count results, so any data from \cite{winters2015} that were included in the other references were removed.

These astrometric offsets with respect to {\it Gaia} are plotted as a function of apparent $G_{RP}$ magnitude in the top panel of Figure~\ref{pubMs_to_Gaia_astrometric_comparison}. As $G_{BP}$ is known to be systematically underestimated for the reddest objects in {\it Gaia} (\citealt{smart2019}) -- which in turn affects the $G_{BP} - G_{RP}$ color -- we instead use absolute {\it Gaia} $G$-band magnitude in the bottom panel as a proxy for color. Colors like $B-R$ directly correlate with $M_G$ (or $M_V$) magnitudes across M and L dwarf spectral types (\citealt{pecaut2013, dieterich2014}). The two panels also show a small bias between these published parallax values and those of {\it Gaia}, and the bias has the same sign as that seen in the {\it Spitzer}-to-{\it Gaia} comparison in Figure~\ref{Gaia_to_K20_astrometric comparison}. Moreover, the two panels in Figure~\ref{pubMs_to_Gaia_astrometric_comparison} suggest that there is a tendency for this bias to increase with fainter apparent magnitude and/or redder color.

The cause for this bias, and whether it highlights an unknown issue with the faintest {\it Gaia} astrometry, is unknown. \cite{smart2019} compared a larger list of previously published parallaxes to those of {\it Gaia} DR2 and also found a difference. They concluded that the discrepancy could be reconciled if either the uncertainties in the (heterogeneous) ground-based parallaxes or the {\it Gaia} uncertainties themselves were increased. Given that our new set of homogeneous {\it Spitzer} astrometry shows a similar discrepancy as previous ground-based measurements suggests that the {\it Gaia} uncertainties are underestimated. 

\subsection{Comparison of {\it Spitzer}+unWISE to Pure-{\it Spitzer} Results}

Above, we hypothesized that the small offset seen in the parallax differences with respect to {\it Gaia} would shrink if the {\it Gaia} uncertainties were found to be underestimated. Another possibility, which we will dispel here, is that our own measurement technique has introduced a small bias.

\begin{figure}
\figurenum{5}
%The script for this plot is pi_comparison_plotter_unWISE_on_off.pro
\includegraphics[scale=0.42,angle=0]{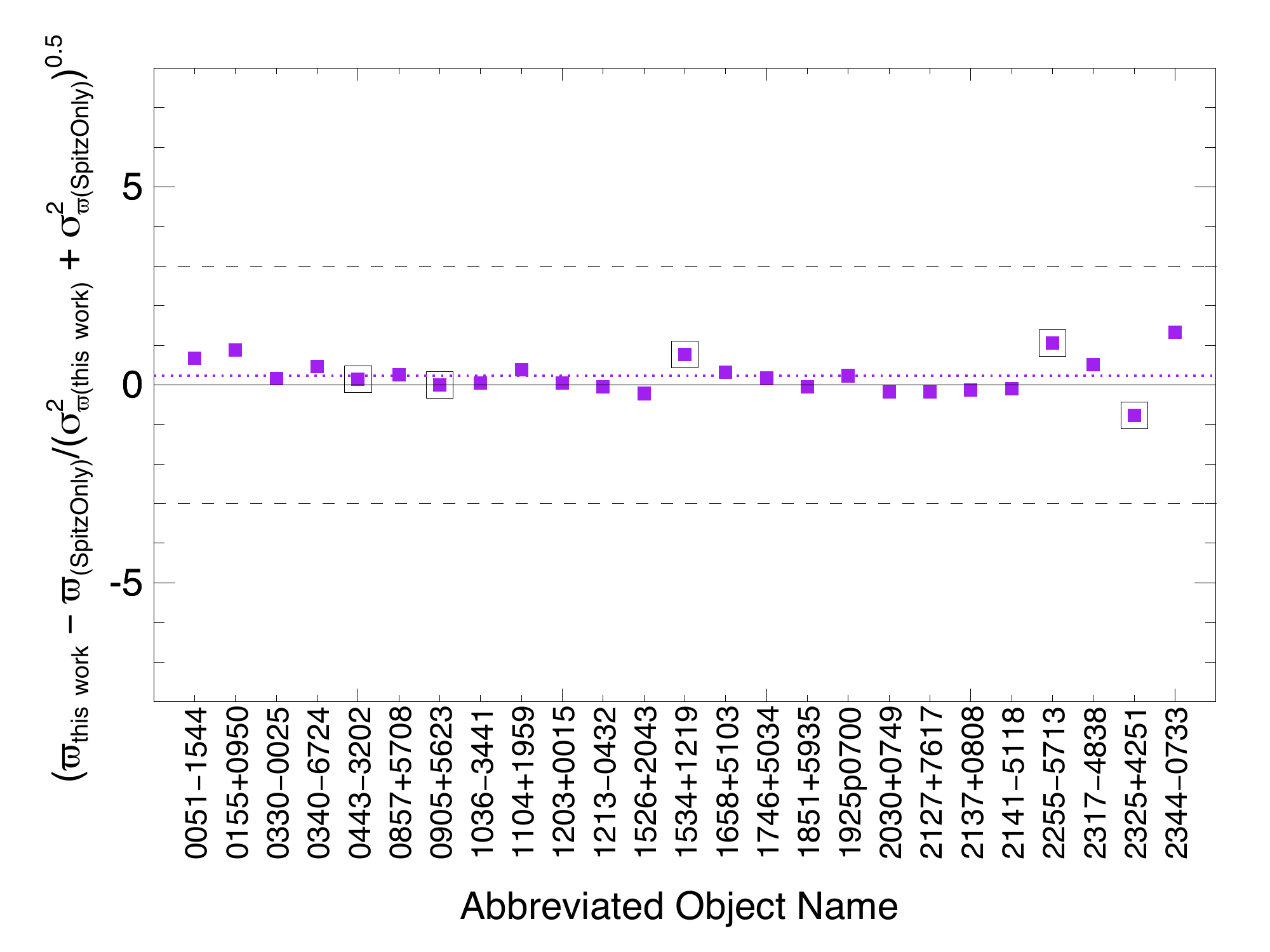}
\caption{Comparison of the astrometric results from this paper to special astrometric runs for which the ancillary unWISE data are not used, for the twenty-five objects (purple squares) in common to this work and {\it Gaia} DR2. Mean offsets along the $y$-axis are shown by the dotted purple line; the dashed black lines show 3$\sigma$ excursions. Objects with $\chi^2_\nu$ values of 1.5 or greater are ringed by an open square and are not included in the computation of the mean.
\label{pi_comparison_unWISE_on_off}}
\end{figure}

The {\it Spitzer} parallax measurements used in Figure~\ref{Gaia_to_K20_astrometric comparison} were supplemented with data from unWISE in order to extend the astrometric time baseline. These objects, although they are among the faintest that {\it Gaia} can measure, are the brightest objects in the {\it Spitzer} program. For this reason, their high-S/N {\it Spitzer} data alone are sufficient to obtain quality parallaxes, so we have performed a special "{\it Spitzer} only" reduction to ascertain whether or not the inclusion of the unWISE data has led to a bias.  A comparison of the reductions with and without the unWISE data is shown in Figure~\ref{pi_comparison_unWISE_on_off}. As expected, no significant difference is present, a bias having been detected only at the 0.2$\sigma$ level.

\subsection{Comparison to Best et al. (2020)}

\begin{figure}
\figurenum{6}
%The script for this plot is pi_comparison_plotter_Best_to_K20.pro
\includegraphics[scale=0.85,angle=0]{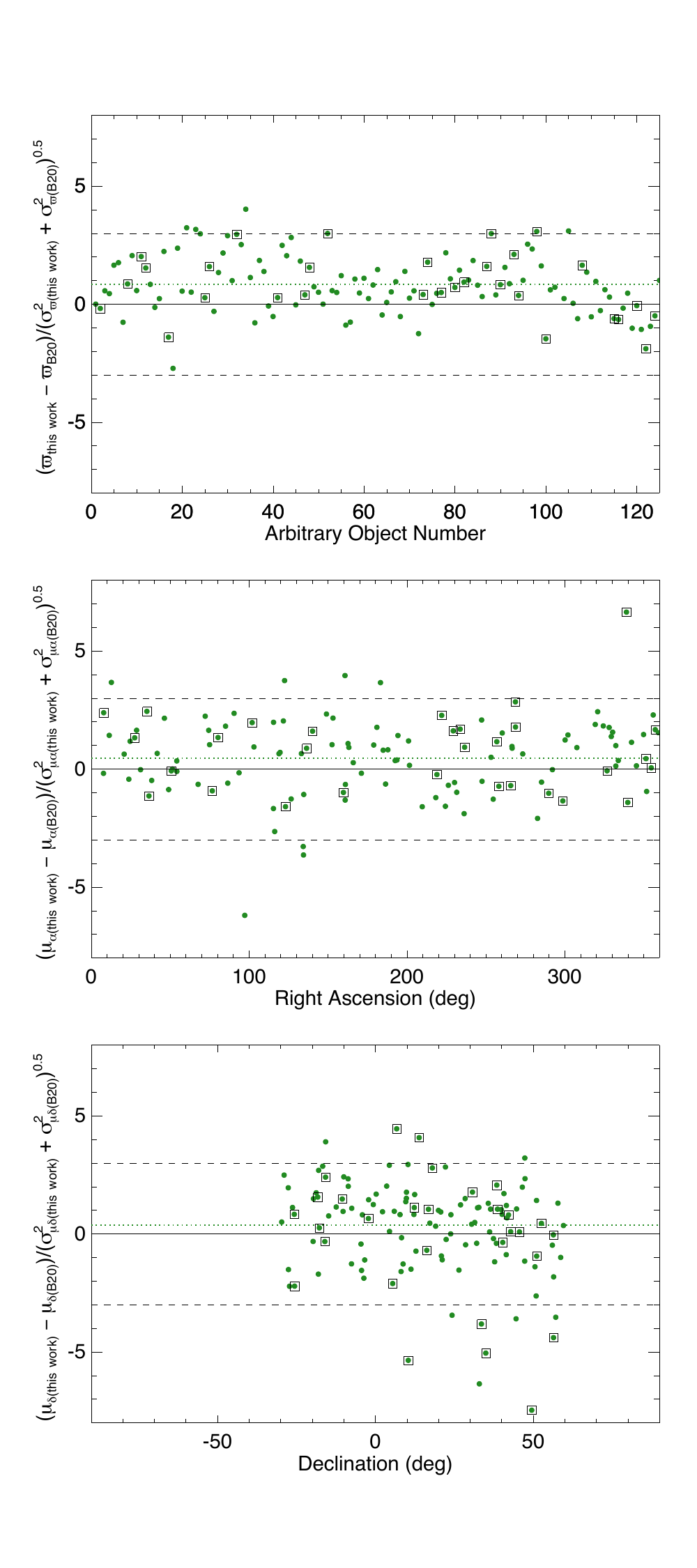}
\caption{Comparison of the astrometric results from this paper to those from \cite{best2020}, for the 124 objects (green points) in common. Mean offsets along the $y$-axis are shown by the dotted green line; the dashed black lines show 3$\sigma$ excursions. Objects with $\chi^2_\nu$ values of 1.5 or greater in either work are marked by squares and are not included in the computation of the mean.
\label{B20_to_K20_astrometric comparison}}
\end{figure}

As this paper was being written, the parallax compilation of \cite{best2020} became available, allowing us to do a comparison of our {\it Spitzer} results to another independent set of astrometry. This comparison is shown in Figure~\ref{B20_to_K20_astrometric comparison}. The offsets seen are at the 0.8$\sigma$ ($\Delta{\varpi_{abs}}=4.3$ mas), 0.5$\sigma$ ($\Delta{\mu_{\alpha}}=1.6$ mas yr$^{-1}$), and 0.4$\sigma$ ($\Delta{\mu_{\delta}}=1.1$ mas yr$^{-1}$) levels for the top, middle, and bottom panels, respectively. Whereas our {\it Spitzer} parallaxes are slightly larger (by 0.8$\sigma$) than those of {\it Gaia}, \cite{best2020} find that their UKIRT parallaxes are slightly smaller (by 1.6$\sigma$) than those of {\it Gaia}. Curiously, \cite{best2020} also conclude that either their parallax uncertainties or those of {\it Gaia} are underestimated, at least the third such case in the recent literature to suggest that {\it Gaia} astrometric uncertainties may be too small for L and T dwarfs.

\section{Supporting Data\label{supporting_data}}

Distance is only one of the important quantities needed when characterizing sources for the mass function analysis. Photometry across the optical through mid-infrared bands is needed to better assess the temperature of each source, which is needed when building a mass function that is tied to $T_{\rm eff}$ as the "observable" parameter. Spectroscopy is another powerful tool, and the most reliable one when assessing the small fraction of sources that have unusual features. These oddities complicate our ability to assign objects to the correct $T_{\rm eff}$ bins because their colors and spectral types follow relations that are different from the bulk of normal, single objects. For example, one oddity identifiable through spectroscopy is low metallicity, which may indicate an older subdwarf (e.g., \citealt{zhang2017}). Another is low-gravity, which may indicate that the object is unusually young since it has yet to contract to its final, equilibrium radius (e.g., \citealt{faherty2016}). Yet another is unresolved binarity, particularly at the L/T transition where spectroscopic blending of features makes composite spectra easier to distinguish (e.g., \citealt{burgasser2010b}). In the subsections that follow, we describe the data acquisition and reduction implemented for our photometric and spectroscopic follow-up campaigns. A compilation of our photometric, spectroscopic, and astrometric data is listed in Table~\ref{table:monster_table}, which is described in the Appendix.

\subsection{Photometry}

\subsubsection{Facilities with 1-2.5 Micron Capability\label{section: JHK}}

The large-area archives searched for existing data were the Two Micron All-Sky Survey (2MASS; \citealt{skrutskie2006}), the various UKIRT-based surveys being done with the Wide-field Camera (WFCAM; \citealt{casali2007}) as part of the UKIRT Infrared Deep Sky Survey (UKIDSS; \citealt{lawrence2007}), and the various surveys being done with the Visible and Infrared Survey Telescope for Astronomy (VISTA; \citealt{emerson2006}) using the VISTA Infrared Camera (VIRCAM; \citealt{dalton2006}). The WFCAM archives searched were those of the UKIDSS Large Area Survey (ULAS), the UKIDSS Galactic Plane Survey (UGPS; \citealt{lucas2008}), the UKIDSS Galactic Clusters Survey (UGCS), and the UKIRT Hemisphere Survey (UHS; \citealt{dye2018}). The VISTA-based survey data searched were those of the VISTA Hemisphere Survey (VHS) and the VISTA Variables in the Via Lactea (VVV; \citealt{minniti2010}). Data were examined using the online WFCAM Science Archive\footnote{See \url{http://wsa.roe.ac.uk:8080/wsa/region_form.jsp}.} and VISTA Science Archive\footnote{See \url{http://horus.roe.ac.uk:8080/vdfs/Vregion_form.jsp}.}.

%Note: Neither this page http://wsa.roe.ac.uk/pubs.html nor the Dye et al. 2018 paper list references for ULAS or UGCS. I couldn't find a reference for VHS, either.

Given the complex spectral energy distributions of L, T, and Y dwarfs, care needs to be taken with regards to filter systems. The two filter systems employed by these near-infrared surveys are those of 2MASS\footnote{See \url{https://old.ipac.caltech.edu/2mass/releases/allsky/doc/sec6_4b.html} for a description of the 2MASS filter system.} and the Maunakea Observatories (MKO; \citealt{tokunaga2002}). Because of bandpass differences between these systems, the magnitude measured in, for example, the 2MASS $J$ filter may differ appreciably from the magnitude of the same L, T, or Y dwarf measured in MKO $J$. As a result, we report $J$ magnitudes in both. The $H$-band filter bandpasses are essentially identical between 2MASS  and MKO, so a single $H$-band magnitude is reported. The 2MASS $K_S$-band and MKO $K$ band are also reported separately. (Note that none of these large-area surveys uses the MKO version of the $K_S$ filter.)

Per the recommendations given at {\url http://horus.roe.ac.uk/vsa/ dboverview.html}, we selected magnitudes with the string {\fontfamily{pcr}\selectfont AperMag3} from both the WFCAM and VISTA Science Archives. For merged catalogs with multiple data sets per band, we chose the individual-epoch {\fontfamily{pcr}\selectfont AperMag3} magnitude with the smallest uncertainty. Magnitudes combined over multiple epochs were avoided; because most of our objects have high motions, these combined magnitudes are generally incorrect because one epoch of blank sky has been averaged into the combined magnitude. That is, the catalog's cross-matching between epochs is done purely on position, not on source identification.

For sources not covered or detected by these large-area surveys, we obtained follow-up photometry using the 2MASS camera (\citealt{milligan1996}) on the 1.5 m Kuiper Telescope on Mt.\ Bigelow, Arizona; the NOAO Extremely Wide Field Infrared Imager (NEWFIRM; \citealt{swaters2009}) at the 4 m Victor M.\ Blanco Telescope on Cerro Tololo, Chile; FLAMINGOS-2 (\citealt{eikenberry2006}) on the 8.1 m Gemini-South Telescope on Cerro Pach\'on, Chile; the Persson's Auxiliary Nasmyth Infrared Camera (PANIC; \citealt{martini2004}) at the 6.5 m Magellan Baade Telescope at Las Campanas Observatory, Chile; the 1.3 m Peters Automated Infrared Imaging Telescope (PAIRITEL; \citealt{bloom2006}) on Mt. Hopkins, Arizona; the Wide-field Infrared Camera (WIRC; \citealt{wilson2003b}) at the 5 m Hale Telescope at Palomar Mountain, California; and the Ohio State Infrared Imager/Spectrometer (OSIRIS) at the 4.1 m Southern Astrophysical Research Telescope (SOAR) located at Cerro Pach\'on, Chile. Data acquisition and reduction from these instruments are described in \cite{kirkpatrick2011} except those for FLAMINGOS-2, which are described in \cite{meisner2020a}.

\subsubsection{Facilities with 3-5 Micron Capability}

In Table~\ref{table:monster_table}, we have used the CatWISE2020 Catalog and Reject Table (\citealt{marocco2020}) as the primary source of photometry in the 3-5 $\mu$m range. Specifically, we used the W1 and W2 magnitudes computed by the moving solutions ({\fontfamily{pcr}\selectfont w1mpro\_pm} and {\fontfamily{pcr}\selectfont w2mpro\_pm}) because these should be more accurate than photometry from the stationary solution given the high motions of our objects and the long, eight-year time baseline covered by the CatWISE2020 data. For comparison, we have also listed photometry (including W3) from the AllWISE Source Catalog and Reject Table. For AllWISE, we selected values from the stationary solution since these should be more stable than the moving solutions, as these were based on fragile motions measured over only a six-month time baseline. (For objects lacking AllWISE detections, the stationary solution from CatWISE2020 was used instead, as noted in the table.)

Table~\ref{table:monster_table} also contains {\it Spitzer}/IRAC photometry in ch1 and ch2. Data from both our photometric follow-up and  astrometric monitoring programs were used. For the latter programs, which had many epochs of ch2 data, the PRF-fit photometry from each individual epoch was used; the reported magnitude is that resulting from the weighted mean flux. We also searched for ancillary data in the {\it Spitzer} Heritage Archive to further supplement our ch1 and ch2 measurements. Those ancillary data sets are listed in Table~\ref{ancillary_spitzer_photometry}. We have reduced those data using the same mosaic portion of our astrometric pipeline, and report the resulting PRF-fit magnitudes in Table~\ref{table:monster_table}. In these reductions, we used the PRF suite applicable to the phase of the mission, either cryogenic or warm, during which the data were taken. For targets in campaigns using IRAC's "sweet spot" (\citealt{ingalls2012}), we took only a portion of the resulting AORs since there is an enormous amount of data available; specfically, we selected a set of nine consecutive AORs toward the beginning of the campaign, another nine toward the middle, and another nine toward the end, and used those to build the mosaic needed for our pipeline.

%\onecolumngrid
\startlongtable
\begin{deluxetable}{lrcll}
\tabletypesize{\scriptsize}
%\tablenum{6}
\tablecaption{Ancillary Spitzer Photometry\label{ancillary_spitzer_photometry}}
\tablehead{
\colhead{Object} &
\colhead{AOR} &
\colhead{Bands} &
\colhead{Program} &
\colhead{PI} \\
\colhead{(1)} &
\colhead{(2)} &
\colhead{(3)} &
\colhead{(4)} &
\colhead{(5)} 
}
\startdata
2MASS 0045+1634&  67432448& 1  & 14019ss&  Vos\\
\nodata        &  67433472& 2  & \nodata&  \nodata\\
WISE 0047+6803 &  58386688& 1  & 12112ss&  Allers \\
2MASS 0103+1935&  43345408& 1  & 80179ss&  Metchev \\
\nodata        &  45626112& 2  & \nodata&  \nodata \\
SDSS 0107+0041 &  10374912& 1,2& 3136*  &  Cruz \\
SIMP 0136+0933 &  21967360& 1,2& 40076* &  Mainzer \\
2MASS 0144-0716&  10375424& 1,2& 3136*  &  Cruz \\
2MASS 0251-0352&  10376448& 1,2& 3136*  &  Cruz \\
WISE 0323+5625 &  32888832& 1  & 61070  &  Whitney \\
\nodata        &  32902912& 2  & \nodata&  \nodata \\
2MASS 0326-2102&  25362944& 1,2& 50059* &  Burgasser \\
2MASS 0340-6724&  53291776& 1  & 11174ss&  Metchev \\
\nodata        &  53291520& 2  & \nodata&  \nodata \\
2MASS 0355+1133&  25363712& 1,2& 50059* &  Burgasser \\
WISE 0401+2849 &  61990912& 1  & 13006  &  Trilling \\
2MASS 0407+1514&  12619008& 1,2& 35*    &  Fazio \\
2MASS 0421-6306&  43338496& 1  & 80179ss&  Metchev \\
\nodata        &  45384960& 2  & \nodata&  \nodata \\
2MASS 0439-2353&  10377472& 1,2& 3136*  &  Cruz \\
2MASS 0443-3202&  25363456& 1,2& 50059* &  Burgasser \\
2MASS 0445-3048&  10377728& 1,2& 3136*  &  Cruz \\
WISE 0457-0207 &  53278464& 1  & 11174ss&  Metchev \\
\nodata        &  53277952& 2  & \nodata&  \nodata \\
PSO 0506+5236  &  67439360& 1  & 14128ss&  Faherty \\
2MASS 0512-2949&  53291008& 1  & 11174ss&  Metchev \\
\nodata        &  53290752& 2  & \nodata&  \nodata \\
2MASS 0523-1403&  10377984& 1,2& 3136*  &  Cruz \\
WISE 0607+2429 &  50990336& 1  & 10167ss&  Gizis \\
\nodata        &  50990080& 2  & \nodata&  \nodata \\
2MASS 0624-4521&  10378240& 1,2& 3136*  &  Cruz \\
2MASS 0641-4322&  50921984& 1,2& 10098  &  Stern \\
2MASS 0700+3157&  10378496& 1,2& 3136*  &  Cruz \\
WISE 0715-1145 &  39058944& 1  & 61071  &  Whitney \\
\nodata        &  39075584& 2  & \nodata&  \nodata \\
2MASS 0755-3259&  38996736& 1  & 61071  &  Whitney \\
\nodata        &  39031808& 2  & \nodata&  \nodata \\
SDSS 0809+4434 &  67435776& 1,2& 14128ss&  Faherty \\
SDSS 0830+4828 &  10379008& 1,2& 3136*  &  Cruz \\
SDSS 0858+3256 &  21984768& 1,2& 40198* &  Fazio \\
SDSS 0909+6525 &  21985280& 1,2& 40198* &  Fazio \\
WISE 0920+4538 &  19064832& 1,2& 30854* &  Uchiyama \\
SIPS 0921-2104 &  10380288& 1,2& 3136*  &  Cruz \\
2MASS 0949-1545&  21985792& 1,2& 40198* &  Fazio \\
2MASS 1022+5825&  10380800& 1,2& 3136*  &  Cruz \\
SDSS 1043+1213 &  43336448& 1  & 80179ss&  Metchev \\
\nodata        &  45622784& 2  & \nodata&  \nodata \\
SDSS 1045-0149 &  10381056& 1,2& 3136*  &  Cruz \\
SDSS 1048+0111 &  10381312& 1,2& 3136*  &  Cruz \\
WISE 1049-5319 &  48640512& 1,2& 90095  &  Luhman \\
2MASS 1051+5613&  10381568& 1,2& 3136*  &  Cruz \\
2MASS 1122-3512&  43331072& 1  & 80179ss&  Metchev \\
\nodata        &  45621504& 2  & \nodata&  \nodata \\
2MASS 1126-5003&  21981952& 1,2& 40198* &  Fazio \\
2MASS 1213-0432&  10382336& 1,2& 3136*  &  Cruz \\
SDSS 1214+6316 &  13778688& 1,2& 244*   &  Metchev \\
SDSS 1219+3128 &  53295104& 1  & 11174ss&  Metchev \\
\nodata        &  53294848& 2  & \nodata&  \nodata \\
Gl 499C        &  53289984& 1  & 11174ss&  Metchev \\
\nodata        &  53289472& 2  & \nodata&  \nodata \\
2MASS 1315-2649&  15033856& 1,2& 20716* &  Gizis \\
2MASS 1324+6358&  13777920& 1,2& 244*   &  Metchev \\
DENIS 1425-3650&  10383104& 1,2& 3136*  &  Cruz \\
2MASS 1448+1031&  10383360& 1,2& 3136*  &  Cruz \\
Gaia 1713-3952 &  45999616& 1  & 80253  &  Whitney \\
\nodata        &  45986304& 2  & \nodata&  \nodata \\ 
VVV 1726-2738  &  21306368& 1,2& 30570* &  Benjamin \\
2MASS 1731+2721&  10384128& 1,2& 3136*  &  Cruz \\
WISE 1741-4642 &  67446272& 1  & 14128ss&  Faherty \\
2MASS 1750-0016&  53283840& 1  & 11174ss&  Metchev \\
\nodata        &  53283072& 2  & \nodata&  \nodata \\ 
SDSS 1750+4222 &  21986048& 1,2& 40198* &  Fazio \\
2MASS 1753-6559&  10384384& 1,2& 3136*  &  Cruz \\
2MASS 1807+5015&  10384640& 1,2& 3136*  &  Cruz \\
WISE 1809-0448 &  54359040& 1  & 11174ss&  Metchev \\
\nodata        &  54358784& 2  & \nodata&  \nodata \\
2MASS 1821+1414&  43343616& 1  & 80179ss&  Metchev \\
\nodata        &  45625344& 2  & \nodata&  \nodata \\
2MASS 1828-4849&  12618496& 1,2& 35*    &  Fazio \\
Gaia 1831-0732 &  12109824& 1  & 146*   &  Churchwell \\
\nodata        &  12105984& 2  & \nodata&  \nodata \\
WISE 1906+4011 &  47929088& 1  & 90152ss&  Gizis \\
\nodata        &  47929344& 2  & \nodata&  \nodata \\
Gaia 1955+3215 &  39262208& 1,2& 61072  &  Whitney \\
2MASS 2002-0521&  67453440& 1  & 14128ss&  Faherty \\
WISE 2030+0749 &  53278208& 1  & 11174ss&  Metchev \\ 
\nodata        &  53277440& 2  & \nodata&  \nodata \\
DENIS 2057-0252&  10385408& 1,2& 3136*  &  Cruz \\
PSO 2117-2940  &  67446784& 1  & 14128ss&  Faherty \\
2MASS 2139+0220&  10385664& 1,2& 3136*  &  Cruz \\
2MASS 2148+4003&  22144256& 1,2& 284*   &  Cruz \\
2MASS 2151-2441&  25364736& 1,2& 50059* &  Burgasser \\
2MASS 2152+0937&  10385920& 1,2& 3136*  &  Cruz \\
2MASS 2209-2711&  35348480& 1,2& 61009  &  Freedman \\ 
DENIS 2252-1730&  42482944& 1,2& 80183  &  Dupuy \\
2MASS 2331-4718&  12619264& 1,2& 35*    &  Fazio \\
\enddata
%\end{deluxetable*}
\onecolumngrid
\tablecomments{Program numbers followed by an asterisk were part of the {\it Spitzer} cryogenic mission and those with a suffix of  "ss" used the IRAC "sweet spot".}
\end{deluxetable}

%\twocolumngrid

\subsection{Spectroscopy\label{section:follow-up_spectra}}

We have obtained near-infrared spectra of some of the objects believed to lie within the 20-pc volume that lacked spectral types in the literature. These are listed in Table~\ref{spectroscopic_followup}. Details on the observing runs and data reduction methods are given in the subsections below.

\begin{deluxetable}{rlll}
\tabletypesize{\scriptsize}
%\tablenum{0}
\tablecaption{Spectroscopic Follow-up\label{spectroscopic_followup}}
\tablehead{
\colhead{Object} &
\colhead{Instrument} &
\colhead{Obs.\ Date (UT)} &
\colhead{Spec.\ Type\tablenotemark{a}} \\
\colhead{(1)} &
\colhead{(2)} &
\colhead{(3)} &
\colhead{(4)} 
}
\startdata
CWISE 0027$-$0121&  Magellan/FIRE&  2018 Dec 01& T9 \\
CWISE 0041$-$4019&  Magellan/FIRE&  2018 Sep 23& T8 pec\\
CWISE 0115$-$4616&  Magellan/FIRE&  2018 Dec 01& T6 \\
CWISE 0119$-$4937&  Magellan/FIRE&  2018 Dec 01& T7 \\
CWISE 0119$-$4502&  Magellan/FIRE&  2018 Dec 01& T8 \\
CWISE 0310$-$5733&  Magellan/FIRE&  2020 Feb 14& T5 \\
Gaia  0412$-$0734&  Keck/NIRES&     2018 Sep 01& L2 pec (composite?)\\
\nodata          &  \nodata   &     2018 Nov 17& \nodata \\
CWISE 0424+0002  &  Magellan/FIRE&  2019 Dec 12& T9:\\
CWISE 0433+1009  &  Keck/NIRES   &  2019 Dec 19& T8 \\
CWISE 0514+2004  &  IRTF/SpeX    &  2018 Nov 25& T0.5 \\
CWISE 0540$-$1802&  CTIO/ARCoIRIS&  2018 Apr 01& T5\\
CWISE 0601+1419  &  IRTF/SpeX    &  2018 Nov 25& T2.5 \\
CWISE 0602$-$4035&  Magellan/FIRE&  2017 Dec 03& T5.5 \\
CWISE 0613+4808  &  LDT/NIHTS&      2019 Nov 13& T5\\
CWISE 0620$-$3006&  Magellan/FIRE&  2017 Dec 06& T2.5\\
Gaia  0623+2631  &  IRTF/SpeX&      2019 Mar 16& L3 pec (composite?)\\
CWISE 0627$-$3730&  Magellan/FIRE&  2017 Dec 03& T6.5\\
CWISE 0630$-$6002&  Magellan/FIRE&  2019 Dec 11& T7 \\
Gaia  0640$-$2352&  Keck/NIRES&     2018 Oct 27& L5 \\
CWISE 0647$-$1600&  Magellan/FIRE&  2017 Dec 03& T6\\
Gaia  0734$-$4330&  Magellan/FIRE&  2020 Feb 13& L7 blue\\
CWISE 0749$-$6827&  Magellan/FIRE&  2017 Dec 03& T8 (pec?)\\
CWISE 0804$-$0000&  CTIO/ARCoIRIS&  2018 Apr 03& T4 \\
CWISE 0845$-$3305&  Magellan/FIRE&  2020 Feb 13& T7 \\
WISE  0902+6708  &  IRTF/SpeX    &  2019 Jan 22& L7 pec (low-g)\\
WISE  0911+2146  &  Magellan/FIRE&  2020 Feb 13& T9\\
CWISE 0917$-$6344&  Magellan/FIRE&  2020 Feb 14& L7\\
CWISE 0953$-$0943&  IRTF/SpeX    &  2019 Jan 23& T5.5\\
%CWISE 1008+2031 & Magellan/FIRE (horrible S/N)& \\
CWISE 1130$-$1158&  CTIO/ARCoIRIS&  2018 Apr 02& sdT5?\\
CWISE 1137$-$5320&  Magellan/FIRE&  2018 Feb 02& T7\\
CWISE 1141$-$2110&  Magellan/FIRE&  2019 Dec 11& T9:\\
CWISE 1152$-$3741&  CTIO/ARCoIRIS&  2018 Apr 02& T7 \\
Gaia  1159$-$3634&  IRTF/SpeX&      2019 Mar 16& M9.5\\
CWISE 1205$-$1802&  CTIO/ARCoIRIS&  2018 Apr 02& T8\\
CWISE 1315$-$4936&  Magellan/FIRE&  2018 Jan 02& T3\\
Gaia  1331$-$6513&  CTIO/ARCoIRIS&  2019 Jun 19& M9\\
WISE  1333$-$1607&  Magellan/FIRE&  2018 Feb 02& T9\\
CWISE 1630$-$0643&  Magellan/FIRE&  2020 Feb 13& T5\\
Gaia  1648$-$2913&  IRTF/SpeX&      2019 Mar 16& L5 pec (composite?)\\
CWISE 1650+5652  &  IRTF/SpeX&      2018 Jun 16& T0\\
CWISE 1726$-$4844&  Magellan/FIRE&  2020 Feb 13& T2.5 \\
Gaia  1807$-$0625&  IRTF/SpeX&      2019 Mar 16& M9 pec (composite?)\\
CWISE 1832$-$5409&  CTIO/ARCoIRIS&  2018 Apr 02& T7\\
Gaia  1836+0315  &  IRTF/SpeX&      2019 Mar 16& L6 v.\ red\\
CWISE 2001$-$4136&  Magellan/FIRE&  2016 Aug 09& T5\\
CWISE 2012+7017  &  LDT/NIHTS&      2019 Nov 13& T4.5\\
CWISE 2058$-$5134&  CTIO/ARCoIRIS&  2019 Jun 19& T0\\
WISE  2126+2530  &  Palomar/DBSP&   2019 Jul 22& M8 \\
CWISE 2138$-$3138&  Keck/NIRES&     2019 Oct 28& T8\\
CWISE 2344$-$4755&  Magellan/FIRE&  2018 Dec 01& T5.5\\
\enddata
\tablenotetext{a}{All are near-infrared spectral types except for that of WISE 2126+2530, which is an optical type.}
\onecolumngrid
\end{deluxetable}

\subsubsection{Palomar/DBSP}

A single object, WISE 2126+2530, was observed on 2019 Jul 22 (UT) using the Double Spectrograph (DBSP; \citealt{oke1982}) at the Hale 5m telescope on Palomar Mountain, California. The D55 dichroic was used to split the light near 5500 \AA\  (0.55 $\mu$m). The blue arm utilized the 600 line mm$^{-1}$ grating blazed at 4000 \AA\ (0.40 $\mu$m), while the red arm utilized the 316 line mm$^{-1}$ grating blazed at 7500 \AA\ (0.75 $\mu$m), producing continuous coverage from 3400 to 10250 \AA\ (0.340 to 1.025 $\mu$m) at a resolving power of $\sim$1500. A 600s exposure was acquired through partly cloudy conditions. Standard reduction procedures, as outlined in section 3.1.1 of \cite{kirkpatrick2016} were employed.

\subsubsection{LDT/NIHTS}

Two objects were observed on 2019 Nov 13 (UT) using the Near-Infrared High Throughput Spectrograph (NIHTS; \citealt{gustafsson2019}) at the 4.3-meter Lowell Discovery Telescope (LDT) at Happy Jack, Arizona. The 1$\farcs$34-wide slit was used providing an average resolving power of 62 over the 0.9--2.5 $\mu$m wavelength range.  A series of ten 120s exposures was obtained of both WISE 0613$+$4808 and WISE 2012$+$7017 at two different positions along the 10$\arcsec$-long slit.  Flats and xenon arcs exposures were taken at the beginning of the night and the A0 V stars, HD 45105 and HD 207646, respectively, were obtained for telluric correction purposes. The data were reduced using the Spextool data reduction package (\citealt{cushing2004}), and telluric correction and flux calibration were achieved following the technique described in \cite{vacca2003}.

\subsubsection{Keck/NIRES}

Four objects were observed over the nights of 2018 Sep 01, Oct 27, and Nov 17, and 2019 Oct 28 and Dec 19 (UT) using the Near-Infrared Echellette Spectrometer (NIRES; see, e.g., \citealt{wilson2004}) at the W.M.\ Keck II telescope on Maunakea, Hawaii. Setup and reductions were identical to those described in \cite{meisner2020b} and covered a spectral range of 0.94-2.45 $\mu$m at a resolving power of $\sim$2700. Note that the spectra for Gaia 0412$-$0734 were combined across nights.

\subsubsection{CTIO/ARCoIRIS}

Eight objects were observed over the nights of 2018 Apr 01-03 and 2019 Jun 19 (UT) using the Astronomy Research using the Cornell Infra Red Imaging Spectrograph (ARCoIRIS) at the Victor Blanco 4m telescope at the Cerro Tololo Inter-American Observatory (CTIO), Chile. Instrument setup and data reductions are identical to those detailed in \cite{greco2019} and covered a spectral range of 0.8-2.4 $\mu$m at a resolving power of $\sim$3500.

\subsubsection{IRTF/SpeX}

Ten objects were observed over the nights of 2018 Jun 16, Nov 25, and 2019 Jan 22/23 and Mar 16 (UT) using SpeX (\citealt{rayner2003}) at the NASA Infrared Telescope Facility (IRTF) on Maunakea, Hawaii. SpeX was used in prism mode with a 0$\farcs$8-wide slit to achieve a resolving power of R = 100-500 over the range 0.8-2.5$\mu$m. All data were reduced using Spextool (\citealt{cushing2004}). A0 stars were used for the telluric correction and flux calibration steps following the technique described in \cite{vacca2003}.

\subsubsection{Magellan/FIRE}

Twenty-five objects were observed over the nights of 2016 Aug 09; 2017 Dec 03 and Dec 06; 2018 Jan 02, Feb 02, Sep 23, and Dec 01; 2019 Dec 11/12; and 2020 Feb 13/14 (UT) using the Folded-port Infrared Echellette (FIRE; \citealt{simcoe2008, simcoe2010}) at the 6.5m Walter Baade (Magellan I) telescope at Las Campanas Observatory, Chile. Observations were done in high-throughput prism mode with the 0$\farcs$6 slit, which gives a resolving power of R$\approx$450 covering 0.8-2.45 $\mu$m.  Reductions followed the steps described in \cite{meisner2020b}.

\subsection{Spectral Classification}

\begin{figure}
\figurenum{7}
%The script for this is nir_spectrum_panel1.pro
\includegraphics[scale=0.57,angle=0]{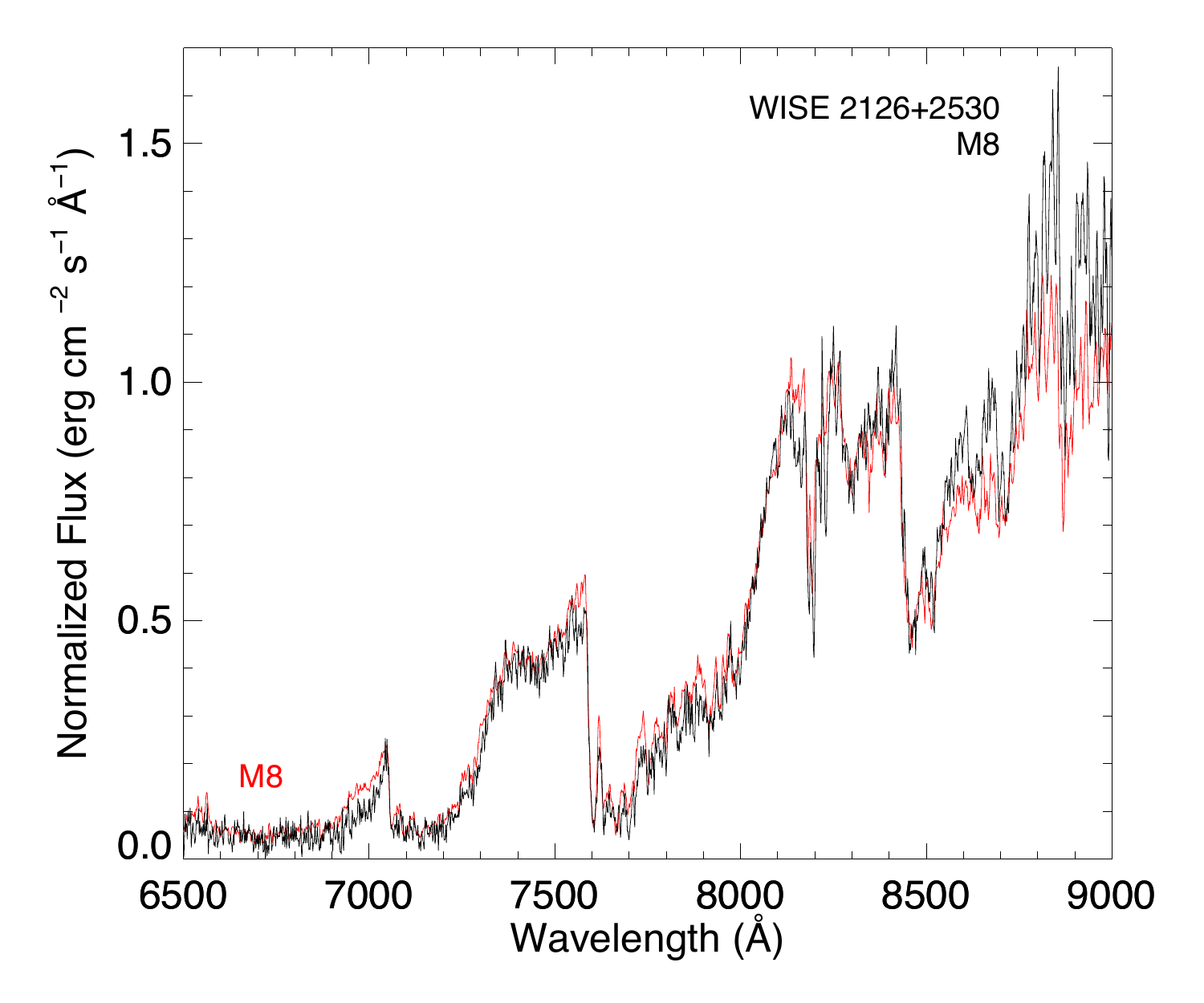}
\caption{Spectrum of WISE 2126+2530 (black) compared to the spectrum of the M8 standard van Biesbroeck 10 (red) from \cite{kirkpatrick2010}. The flux of both objects is normalized to one at 8250 \AA, a high-S/N portion of the spectrum free from telluric absorption.
\label{spectra-panel0}}
\end{figure}

\begin{figure}
\figurenum{8}
%The script for this is nir_spectrum_panel1.pro
\includegraphics[scale=0.70,angle=0]{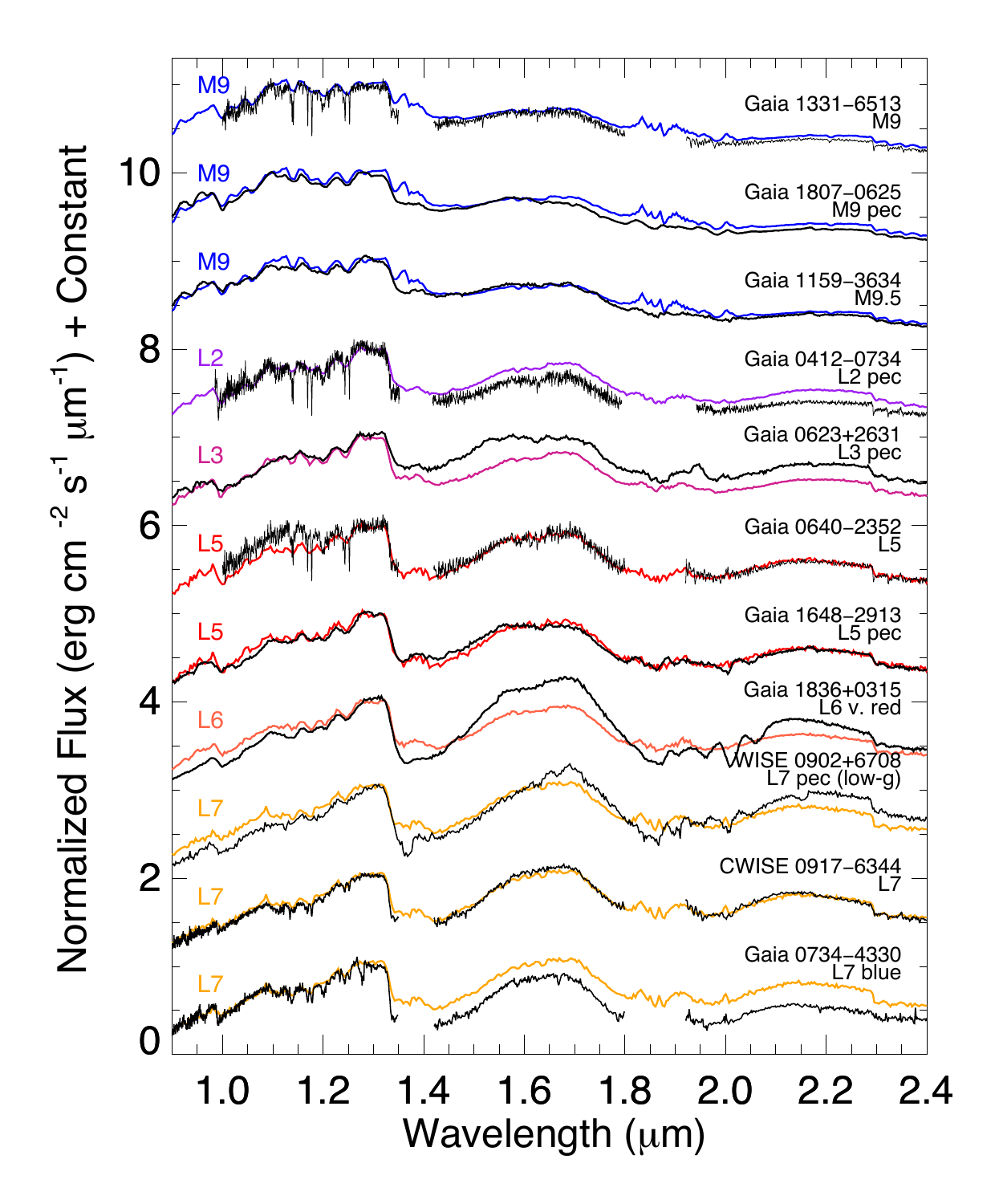}
\caption{Spectra of M- and L-type dwarfs compared to the spectrum of the standard nearest in type. These near-infrared standards are taken from \cite{kirkpatrick2010}. The flux of all objects is normalized to one at 1.28 $\mu$m and offset by integral increments to ease comparison. Spectra of the target objects are in black and those of the standards in other colors. Our spectral classification of each target object is also shown in black and that of the nearest standard in other colors. Smoothing has been applied for some of the noisier target spectra.
\label{spectra-panel1}}
\end{figure}

\begin{figure*}
\figurenum{9}
%The script for this is nir_spectrum_panel2.pro
\includegraphics[scale=0.60,angle=0]{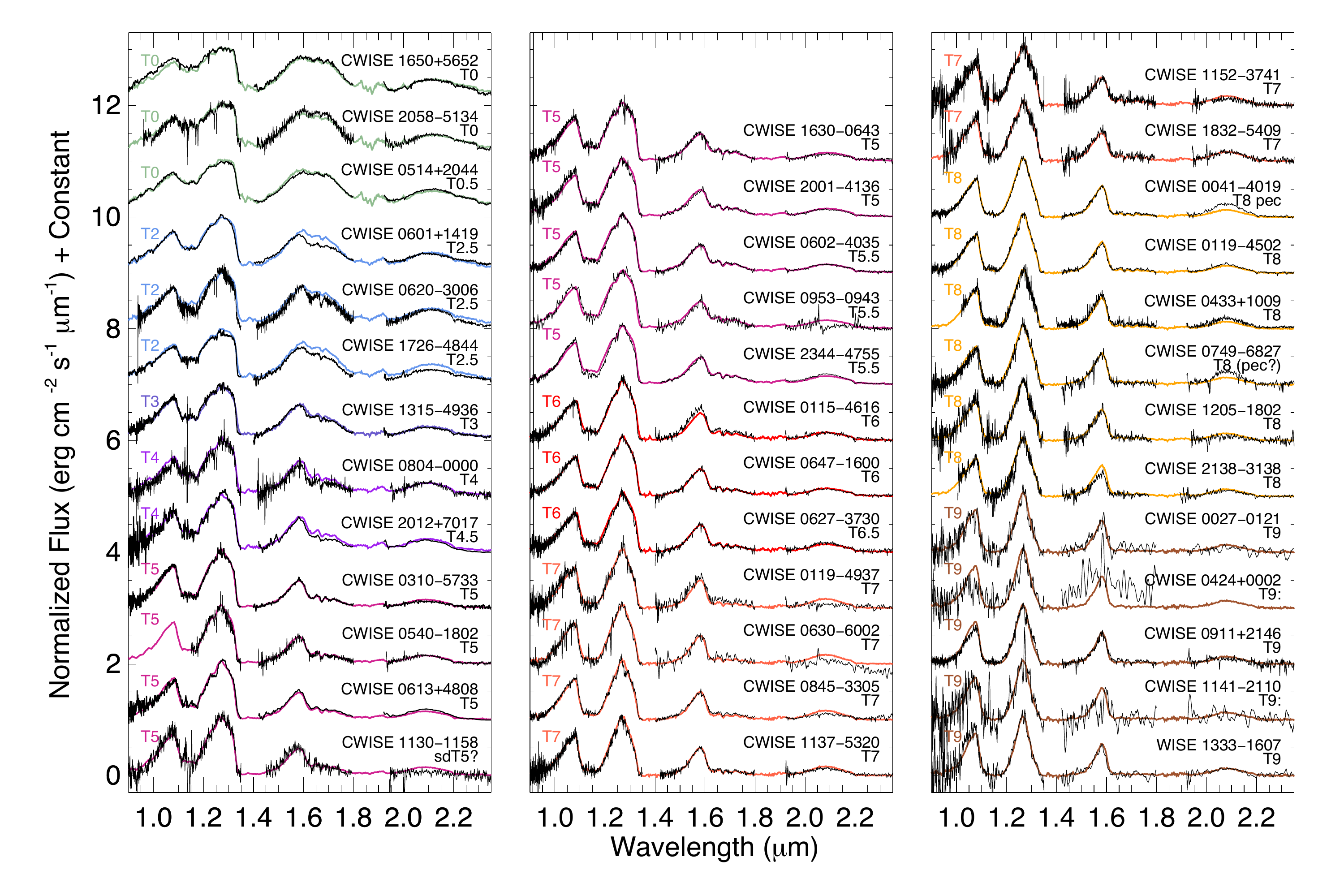}
\caption{Spectra of {\it WISE}-selected objects compared to the spectrum of the standard nearest in type. These near-infrared standards are taken from \cite{burgasser2006} and \cite{cushing2011}. See the caption to Figure~\ref{spectra-panel1} for other details.
\label{spectra-panel2}}
\end{figure*}

The spectra were classified as follows. For the single optical spectrum of WISE 2126+2530 in Figure~\ref{spectra-panel0}, we overplotted spectral standards from \cite{kirkpatrick1997}, which are based on the optical classification system of \cite{kirkpatrick1991} and looked for the best match over the entirety of the spectral range. For near-infrared spectra in Figures~\ref{spectra-panel1} and \ref{spectra-panel2}, we also performed a best by-eye fit, but using the near-infrared standards established by \cite{kirkpatrick2010} for the L dwarfs, \cite{burgasser2006} for early-T through late-T, and \cite{cushing2011} for late-T through early-Y. In total, we classify four objects as M dwarfs, eight as L dwarfs, and 38 as T dwarfs.

\section{Building the 20-pc Census\label{section:data_analysis}}

\subsection{Objects to consider}

The newly reduced {\it Spitzer} astrometry, along with published literature values, now enables a refinement of the 20-pc census. If an object has a trigonometric parallax measurement with an uncertainty $\le$12.5\%, we take that parallax at face value and retain the object if $\varpi_{obs} \ge 50$ mas. In this group there are a few objects that are worthy of special mention:

\begin{itemize}
    \item CWISE 0536$-$3055: Based on the data available to \cite{meisner2020a}, those authors were unable to confirm the motion of this candidate. Using the {\it Spitzer} ch1 and ch2 magnitudes and color, our type and distance estimates suggest a [T9.5]\footnote{We use brackets to denote estimates for spectral types not yet measured.} dwarf at $\sim$17.4 pc. Our {\it Spitzer} astrometry from Table~\ref{spitzer_results_highq} gives a total proper motion of 37.4$\pm$13.7 mas yr$^{-1}$, which is different from zero only at the 2.7$\sigma$ level. More telling, however, is the high-quality absolute parallax, which is 78.1$\pm$3.8 mas (only 5\% uncertainty; Table ~\ref{spitzer_results_highq}) based on {\it Spitzer} astrometric sampling with good coverage over the parallactic ellipse (Figure Set 1). CWISE 0536$-$3055 is therefore confirmed to be nearby and to fall within 20 pc of the Sun. This object represents a rare case in which the six-month parallactic motion (156.2 mas) is far (8.4$\times$) larger than the six-month proper motion (18.7 mas). Obtaining a radial velocity of this object would inform us whether CWISE 0536$-$3055 is coming toward our Solar System or away, and how that translates into a closest approach distance.
    
    \item WISE 0546$-$0959: As with CWISE 0536$-$3055 above, this T5 dwarf has an exceptionally small proper motion of 11.8$\pm$3.5 mas yr$^{-1}$ according to \cite{best2020} or 10.3$\pm$2.5 mas yr$^{-1}$ according to our {\it Spitzer} astrometry, despite its large parallax of 50.4$\pm$3.6 (\citealt{best2020}) or 57.5$\pm$3.9 mas (our {\it Spitzer} measurement). In this case, the six-month parallactic motion is $\sim$20$\times$ larger than the six-month proper motion. 
    
    \item CWISE 1411$-$4811: Despite a robust {\it Spitzer} parallax value of 58.2$\pm$4.7 mas, this object has no measured spectral type. Its values of W1$-$W2 = 2.28$\pm$0.04 mag and M$_{ch2}$ = 13.10$\pm$0.18 mag (Table~\ref{table:monster_table}) suggest a type of [T6.5].
    
    \item WISE 1600$-$4543: This object has no measured spectral type, despite a robust parallax measurement of 74.7951$\pm$0.9190 mas from {\it Gaia} DR2. Using data in Table~\ref{table:monster_table}, we find that this source has M$_{W2}$ = 11.74$\pm$0.06 mag, which suggests [L9]. The color of $J_{2MASS}-$W2 = 2.62$\pm$0.04 mag suggests a type between mid-L and early-T. 
    
    \item CWISE 1926$-$3429: Despite a robust {\it Spitzer} parallax value of 51.6$\pm$3.9 mas, this object has no measured spectral type. Values of ch1$-$ch2 = 0.98$\pm$0.03 mag and M$_{ch2}$ = 12.67$\pm$0.17 mag (Table~\ref{table:monster_table}) suggest a type of [T5.5].

\end{itemize}

In addition to objects with well measured parallaxes, there is another set of potential 20-pc members with poorer or non-existent parallax measurements that need additional scrutiny. The objects are listed in Table~\ref{table:distance_estimates} and are (a) pulled from Table~\ref{new_discoveries} or Tables~\ref{spitzer_results_lowq}-\ref{spitzer_results_poorq}, (b) are objects originally included in {\it Spitzer} program 14224 but dropped because of time restrictions, or (c) are previously published objects rediscovered by the CatWISE or Backyard Worlds teams for which initial estimates indicated distances within 23 pc of the Sun. We use a combination of photometric and spectrophotometric distance estimates to determine whether each object should be included in the 20-pc census. Namely, we use data from 20-pc census members with robust parallax measurements (uncertainties $\le12.5\%$) to construct three independent relations of M$_J$ vs.\ $J-$W2 (valid for $J-$W2 $\ge$ 4.0 mag, or for 2.0 $ \le J-$W2 $<$ 4 mag if W1$-$W2 $\ge$ 2.2 mag), M$_H$ vs.\ near-infrared spectral type (valid for all L, T, and Y spectral types), and M$_{ch2}$ vs.\ ch1$-$ch2 (valid for ch1$-$ch2 $\ge$ 0.4 mag). Using data provided in Table~\ref{table:monster_table}, we use the apparent magnitudes and colors of each object in Table~\ref{table:distance_estimates} to estimate a distance from each relation, and then average the results to provide a final distance estimate. For some objects, there is not sufficient observational data for any of these relations -- or the object has colors outside the range for which the relations are valid -- so instead we use a M$_{W2}$ vs.\ W1$-$W2 relation (valid for W1$-$W2 $\ge$ 0.5 mag), also constructed from 20-pc members with robust parallax measurements, to provide a distance estimate. 

\startlongtable
\begin{deluxetable*}{rcccccccccccl}
\tabletypesize{\footnotesize}
\tablecaption{Other Objects Considered for the 20-pc Census\label{table:distance_estimates}}
\tablehead{
\colhead{Object} & 
\colhead{Spec.} &
\colhead{Spec.} &
\colhead{Our} &
\colhead{Pub.} &
\colhead{Pub.} &
\colhead{d$_{est}$}&
\colhead{d$_{est}$}&
\colhead{d$_{est}$}&
\colhead{d$_{est}$}&
\colhead{Final} &
\colhead{Include} &
\colhead{Remarks} \\
\colhead{name} & 
\colhead{type} &
\colhead{type} &
\colhead{$\varpi_{abs}$} &
\colhead{$\varpi_{abs}$} &
\colhead{$\varpi_{abs}$} &
\colhead{$J$,W2}&
\colhead{$H$,type}&
\colhead{ch1,ch2}&
\colhead{W1,W2}&
\colhead{d$_{est}$} &
\colhead{in 20-pc} &
\colhead{} \\
\colhead{} & 
\colhead{} &
\colhead{ref.} &
\colhead{(mas)} &
\colhead{(mas)} &
\colhead{ref.} &
\colhead{(pc)}&
\colhead{(pc)}&
\colhead{(pc)}&
\colhead{(pc)}&
\colhead{(pc)} &
\colhead{census?} &
\colhead{} \\
\colhead{(1)} &                          
\colhead{(2)} &  
\colhead{(3)} &  
\colhead{(4)} &
\colhead{(5)} &
\colhead{(6)} &
\colhead{(7)} &
\colhead{(8)} &
\colhead{(9)} &                          
\colhead{(10)} &
\colhead{(11)} &
\colhead{(13)} &
\colhead{(12)} 
}
\startdata
CWISE 0027$-$0121& T9     & T& 54.2$\pm$7.9 &  \nodata     & - &     16.00&    \nodata&      17.47&   \nodata&    16.73 &    yes\\
2MASS 0034$-$0706& L4.3   & D& \nodata      &  55.8$\pm$12.3&r &   \nodata&      32.03&    \nodata&   \nodata&    32.03 &     no\\
CWISE 0043$-$3822& [T8.5] & T& 38.1$\pm$15.7&  \nodata     & - &   \nodata&    \nodata&      19.24&   \nodata&    19.24 &    yes\\
WISE  0048+2508  & [T8.5] & m& \nodata      &  \nodata     & - &     12.48&    \nodata&      14.54&   \nodata&    13.51 &    yes\\
2MASS 0051$-$1544& L5     & B& 34.0$\pm$6.6 &  29.1$\pm$1.4& G &   \nodata&    \nodata&    \nodata&   \nodata&    34.36 &     no\\
2MASS 0103+1935  & L6(o)  & K& 35.9$\pm$5.7 &  46.9$\pm$7.6& F &   \nodata&    \nodata&    \nodata&   \nodata&    21.32 &     no&    see text\\
CWISE 0115$-$4616& T6     & T& \nodata      &  \nodata     & - &     21.04&      31.31&      19.97&   \nodata&    24.11 &     no\\
CWISE 0119$-$4937& T7     & T& \nodata      &  \nodata     & - &     18.11&      39.29&      20.47&   \nodata&    25.96 &     no\\
CWISE 0119$-$4502& T8     & T& \nodata      &  \nodata     & - &     20.50&      11.51&      18.75&   \nodata&    16.92 &    yes\\
WISE  0132$-$5818& [T9]   & m& 27.2$\pm$7.3 &  \nodata     & - &     22.40&    \nodata&      21.67&   \nodata&    22.04 &     no\\
WISE  0135+1715  & T6     & k& 65.3$\pm$10.0&  46.7$\pm$3.5& W &   \nodata&    \nodata&    \nodata&   \nodata&    21.41 &     no\\
WISE  0138$-$0322& T3     & J& 38.5$\pm$6.4 &  43.9$\pm$2.9& W &   \nodata&    \nodata&    \nodata&   \nodata&    22.78 &     no\\
\enddata
\tablecomments{(This table is available in its entirety in a machine-readable form in the online journal. A portion is shown here for guidance regarding its form and content.)}
\tablenotetext{a}{This object is excluded for our 20-pc L, T, and Y dwarf census because its type is earlier than L0.}
\tablecomments{Reference code for infrared spectral type. Values in brackets are estimates, and the types for 2MASS 0103+1935 and 2MASS 0639$-$7418 are based on optical spectra: 
(B) \citealt{burgasser2010}, 
(b) \citealt{burgasser2010b}, 
(C) \citealt{cruz2007},
(D) \citealt{bardalez2014}, 
(F) \citealt{faherty2016},
(G) \citealt{mace2013}, 
(g) \citealt{greco2019},
(i) \citealt{kirkpatrick2016},
(J) \citealt{kirkpatrick2011}, 
(j) \citealt{kirkpatrick2010},
(K) \citealt{kirkpatrick2000}, 
(k) \citealt{kirkpatrick2012}, 
(M) \citealt{meisner2020a}, 
(m) \citealt{meisner2020b},
(R) \citealt{reyle2014},
(S) \citealt{schneider2017},
(T) this paper,
(U) \citealt{burningham2013},
(u) \citealt{burningham2010},
(W) \citealt{best2015}.}
\tablecomments{Reference code for published parallax: 
(c) \citealt{theissen2020},
(F) \citealt{faherty2012},
(G) \citealt{gaia2018}, 
(g) \citealt{gaia2018} parallax for the primary is cited,
(r) \citealt{smart2018},
(W) \citealt{best2020}.}
\end{deluxetable*}
%\end{center}

We also provide spectral types in Table ~\ref{table:distance_estimates}. For objects without measured spectral types, we provide type estimates by using the final distance estimate in the table combined with the object's ch2 magnitude to provide an estimate of $M_{ch2}$. We then take data from 20-pc census members having robust parallax measurement (uncertainties $\le12.5\%$) to construct a relation of spectral type vs.\ $M_{ch2}$ (valid over the entire range needed, 10.5 $< M_{ch2} <$ 16.0 mag), and use this to predict the type. (A value of $M_{W2}$ is used as a proxy for $M_{ch2}$ when no ch2 magnitude is available.) These estimated types are enclosed within brackets in the table.

Several objects requiring special consideration are noted by "see text" under the Remarks column in Table ~\ref{table:distance_estimates}. Those objects are discussed below:

\begin{itemize}
   
   \item 2MASS 0103+1935: This optical L6 dwarf (\citealt{kirkpatrick2000}) has two independent parallax measurements, both low quality, of 35.9$\pm$5.7 mas (Table~\ref{spitzer_results_lowq}) and 46.9$\pm$7.6 mas (\citealt{faherty2012}). Given that both measures suggest a parallax below 50 mas, we consider this object to fall outside of 20 pc.

   \item CWISE 0212+0531: This object was announced in \cite{meisner2020a}, although those authors were not able to confirm the object's motion.  Based on the {\it Spitzer} ch1 and ch2 magnitudes and color, our spectral type and distance estimates suggest [$\ge$Y1] at $<$13.3 pc. Our {\it Spitzer} astrometry from Table~\ref{spitzer_results_poorq} gives a total proper motion of 82.6$\pm$52.7 mas yr$^{-1}$, which is different from zero at only the 1.6$\sigma$ level. The resulting parallax is 24.7$\pm$16.3 mas, with one parallax factor being represented by only a single {\it Spitzer} data point (Figure Set 1). Because both the motion and parallax are insignificantly different from zero, and because the measured parallax is much smaller than the expected value, we consider this to be a background object.

   \item CWISE 0423$-$4019: Our {\it Spitzer} photometry suggests a [T9] dwarf at $\sim$16.5 pc. Our {\it Spitzer} parallax measurement of $-$11.7$\pm$6.9 mas and total proper motion of 3.8$\pm$3.3 mas yr$^{-1}$, however, show that this is a background object and not a nearby brown dwarf.

   \item CWISE 0424+0002: This object was announced in \cite{meisner2020a}, although those authors were not able to confirm the object's motion. Our {\it Spitzer} astrometry from Table~\ref{spitzer_results_poorq} gives a total proper motion of 208.7$\pm$35.0 mas yr$^{-1}$, which is different from zero at the 6.0$\sigma$ level. The resulting parallax is 37.4$\pm$11.7 mas, representing a 31\% uncertainty, and there is only a single {\it Spitzer} data point at one of the maximum parallax factors (Figure Set 1). Our spectrum from Figure~\ref{spectra-panel1} confirms that it is nearby. Because the motion is confirmed but the trigonometric parallax is not yet credible, we use our (spectro)photometric distance estimates to place this object within 20 pc.

   \item CWISE 0442$-$3855: Our {\it Spitzer} photometry suggests a [T8.5] dwarf at $\sim$16.8 pc. Our {\it Spitzer} parallax measurement of $-$12.4$\pm$4.9 mas and total proper motion of 3.6$\pm$2.6 mas yr$^{-1}$, however, show that this is a background object and not a nearby brown dwarf.

   \item CWISE 0617+1945: Using the colors of this object in Table~\ref{table:monster_table}, we are unable to provide a distance estimate using any of our four preferred absolute magnitude relations. Using the MKO-based $JHK$ magnitudes from Table~\ref{table:monster_table}, the color-color plots presented in section~\ref{section:plot_analysis} suggest that this is a late-L dwarf, which would indicate M$_H$ = 13.8 mag and a distance of $\sim$7.5 pc. As further discussed in section~\ref{section:known_multiples}, the object appears to have a co-moving companion to its north-east, which is faint enough that it does not strongly affect the distance estimate. We consider this pair to fall within 20 pc.
   
   \item ULAS 0745+2332: This object, discovered by \cite{burningham2013}, lies in very close proximity to a background star that complicated our {\it Spitzer} astrometric measurements, leading to a false, negative parallax (Table~\ref{spitzer_results_poorq}). This object is not detected in any of the various {\it WISE} catalogs consulted for Table~\ref{table:monster_table}. The discovery paper lists a T8.5 spectral type and estimated distance of $<$19.4 pc, so we include this object in the 20-pc census.
   
   \item WISE 0830+2837: This candidate Y dwarf from \cite{bardalez2020} is sufficiently red in its {\it Spitzer} colors to be a possible bridge source in T$_{eff}$ between spectroscopically verified early-Y dwarfs and WISE 0855$-$0714. Given its estimated distance of $\sim$8.2 pc and our low-quality parallax of 90.6$\pm$13.7 mas, we consider this object to be well within 20 pc.

   \item CWISE 1008+2031: This object was announced in \cite{meisner2020a}, although those authors were not able to confirm the object's motion. Our {\it Spitzer} astrometry from Table~\ref{spitzer_results_poorq} gives a total proper motion of 215.3$\pm$51.5 mas yr$^{-1}$, which is different from zero at the 4.2$\sigma$ level. The resulting parallax is 37.1$\pm$15.1 mas, representing a 41\% uncertainty, with the {\it Spitzer} astrometric sampling providing only a single point at one of the maximum parallax factors (Figure Set 1). Because the motion of this object confirms it as being nearby and our photometric distance estimates place it within 20 pc, we include it in the 20-pc census.

%   \item CWISE 1022+1455: Based on {\it Spitzer} photometry, this object was estimated to be a [T8.5] at a distance of $\sim$26.2 pc by \cite{meisner2020a}. Our {\it Spitzer} parallax value of 76.0$\pm$16.4 mas suggests a much closer distance; however, this value is based on data that samples one extreme of the parallactic ellipse with only a single data point. Given the discrepancy in these distances, a closer look at other colors is warranted. From Table~\ref{table:monster_table}, we find that the source has been detected only in {\it Spitzer} ch1 and ch2 and {\it WISE W2}, although there is a $J$-band limit from \cite{meisner2020a} and a W1 limit from CatWISE2020. These result in colors of ch1$-$ch2 = 1.94 $\pm$0.09 mag, $J_{MKO}-$ch2 $>$ 3.9 mag, and W1$-$W2 $>$ 3.7 mag. The first of these suggests the [T8.5] type mentioned above; the second suggests [$>$T9] and the third [$>$Y0] based on plots of these colors versus spectral type in \cite{kirkpatrick2011} and \cite{kirkpatrick2019}. The {\it Spitzer} parallax value and ch1 magnitude are consistent with a normal dwarf of type [Y0.5]. However, the ch1$-$ch2 color is $\sim$0.8 mag too blue for this type. We therefore conclude either that this is a $<$20 pc brown dwarf of unusual type or that the {\it Spitzer} parallax is simply not credible. Until additional astrometry is available, we prefer the latter hypothesis and exclude this object from the 20-pc sample.
   
   \item WISE 1040+4503: This object was announced in \cite{meisner2020a}, although those authors were not able to confirm the object's motion. Our {\it Spitzer} astrometry from Table~\ref{spitzer_results_poorq} gives a total proper motion of 91.7$\pm$32.3 mas yr$^{-1}$, which is different from zero at the 2.8$\sigma$ level. The resulting parallax is 18.8$\pm$9.8 mas, representing a 52\% uncertainty, with the {\it Spitzer} astrometric sampling providing only a single point at one of the maximum parallax factors (Figure Set 1). Given that the photometric distance estimate is outside of 20 pc and that a distance within 20 pc is not suggested by the available astrometry, we exclude this object from the 20-pc census. It may, in fact, be a background object.

   \item CWISE 1047+5457: \cite{meisner2020a} estimated that this is a [Y0] dwarf at $\sim$21.7 pc. Our low-quality parallax value of 75.2$\pm$12.8 suggests that it is closer. One of the maximum parallax factors is sampled with only one {\it Spitzer }data point (Figure Set 1), but this together with the other data samples strongly suggest a parallax $>$50 mas. We consider this object to lie within 20 pc, although higher quality astrometry is clearly needed.

   \item CFBDS 1118$-$0640: This object, which is a common proper motion companion to the mid-M dwarf 2MASS J11180698$-$0640078, was included in our {\it Spitzer} parallax program through a mistake. Its spectral type of T2 was paired up incorrectly with the {\it WISE} magnitudes of the primary, resulting in a photometric distance of $<$20 pc. The {\it Gaia} DR parallax of the primary is 9.90$\pm$0.15 mas, and our {\it Spitzer} parallax of the companion T dwarf is 1.4$\pm$5.2 mas. This object is therefore excluded from the 20-pc census.

   \item CWISE 1130$-$1158: This object has wildly discrepant distance estimates, with those using colors predicting a value within 20 pc and the one using spectral type indicating a value well outside 20 pc. Our spectroscopic follow-up from section~\ref{section:follow-up_spectra} suggests that this object has a peculiar spectrum, particularly a depressed $K$-band spectrum similar to that seen in other T-type subdwarfs (e.g., \citealt{pinfield2014a}). We therefore classify this object as an sdT5?. Given its possible subdwarf status, neither the color-based nor type-based relations may be accurate. For now, we consider this object to fall outside 20 pc, but additional astrometry is needed.
   
   \item 2MASS 1158+0435: This is an optical and near-infrared sdL7 (\citealt{kirkpatrick2010}) placed on the parallax program because distance estimates for L subdwarfs are not yet well established. Our {\it Spitzer} parallax value of 39.2$\pm$6.2 mas is based on a well-sampled parallactic ellipse (Figure Set 1), so we consider this object to lie outside of 20 pc.

   \item ULAS 1319+1209: \cite{burningham2010} classify this object as T5 pec based on a T5 fit in the $J$ band and a T3 fit in the $H$ band. In preparing our list of target objects for the {\it Spitzer} parallax program, we mistook this object to be the bright proper motion star immediately to its north, which has an AllWISE value of W2 = 12.56$\pm$0.03 mag. This led to an incorrect distance estimate of $\sim$9 pc. Our {\it Spitzer} parallax (7.8$\pm$6.5 mas) was measured for this brighter star, Gaia DR2 3739496602924096000, not of the T dwarf\footnote{Because our measurements are not of a brown dwarf or even of an object within 20 pc, we have excluded this source from Table~\ref{table:monster_table}.}. Investigating this further, we find that the {\it Gaia} star, which is not listed in SIMBAD, has a {\it Gaia} DR2 parallax of 9.22$\pm$0.11 mas and motions of $\mu_{RA}$ = $-$135.2$\pm$0.2 mas yr$^{-1}$ and $\mu_{Dec}$ = 3.8$\pm$0.2 mas yr$^{-1}$. The motion measured by \cite{burningham2013} for the T dwarf is $\mu_{RA}$ = $-$120.9$\pm$16.0 mas yr$^{-1}$ and $\mu_{Dec}$ = $-$22.9$\pm$14.6 mas yr$^{-1}$ which is consistent within the uncertainties to those of the {\it Gaia} star. \cite{murray2011} estimate the distance of ULAS 1319+1209 to be 75$\pm$12 pc and note that it might be a halo T dwarf, although \cite{liu2011} contend that thick disk membership is more likely. \cite{burningham2013} estimate that the T dwarf falls between 58.6 and 99.1 pc if it is a single object, and could be as distant as 140.0 pc if a binary. These higher values are consistent with the distance to the {\it Gaia} object at 108.5 pc. The {\it Gaia} star has {\fontfamily{pcr}\selectfont teff\_val} = 3974K, which would correspond to a late-K dwarf, whose metallicity should be easily measurable. We believe that this may be a new benchmark system and a particularly valuable one since the T dwarf shows peculiarities that may or may not be linked to a lower metallicity. 

   \item Gaia 1331$-$6513: This is another object, like CWISE 0536$-$3055 discussed above, that has a very low motion value given its proximity to the Sun ($\sim$16.0 pc). The total motion from {\it Gaia} DR2 is 21.2$\pm$0.3 mas yr$^{-1}$, meaning that the parallactic motion over six months is twelve times larger than the proper motion. A measurement of the radial velocity would inform us whether this object is coming toward the Sun or away from it and the timescale for closest approach to the Solar System.
   
   \item WISE 1355$-$8258: This object was announced in \cite{schneider2016}, and \cite{kirkpatrick2016} noted its unusual near-infrared spectrum, which they tentatively interpreted to be an sdL5. \cite{bardalez2018} attempted to explain the spectrum as that of an unresolved binary but were unable to find a binary fit that provided a convincing explanation. They noted, however, a possible kinematic association with the AB Doradus Moving Group, despite finding no spectroscopic evidence of low-gravity. Their best guess for the distance is 27-33 pc. Using {\it WISE} astrometry, \cite{theissen2020} measure a fragile parallax of 60$\pm$19 mas (32\% error). Using a combination of 2MASS and {\it WISE} astrometry, E.\ L.\ Wright (priv.\ comm.) finds a still fragile parallax of 73$\pm$16 mas (22\% error). For now, we consider this object to lie outside of 20 pc but encourage future astrometric monitoring in an effort to better understanding this intriguing object.
   
   \item CWISE 1446$-$2317: \cite{marocco2020} show that the {\it Spitzer} colors of this object place it among the coldest Y dwarfs currently known. Our {\it Spitzer} parallax measurements of 95.6$\pm$13.9 mas, though somewhat fragile based on its poorly sampled parallactic ellipse (Figure Set 1), nonetheless strongly suggests proximity to the Sun. We include this object within the 20 pc census.

   \item CWISE 1458+1734: This object is from \cite{meisner2020a}, who suggest a spectral type of [T8] and distance of $\sim$21.6 pc. Our {\it Spitzer} parallax measurement of 1.3$\pm$7.2 mas (Table~\ref{spitzer_results_poorq}) is based on a fit to a well-sampled parallactic ellipse. The proper motion of this source is measured at high significance, 503.6$\pm$26.1 mas yr$^{-1}$ (Table~\ref{spitzer_results_poorq}), so the lack of a measurable parallax is puzzling. We have compared the UHS $J$-band image from 2013 May to our own $J$-band image taken from Palomar/WIRC in 2020 Jul (Figure~\ref{figure:cwise1458}) and confirm a motion along nearly the same position angle indicated by our astrometric fit in Figure Set 1. We note, however that the position angle of the motion vector is almost perfectly aligned with the major axis of the parallactic ellipse, meaning that an incorrect motion magnitude could easily erase the parallactic signature. We have performed a test of this hypothesis by determining what value of the total motion is needed to create a parallactic signature matching the distance estimate in Table~\ref{table:distance_estimates} while also assuming that the motion {\it direction} measured by our {\it Spitzer}+unWISE astrometry is correct. We get the correct result if the total proper motion is reduced from 504 mas yr$^{-1}$ to $\sim$300 mas yr$^{-1}$. This hypothesis is supported by the fact that CWISE 1458+1734 is moving between -- and is bracketed by -- two background objects that themselves fall along nearly the same position angle as the proper motion, and it is thus conceivable the unWISE astrometry of the T dwarf is pulled southeastward at early epochs by blending from the southeast source and northwestward at later epochs by blending from the northwest source, thereby inflating the true motion value. Crude measurements of the astrometry from the images in Figure~\ref{figure:cwise1458} give a proper motion of $\sim$305 mas yr$^{-1}$, confirming our hypothesis. Nevertheless, the photometric distance of this source places it just outside 20 pc, so it is not included in our 20-pc census. 
   
\begin{figure}
\figurenum{10}
\includegraphics[scale=0.40,angle=0]{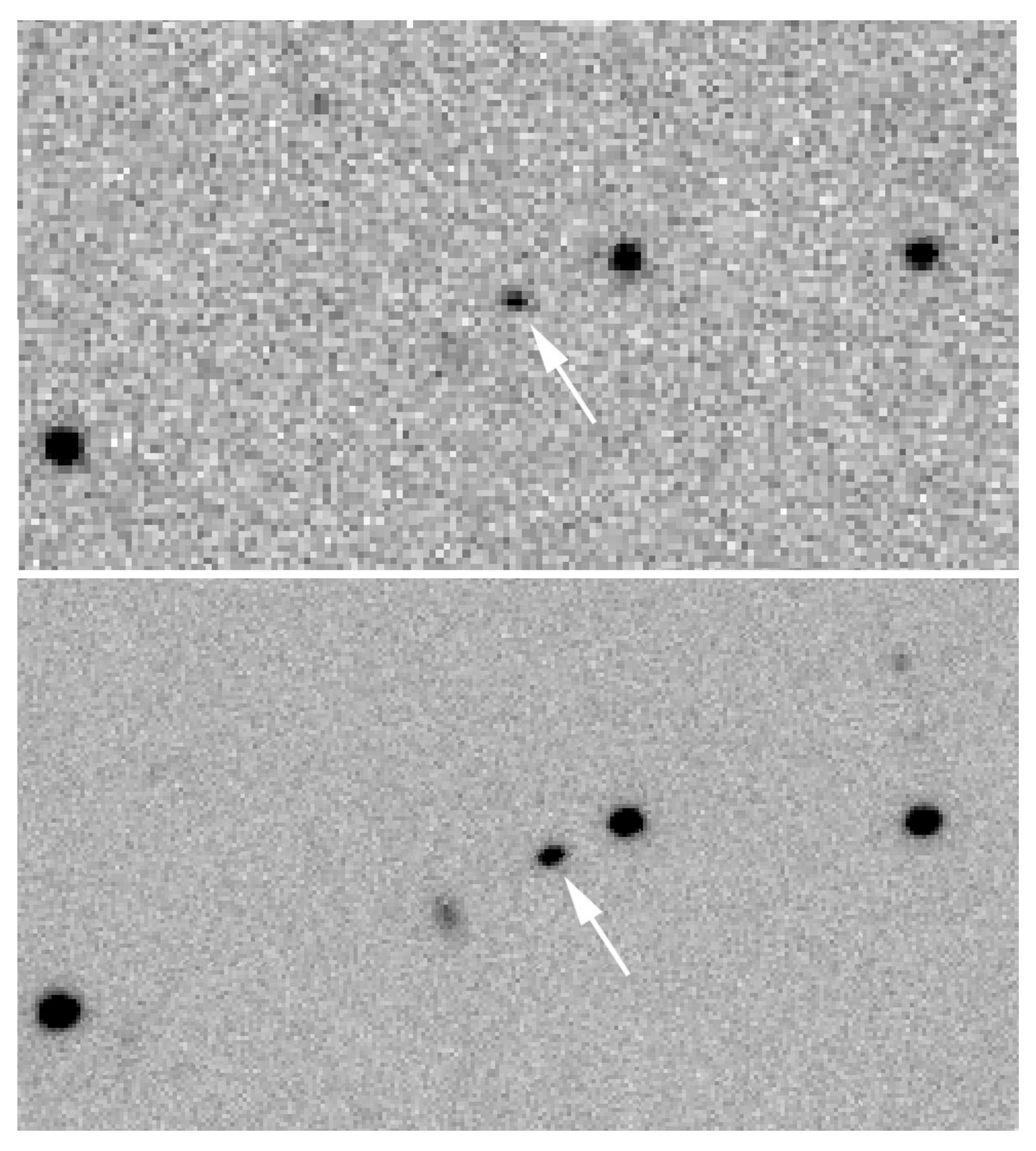}
\caption{Images at $J$-band for CWISE 1458+1734. (Top) The 2013 May image from UHS. (Bottom) Our 2020 Jul image from Palomar/WIRC. These images are 1$\times$0.5 arcmin with north up and east to the left. Arrows mark the location of CWISE1458+1734 and illustrate its motion over the 7.2-yr baseline.
\label{figure:cwise1458}}
\end{figure}

   \item WISE 1534$-$1043: This object is from \cite{meisner2020a}, who note that its placement on the $J-$ch2 vs.\ ch1$-$ch2 color plot suggest it is a mid- to late-T subdwarf. As such, deriving a photometric distance estimate from relations that assume solar metallicity is useless. Too few late-T subdwarfs are known to enable a better distance estimate, particularly since we do not know if the object's metallicity is similar to or more extreme than known T subdwarfs, so our {\it Spitzer } trigonometric distance measurement of 47.8$\pm$14.3 mas (Table~\ref{spitzer_results_poorq}) is the best current indicator, despite the large relative uncertainty of 30\%. The object's high proper motion, 2772.7$\pm$57.3 mas yr$^{-1}$, also points to an old, kinematically heated object. (At 20 pc, this would correspond to a tangential velocity of 263 km s$^{-1}$.) The $<$50mas parallax suggests that we exclude this object from the 20-pc census as we await additional astrometric measurements.

   \item WISE 1619+1347: This object was announced in \cite{meisner2020b}, although those authors were not able to confirm the object's motion. Our {\it Spitzer} astrometry from Table~\ref{spitzer_results_poorq} gives a total proper motion of 29.1$\pm$16.6 mas yr$^{-1}$, which is different from zero at only the 1.8$\sigma$ level. The resulting, negative parallax of $-$9.1$\pm$4.3 mas, is based on {\it Spitzer} astrometric data that sample the parallactic ellipse well (Figure Set 1). We therefore consider this to be a background object.
   
   \item CWISE 1827+5645: This object was re-discovered by high school student Justin Hong as part of the Summer Research Connection at Caltech in the summer of 2020. The object was first discovered during the original {\it WISE} mission and chosen for {\it Spitzer} follow-up in program 70062, where it was measured to have a ch1$-$ch2 color indicative of a late-T dwarf. Subsequent Palomar/WIRC $J$-band imaging indicated a magnitude of $\sim$19.0 mag, ruling out the possibility of its being a late-T dwarf. The object was rediscovered again by the Backyard Worlds team but was paired up with a $J$ = 19.33$\pm$0.17 mag UHS object -- the same object seen in the Palomar imaging -- and believed to be a more distant early-T dwarf based on its implied $J-$W2 color. This $J$-band source is, however, an interloper in the field and not the brown dwarf candidate itself. (The same background object also contaminates the proper motion measure from CatWISE2020.) The {\it Spitzer} photometry from 2012 is clean; this color, together with clear evidence of motion through the epochal coverage of {\it WISE} and {\it NEOWISE} images, indicates a [T9.5] dwarf just outside of the 20-pc census.

   \item CWISE 2058$-$5134: We are unable to provide a distance estimate to this object using any of our four preferred absolute magnitude relations. Our spectroscopic follow-up (Table~\ref{spectroscopic_followup}) shows that this is a T0 dwarf, which would indicate M$_{JMKO}$ = 14.5 mag using plots illustrated in the following section. This suggests a distance of $\sim$33.9 pc. We consider this object to fall outside of 20 pc.
   
\end{itemize}

\subsection{The Resulting Census and Final Checks}

Our final, full-sky census of L, T, and Y dwarfs within 20 pc of the Sun is presented in Table~\ref{20pc_sample}. This includes not only solivagant dwarfs within that distance but also all known L, T, and Y dwarf companions to earlier type stars within 20 pc. For objects confirmed or believed to be double or triple systems, each component that is an L, T, or Y dwarf is listed. The table lists each object's discovery name, discovery reference, and optical and near-infrared spectral types (with reference), if measured. The table also lists the absolute parallax from Table~\ref{table:monster_table} and the total proper motion, along with a reference for the astrometry. For cases in which either a spectral type or parallax is estimated, the estimated value is shown in brackets. (For the $T_{\rm eff}$ values listed in the penultimate column, the reader is referred to section~\ref{subsection:assigning_temps}.) The last column of the table is reserved for special notes. If a note of "[]" is listed, then that object's listed parallax should be ignored in favor of the spectrophotometric estimate shown in brackets. If a note of "yng" or "sd" is listed, that object is discussed further in section~\ref{section:characterization}.

\startlongtable

\onecolumngrid
\tablenotetext{a}{The astrometry listed is for the primary star in the system.}
\tablenotetext{b}{Values in brackets are estimates only.}
\tablenotetext{c}{This object's parallax has been converted from relative to absolute by adding 0.9$\pm$0.3 mas, per the discussion in section 8 of \cite{kirkpatrick2019}.}
\tablenotetext{d}{A "yng" entry indicates that the spectrum of this object suggests low gravity and youth. An "sd" entry indicates that the spectrum of this object suggests low metallicity and hence, old age. A value in brackets indicates that the value of the parallax in the $\varpi_{abs}$ column is uncertain and that our distance estimate from Table~\ref{table:distance_estimates} suggests the bracketed value be considered as the parallax instead.}
\tablenotetext{e}{Analysis in section~\ref{sec:suspected_multiples} shows that this object is probably a late-M dwarf. It has been dropped from subsequent analysis and not considered a member of the L, T, and Y dwarf census.}
\tablenotetext{f}{The {\it Gaia} DR2 identifications for these sources are given in Table~\ref{table:monster_table} and are: 
Gaia J041246.85$-$073416.8 = Gaia DR2 3195979005694112768,
Gaia J171340.47$-$395211.8 = Gaia DR2 5972124644679705728,
Gaia J183118.29$-$073227.6 = Gaia DR2 4159791176135290752,
Gaia J183610.72+031524.6   = Gaia DR2 4283084190940885888,
Gaia J195557.27+321518.2   = Gaia DR2 2034222547248988032.}
\tablecomments{References:
(0) This paper,
(1) \citealt{kirkpatrick2019},
(2) \citealt{kirkpatrick2012},
(3) \citealt{burgasser2004},
(4) \citealt{kirkpatrick2011},
(5) \citealt{albert2011},
(6) \citealt{tinney2018},
(7) \citealt{pinfield2014b},
(8) \citealt{mace2013},
(9) \citealt{schneider2015},
(10) \citealt{cushing2011},
(11) \citealt{mainzer2011},
(12) \citealt{kirkpatrick2013-0647},
(13) \citealt{pinfield2014a},
(14) \citealt{luhman2014-solar_comp},
(15) \citealt{kirkpatrick2014},
(16) \citealt{luhman2014-0855},
(17) \citealt{tinney2012},
(18) \citealt{cushing2014},
(19) \citealt{burningham2013},
(20) \citealt{thompson2013},
(21) \citealt{tinney2014},
(22) \citealt{martin2018},
(23) \citealt{lodieu2012},
(24) \citealt{looper2007},
(25) \citealt{mace2013b},
(26) \citealt{scholz2010b},
(27) \citealt{warren2007},
(28) \citealt{mugrauer2006},
(29) \citealt{tinney2005},
(30) \citealt{delorme2008},
(31) \citealt{burningham2010},
(32) \citealt{burgasser2002},
(33) \citealt{scholz2011},
(34) \citealt{burgasser2003b},
(35) \citealt{bihain2013},
(36) \citealt{nakajima1995},
(37) \citealt{lucas2010},
(38) \citealt{artigau2010},
(39) \citealt{lodieu2007},
(40) \citealt{leggett2012},
(41) \citealt{luhman2012},
(42) \citealt{burningham2008},
(43) \citealt{burgasser1999},
(44) \citealt{wright2013},
(45) \citealt{goldman2010},
(46) \citealt{burningham2011},
(47) \citealt{tsvetanov2000},
(48) \citealt{cardoso2015},
(49) \citealt{scholz2010a},
(50) \citealt{pinfield2012},
(51) \citealt{burgasser2000},
(52) \citealt{burgasser2003a},
(53) \citealt{chiu2006},
(54) \citealt{murray2011},
(55) \citealt{strauss1999},
(56) \citealt{gelino2011},
(57) \citealt{knapp2004},
(58) \citealt{biller2006},
(59) \citealt{burningham2009},
(60) \citealt{scholz2003},
(61) \citealt{burgasser2006},
(62) \citealt{burgasser2010},
(63) \citealt{kasper2007}.,
(64) \citealt{pinfield2008},
(65) \citealt{luhman2011},
(66) \citealt{faherty2012},
(67) \citealt{dupuy2012},
(68) \citealt{vrba2004},
(69) \citealt{subasavage2009},
(70) \citealt{marocco2010},
(71) \citealt{burgasser2008},
(72) \citealt{gaia2018},
(73) \citealt{cushing2016},
(74) \citealt{smart2013},
(75) \citealt{vanaltena1995},
(76) \citealt{tinney2003},
(77) \citealt{manjavacas2013},
(78) \citealt{henry2006},
(79) \citealt{harrington1980},
(80) \citealt{deacon2017}, 
(81) \citealt{martin1999}, 
(82) \citealt{day-jones2013},
(83) \citealt{smith2014},
(84) \citealt{loutrel2011},
(85) \citealt{scholz2014},
(86) \citealt{best2013},
(87) \citealt{kellogg2018},
(88) \citealt{deacon2012b},
(89) \citealt{aberasturi2011},
(90) \citealt{luhman2013},
(91) \citealt{sheppard2009},
(92) \citealt{reyle2014},
(93) \citealt{freed2003},
(94) \citealt{robert2016},
(95) \citealt{andrei2011},
(96) \citealt{leggett2000},
(97) \citealt{gauza2015},
(98) \citealt{hall2002},
(99) \citealt{deacon2011},
(100) \citealt{bowler2010},
(101) \citealt{goldman1999}, 
(102) \citealt{dupuy2012}, 
(103) \citealt{kirkpatrick2008}, 
(104) \citealt{marocco2013}, 
(105) \citealt{kirkpatrick2000}, 
(106) \citealt{bardalez2014}, 
(107) \citealt{reid2000}, 
(108) \citealt{knapp2004}, 
(109) \citealt{wilson2003}, 
(110) \citealt{cruz2009}, 
(111) \citealt{kendall2007}, 
(112) \citealt{reid2008}, 
(113) \citealt{liebert2003}, 
(114) \citealt{schneider2014}, 
(115) \citealt{cruz2003}, 
(116) \citealt{golimowski2004}, 
(117) \citealt{reid2006}, 
(118) \citealt{allers2013}, 
(119) \citealt{castro2013}, 
(120) \citealt{cruz2007}, 
(121) \citealt{fan2000}, 
(122) \citealt{geballe2002}, 
(123) \citealt{salim2003}, 
(124) \citealt{burgasser2010b}, 
(125) \citealt{thorstensen2003}, 
(126) \citealt{kirkpatrick2014}, 
(127) \citealt{reid2001}, 
(128) \citealt{bouy2004}, 
(129) \citealt{phan-bao2008}, 
(130) \citealt{scholz2002}, 
(131) \citealt{deacon2005}, 
(132) \citealt{burgasser2008b}, 
(133) \citealt{kirkpatrick2010}, 
(134) \citealt{schmidt2010}, 
(135) \citealt{gizis2002}, 
(136) \citealt{hawley2002}, 
(137) \citealt{kendall2004}, 
(138) \citealt{delfosse1997}, 
(139) \citealt{kirkpatrick1999}, 
(140) \citealt{bouy2003}, 
(141) \citealt{geissler2011}, 
(142) \citealt{gizis2000}, 
(143) \citealt{folkes2007}, 
(144) \citealt{ruiz1997}, 
(145) \citealt{liu2005}, 
(146) \citealt{koen2017}, 
(147) \citealt{gomes2013}, 
(148) \citealt{luhman2014-solar_comp}, 
(149) \citealt{kirkpatrick2016}, 
(150) \citealt{schmidt2007}, 
(151) \citealt{forveille2004}, 
(152) \citealt{potter2002}, 
(153) \citealt{goto2002}, 
(154) \citealt{burgasser2007c}, 
(155) \citealt{west2008}, 
(156) \citealt{burgasser2004}, 
(157) \citealt{beamin2013}, 
(158) \citealt{schneider2017}, 
(159) \citealt{metodieva2015}, 
(160) \citealt{folkes2012}, 
(161) \citealt{looper2008}, 
(162) \citealt{gizis2011}, 
(163) \citealt{liu2002}, 
(164) \citealt{menard2002}, 
(165) \citealt{gizis2003}, 
(166) \citealt{cushing2005}, 
(167) \citealt{reid2008b}, 
(168) Gaia Data Release 2: \citealt{gaia2016} and \citealt{gaia2018}, 
(169) \citealt{bartlett2017}, 
(170) \citealt{dahn2002}, 
(171) \citealt{dieterich2014}, 
(172) \citealt{dahn2017}, 
(173) Hipparcos: \citealt{vanleeuwen2007}, 
(174) \citealt{kirkpatrick2019}, 
(175) \citealt{faherty2016}, 
(176) \citealt{gagne2015b}, 
(177) \citealt{scholz2018}, 
(178) \citealt{cushing2018},
(179) \citealt{lazorenko2018},
(180) \citealt{dupuy2017},
(181) \citealt{best2015}, 
(182) \citealt{gizis2012}, 
(183) \citealt{lodieu2002}, 
(184) \citealt{luhman2014}, 
(185) \citealt{lodieu2005}, 
(186) \citealt{artigau2006},
(187) \citealt{artigau2011},
(188) \citealt{metchev2008},
(189) \citealt{burgasser2003c},
(190) \citealt{thalmann2009},
(191) \citealt{marocco2019},
(192) \citealt{luhman2007},
(193) \citealt{looper2008},
(194) \citealt{liebert2006},
(195) \citealt{ellis2005},
(196) \citealt{kellogg2015},
(197) \citealt{burgasser2003-opt_spec},
(198) \citealt{liu2016},
(199) \citealt{smart2018},
(200) NPARSEC unpublished (Smart, priv.\ comm.),
(201) \citealt{gizis2015},
(202) \citealt{bouy2005},
(203) \citealt{koerner1999},
(204) \citealt{casewell2008},
(205) \citealt{marocco2015},
(206) \citealt{dupuy2015},
(207) \citealt{manjavacas2019},
(208) \citealt{burgasser2005},
(209) \citealt{liu2010},
(210) \citealt{burgasser2013},
(211) \citealt{kniazev2013},
(212) \citealt{liu2012},
(213) \citealt{burgasser2011b},
(214) \citealt{pineda2016},
(215) \citealt{law2006},
(216) \citealt{gizis2015b},
(217) \citealt{deacon2017b},
(218) \citealt{nilsson2017},
(219) \citealt{faherty2018b},
(220) \citealt{reyle2018},
(221) \citealt{marocco2019},
(222) \citealt{king2010},
(223) \citealt{reid2006b},
(224) \citealt{torres2019},
(225) \citealt{aberasturi2014},
(226) \citealt{ireland2008},
(227) \citealt{pravdo2005},
(228) \citealt{mamajek2018},
(229) \citealt{scholz2020},
(230) \citealt{deacon2012},
(231) \citealt{marocco2020},
(232) \citealt{meisner2020a},
(233) \citealt{meisner2020b},
(234) \citealt{bardalez2020},
(235) \citealt{volk2003},
(236) \citealt{greco2019},
(237) \citealt{best2020},
(238) \citealt{burgasser2015},
(239) \citealt{burgasser2015b},
(240) \citealt{dupuy2019}.
}
\end{deluxetable*}

Having now compiled the census, it is instructive to look back to previous attempts at assembling lists of nearby L, T, and Y dwarfs. These comparisons show how quickly our knowledge of this sample has evolved in just over fifteen years.

\cite{kendall2004} published a list of the sixteen nearest L dwarfs, out to $\sim$11 pc. Of those, fourteen appear in Table~\ref{20pc_sample}, the two exceptions being objects now considered to be late-M dwarfs: SDSS J143808.31+640836.3, which \cite{cruz2003} classify as M9.5 in the optical, and 2MASSW J2306292$-$050227\footnote{\citealt{kendall2004} mistakenly list this as 2MASSW J2306292+154905.} (aka TRAPPIST 1), which \cite{gizis2000} type as an optical M7.5. 

%\cite{reid2008} published a list of late-M through T0 dwarfs believed to lie within 20 pc. Of the twenty-seven systems listed with optical types between L0 and T0, twenty-three appear in Table~\ref{20pc_sample}. The exceptions are 2MASS J02284243+1639329, 2MASS J09111297+7401081, 2MASS J19360187$-$5502322, and 2MASS J20360316+1051295. These objects are now known to fall just outside of the 20-pc volume, as the {\it Gaia} DR2 parallaxes for all four fall between 40 and 50 mas. 

\cite{looper2008} published a list of L dwarfs believed to fall within 10 pc. All ten of those objects appear in Table~\ref{20pc_sample}. 

\cite{reid2008} published a list of ninety-four L dwarf systems believed to lie within 20 pc. Eighty-four of these appear in Table~\ref{20pc_sample}. The exceptions are eight systems -- 2MASS J01550354+0950003, 2MASS J02284243+1639329, DENIS J061549.3$-$010041, SDSS J080531.84+481233.0, DENIS J082303.1$-$491201, 2MASS J09111297+7401081, 2MASS J19360187$-$5502322, and 2MASS J20360316+1051295 -- that are now known to fall just outside of the 20-pc volume according to {\it Gaia} DR2, and two objects -- DENIS J065219.7$-$253450 (M9.2 near-infrared; \citealt{bardalez2014}) and 2MASSW J1421314+182740 (M8.9 near-infrared; \citealt{bardalez2014}) -- that we consider to be late-M dwarfs.

\cite{kirkpatrick2012} published the full stellar census within 8 pc, using a combination of preliminary trigonometric parallaxes and spectrophotometric distance estimates for the L, T, and Y dwarfs. All thirty-three of those L, T, and Y dwarfs appear in Table~\ref{20pc_sample}.

\cite{kirkpatrick2019} gave a listing of 235 L0-L5.5 and T6-Y1+ dwarfs within 20 pc but missed a few objects, discovered prior to their paper, that this new census now includes. In the L0-L5.5 range, a handful of component objects in systems known to be binaries or triples were overlooked -- DENIS 0205$-$1159A ([L5]; \citealt{bouy2005}),  2MASS 1315$-$2649A (L5.5/L5; \citealt{burgasser2011b}), LSPM 1735+2634B (L0:; \citealt{law2006}), Gl 802B ([L5-L7]; \citealt{ireland2008}), and DENIS 2252$-$1730 ([L4:]; \citealt{reid2006b}). Several previously published objects near the L0 or L5.5 edges are now considered to fall within the L0-L5.5 range based on published spectral types -- 2MASS 0413+3709 ([L1]; \citealt{kirkpatrick2010}), 2MASS 0421$-$6306 (L5$\beta$; \citealt{cruz2007}), 2MASS 0835$-$0819 (L5; \citealt{cruz2003}), 2MASS 0908+5032 (L5/L6; \citealt{cruz2003}), and 2MASS 1010$-$0406 (L6/L5; \citealt{cruz2003}). Two objects in the middle of the L0-L5.5 range were also overlooked: WISE J0508+3319 (L2; \citealt{kirkpatrick2016}) and DENIS J1013$-$7842 (L3; \citealt{aberasturi2014}). Finally, one object (2MASS J21580457$-$1550098; L4:/L5; \citealt{gizis2003}) has now been dropped from the \cite{kirkpatrick2019} list because a definitive parallax from \cite{smart2018} shows that it likely lies beyond 20 pc. In the T6-Y1+ range, two objects near T6 were overlooked -- UGPS 0355+4743 (T6:; \citealt{smith2014}) and 2MASS 2154+5942 (T6; \citealt{looper2007}) -- along with two later T dwarfs  -- 2MASS 1315$-$2649B (T7; \citealt{burgasser2011b}) and Gl 758B (T7; \citealt{thalmann2009}).

\cite{bardalez2019} published a list of 472 dwarfs of type M7 through L5 within 25 pc, of which 283 fall within 1$\sigma$ of 20 pc. Three of the L dwarfs do not appear in our Table~\ref{20pc_sample} because we consider them to have late-M spectral types: DENIS J065219.7$-$253450 (see above), 2MASS J14213145+1827407 (see above), and 2MASSI J1438082+640836 (M9.5 optical; \citealt{cruz2003}). Several other L dwarfs are now known (or are likely, within the uncertainties) to be outside of the 20-pc volume based on accurate parallaxes: DENIS J1228.2$-$1547AB (\citealt{dupuy2017}), SDSS J133148.92$-$011651.4 (\citealt{smart2018}), SDSS J144600.59+002451.9 (\citealt{faherty2012}), SDSS J153453.33+121949.2 (\citealt{gaia2018}), 2MASS J21580457-1550098 (\citealt{smart2018}), and 2MASS J23512200+3010540 (\citealt{liu2016}). Two other L dwarfs, 2MASS J04474307-1936045 ($\sim$26 pc; \citealt{faherty2012}), and SDSS J092308.70+234013.6 ($\sim$21pc; \citealt{schmidt2010}), have published spectrophotometric distance estimates placing them outside of 20 pc, so they are not included in our table.

Finally, there are two objects noted in \cite{best2020} as falling within 20 pc that are nonetheless excluded from Table~\ref{20pc_sample}. 2MASS J05160945$-$0445499 has a parallax listed by \cite{best2020} as 54.2$\pm$4.3 mas, but a more accurate parallax of 47.83$\pm$2.85 mas from NPARSEC (Smart, priv.\ comm.), places this object just outside of 20 pc. WISEA J055007.94+161051.9 has a \cite{best2020} parallax of 53.9$\pm$2.8 mas, but a more accurate {\it Gaia} DR2 parallax of 49.1169$\pm$0.8467 places it beyond 20 pc. 

%As this paper was being written, \cite{best2020} published a list of 348 parallaxes of L0 through T8 dwarfs thought to lie within 25 pc and falling between declinations of $-$30$^\circ$ and +60$^\circ$. Two objects from that list purported to lie within 20 pc are not originally given in our Table~\ref{20pc_sample}: 2MASSW J1326201$-$272937, has a \cite{best2020} parallax of 54.7$\pm$5.9 mas but our value of 39.80$\pm$8.66 mas from \cite{smart2018} places this object outside 20 pc. Because the \cite{best2020} formal uncertainty is smaller, we now include this object in our table.

The above checks are illustrative of the fact that our knowledge of the nearby census is constantly changing. New objects are still being discovered. Some objects already known within the census are found to be binary (or triple), and some higher mass stars within 20 pc are found to have L, T, or Y companions. Some objects originally thought to lie within the volume are found, once better astrometry is available, to fall outside. And objects are sometimes discovered then forgotten simply because there does not exist a living, publicly available database that adequately captures this information. Nonetheless, our knowledge -- and our completeness -- of this census is improving with time, thereby enabling a more robust look into the low-mass products of star formation.

\section{Characterizing the 20-pc Census\label{section:characterization}}

With the census of L, T, and Y dwarfs within 20 pc now compiled, we can begin to study the field mass function. As described in section~\ref{section:temps_densities} below, we must compute space densities binned by effective temperature so that we can compare the empirical data to mass function simulations. This requires us to calculate an effective temperature for each individual object. Most objects can be assigned temperatures using relations typical of old, solar-metallicity field objects, but some objects within the census are young or low metallicity. To handle these properly, we first need to identify which objects they are. Moreover, because we want to assign temperatures to individual objects, this means recognizing when objects are unresolved multiple systems, to the extent that our existing data can help to address that. In the next subsections we delve into this characterization of the census, as a prelude to determining the space densities we need.

\subsection{Low-gravity (Young) Objects\label{section:known_youngs}}

Brown dwarfs with ages less than $\sim$100 Myr have not yet fully contracted to their final, equilibrium radius (\citealt{kirkpatrick2008}) and are identifiable through spectroscopic and photometric signatures that indicate a lower gravity than normal, old brown dwarfs that {\it have} fully contracted. These young brown dwarfs represent a challenge to determining the mass function via our methodology because the standard mapping of spectral type, absolute magnitude, or color into effective temperature does not apply to them (\citealt{faherty2016}). Young objects that fall within the 20-pc census need to be identified so that they can be placed into the correct bins of T$_{eff}$.

On the other hand, these same objects also represent an opportunity to probe the low-mass cutoff. Objects below a few Jupiter masses are generally very difficult to find if formed billions of years ago because of the intrinsic faintness resulting from their long cooling times. However, objects of similar mass can be much more easily detected when they are younger because they will be much warmer and brighter. An isolated brown dwarf that shows signs of low gravity, if it can be associated kinematically to a moving group or young association of known age, can be placed on theoretical isochrones to produce a mass estimate. Although it was once believed that a large reservoir of rogue planets -- objects that escaped their original protoplanetary disks -- existed in the Milky Way (\citealt{sumi2011}), microlensing results with more robust statistics have shown that the population of field objects having masses down to at least a few Jupiter masses appears to be drawn from the same population as higher-mass brown dwarfs and stars (\citealt{mroz2017}). Thus, such young brown dwarfs can serve as independent probes of the low-mass cutoff value of star formation itself.

Spectroscopic signatures of youth have been noted in late-M, L, and even some T dwarfs (e.g., \citealt{cruz2009, allers2013, gagne2015}), and classification systems have been developed to incorporate these. The most commonly used system (\citealt{kirkpatrick2005}) assigns a suffix of $\beta$, $\gamma$, or $\delta$ to the core type to indicate the degree to which low-gravity signatures are evident, with the infrequently used $\alpha$ suffix assigned to spectra with gravities typical of old field objects. \cite{faherty2016} note that a fraction of objects assigned $\beta$ designations seem not to belong to any known, young moving groups, and some young associations of presumably fixed age can contain objects with both $\beta$ and $\gamma$ designations. \cite{sengupta2010} point out that the rotation rates of some brown dwarfs can make them oblate, but non-sphericity in an old object seen equator-on is unlikely to produce the radius inflation needed to turn an $\alpha$ classification into a $\beta$ classification, for example. The differences between the two classifications is thought to be around 0.5 dex in log($g$) (see Figure 9 of \citealt{burrows1997}), so a simple calculation shows that a radius increase of 10$\times$ would be needed to achieve the effect.  \cite{gonzales2019} has further noted that the late-M dwarf TRAPPIST-1, though presumably of field age, nonetheless has near-infrared spectral indices indicating an intermediate gravity. If this star's radius is truly inflated, it could be due to magnetic activity or to tidal interactions by the numerous planets in its solar system. (It has also been shown that low-gravity indices can sometimes be incorrectly assigned in the near-infrared for subdwarfs [\citealt{aganze2016}], although a more careful analysis of the overall spectral energy distribution can eliminate this problem.) For the remainder of our analysis, we will regard $\beta$ designations to be true indicators of low gravity even if youth cannot confidently be assigned through moving group membership.

Several L, T, and Y dwarfs in the 20-pc census (Table~\ref{20pc_sample}) are known to have low-gravity features. Given that our {\it Spitzer} monitoring has improved the astrometry for many of these targets, we can now run analyses to determine if there are any objects found to be high-probability members of any known moving groups but lacking spectra or having spectra where gravity diagnostics are less clear. For this exercise, we consider only those objects in the 20-pc census having measured trigonometric parallaxes, and we use two separate tools that can assess membership probabilities based on the subset of kinematic data we have -- positions, distances, and motions, but not radial velocities. The first tool is Banyan~$\Sigma$ (\citealt{gagne2018}), which uses Bayesian inference to compute the membership probabilities for twenty-nine different associations within 150 pc of the Sun. For our set of input parameters (RA, Dec, $\varpi_{abs}$, $\mu_\alpha$, $\mu_\delta$), Banyan~$\Sigma$ uses Bayes' theorem to marginalize over radial velocity, and the Bayesian priors are set so that a probability threshold of 90\% will recover 82\% of true members. The second tool is LACEwING (\citealt{riedel2017}), which determines the membership probabilities in 16 different young associations within 100 pc of the Sun. Unlike Banyan~$\Sigma$, the LACEwING code takes a frequentist approach and works directly in observable space (proper motion, sky position, etc.) rather than in $XYZ$ and $UVW$ for its probability computations.

Table~\ref{table:young_objects} shows the results of our Banyan~$\Sigma$ and LACEwING runs. The table retains only those objects that have $\beta$ or $\gamma$ spectral classifications ("Sp.Type Opt" or "Sp.Type NIR", copied from Table~\ref{20pc_sample}) in the literature, have a Banyan~$\Sigma$ probability of $\ge$90\% for young association membership, or have a non-zero LACEwING probability for membership. Other columns list the possible associations assigned by Banyan~$\Sigma$ and LACEwING. When there are multiple moving groups that match, the relative probabilities are listed for those groups having at least a 5\% probability. The final columns list whether or not the spectrum shows low-gravity features ("Low-g?"), whether the results suggest possible membership in a moving group ("Assoc.\ Memb.?"), the published reference first noting the object's possible youth ("Youth Ref"), and the mass estimate and its published reference ("Mass" and "Mass Ref.") for any objects with established membership. 

\movetabledown=25mm
\begin{longrotatetable}
\begin{deluxetable*}{cccccccccccc}
\tabletypesize{\footnotesize}
\tablecaption{Potentially Young L, T, and Y Dwarfs within 20 pc of the Sun\label{table:young_objects}}
\tablehead{
\colhead{Object} &                          
\colhead{Sp.Type} &
\colhead{Sp.Type} &
\colhead{Banyan~$\Sigma$} &
\colhead{Banyan~$\Sigma$} &
\colhead{LACEwING} &
\colhead{LACEwING} &
\colhead{Low-g?} &                          
\colhead{Assoc.} &
\colhead{Youth} &
\colhead{Mass} &
\colhead{Mass} \\
\colhead{} &                          
\colhead{Opt.} &
\colhead{NIR} &
\colhead{Prob.} &
\colhead{Assoc.\tablenotemark{a}} &
\colhead{Prob.} &
\colhead{Assoc.\tablenotemark{a}} &
\colhead{} &                          
\colhead{Memb.?} &
\colhead{Ref.} &
\colhead{(M$_{Jup}$)} &
\colhead{Ref.} \\
\colhead{(1)} &                          
\colhead{(2)} &  
\colhead{(3)} &  
\colhead{(4)} &
\colhead{(5)} &
\colhead{(6)} &
\colhead{(7)} &
\colhead{(8)} &
\colhead{(9)} &                          
\colhead{(10)} &
\colhead{(11)} &
\colhead{(12)} \\
}
\startdata
   WISE  0031+5749& \nodata      & L8          & 97.62&                       CarN&  0&  field& no& no& \nodata& \nodata& \nodata\\
   2MASS 0034+0523& \nodata      & T6.5        &  1.97&                      field& 34&    Arg(81),$\beta$Pic(19)& no& no& \nodata& \nodata& \nodata\\
   2MASS 0045+1634& L2$\beta$    & L2$\gamma$  & 99.61&                        Arg& 20&    Arg(77),$\beta$Pic(23)&   yes  &  yes & R& 24.98$\pm$4.62& Q\\
   WISE  0047+6803& L7($\gamma$?)& L6-8$\gamma$& 99.67&                        ABD& 25&    ABD(69),Arg(31)&   yes  &  yes & T& 11.84$\pm$2.63& Q\\
   SIMP  0136+0933& T2           & T2          & 97.46&                       CarN&  0&  field& yes& yes& B& 12.7$\pm$1.0& B\\                  
   2MASS 0144-0716& L5           & L4.5        &  0.01&                      field& 22&    Arg& no& no& \nodata& \nodata& \nodata\\
   WISE  0206+2640& \nodata      & L9 pec (red)& 33.14&                      field& 33&   Hyad& no& no?& \nodata& \nodata& \nodata\\
   CWISE 0238-1332& \nodata      & [$\ge$Y1]   & 95.55&                        Arg& 51&    ABD(72),$\beta$Pic(28)& \nodata& maybe?& \nodata& \nodata& \nodata\\\
   WISE  0241-3653& \nodata      & T7          & 95.55&                        Arg&  0&  field& no& no?& \nodata& \nodata& \nodata\\
   WISE  0316+4307& \nodata      & T8          & 95.26&                       CarN&  0&  field& no& no?& \nodata& \nodata& \nodata\\
   2MASS 0318-3421& L7           & L6.5        &  0.01&                      field& 32&    ABD(76),Arg(24)& no& no& \nodata& \nodata& \nodata\\
   WISE  0323+5625& \nodata      & L7          &  3.85&                      field& 26&   Hyad& no& no& \nodata& \nodata& \nodata\\
%   LP 944-20(0339-3525)& M9$\beta$& L0$\beta$  & \\                                                                
   2MASS 0355+1133& L5$\gamma$   & L3-6$\gamma$& 99.64&                        ABD& 48&    ABD&   yes  &  yes & R& 21.62$\pm$6.14& Q\\
   UGPS  0355+4743& \nodata      & [T6]        &  0.00&                      field& 26&   Hyad& \nodata& no?& \nodata& \nodata& \nodata\\
   2MASS 0407+1514& \nodata      & T5.5        & 97.38&           Arg(51),CarN(49)& 25& $\beta$Pic& no& maybe?& \nodata& \nodata& \nodata\\
   2MASS 0421-6306& L5$\beta$    & L5$\gamma$  & 99.74&           Arg(82),CarN(18)& 24&    Arg(71),CarN(29)&   yes  &  yes & C& \nodata& \nodata\\
   CWISE 0424+0002& \nodata      & T9:         & 14.93&                      field& 32&    ABD(64),$\beta$Pic(20),Col(16)& no?& no& \nodata& \nodata& \nodata\\
   WISE  0513+0608& \nodata      & T6.5        & 47.78&                      field& 23&    ABD& no& no& \nodata& \nodata& \nodata\\
   2MASS 0523-1403& L2.5         & L5          & 94.97&                        Arg&  0&  field& no& no& \nodata& \nodata& \nodata\\
AB Dor C[b](0528-6526)& \nodata    & \nodata     &\nodata&                   \nodata& \nodata& \nodata& \nodata& yes\tablenotemark{b}& U& 14$\pm$1& U\\
   CWISE 0536-3055& \nodata      & [T9.5]      & 98.95&                 $\beta$Pic&  0&  field& \nodata& no?& \nodata& \nodata& \nodata\\                
   2MASS 0559-1404& T5           & T4.5        &  0.00&                      field& 22&    ABD& no& no& \nodata& \nodata& \nodata\\
   LSR   0602+3910& L1           & L1$\beta$   &  0.03&                      field&  0&  field& yes& no& E& \nodata& \nodata\\
%28$\pm$16& Q\\ for LSR 0602+3910 mass estimate
   2MASS 0624-4521& L5           & L5          & 95.12&                        Arg&  0&  field& no& no& \nodata& \nodata& \nodata\\
   WISE  0627-1114& \nodata      & T6          & 98.91&                        ABD& 29&    ABD& no& maybe?& \nodata& \nodata& \nodata\\
   WISE  0642+4101& \nodata      & extr.\ red  & 95.04&             ABD(90),Col(9)& 22&    ABD&  maybe?&  yes?& G& \nodata& \nodata\\
WISE  0700+3157ABC& L3+L6.5+L6.5::& L3:+L6.5:+?&  0.01&                      field& 22&    ABD& no& no& \nodata& \nodata& \nodata\\
   WISE  0701+6321& \nodata      & T3          & 94.77&            Col(83),Arg(15)&  0&  field& no& no& \nodata& \nodata& \nodata\\
   SDSS  0758+3247& T3           & T2.5        & 99.37& Arg(70),CarN(18),$\beta$Pic(12)& 0& field&   no&   no?& \nodata& \nodata& \nodata\\
   WISE  0759-4904& \nodata      & T8          &  0.00&                      field& 29&    Arg& no& no& \nodata& \nodata& \nodata\\
   DENIS 0817-6155& \nodata      & T6          &  0.00&                      field& 23&    Arg(74),ABD(26)& no& no& \nodata& \nodata& \nodata\\
   WISE  0820-6622& \nodata      & L9.5        & 99.80&                       CarN& 35&   CarN&   no&  maybe?& \nodata& \nodata& \nodata\\
   2MASS 0859-1949& L7:          & L8          &  0.04&                      field& 21&    Arg& no& no& \nodata& \nodata& \nodata\\
   2MASS 0908+5032& L5           & L6          & 79.86&                       CarN& 23&    ABD& no& no?& \nodata& \nodata& \nodata\\
   2MASS 1010-0406& L6           & L5          & 99.65&                       CarN&  0&  field&   no&  no& \nodata& \nodata& \nodata\\
   2MASS 1022+5825& L1$\beta$    & L1$\beta$   &  0.00&                      field&  0&  field&   yes&  no& R& \nodata& \nodata\\
%28$\pm$16& Q\\ mass estimate for 1022+5825
 WISE  1049-5319AB& L8:+T1.5::   & L7.5+T0.5:  & 94.85&                        Arg& 33&    Arg(66),ABD(34)& no&  no& \nodata& \nodata& \nodata\\                   
   DENIS 1058-1548& L3           & L3          & 96.16&                        Arg&  0&  field& no&  no& \nodata& \nodata& \nodata\\
   2MASS 1108+6830& L1$\gamma$   & L1$\gamma$  & 97.97&                        ABD&  0&  field&   yes&  yes& J& \nodata& \nodata\\
   SDSS  1110+0116& \nodata      & T5.5        & 99.25&                        ABD&  0&  field&   yes&  yes& P& 10-12& J\\
LHS 2397aB(1121-1313)& \nodata   & [L7.5]      & 95.43&                       CarN&  0&  field& no&  no& \nodata& \nodata& \nodata\\
   2MASS 1213-0432& L5           & L4.2        & 99.17&           CarN(69),Arg(31)&  0&  field& no&  no& \nodata& \nodata& \nodata\\
   SDSS  1219+3128& \nodata      & L9.5        & 94.18&                        Arg&  0&  field& no&  no& \nodata& \nodata& \nodata\\
%   VHS  1256-1257B& \nodata      & L7: VL-G    & 64.81&      $\beta$Pic(93),Col(7)&  0&  field& yes& no& Z& 19$\pm$5& W\\
 Gl 494C(1300-1221)& \nodata     & T8          & 99.28&                       CarN&  0&  field& no& no& \nodata& \nodata& \nodata\\
   ULAS  1302+1308& \nodata      & T8          & 98.93&                       CarN&  0&  field& no& no& \nodata& \nodata& \nodata\\ 
Kelu-1AB(1305-2541)& L3+L3       & L2:+L4:     & 99.32&                        Arg&  0&  field& no& no& \nodata& \nodata& \nodata\\
   2MASS 1324+6358& \nodata      & T2: pec     & 98.60&             ABD(92),Col(8)&  0&  field& yes?& yes& D& 11-12& D\\
   2MASS 1326-2729& L5           & L6.6:       & 96.18&                       CarN&  0&  field& no& no?& \nodata& \nodata& \nodata\\
   DENIS 1425-3650& L3           & L4$\gamma$  & 99.49&                        ABD& 26&    ABD&   yes  &  yes & J& 22.52$\pm$6.07& Q\\
   WISE  1612-3420& \nodata      & T6.5        & 64.61&                        ABD& 20&    ABD& no?& yes?& \nodata& \nodata& \nodata\\
   SDSS  1624+0029& T6           & T6          & 98.98&                       CarN&  0&  field& no& no& \nodata& \nodata& \nodata\\     
   WISE  1741-4642& \nodata      & L6-8$\gamma$& 99.01&                        ABD& 25&    ABD& yes& yes& S& \nodata& \nodata\\
   WISE  1753-5904& \nodata      & [T8.5]      &  0.11&                      field& 38& Arg(62),ABD(16),$\beta$Pic(13),CarN(8)& \nodata& no& \nodata& \nodata& \nodata\\
   2MASS 1753-6559& L4           & \nodata     & 99.54&     Arg(88),CarN(7),ABD(6)& 39&    Arg(66),ABD(34)& no&  no?& \nodata& \nodata& \nodata\\
   WISE  1818-4701& \nodata      & [T8.5]      & 95.99&                       CarN&  0&  field& \nodata& no& \nodata& \nodata& \nodata\\
 Gl 758B(1923+3313)& \nodata     & T7:         & 99.05&                        Arg&  0&  field& no?&  no&  \nodata& \nodata& \nodata\\
   WISE  1926-3429& \nodata      & [T5.5]      & 99.41&      $\beta$Pic(91),Arg(8)& 25& $\beta$Pic(64),Arg(36)& \nodata&  maybe?& \nodata& \nodata& \nodata\\
   2MASS 2002-0521& L5$\beta$    & L5-7$\gamma$&  0.00&                      field&  0&  field&   yes&  no& J& \nodata& \nodata\\
   DENIS 2057-0252& L1.5         & L2$\beta$   &  0.00&                      field&  0&  field&   yes&  no& Q& \nodata& \nodata\\
% 70$\pm$13& Q\\ mass estimate for 2057-0252
   WISE  2121-6239& \nodata      & T2          & 68.54&                        Arg& 37&    ABD& no& no?& \nodata& \nodata& \nodata\\
   WISE  2236+5105& \nodata      & T5.5        & 98.03&                       CarN&  0&  field& no& no?& \nodata& \nodata& \nodata\\                
   2MASS 2244+2043& L6.5 pec     & L6-8$\gamma$& 99.71&                        ABD&  0&  field& yes& yes& L& 10.46$\pm$1.49& Q\\
   WISE  2255-3118& \nodata      & T8          & 99.12&                 $\beta$Pic& 31& $\beta$Pic(86),Arg(14)&  no?&  maybe?& \nodata& \nodata& \nodata\\
   WISE  2313-8037& \nodata      & T8          &  0.00&                      field& 40&    ABD(75),$\beta$Pic(25)& no?& no& \nodata& \nodata& \nodata\\
   2MASS 2317-4838& L4 pec       & L6.5 pec (red)&0.00&                      field& 22& $\beta$Pic& yes?& no?& \nodata& \nodata& \nodata\\
   ULAS  2321+1354& \nodata      & T7.5        &  0.00&                      field& 23& $\beta$Pic& no?& no&  \nodata& \nodata& \nodata\\
   2MASS 2322-3133& L0$\beta$    & L2$\beta$   &  0.00&                      field&  0&  field&   yes&  no& F& \nodata& \nodata\\
   WISE  2332-4325& \nodata      & T9:         & 99.68&                        ABD& 56&    ABD&   no?&  maybe?& \nodata& \nodata& \nodata\\
   WISE  2343-7418& \nodata      & T6          &  0.00&                      field& 43&    ABD(80),Arg(20)& no?& no?& \nodata& \nodata& \nodata\\
   WISE  2357+1227& \nodata      & T6          &  0.00&                      field& 47&    ABD(66),$\beta$Pic(34)& no?& no?& \nodata& \nodata& \nodata\\
\enddata
\tablecomments{Reference code for Youth Ref.: 
%A = \citealt{allers2013},
B = \citealt{gagne2017},
C = \citealt{cruz2009},
D = \citealt{gagne2018b},
E = \citealt{gagne2015b},
F = \citealt{faherty2012},
G = \citealt{gagne2014},
%H = \citealt{thalmann2009},
J = \citealt{gagne2015},
%K = this paper,
L = \citealt{looper2008},
%M = \citealt{goldman2010},
P = \citealt{knapp2004},
Q = \citealt{faherty2016},
R = \citealt{reid2008},
S = \citealt{schneider2014},
T = \citealt{thompson2013},
U = \citealt{climent2019}.}
%W = \citealt{dupuy2020},
%Z = \citealt{gauza2015}.}
\tablenotetext{a}{Code for moving groups and young associations: 
ABD = AB Doradus Moving Group (age 120$\pm$10 Myr; \citealt{barenfeld2013}), 
Arg = Argus Association (age 45$\pm$5 Myr; \citealt{zuckerman2019}), 
$\beta$Pic = $\beta$ Pictoris Moving Group (age 
26$\pm$3 Myr \citealt{malo2014};
24$\pm$3 Myr \citealt{bell2015};
22$\pm$6 Myr \citealt{shkolnik2017};
$18.5_{-2.4}^{+2.0}$ Myr \citealt{miret-roig2020}; 
17.8$\pm$1.2 Myr \citealt{crundall2019}), 
CarN = Carina-Near Moving Group (age 200$\pm$50 Myr; \citealt{zuckerman2006}), 
Col = Columba Association (age 42$^{+6}_{-4}$ Myr; \citealt{bell2015}), 
Hyad = Hyades (age 625$\pm$50 Myr; \citealt{lodieu2020} and references therein).}
\tablenotetext{b}{By definition, this member of the AB Doradus multiple star system is a member of the AB Doradus Moving Group. Because this companion to the C component of the system has not been independently confirmed, it is not included in subsequent analysis.}
\end{deluxetable*}
\end{longrotatetable}

Objects in Table~\ref{table:young_objects} that have "yes" under the "Low-g?" column are ones for which a low-gravity classification exists. For these we assign their T$_{eff}$ values using each object's measured near-infrared spectral type and the relation from spectral type to effective temperature applicable to young objects, as given in Table 19 of \cite{faherty2016}. For all other objects in the table, we assume that relations applicable to objects of normal gravity apply.

A number of objects in this table have full space motions available and have been confidently assigned membership in a young moving group. This has allowed researchers to identify several members of the 20-pc census that have masses below 25 M$_{Jup}$. Presently, there are no young moving group members within 20 pc that push below 10 M$_{Jup}$, although other members of lower mass have been identified at larger distances from the Sun. Three such examples are (1) PSO J318.5338-22.8603, a late-L dwarf member of the $\beta$ Pic Moving Group, which has a mass of 6.5$^{+1.3}_{-1.0}$ M$_{Jup}$ (\citealt{liu2013}), (2) 2MASSW J1207334$-$393254b, a late-L dwarf member of the TW Hya Association, which has a mass of 5$\pm$2 M$_{Jup}$ (\citealt{chauvin2004}), and (3) 2MASS J11193254$-$1137466AB, another late-L dwarf member of the TW Hya Association (\citealt{kellogg2016}), which \cite{best2017} show is an equal-mass system comprised of two $3.7_{-0.9}^{+1.2}$ M$_{Jup}$ brown dwarfs.

With the possible exception of 2MASS 1119$-$1137AB, none of these push below the 5 M$_{Jup}$ value established as the upper bound of the low-mass cutoff by \cite{kirkpatrick2019}, but there are several intriguing objects in Table~\ref{table:young_objects} that could.  These objects are labeled with "maybe?" under "Assoc.\ Memb.?" in the table and include CWISE 0238$-$1332, 2MASS 0407+1514, WISE 0627$-$1114, WISE 0820$-$6622, WISE 1926$-$3429, WISE 2255$-$3118, and WISE 2332$-$4325. Specifically, if the [$\ge$Y1] dwarf CWISE 0238$-$1332 were confirmed as an AB Dor, Argus, or $\beta$ Pic member, it would have a mass of below $\sim$4 M$_{Jup}$. Similarly, if the T8 dwarf WISE 2255$-$3118 were confirmed as a $\beta$ Pic member, it would have a mass of $\sim$2 M$_{Jup}$. (We find that the latter object, however, has a spectrum from \citealt{kirkpatrick2011} that is not noted for any peculiarities.) For these potentially young objects, obtaining radial velocities to determine robust membership may be quite difficult, but establishing new ultra-low-mass objects in the 20-pc census would provide extremely valuable knowledge.

Finally, we note that the \cite{faherty2016} young relations show that young M9 and M9.5 dwarfs fall into the same 2100-2250 K bin as early-L dwarfs of normal gravity. This means that such objects need to be included in our present census so that this temperature bin is complete. The only known low-gravity dwarf in \cite{faherty2016} that matches this criterion and falls within 20 pc is LP 944-20, but that object is believed to be somewhat older (475-650 Myr; \citealt{tinney1998}) than the low-gravity dwarfs needing special T$_{eff}$ estimates and therefore is not considered further here.

\subsection{Low-metallicity (Old Subdwarf) Objects\label{section:known_sds}}

There is a sizable number of objects in the 20-pc L, T, Y dwarf census of Table~\ref{20pc_sample} that have subdwarf spectral types or peculiar spectra whose features are attributed to low metallicity. See \cite{zhang2017, zhang2018, zhang2019} for comprehensive lists of known sdL and sdT dwarfs. Because subdwarfs are generally older objects, it is no surprise that our volume-limited census has few subwarfs of type sdL (two) but many of type sdT (thirteen): unless the object is very near the stellar/substellar mass boundary, it will have cooled to later types given its long lifetime. These low-metallicity objects are listed below:

\begin{itemize}

    \item WISE 0448-1935: This T5 pec dwarf was discovered by \cite{kirkpatrick2011}, who noted an excess of flux at $Y$-band and a flux deficit at $K$-band relative to the T5 spectral standard. They note that such features are common to other known or suspected low-metallicity T dwarfs.

%Note: WISE 0614+3912 is listed in Zhang et al. 2019 as low-metallicity and they quote my 2011 paper, but I label this is a normal T dwarf there and include no additional commentary on it. So I'm dropping this object from the list here.
    
    \item 2MASS 0645$-$6646: This object had the highest proper motion of all new discoveries listed in the 2MASS motion survey of \cite{kirkpatrick2010}, who classified it as an sdL8. It is one of only two L-type subdwarfs within the 20-pc census. Likely due to its very southerly declination, it has received far less follow-up than many of the more distant L-type subdwarfs known. 
    
    \item 2MASS 0729-3954: This T8 pec dwarf was discovered by \cite{looper2007}, who noted excess $Y$-band flux and depressed $H$- and $K$-band fluxes relative to the T8 standard. They noted that such features are seen in other T dwarfs suspected of low metallicity and/or high gravity.
    
    \item WISE 0833+0052: This object was discovered by \cite{pinfield2014a}, who classified it as a T9 with a suppressed $K$-band flux. They note that the blue $Y-J$ color was not evident in the confirmation spectrum, but would otherwise point at a $Y$-band excess like that seen in other T dwarfs suspected of having a low metallicity.
    
%Note: ULAS 0901-0306: is listed in Zhang et al. 2019 as low-metallicity and they quote Lodieu et al. (2007). However, that paper only notes a suppressed K-band and high gravity but not specifically low metallicity. A higher Y-band flux isn't noted, and the evidence for K-band flattening is a bit weak given the S/N of the spectrum. I'm dropping this object from the list of low-Z objects here.

    \item 2MASS 0937+2931: This T6 pec dwarf was discovered by \cite{burgasser2002}, who noted the highly suppressed $K$-band peak in its spectrum. Those authors argued that for a fixed effective temperature and composition, an older and more massive T dwarf would necessarily have a higher photospheric pressure than a younger object of lower mass, which would increase the relative importance of the collision induced absorption (CIA) by H$_2$. Another possible hypothesis for the deficit of flux at $K$-band, they argued, is decreased metallicity, which also increases the relative importance of CIA H$_2$. Of course, a combination of both effects -- both a lower metallicity and an extreme age/high mass -- could be contributing to the suppression of the $K$-band flux by CIA H$_2$. We will also note here that theoretical models of CIA H$_2$ by \cite{borysow1997} demonstrate that this absorption in T dwarf atmospheres is strong across the $J$, $H$, and $K$ bands, although stronger at $K$ than at $H$ and stronger at $H$ than at $J$. This would have the additional effect of enhancing the $Y$-band flux relative to $J$ while flattening the $K$-band flux peak.
    
    \item 2MASS 0939-2448: \cite{burgasser2006} note a broader $Y$-band peak in this object along with a depressed $K$-band peak. Those authors found that the $K$-band depression is much greater than that allowed by models that cover a physical range of gravities, leading them to conclude that a lower metallicity was the primary cause. 
    
    \item LHS 6176B (0950+0117): This object was discovered by \cite{burningham2013}, who established its companionship with the M dwarf LHS 6176A, which has a metallicity of [Fe/H] = $-0.30{\pm}0.1$ dex. The published near-infrared spectrum in that paper appears to show a depressed $K$-band and what may be a broader $Y$-band peak as well, although the spectrum only samples part of the $Y$-band itself.
    
%NOTE    \item WISE 1042-3842: is listed in Zhang et al. 2019 as low-metallicity and they quote my 2011 paper, but I label this is a normal T dwarf there and include no additional commentary on it. So I'm dropping this object from the list here.
    
%NOTE    \item 2MASS 1114-2618: In Burgasser paper, same arguments as 0939. However, the spectrum of this object is far less extreme.

%NOTE    \item WISE 1150+6302: is listed in Zhang et al. 2019 as low-metallicity and they quote my 2011 paper, but I label this is a normal T dwarf there and include no additional commentary on it. So I'm dropping this object from the list here.
    
%NOTE    \item WISE 1217+1626AB: is listed in Zhang et al. 2019 as low-metallicity and they quote my 2011 paper, but I label this is a normal T dwarf there and include no additional commentary on it. So I'm dropping this object from the list here.
    
    \item SDSS 1416+1348 ("A") and ULAS 1416+1348 ("B"): This is a close, common-proper-motion pair. The brighter, SDSS object is commonly typed as an sdL7 (\citealt{kirkpatrick2016, zhang2017}) and the fainter, ULAS object as an (sd)T7.5 (\citealt{burgasser2010}). \cite{gonzales2020} show through spectral retrieval methods that both objects are slightly subsolar in metallicity, with [M/H] $\approx$ $-$0.3 dex. 
     
    \item Gl 547B (1423+0116): Also known as BD+01 2920B, this T8 dwarf is the companion to an early-G dwarf. The discovery spectrum from \cite{pinfield2012} shows a broader $Y$-band peak and more depressed $K$-band peak than the spectral standard of the same type. Those authors list the metallicity of the primary star as [Fe/H] = $-0.38{\pm}0.06$ dex, which directly links the $Y$- and $K$-band peculiarities of this companion and other objects in this list to a lower metallicity cause.
    
%NOTE:    \item WISE 1436-1814: This object was discovered by \cite{kirkpatrick2011}. Their spectrum shows excess flux at $Y$ and depressed flux at $K$ that they noted could suggest lower metal content. -- NO! Our spectrum in that paper show EXCESS K-band *and* Y-band flux!
     
    \item Gl 576B (1504+0537): Also known as HIP 73786B, this object was uncovered as a common-proper-motion companion by \cite{scholz2010b}. \cite{murray2011} found that the primary star has a metallicity of [Fe/H] = $-0.30{\pm}0.1$ dex, and that the spectrum of the secondary has depressed $H$- and $K$-band peaks. (Their spectrum does not fully sample the $Y$-band peak.)  \cite{zhang2019} classify this companion as an sdT5.5. 

    \item WISE 1523+3125: \cite{mace2013} discovered this object and noted that it has the same $Y$- and $K$-band peculiarities noted for known subdwarfs.
    
%NOTE    \item WISE 1707-1744: is listed in Zhang et al. 2019 as low-metallicity and they quote Greg's 2013 paper, but there is no additional commentary on it. So I'm dropping this object from the list here.
     
    \item WISE 2005+5424: This is an sdT8 from \cite{mace2013b} and a companion to Wolf 1130A, whose metallicity is known ([Fe/H] = $-0.64{\pm}0.17$; \citealt{rojas-ayala2012}). \cite{mace2018} have measured a refined value of [Fe/H] = $=-0.70{\pm}0.12$. \cite{zhang2019} have suggested that this object may eventually require a more extreme classification (esdT8) once other T subdwarfs are identified.
    
    \item WISE 2134-7137: This object was discovered by \cite{kirkpatrick2011}. As they note, the spectrum of this object exhibits excess flux at $Y$ and depressed flux at $K$, which could suggest lower metal content.
    
%NOTE    \item WISE 2319-1844: is listed in Zhang et al. 2019 as low-metallicity and they quote my 2011 paper, but I label this is a normal T dwarf there and include no additional commentary on it. So I'm dropping this object from the list here.
    
    \item WISE 2325-4105: This object, which was discovered by \cite{kirkpatrick2011}, has a spectrum exhibiting excess flux at $Y$ and depressed flux at $K$. Both of these traits are common to most of the objects on this list.

\end{itemize}

%Despite this, the available models provide poor matches to the overall spectral energy distribution of the source and only confine the effective temperature to a broad range of 600-900K, which is not inconsistent with the T$_{eff}$ range of solar metallicity T8. In the absence of other knowledge, we will use the spectral type to temperature relations derived for normal L and T dwarfs to provide the effective temperatures for the subdwarfs as well.

%2MASS 1217+1626B may even be the first sdY identified, although given the wide range of photometric properties seen within the Y dwarf population, other sdY dwarfs likely already populate our census and will be more clearly identified with the higher S/N and wider wavelength coverage provided by {\it JWST}.

%For Teff values of sdL's, see equation on page 1387 of Zhang-Primeval-Paper3.

A few other suspected subdwarfs within the 20-pc census are listed in section~\ref{sec:suspected_subdwarfs} below.

\subsection{Confirmed L, T, and Y Multiples\label{section:known_multiples}}

\cite{kirkpatrick2019} listed a number of known L, T, and Y multiples falling within the 20-pc census: WISE 0146+4234AB,  WISE 0226$-$0211AB, WISE 0458+6434AB, WISE 0614+3912AB, WISE 1217+1626AB, 2MASS 1225$-$2739AB, SDSS/ULAS 1416+1348AB, and 2MASS 1553+1532AB. All of these are confirmed via high-resolution imaging observations and/or common proper motion.

A number of other L, T, and Y multiples in the 20-pc census are further discussed below. Each of these has likewise been confirmed via imaging and/or motion. (For systems with a suspected, but not confirmed, tertiary component, the component's suffix is shown in brackets.)

\begin{itemize}

    \item GJ 1001BC (0004-4044): Using multiple instruments on the {\it Hubble Space Telescope} ({\it HST}), \cite{golimowski2004} discovered that the mid-L dwarf GJ 1001B was a binary. The multiple observations over different epochs confirmed that the binary is a common-proper-motion pair. 

    \item DENIS 0205$-$1159AB[C]: The host object in this system was discovered by \cite{delfosse1997}. The B component, which was discovered through Keck Observatory imaging by \cite{koerner1999}, was found by \cite{bouy2005} through {\it Hubble Space Telescope} imaging to be elongated, leading to speculation that B is a close binary. It appears that the C component has never been independently verified.
    
    \item SDSS 0423$-$0414AB: The primary in this system was discovered by \cite{geballe2002}. The companion was discovered by \cite{burgasser2005} using imaging from the {\it Hubble Space Telescope}.
    
    \item CWISE 0617+1945AB: This object is new to this paper. Publicly available UGPS $K$-band images from 2010 Nov 16 UT and 2013 Apr 03 UT, which clearly show the source's motion to the WSW, also show a common-proper-motion companion 1$\farcs$3 arcsec to the NW (Figure~\ref{figure:cwise0617}). The CatWISE2020 Catalog gives motions of $\mu_\alpha = -103.80{\pm}4.0$ mas yr$^{-1}$ and $\mu_\delta = -59.80{\pm}3.8$ mas yr$^{-1}$ for the A component. Only the A component is listed in {\it Gaia} DR2, but it has no parallax or proper motion measurements reported there. Null information in these columns is generally taken to mean that the five-parameter astrometric solution of position, parallax, and proper motion could not converge over the small time baseline of {\it Gaia} data available for DR2. This may be evidence that the source is an unresolved physical double whose orbital motion was confounding the {\it Gaia} fit. It is also possible that the A component is confused by an object in the background except that POSS-II F (red) and N (near-infrared) plates from the mid-1990s do not show any comparably bright background source at the present position that would be compromising {\it Gaia}'s astrometry. A plot of $J_{MKO}-K_{MKO}$ vs.\ $J_{MKO} -$ W2 using the data presented in Table~\ref{table:monster_table} shows that the A component falls squarely in the locus of other mid- to late-L dwarfs. Using an estimate of the $J$-band magnitude of B and assuming it is equidistant with A, we determine a spectral type for B of [T8:].
  
\begin{figure}
\figurenum{11}
\includegraphics[scale=0.26,angle=0]{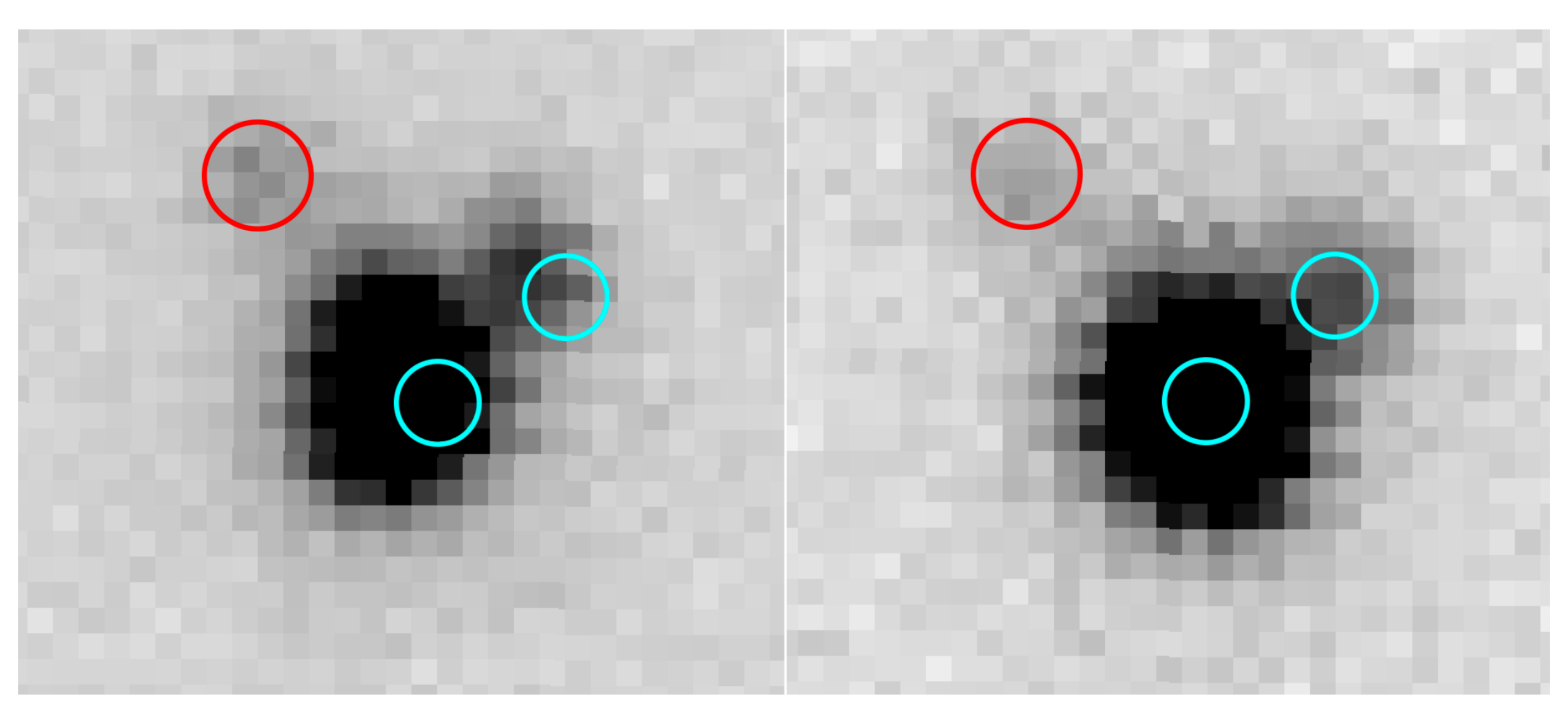}
\caption{UGPS $K$-band images for CWISE 0617+1945. (Left) The 2011 Nov image. (Right) The 2013 Apr image. Both images are five arcsec on a side. Circles on both images mark the 2013 positions of CWISE 0617+1945A and CWISE 0617+1945B (cyan) and the non-moving background source (red). North is up and east is to the left.
\label{figure:cwise0617}}
\end{figure}
    
    \item 2MASS 0700+3157AB[C]: This system was discovered serendipitously by \cite{thorstensen2003} when performing astrometric measurements of the unrelated nearby DC10 white dwarf LHS 1889. Using imaging observations with the {\it Hubble Space Telescope}, \cite{reid2006} discovered a faint companion. \cite{dupuy2017} have performed high-resolution astrometric monitoring of the system and found that the L3: primary is marginally less massive (68.0$\pm$2.6 M$_{Jup}$) than the L6.5: secondary ($73.3^{+2.9}_{-3.0}$ M$_{Jup}$) despite the large difference in their luminosities. This led those authors to surmise that the B component was comprised of two lower-mass brown dwarfs, although they were unable to find a three-body solution in which theoretical evolutionary models could self consistently apportion the masses and luminosities at a single coeval age. For now, we consider the C component likely, but not confirmed.
    
    \item 2MASS 0746+2000AB: Based on its location on the color-magnitude diagram, 2MASS 0746 was suspected to be an unresolved binary by \cite{reid2000}. \cite{reid2001a} confirmed this hypothesis with imaging from {\it HST} and verified common proper motion of the components using earlier observations from the W.\ M.\ Keck Observatory.

    \item 2MASS 0915+0422AB: This object was discovered by \cite{reid2006}, who also found it to be a binary using imaging from {\it HST}.
    
    \item WISE 1049$-$5319AB: This object, commonly referred to as Luhman 16AB, is the third closest system to the Sun and has been known as a binary since its discovery (\citealt{luhman2013}).
    
    \item Kelu-1AB (1305$-$2541): The overluminosity of this object relative to L dwarfs of similar spectral type had been noted after its trigonometric parallax was measured by \cite{dahn2002} and \cite{vrba2004}. \cite{liu2005} imaged the companion using the W.\ M.\ Keck Observatory and used earlier observations from {\it HST} to confirm common proper motion between the components.

    \item 2MASS 1315$-$2649AB: This highly active L dwarf was discovered serendipitously by \cite{hall2002} and identified as a binary via high-resolution imaging at the W.\ M.\ Keck Observatory by \cite{burgasser2011b}.
    
    \item Gl 564BC (1450+2354): \cite{potter2002} discovered this close pair as companion binary to the G2 V star Gl 564A using the Gemini North Telescope. Their subsequent observations at Gemini along with spectroscopy from the W.\ M.\ Keck Observatory confirmed the physical association of the L dwarf pair with the G dwarf primary.

    \item 2MASS 1520$-$4422AB: Observations of this object with the New Technology Telescope by \cite{kendall2007} revealed that the object is a double and that the two components are both L dwarfs. The difference in magnitude between the objects matches expectations if two objects are equidistant.

    \item 2MASS 1534$-$2952AB: This mid-T dwarf was discovered by \cite{burgasser2002} and found to be a binary through {\it HST} imaging by \cite{burgasser2003c}.
    
    \item 2MASS 2152+0937AB: This mid-L dwarf was discovered by \cite{reid2006}, who also identified it as an equal-magnitude binary through {\it HST} imaging.
    
    \item Gl 845BC (2204$-$5646): This object is the companion to the nearby K dwarf $\epsilon$ Ind. It was discovered by \cite{scholz2003} and further identified through imaging as a likely pair of T dwarfs by \cite{volk2003}. \cite{mccaughrean2004} acquired individual spectroscopy to confirm this as a physical pair of T dwarfs.
    
    \item DENIS 2252$-$1730AB: \cite{kendall2004} discovered this object, and it was identified as a binary system by \cite{reid2006b} through {\it HST}/NICMOS imaging.
    
    \item 2MASS 2255$-$5713AB: This object was discovered by \cite{kendall2007} and identified as a binary system through {\it HST}/NICMOS imaging by \cite{reid2008b}.
    
\end{itemize}

Previously suspected multiple systems and new ones identified here for the first time are addressed in section~\ref{sec:suspected_multiples}.

\subsection{Analysis of Color-Magnitude and Color-Color Plots\label{section:plot_analysis}}

In order to identify other unresolved binaries or subdwarfs in the 20-pc census, we examine color-magnitude and color-color diagrams built from the photometric, astrometric, and spectroscopic data compiled in Table~\ref{table:monster_table}. On these we highlight known multiple systems, low-gravity objects, and low-metallicity subdwarfs, as discussed above.

As mentioned in section~\ref{section: JHK}, the data presented in Table~\ref{table:monster_table} are drawn from a variety of sources, leading to heterogeneity, particularly in the photometric values. For example, although 2MASS covers the entire sky, it is not deep enough to detect many of the late-T and Y dwarfs. For those objects, the hemispheric surveys of UHS in the north and VHS in the south can provide deeper data. Although $H$-band filters are largely invariant across surveys, the same is not true of $J$ and $K$. As shown in Figure 3 of \cite{gonzalez2018}, the 2MASS filters $J_{\rm 2MASS}$ and $K_S$ are markedly different from the $J_{\rm MKO}$ and $K_{\rm MKO}$ filters used by WFCAM. Furthermore, although the VISTA employs the same $J_{\rm MKO}$ as WFCAM, its $K_S$ filter is much closer to the $K_S$ filter used by 2MASS. Similarly, although {\it WISE} data in bands W1 and W2 cover the entire sky, deeper observations by {\it Spitzer} are done with complementary, though not identical, ch1 and ch2 filters, as shown in Figure 2 of \cite{mainzer2011}.  

\begin{figure*}
\figurenum{12}
\gridline{\fig{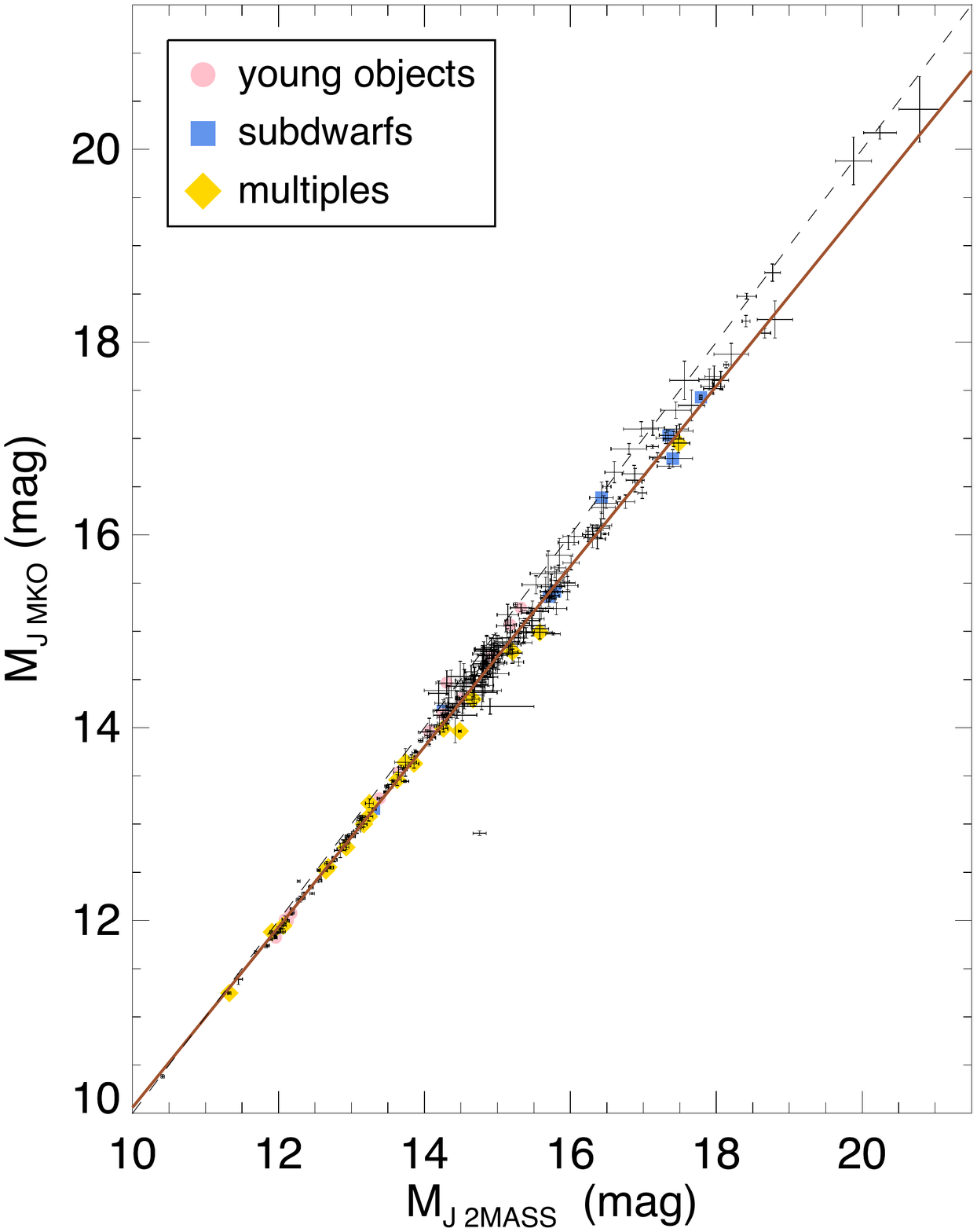}{0.25\textwidth}{(a)}
          \fig{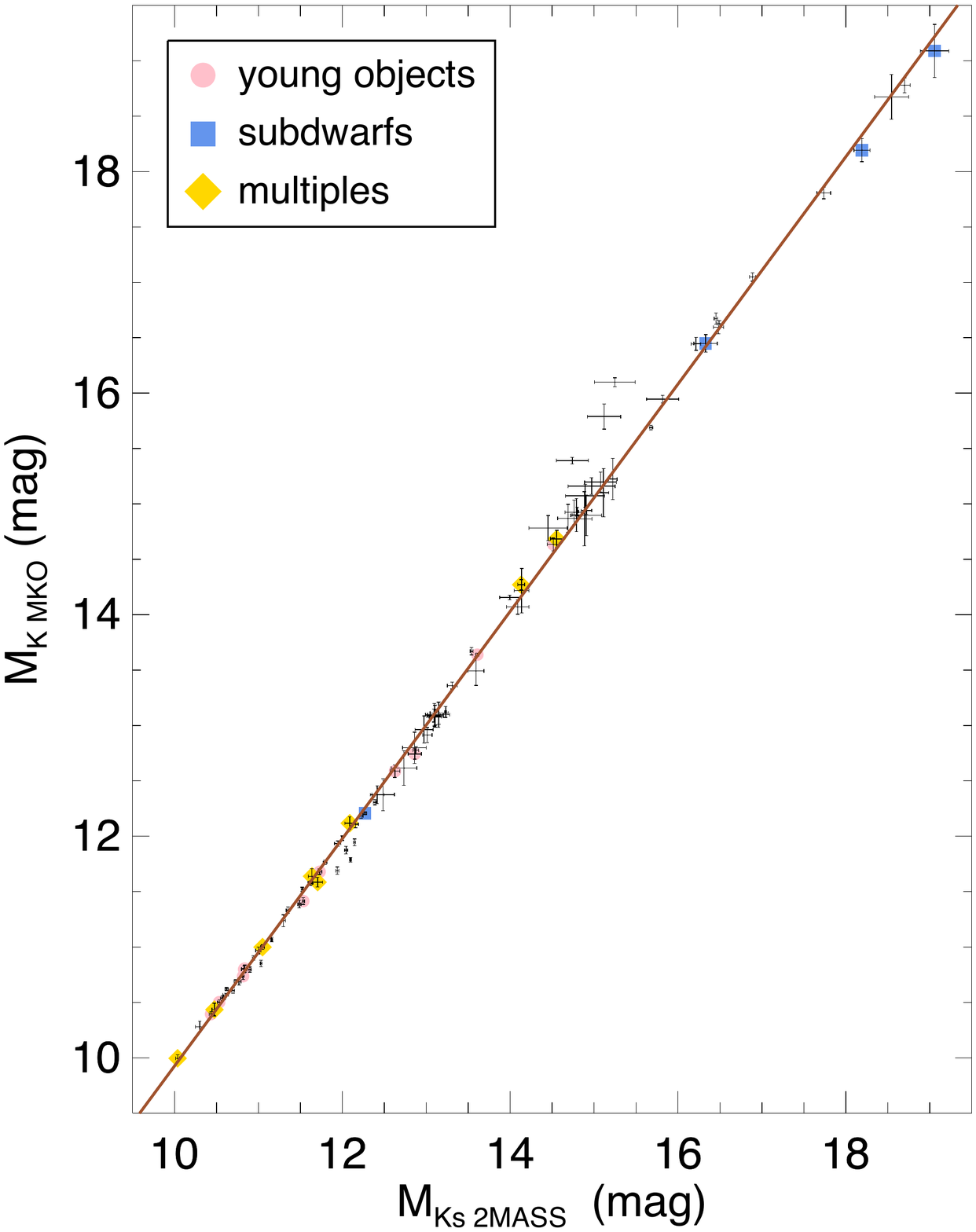}{0.25\textwidth}{(b)}
          \fig{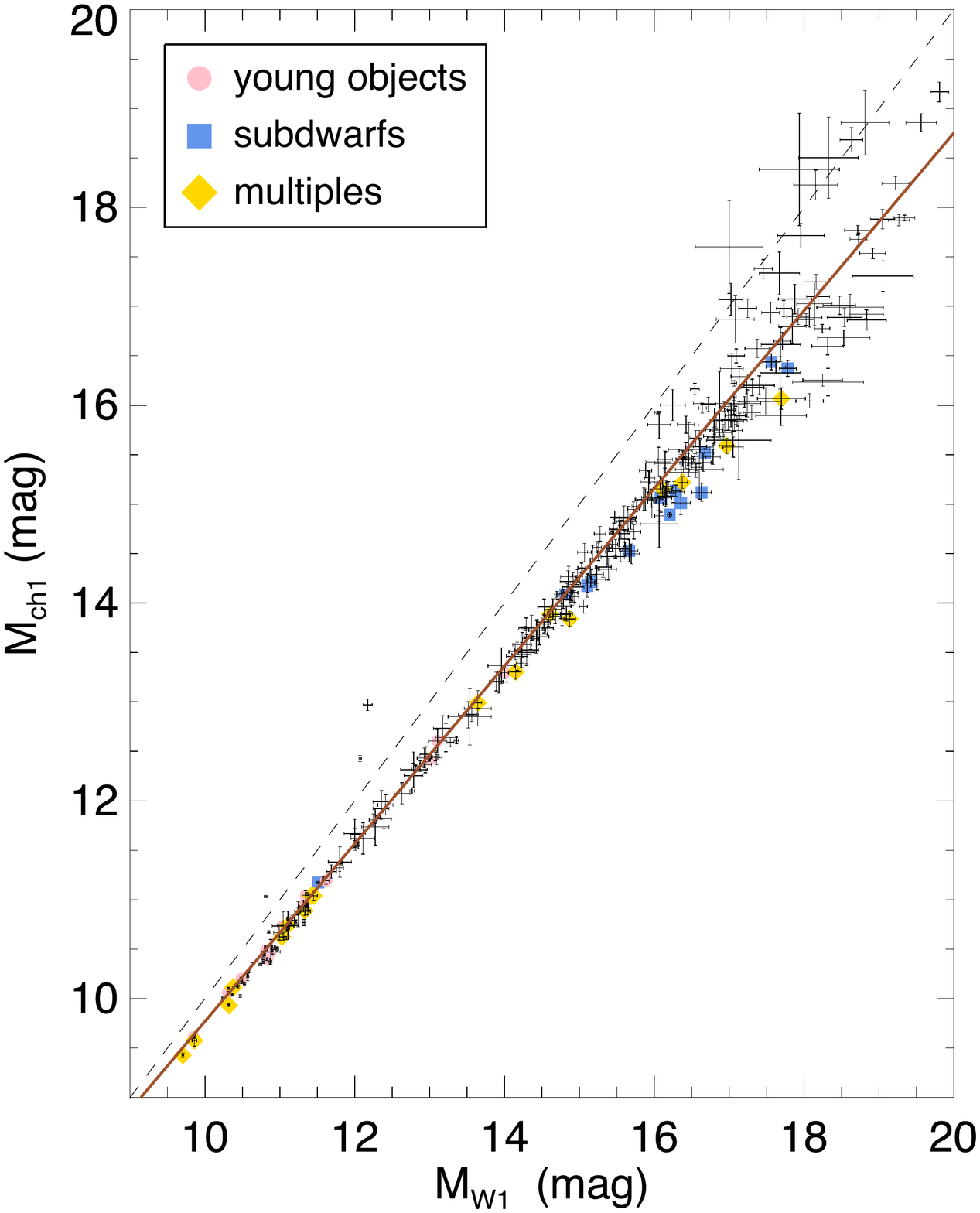}{0.25\textwidth}{(c)}
          \fig{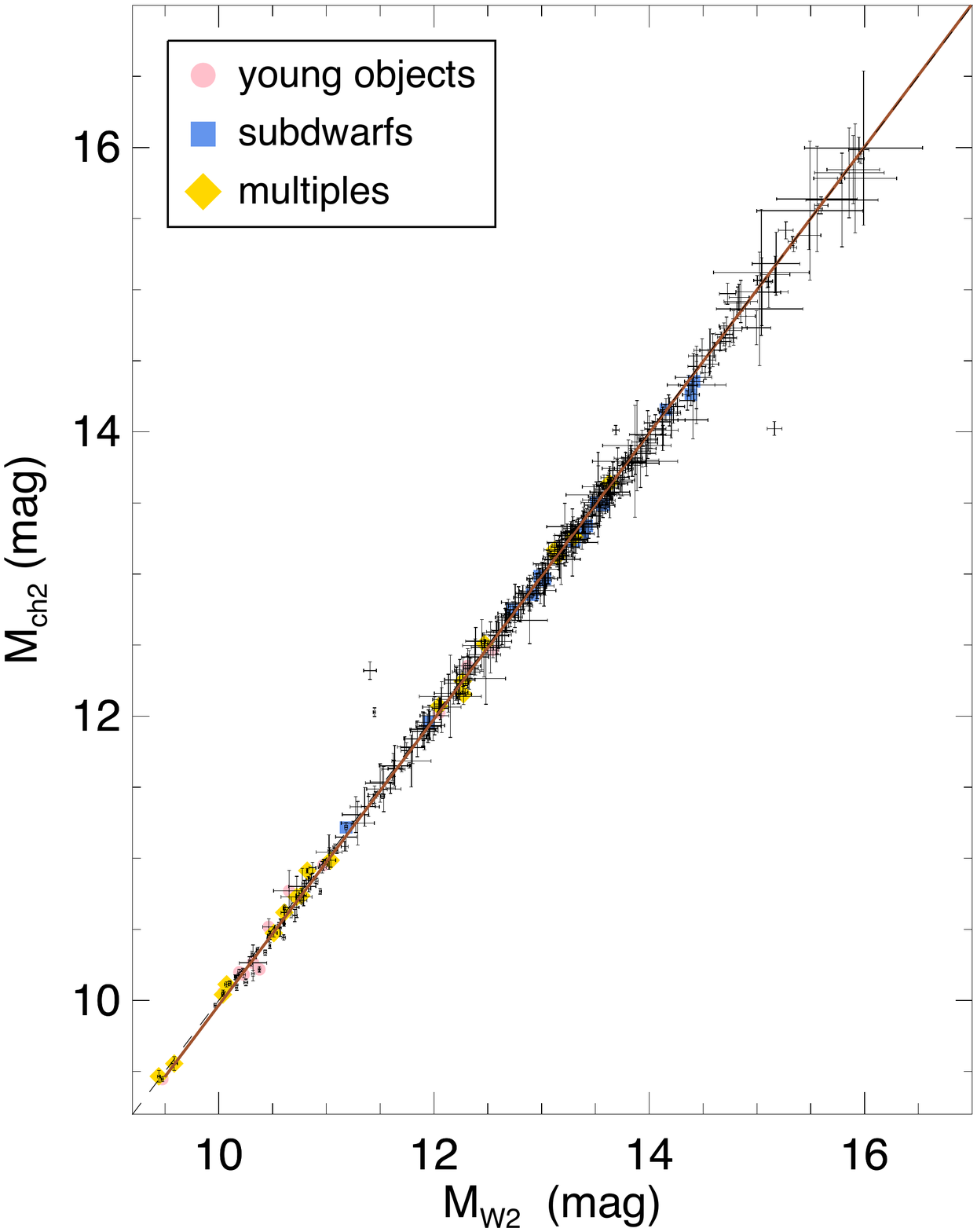}{0.25\textwidth}{(d)}}
\caption{Plots showing the comparison of absolute magnitudes for objects within the 20-pc census that have measurements in both bands and robust parallax measurements: (a) $M_{J \rm 2MASS}$ vs.\ $M_{J \rm MKO}$, (b) $M_{Ks \rm 2MASS}$ vs.\ $M_{K\rm MKO}$, (c) $M_{\rm W1}$ vs.\ $M_{\rm ch1}$, and (d) $M_{\rm W2}$ vs.\ $M_{\rm ch2}$. Objects identified as low-gravity, low-metallicity, or having unresolved multiplicity are color coded, per the legend. The brown line shows the linear least-squares fit to the data, excluding color-coded objects. The parameters for these fits are given in Table~\ref{table:equations}. The one-to-one line is shown by the black dashed line.
\label{figure:filter_transformations}}
\end{figure*}

Ideally, transforming magnitudes in one filter to the complimentary filter in the other survey(s) would allow us to examine homogenized color-color and color-magnitude diagrams using as much data as has been currently collected for the 20-pc L, T, and Y dwarfs. Figure~\ref{figure:filter_transformations} shows the relation in absolute magnitude between $J_{\rm 2MASS}$ and $J_{\rm MKO}$, $K_S$ and $K_{\rm MKO}$, W1 and ch1, W2 and ch2. Linear least squares fits to the trends are illustrated in the plots and listed in Table~\ref{table:equations}. The line of one-to-one correspondence is shown by the black dashed line on each panel.

\begin{figure}
\figurenum{13}
\includegraphics[scale=0.35,angle=0]{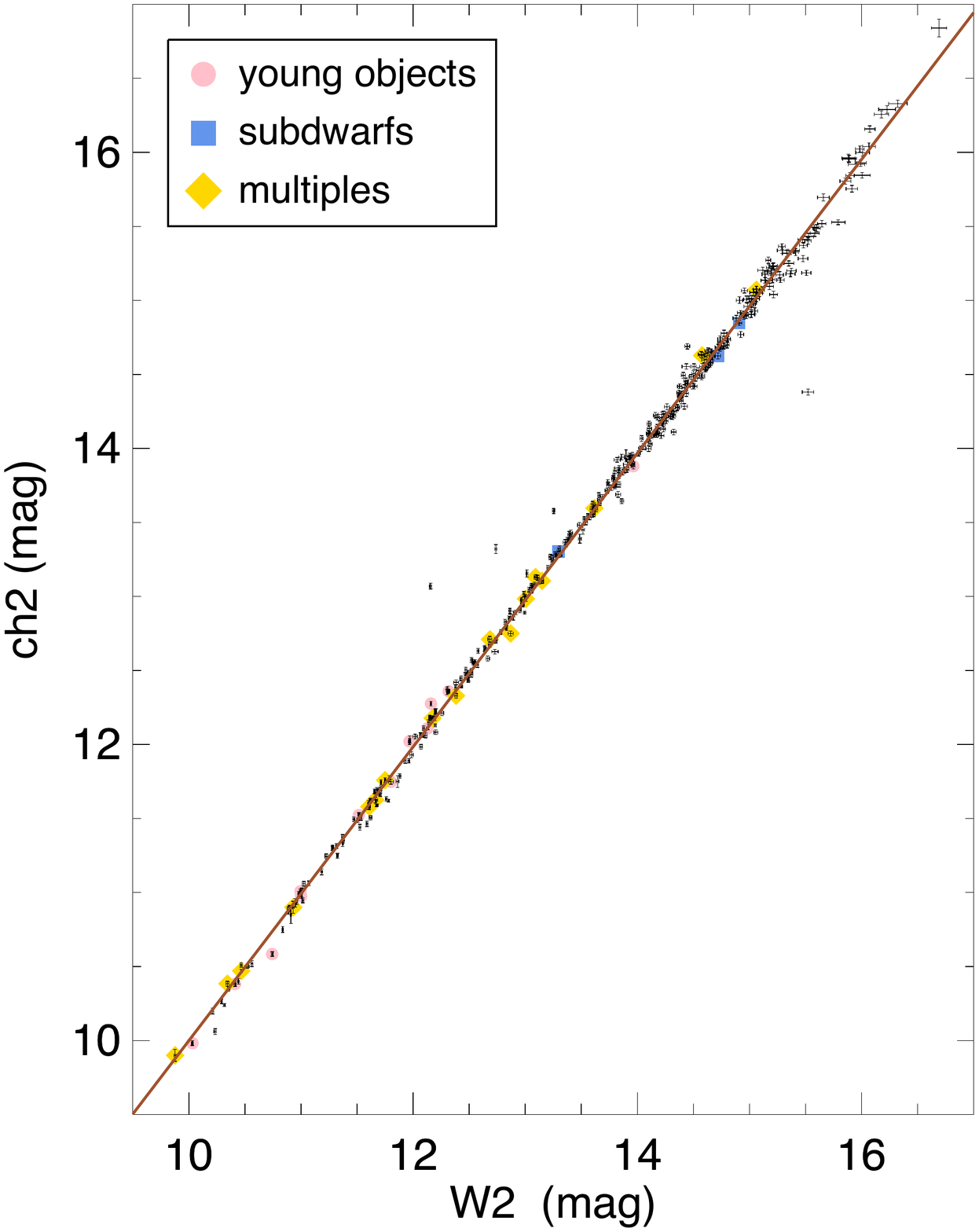}
\caption{Comparison of the W2 vs.\ ch2 apparent magnitudes for objects within the 20-pc census that have measurements in both bands. Objects identified as low-gravity, low-metallicity, or having unresolved multiplicity are color coded, per the legend. The brown line shows the linear least-squares fit to the data, excluding color-coded objects. The parameters for this fit are given in Table~\ref{table:equations}.
\label{figure:ch2_vs_W2}}
\end{figure}

The fits to these trends show significant deviations from the one-to-one line for all of these plots except $M_{\rm W2}$ vs.\ $M_{\rm ch2}$. Transforming between the W2 and ch2 magnitudes is thus an easy transformation (Figure~\ref{figure:ch2_vs_W2}; Table~\ref{table:equations}) not requiring a color or spectral type term. However, transforming between the three other pairs of bands would involve such terms. For these, the uncertainties in the fits as well as uncertainties in the magnitude and color/type measurements would result in a transformed value with a necessarily large uncertainty. Therefore, in the following plots, the only transformations we will include are converting W2 magnitudes into ch2 magnitudes for objects that lack a ch2 measurement.

\begin{longrotatetable}
\begin{deluxetable*}{lcccccccccccc}
\tabletypesize{\footnotesize}
\tablecaption{Polynomial Fits to Trends Shown in Figures~\ref{figure:filter_transformations}-\ref{figure:yaxis_teff}\label{table:equations}}
\tablehead{
\colhead{$x$} & 
\colhead{$y$} &
\colhead{$c_0$} &
\colhead{$c_1$} &
\colhead{$c_2$} &
\colhead{$c_3$} &
\colhead{$c_4$} &
\colhead{$c_5$} &
\colhead{$c_6$} &
\colhead{$c_7$} &
\colhead{Valid Range} &
\colhead{RMS\tablenotemark{a}} &
\colhead{Fig\#} \\
\colhead{} & 
\colhead{} &
\colhead{($\sigma_{c0}$)} &
\colhead{($\sigma_{c1}$)} &
\colhead{($\sigma_{c2}$)} &
\colhead{($\sigma_{c3}$)} &
\colhead{($\sigma_{c4}$)} & 
\colhead{($\sigma_{c5}$)} &
\colhead{($\sigma_{c6}$)} &
\colhead{($\sigma_{c7}$)} &
\colhead{} &
\colhead{} &
\colhead{} \\
\colhead{(1)} &                          
\colhead{(2)} &  
\colhead{(3)} &  
\colhead{(4)} &
\colhead{(5)} &
\colhead{(6)} &
\colhead{(7)} &
\colhead{(8)} &                          
\colhead{(9)} &  
\colhead{(10)} &  
\colhead{(11)} &
\colhead{(12)} &
\colhead{(13)} \\
}
\startdata
$M_{J \rm 2M}$& $M_{J \rm MKO}$&
     7.0584e-01 &
     9.3542e-01 &
     \nodata &
     \nodata &
     \nodata &
     \nodata &
     \nodata &
     \nodata &
     $10 < M_{J \rm 2M} < 20$ &
     0.17 &
     \ref{figure:filter_transformations}a\\
& & (1.3556e-02)& (9.9161e-04)& \nodata & \nodata & \nodata & \nodata & \nodata & \nodata & \nodata & \nodata & \nodata \\
$M_{Ks}$& $M_{K \rm MKO}$&
     -4.5368e-01 &
      1.0358 &
     \nodata &
     \nodata &
     \nodata &
     \nodata &
     \nodata &
     \nodata &
     $9.5 < M_{Ks} < 18.5$ &
     0.13 &
     \ref{figure:filter_transformations}b\\
& & (2.5737e-02)& (2.1127e-03)& \nodata & \nodata & \nodata & \nodata & \nodata & \nodata & \nodata & \nodata & \nodata \\
$M_{\rm W1}$& $M_{\rm ch1}$&
     7.8505e-01 &
     8.9844e-01 &
     \nodata &
     \nodata &
     \nodata &
     \nodata &
     \nodata &
     \nodata &
     $9 < M_{\rm W1} < 20$ &
     0.33 &
     \ref{figure:filter_transformations}c\\
& & (1.0156e-02)& (8.1555e-04)& \nodata & \nodata & \nodata & \nodata & \nodata & \nodata & \nodata & \nodata & \nodata \\
$M_{\rm W2}$& $M_{\rm ch2}$&
     -9.6870e-02 &
     1.0063 &
     \nodata &
     \nodata &
     \nodata &
     \nodata &
     \nodata &
     \nodata &
     $9 < M_{\rm W2} < 17$ &
     0.11 &
     \ref{figure:filter_transformations}d\\
& & (1.5392e-02)& (1.2866e-03)& \nodata & \nodata & \nodata & \nodata & \nodata & \nodata & \nodata & \nodata & \nodata \\
W2& ch2&
     7.0888e-02 &
     9.9268e-01 &
     \nodata &
     \nodata &
     \nodata &
     \nodata &
     \nodata &
     \nodata &
     $9.5 < {\rm W2} < 17.5$ &
     0.10 &
     \ref{figure:ch2_vs_W2}\\
& & (8.9087e-03)& (6.5958e-04)& \nodata & \nodata & \nodata & \nodata & \nodata & \nodata & \nodata & \nodata & \nodata \\
SpT& $M_{J \rm MKO}$&
1.1808e+01&  3.3790e-01&  -1.9013e-01&   7.1759e-02&  -9.9829e-03&   6.3147e-04&  -1.8672e-05&   2.1526e-07&
  $0 \le {\rm SpT} \le 22$ &
  0.60 &
  \ref{figure:xaxis_spectype}a \\
& & (5.5134e-03)&   (8.8293e-03)&   (6.1245e-03)&   (1.8277e-03)&   (2.6582e-04)&   (1.9917e-05)&   (7.3940e-07)&   (1.0762e-08)& \nodata & \nodata & \nodata \\
SpT& $M_H$ &
1.0966e+01&   6.0330e-01&  -3.5647e-01&   1.1696e-01&  -1.6688e-02&   1.1719e-03&  -4.0259e-05&   5.4808e-07&
  $0 \le {\rm SpT} \le 22$ &
  0.57 &
  \ref{figure:xaxis_spectype}b \\
& & (1.2196e-02)&   (2.0708e-02)&   (1.2508e-02)&   (3.4795e-03)&   (4.9038e-04)&   (3.6092e-05)&   (1.3227e-06)&   (1.9054e-08)& \nodata & \nodata & \nodata \\
SpT& $M_{\rm ch1}$&
 9.9434e+00&   3.4919e-01&  -1.0725e-01&   1.7669e-02&  -7.1402e-04&  -5.0487e-05&   4.7761e-06&  -9.9981e-08&
  $0 \le {\rm SpT} \le 22$ &
  0.38 &
  \ref{figure:xaxis_spectype}c \\
& & (9.9630e-03)&   (1.8691e-02)&   (1.1499e-02)&   (3.1349e-03)&   (4.2974e-04)&   (3.0823e-05)&   (1.1039e-06)&   (1.5571e-08)& \nodata & \nodata & \nodata \\
SpT& $M_{\rm ch2}$&      
1.0071e+01&   1.8897e-01&  -6.2186e-02&   1.9711e-02&  -2.3844e-03&   1.3230e-04&  -3.3136e-06&   2.9971e-08&
  $0 \le {\rm SpT} \le 22$ &
  0.31 &
  \ref{figure:xaxis_spectype}d \\
& & (7.3148e-03)&   (1.1131e-02)&   (7.2528e-03)&   (2.1162e-03)&   (3.0323e-04)&   (2.2351e-05)&   (8.1392e-07)&   (1.1594e-08)& \nodata & \nodata & \nodata \\
SpT& $J - {\rm ch2}$& 
1.8153e+00&   1.8527e-01&  -1.7678e-01&   7.2989e-02&  -1.1351e-02&   8.1335e-04&  -2.7663e-05&   3.6779e-07&
   $0 \le {\rm SpT} \le 22$ &
   0.44 &
  \ref{figure:xaxis_spectype}e \\
& & (7.9722e-03)&   (1.1627e-02)&   (7.0547e-03)&   (1.9164e-03)&   (2.6068e-04)&   (1.8578e-05)&   (6.6362e-07)& (9.3747e-09)& \nodata & \nodata & \nodata \\
SpT& $H - {\rm ch2}$& 
1.1150e+00&   6.7204e-02&  -7.3996e-02&   4.0283e-02&  -7.1625e-03&   5.7419e-04&  -2.1740e-05&   3.2110e-07&
   $0 \le {\rm SpT} \le 22$ &
   0.42 &
  \ref{figure:xaxis_spectype}f \\
& & (1.5068e-02)&   (2.3813e-02)&   (1.3338e-02)&   (3.4641e-03)&   (4.6215e-04)&   (3.2659e-05)&   (1.1624e-06)&   (1.6400e-08)& \nodata & \nodata & \nodata \\
SpT& ch1 $-$ ch2&
2.6662e-02&  -2.6015e-02&   8.1897e-04&   3.2520e-04&
     \nodata &
     \nodata &
     \nodata &
     \nodata &
   $0 \le {\rm SpT} \le 22$ &
   0.19 &
  \ref{figure:xaxis_spectype}g \\
& & (8.2684e-03)&   (3.1254e-03)&   (3.3481e-04)&   (1.0531e-05)& \nodata & \nodata & \nodata & \nodata & \nodata & \nodata & \nodata \\
SpT& W1 $-$ W2&
2.2668e-01&   2.9069e-02&  -4.6379e-03&   5.7825e-04&
     \nodata &
     \nodata &
     \nodata &
     \nodata &
   $0 \le {\rm SpT} \le 22$ &
   0.28 &
  \ref{figure:xaxis_spectype}h \\
& & (4.6612e-03)&   (2.0308e-03)&   (2.5436e-04)&   (9.1089e-06)& \nodata & \nodata & \nodata & \nodata & \nodata & \nodata & \nodata \\
$M_{J \rm MKO}$& SpT&
-7.7784e+01&   1.3260e+01& -6.1185e-01&   9.6221e-03&
     \nodata &
     \nodata &
     \nodata &
     \nodata &
    $14.3 \le M_{J \rm MKO} \le 24.0$\tablenotemark{e} &
    0.53 &
    \ref{figure:yaxis_spectype}a \\
& & (1.5730e+01)&   (2.6086e+00)&   (1.4247e-01)&   (2.5610e-03)& \nodata & \nodata & \nodata & \nodata & \nodata & \nodata & \nodata \\
$M_H$& SpT&
-6.9184e+01&   1.1863e+01&  -5.4084e-01&   8.4661e-03&
     \nodata &
     \nodata &
     \nodata &
     \nodata &
     $14.5 \le M_H \le 24.0$\tablenotemark{e}& 
     0.51 &
    \ref{figure:yaxis_spectype}b \\
& & (1.5192e+01)&   (2.5100e+00)&   (1.3661e-01)&   (2.4473e-03)& \nodata & \nodata & \nodata & \nodata & \nodata & \nodata & \nodata \\
$M_{\rm ch1}$& SpT&
-1.2682e+02&   2.1824e+01&  -1.0888e+00&   1.8362e-02&
     \nodata &
     \nodata &
     \nodata &
     \nodata &
     $10.0 \le M_{\rm ch1} \le 19.0$ &
     0.89 &
    \ref{figure:yaxis_spectype}c \\
& & (1.3047e+01)&   (2.8362e+00)&   (2.0216e-01)&   (4.7350e-03)& \nodata & \nodata & \nodata & \nodata & \nodata & \nodata & \nodata \\
$M_{\rm ch2}$& SpT&
1.4559e+03&  -4.6516e+02&   5.4301e+01&  -2.7423e+00&   5.0950e-02&
     \nodata &
     \nodata &
     \nodata &
     $10.0 \le M_{\rm ch2} \le 16.0$ &
     1.26 &
    \ref{figure:yaxis_spectype}d \\
& & (3.8543e+02)&   (1.2245e+02)&   (1.4499e+01)&   (7.5845e-01)&   (1.4790e-02)& \nodata & \nodata & \nodata & \nodata & \nodata & \nodata \\
$J - {\rm ch2}$& SpT&  
1.1022e+01&   3.4335e+00&  -4.8308e-01&   2.6036e-02&
     \nodata &
     \nodata &
     \nodata &
     \nodata &
     $2.1 \le J - {\rm ch2} \le 8.5$\tablenotemark{f} & 
     0.53 &
    \ref{figure:yaxis_spectype}e \\
& & (6.2228e-01)&   (4.6900e-01)&   (1.0643e-01)&   (7.2766e-03)& \nodata & \nodata & \nodata & \nodata & \nodata & \nodata & \nodata \\
$H- {\rm ch2}$& SpT&  
1.0280e+01&   3.5828e+00&  -5.0032e-01&   2.7292e-02&
     \nodata &
     \nodata &
     \nodata &
     \nodata &
     $2.4 \le H - {\rm ch2} \le 8.3$\tablenotemark{f} & 
     0.54 &
    \ref{figure:yaxis_spectype}f \\
& & (8.4671e-01)&   (6.1648e-01)&   (1.3683e-01)&   (9.2515e-03)&
\nodata & \nodata & \nodata & \nodata & \nodata & \nodata & \nodata \\
ch1 $-$ ch2& SpT& 
5.4614e+00&   2.1717e+01&  -1.6691e+01&   6.1763e+00&  -8.1737e-01&
     \nodata &
     \nodata &
     \nodata &
     $0.1 \le {\rm ch1} - {\rm ch2} \le 3.0$&
     1.26 &
    \ref{figure:yaxis_spectype}g \\
& & (1.5994e-01)&   (1.0797e+00)&   (1.7084e+00)&   (8.9977e-01)&   (1.5127e-01)&  \nodata & \nodata & \nodata & \nodata & \nodata & \nodata \\
W1 $-$ W2& SpT& 
-3.9840e+00&   3.4029e+01&  -2.5352e+01&   1.0073e+01&  -1.9779e+00&   1.5181e-01&
     \nodata &
     \nodata &
     $0.4 \le {\rm W1} - {\rm W2} \le 4.0$&
     1.16 &
    \ref{figure:yaxis_spectype}h \\
& & (5.8726e-01)&   (2.7025e+00)&   (3.8349e+00)&   (2.3059e+00)&   (6.1348e-01)&   (5.9415e-02)& \nodata & \nodata & \nodata & \nodata & \nodata \\
ch1 $-$ ch2& $M_{J \rm MKO}$&  
1.4839e+01&  -1.5369e+00&   1.3741e+00&   4.7706e-02&
     \nodata &
     \nodata &
     \nodata &
     \nodata &
   $0.2 \le {\rm ch1} - {\rm ch2} \le 3.7$ &
   0.82 &
  \ref{figure:xaxis_ch1ch2}a \\
& & (1.9346e-02)&   (4.9810e-02)&   (3.7680e-02)&   (8.4926e-03)& \nodata & \nodata & \nodata & \nodata & \nodata & \nodata & \nodata \\
ch1 $-$ ch2& $M_H$&
1.3650e+01&   4.2277e-01&   5.9475e-01&   1.4662e-01& 
     \nodata &
     \nodata &
     \nodata &
     \nodata &
   $0.2 \le {\rm ch1} - {\rm ch2} \le 3.7$ &
   0.73 &
  \ref{figure:xaxis_ch1ch2}b \\
& &  (2.9219e-02)&   (7.3179e-02)&   (5.4990e-02)&   (1.2372e-02)& \nodata & \nodata & \nodata & \nodata & \nodata & \nodata & \nodata \\
ch1 $-$ ch2& $M_{\rm ch2}$&
1.1685e+01&   1.2405e+00&  -2.6707e-01&   9.7851e-02& 
     \nodata &
     \nodata &
     \nodata &
     \nodata &
   $0.2 \le {\rm ch1} - {\rm ch2} \le 3.7$ &
   0.37 &
  \ref{figure:xaxis_ch1ch2}c \\
& &  (2.1383e-02)&   (4.3465e-02)&   (2.6277e-02)&   (4.6742e-03)& \nodata & \nodata & \nodata & \nodata & \nodata & \nodata & \nodata \\
ch1 $-$ ch2& $J - {\rm ch2}$& 
3.2442e+00&  -3.3515e+00&   2.2401e+00&  -2.1036e-01&
     \nodata &
     \nodata &
     \nodata &
     \nodata &
   $0.4 \le {\rm ch1} - {\rm ch2} \le 3.7$ &
   0.59 &
  \ref{figure:xaxis_ch1ch2}d \\
& & (1.0321e-02)&   (2.9530e-02)&   (2.4059e-02)&   (5.7791e-03)&
\nodata & \nodata & \nodata & \nodata & \nodata & \nodata & \nodata \\
ch1 $-$ ch2& $H - {\rm ch2}$&
1.8968e+00&  -3.8478e-01&   5.4798e-01&   1.0067e-01&
     \nodata &
     \nodata &
     \nodata &
     \nodata &
   $0.4 \le {\rm ch1} - {\rm ch2} \le 3.7$ &
   0.54 &
  \ref{figure:xaxis_ch1ch2}e \\
& & (1.5548e-02)&   (4.6766e-02)&   (3.8685e-02)&   (9.2812e-03)& \nodata & \nodata & \nodata & \nodata & \nodata & \nodata & \nodata \\
ch1 $-$ ch2& W1 $-$ W2&
3.6295e-01&   1.4472e+00&  -9.0895e-02&
     \nodata &
     \nodata &
     \nodata &
     \nodata &
     \nodata &
   $0.0 \le {\rm ch1} - {\rm ch2} \le 3.7$ &
   0.26 &
  \ref{figure:xaxis_ch1ch2}f \\
& & (1.7547e-03)&   (6.3917e-03)&   (3.7513e-03)& \nodata & \nodata & \nodata & \nodata & \nodata & \nodata & \nodata & \nodata \\
W1 $-$ W2& $M_{J \rm MKO}$&
1.5375e+01&  -1.8851e+00&   8.6518e-01&
     \nodata &
     \nodata &
     \nodata &
     \nodata &
     \nodata &
   $1.0 \le {\rm W1} - {\rm W2} \le 4.5$ &
   1.06 &
   \ref{figure:xaxis_W1W2}a \\
& &  (3.0925e-02)&   (2.6582e-02)&   (5.4656e-03)& \nodata & \nodata & \nodata & \nodata & \nodata & \nodata & \nodata & \nodata \\
W1 $-$ W2& $M_H$&
1.3974e+01&  -5.0420e-01&   6.1351e-01&
     \nodata &
     \nodata &
     \nodata &
     \nodata &
     \nodata &
   $1.0 \le {\rm W1} - {\rm W2} \le 4.5$ &
   0.97 &
   \ref{figure:xaxis_W1W2}b \\
& & (5.0500e-02)&   (4.2061e-02)&   (8.3712e-03)& \nodata & \nodata & \nodata & \nodata & \nodata & \nodata & \nodata & \nodata \\
W1 $-$ W2& $M_{\rm ch2}$&
1.1923e+01&   3.2350e-01&   1.1564e-01&
     \nodata &
     \nodata &
     \nodata &
     \nodata &
     \nodata &
   $1.0 \le {\rm W1} - {\rm W2} \le 4.5$ &
   0.46 &
   \ref{figure:xaxis_W1W2}c \\
& & (4.6335e-02)&   (3.6466e-02)&   (6.8309e-03)& \nodata & \nodata & \nodata & \nodata & \nodata & \nodata & \nodata & \nodata \\
W1 $-$ W2& $J - {\rm W2}$&
3.4006e+00&  -2.0109e+00&   6.8777e-01&
     \nodata &
     \nodata &
     \nodata &
     \nodata &
     \nodata &
   $1.0 \le {\rm W1} - {\rm W2} \le 4.5$ &
   0.77 &
   \ref{figure:xaxis_W1W2}e \\
& & (1.1512e-02)&   (1.1197e-02)&   (2.5620e-03)& \nodata & \nodata & \nodata & \nodata & \nodata & \nodata & \nodata & \nodata \\
W1 $-$ W2& $H - {\rm W2}$&
1.0705e+00&   8.5334e-01&  -3.0341e-01&   1.1371e-01&
     \nodata &
     \nodata &
     \nodata &
     \nodata &
   $1.0 \le {\rm W1} - {\rm W2} \le 4.5$ &
   0.67 &
   \ref{figure:xaxis_W1W2}f \\
& & (4.6358e-02)&   (6.8836e-02)&   (3.2019e-02)&   (4.6830e-03)& \nodata & \nodata & \nodata & \nodata & \nodata & \nodata & \nodata \\
$J_{\rm MKO} - {\rm ch2}$& $M_{J \rm MKO}$&  
1.1915e+01&   1.5841e+00&  -1.6137e-02& 
     \nodata &
     \nodata &
     \nodata &
     \nodata &
     \nodata &
   $1.0 \le J_{\rm MKO} - {\rm ch2} \le 12.0$\tablenotemark{b} &
   0.39 &
   \ref{figure:xaxis_Jch2}a \\   
& & (4.9157e-02)&   (2.3193e-02)&   (2.4580e-03)& \nodata & \nodata & \nodata & \nodata & \nodata & \nodata & \nodata & \nodata \\
$J_{\rm MKO} - {\rm ch2}$& $M_H$&  
1.1583e+01&   1.9032e+00&  -5.0925e-02&
     \nodata &
     \nodata &
     \nodata &
     \nodata &
     \nodata &
   $1.0 \le J_{\rm MKO} - {\rm ch2} \le 12.0$\tablenotemark{c} &
   0.42 &
   \ref{figure:xaxis_Jch2}b \\   
& & (3.3133e-02)&   (1.7627e-02)&   (2.0930e-03)& \nodata & \nodata & \nodata & \nodata & \nodata & \nodata & \nodata & \nodata \\
$J_{\rm MKO} - {\rm ch2}$& $M_{\rm ch2}$&
1.2404e+01&   3.6423e-01&   4.5527e-03& 
     \nodata &
     \nodata &
     \nodata &
     \nodata &
     \nodata &
   $1.0 \le J_{\rm MKO} - {\rm ch2} \le 12.0$\tablenotemark{d} &
   0.34 &
   \ref{figure:xaxis_Jch2}c \\   
& & (1.9595e-02)&   (7.9067e-03)&   (6.5360e-04)& \nodata & \nodata & \nodata & \nodata & \nodata & \nodata & \nodata & \nodata \\
$H - {\rm ch2}$& $M_{J \rm MKO}$&  
1.1777e+01&   1.4108e+00&  -4.6998e-05& 
     \nodata &
     \nodata &
     \nodata &
     \nodata &
     \nodata &
   $2.5 \le H - {\rm ch2} \le 12.0$\tablenotemark{b} &
   0.50 &
   \ref{figure:xaxis_Hch2}a \\   
& & (5.9763e-02)&   (2.5921e-02)&   (2.5617e-03)& \nodata & \nodata & \nodata & \nodata & \nodata & \nodata & \nodata & \nodata \\
$H - {\rm ch2}$& $M_H$&  
1.1454e+01&   1.6462e+00&  -1.7633e-02& 
     \nodata &
     \nodata &
     \nodata &
     \nodata &
     \nodata &
   $2.0 \le H - {\rm ch2} \le 12.0$\tablenotemark{c} &
   0.35 &
   \ref{figure:xaxis_Hch2}b \\   
& & (4.2250e-02)&   (2.0841e-02)&   (2.3369e-03)& \nodata & \nodata & \nodata & \nodata & \nodata & \nodata & \nodata & \nodata \\
$H - {\rm ch2}$& $M_{\rm ch2}$& 
1.2709e+01&   1.4789e-01&   2.7211e-02& 
     \nodata &
     \nodata &
     \nodata &
     \nodata &
     \nodata &
   $3.0 \le H - {\rm ch2} \le 12.0$\tablenotemark{d} &
   0.30 &
   \ref{figure:xaxis_Hch2}c \\   
& & (2.6042e-02)&   (1.0488e-02)&   (9.1350e-04)& \nodata & \nodata & \nodata & \nodata & \nodata & \nodata & \nodata & \nodata \\
$M_H$& $T_{\rm eff}$& 
1.2516e+04&  -1.5666e+03&   6.7502e+01&  -9.2430e-01&  -1.9530e-03&
     \nodata &
     \nodata &
     \nodata &
     $9.5 \le M_H \le 25.0$ &
     88.1 &
     \ref{figure:yaxis_teff}a \\
& & (1.0770e+03)&   (2.7058e+02)&   (2.4638e+01)&   (9.6594e-01)&   (1.3793e-02)& \nodata & \nodata & \nodata & \nodata & \nodata & \nodata \\
SpT& $T_{\rm eff}$&  
2.2375e+03&  -1.4496e+02&   4.0301e+00&
     \nodata &
     \nodata &
     \nodata &
     \nodata &
     \nodata &
     $0.0 \le SpT \le 8.75$ &
     134 &
     \ref{figure:yaxis_teff}b \\ 
& & (1.1342e+01)&   (4.2745e+00)&   (8.8587e-01)& \nodata & \nodata & \nodata & \nodata & \nodata & \nodata & \nodata & \nodata \\
SpT& $T_{\rm eff}$& 
1.4379e+03& -1.8309e+01& 
     \nodata &
     \nodata &
     \nodata &
     \nodata &
     \nodata &
     \nodata &
     $8.75 \le SpT \le 14.75$ &
     79 &
     \ref{figure:yaxis_teff}b \\ 
SpT& $T_{\rm eff}$& 
5.1413e+03&  -3.6865e+02&   6.7301e+00&
     \nodata &
     \nodata &
     \nodata &
     \nodata &
     \nodata &
     $14.75 \le SpT \le 22.0$ &
     79 &
     \ref{figure:yaxis_teff}b \\ 
& & (7.9271e+02)&   (8.7788e+01)&   (2.4174e+00) & \nodata & \nodata & \nodata & \nodata & \nodata & \nodata & \nodata & \nodata \\
\enddata
\tablenotetext{a}{The units are those of the $x$ coordinate: magnitude for apparent magnitudes, absolute magnitudes, and colors; spectral subclass for SpT; and K for $T_{\rm eff}$.}
\tablenotetext{b}{Relation should be used only for sources having separate indications of $M_{J \rm MKO} \ge 16.0$ mag.}
\tablenotetext{c}{Relation should be used only for sources having separate indications of $M_H \ge 15.0$ mag.}
\tablenotetext{d}{Relation should be used only for sources having separate indications of $M_{\rm ch2} \ge 13.0$ mag.}
\tablenotetext{e}{Relation should be used only for sources having separate indications that SpT $\ge$ T4.}
\tablenotetext{f}{Relation should be used only for sources having separate indications that SpT $\ge$ T6.5.}
\tablecomments{These are simple polynomial equations of the form $$y = \sum_{i=0}^{n}c_ix^i.$$ For spectral types, SpT = 0 for L0, SpT = 5 for L5, SpT = 10 for T0, SpT = 15 for T5, SpT = 20 for Y0, etc.}
\end{deluxetable*}
\end{longrotatetable}

\begin{figure*}
\figurenum{14}
\gridline{\fig{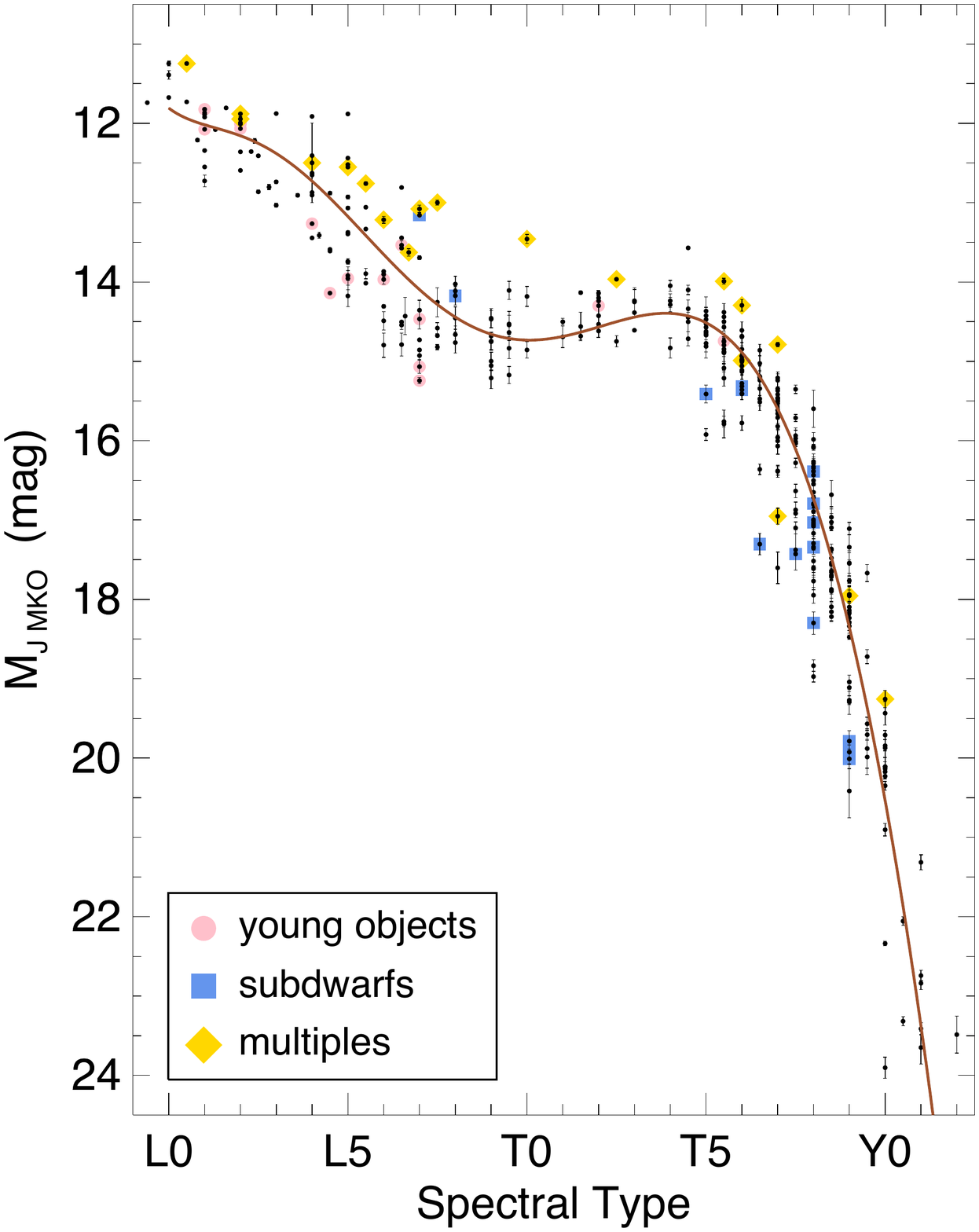}{0.25\textwidth}{(a)}
          \fig{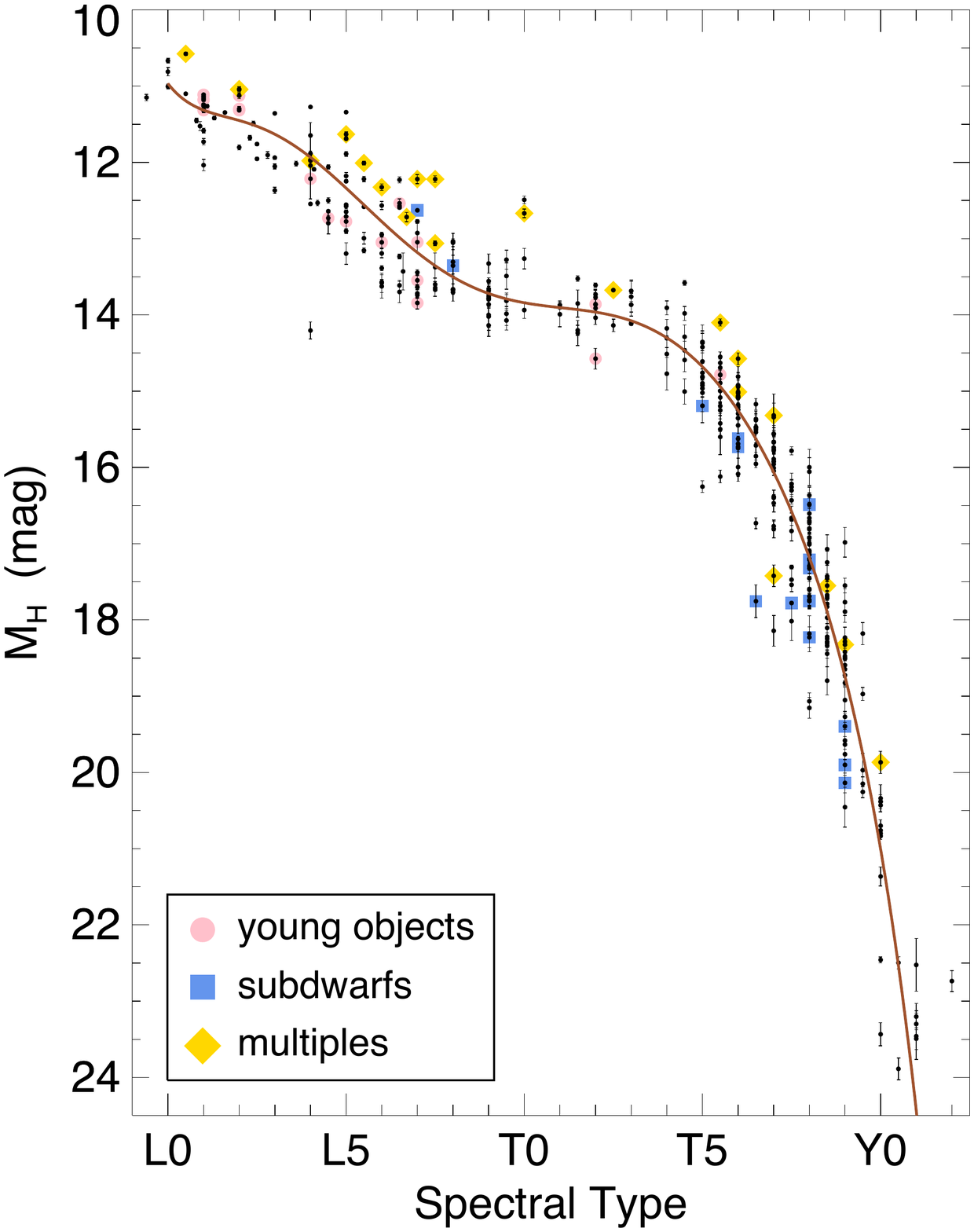}{0.25\textwidth}{(b)}
          \fig{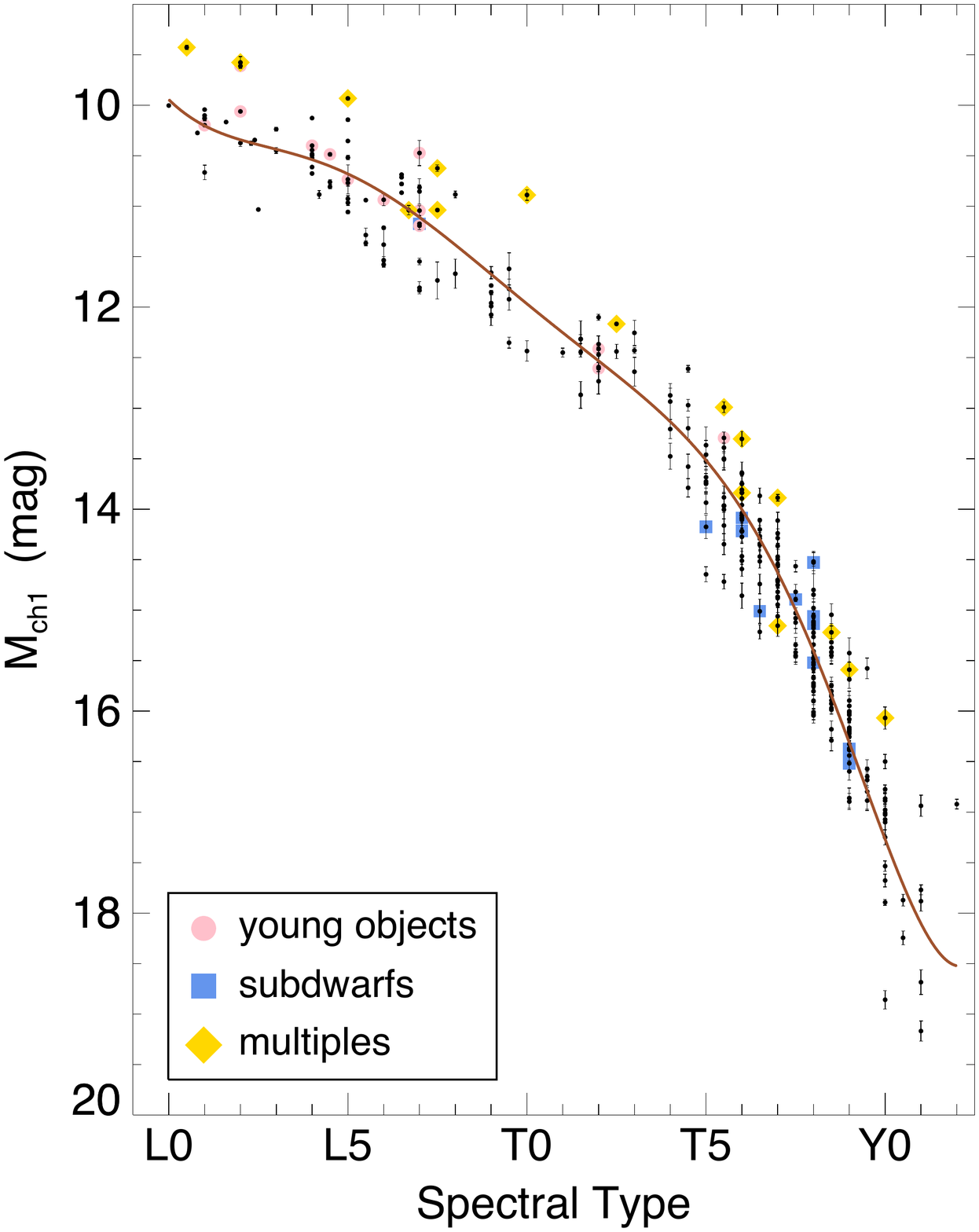}{0.25\textwidth}{(c)}
          \fig{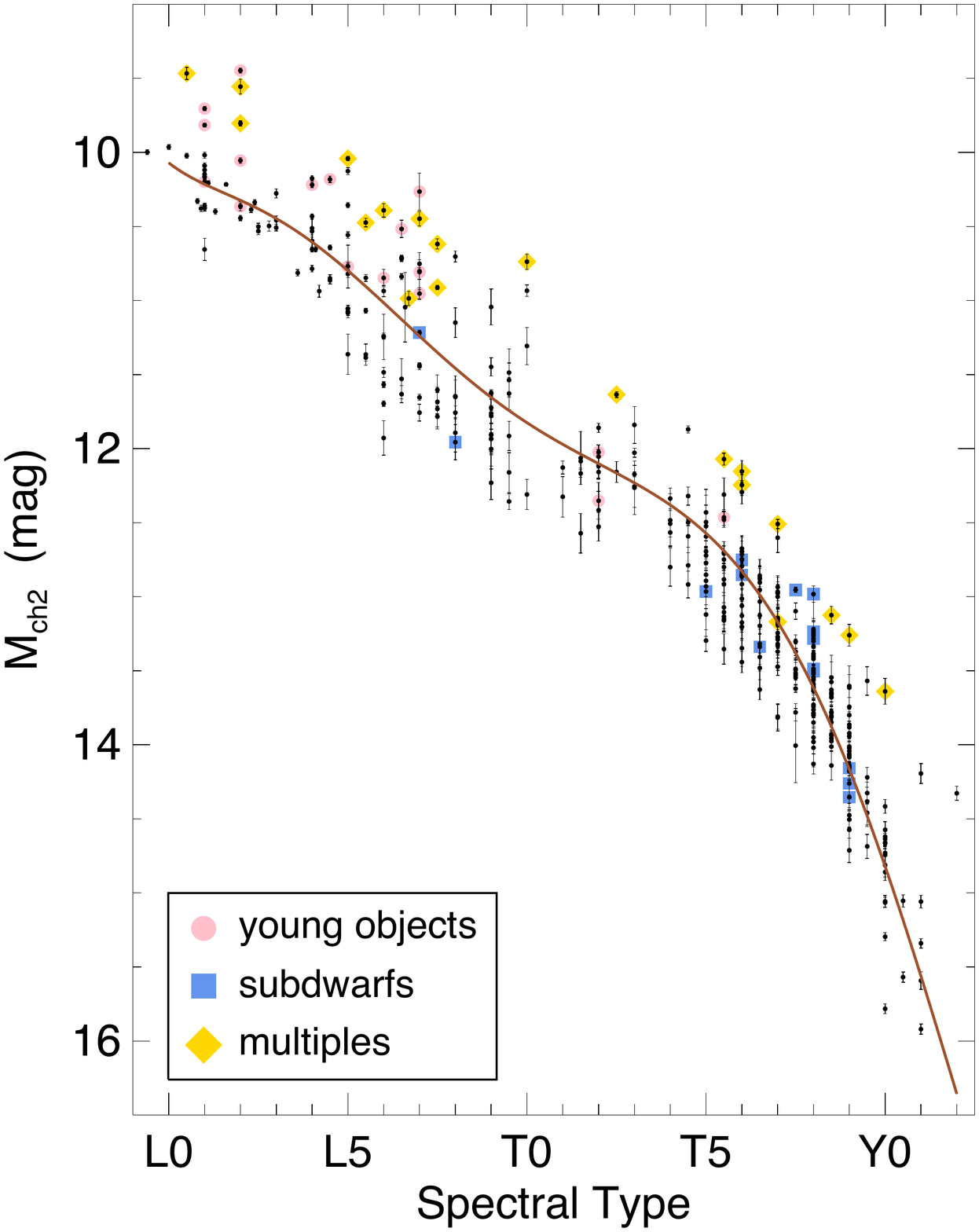}{0.25\textwidth}{(d)}}
\gridline{\fig{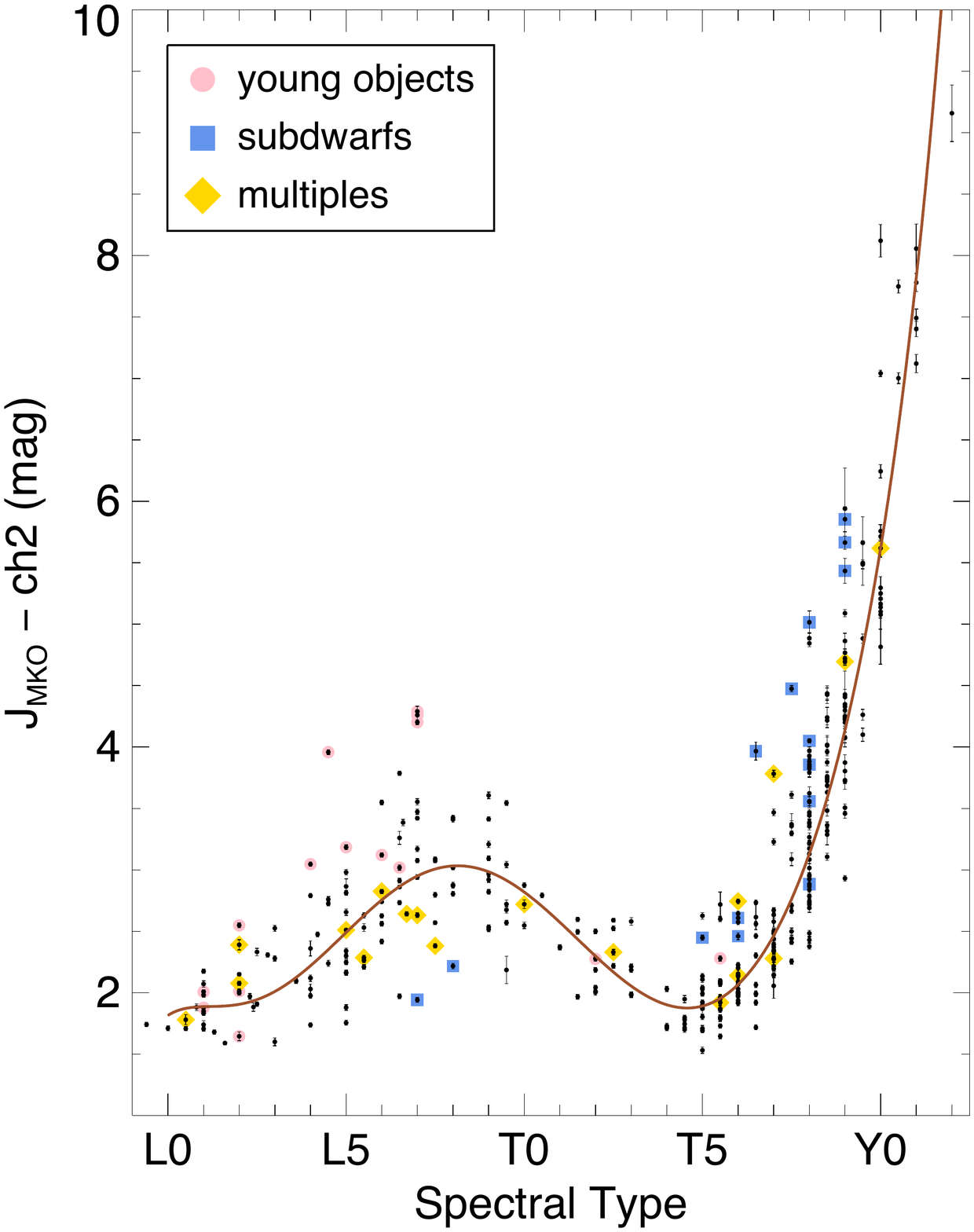}{0.25\textwidth}{(e)}
          \fig{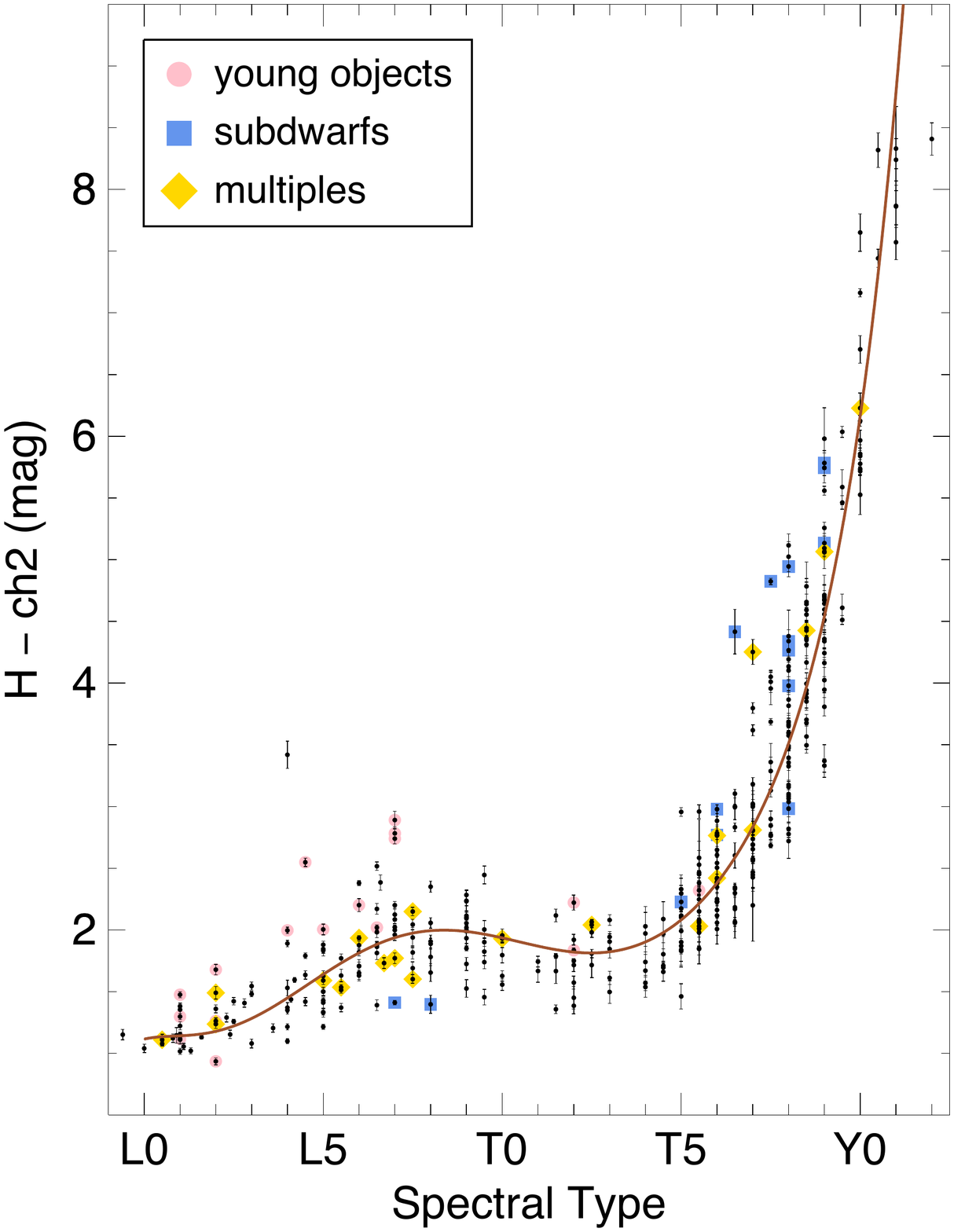}{0.25\textwidth}{(f)}
          \fig{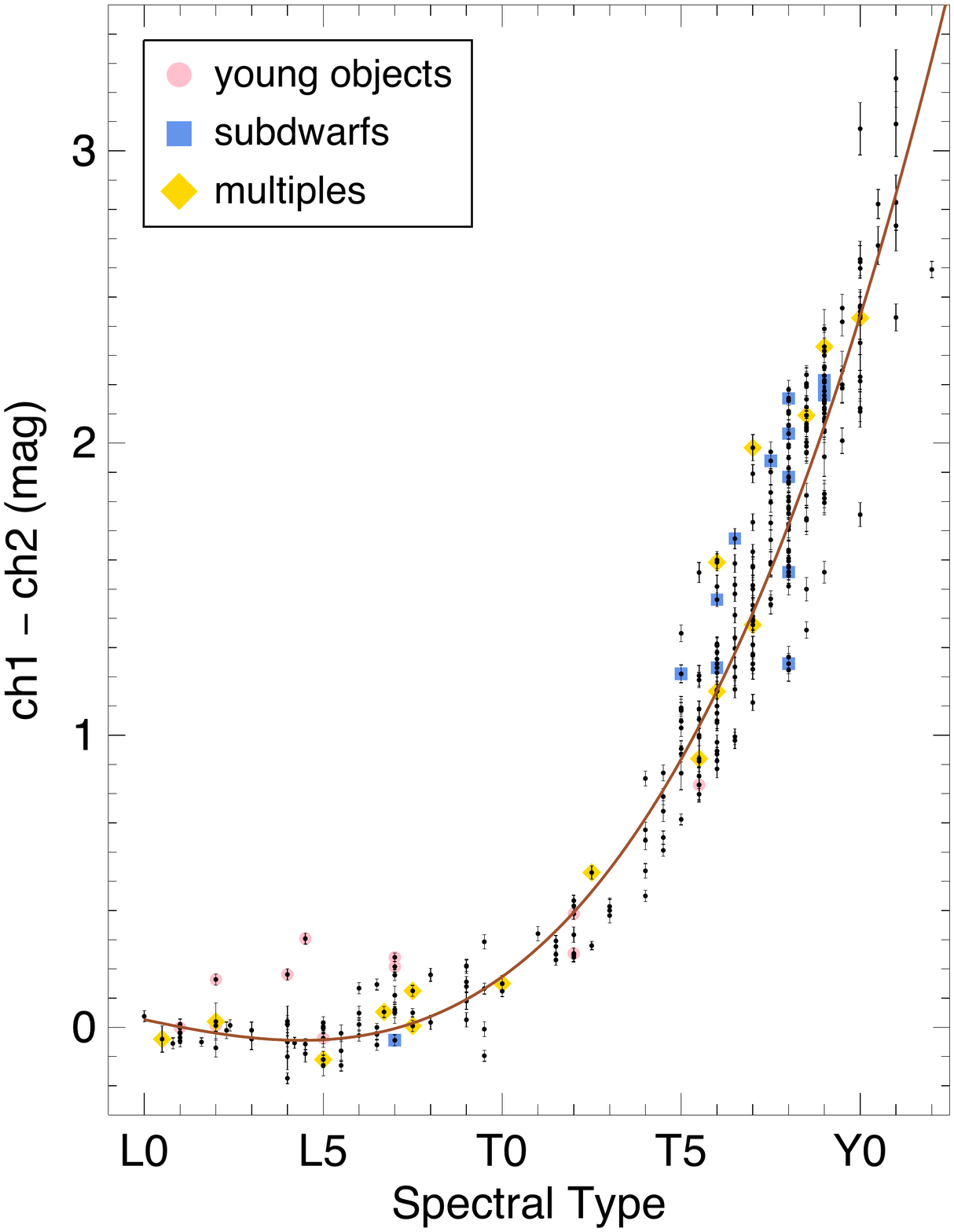}{0.25\textwidth}{(g)}
          \fig{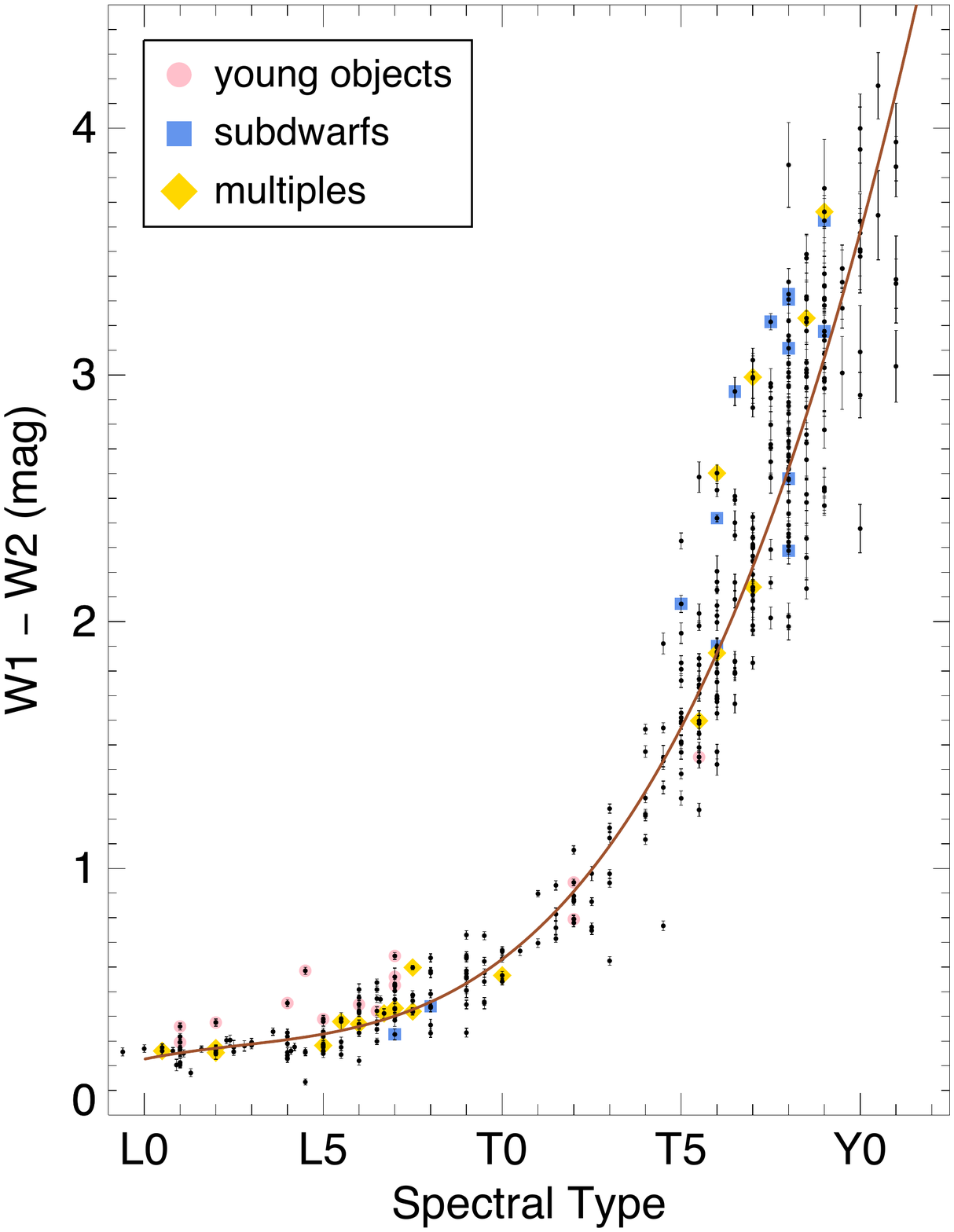}{0.25\textwidth}{(h)}}
\caption{Plots of various absolute magnitudes (a-d) and colors (e-h) as a function of near-infrared spectral type. Only members of the 20-pc census are shown, and plots a-d show only the subset of 20-pc objects having parallaxes measured to better than 12.5\%. Plots of $M_{\rm ch2}$, $J_{\rm MKO} - {\rm ch2}$, and $H - {\rm ch2}$ are supplemented with W2 magnitudes when ch2 magnitudes are not available, as described in section~\ref{section:plot_analysis}. Polynomial fits that exclude known young objects (pink circles, section~\ref{section:known_youngs}), subdwarfs (blue squares, section~\ref{section:known_sds}), and multiple systems (yellow diamonds, section~\ref{section:known_multiples}) are shown in brown and described in Table~\ref{table:equations}.
\label{figure:xaxis_spectype}}
\end{figure*}

Trends of absolute magnitude with spectral type are illustrated in Figure~\ref{figure:xaxis_spectype}(a-d). Most of the known, unresolved doubles have components of nearly equal magnitudes, and not surprisingly, most of these objects stand out as overluminous for their types relative to the main trends. On the plots of $M_{J \rm MKO}$ and $M_H$, T-type subdwarfs tend to be overluminous with respect to the mean trend, whereas L-type subdwarfs are underluminous, although for the latter there are only two examples with which to judge. At $M_{\rm ch1}$ and $M_{\rm ch2}$, the subdwarfs are indistiguishable from objects of solar metallicity. Young L dwarfs within 20 pc tend to be overluminous with respect to the mean trend in all four absolute magnitudes, whereas young T dwarfs -- at least for the three known examples -- do not distinguish themselves from the run of older T dwarfs.

Trends of colors with spectral type are illustrated in Figure~\ref{figure:xaxis_spectype}(e-h). The two known L subdwarfs are much bluer than the mean trend in $J_{\rm MKO} - {\rm ch2}$ and $H - {\rm ch2}$ colors, though indistinguishable from the mean trend in ${\rm ch1} - {\rm ch2}$ and ${\rm W1} - {\rm W2}$. The T subdwarfs tend to lie redward of the mean trend in all four colors. Young L dwarfs are markedly redder than the trend in all four colors, whereas the few young T dwarfs known do not clearly differentiate themselves.

\begin{figure*}
\figurenum{15}
\gridline{\fig{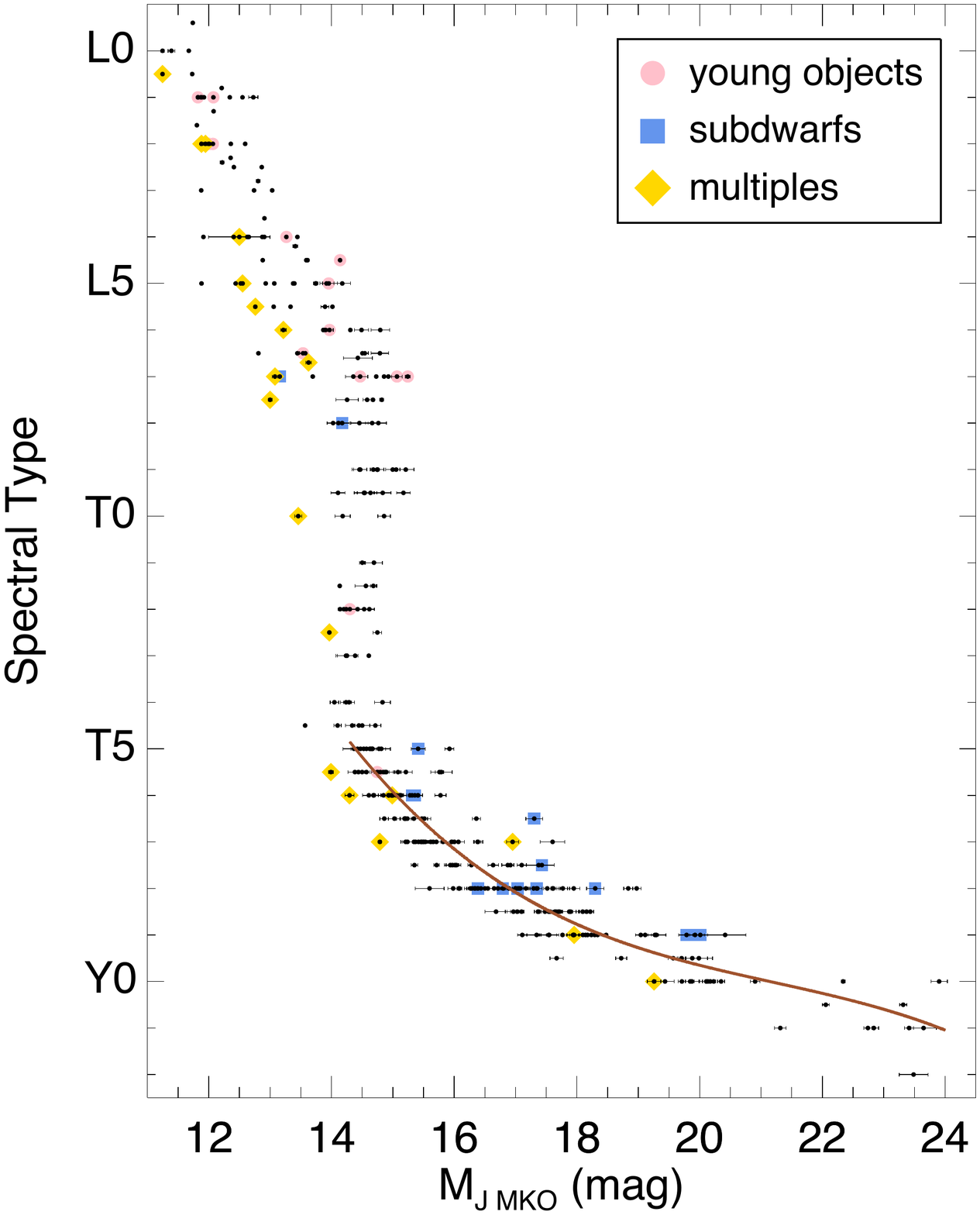}{0.25\textwidth}{(a)}
          \fig{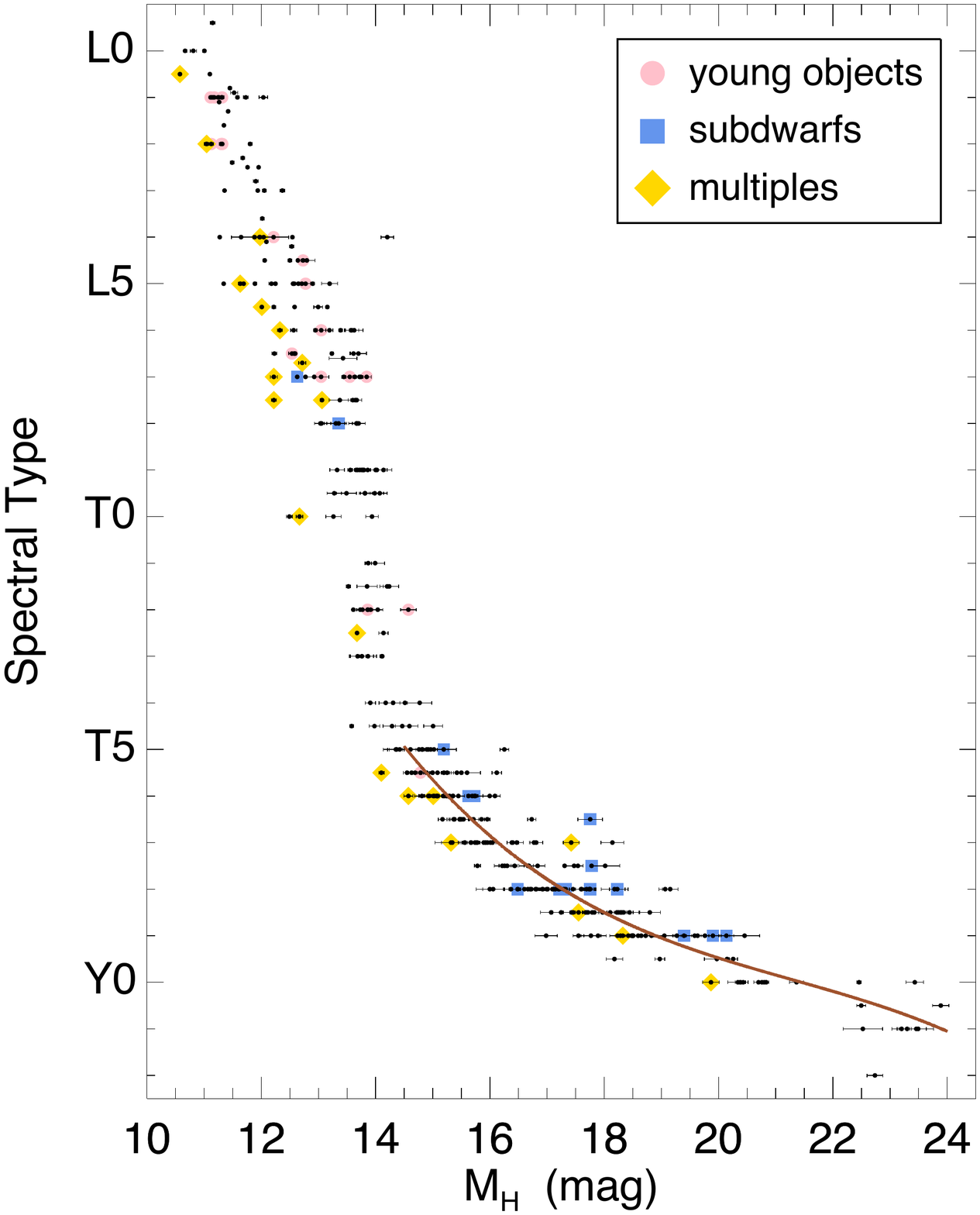}{0.25\textwidth}{(b)}
          \fig{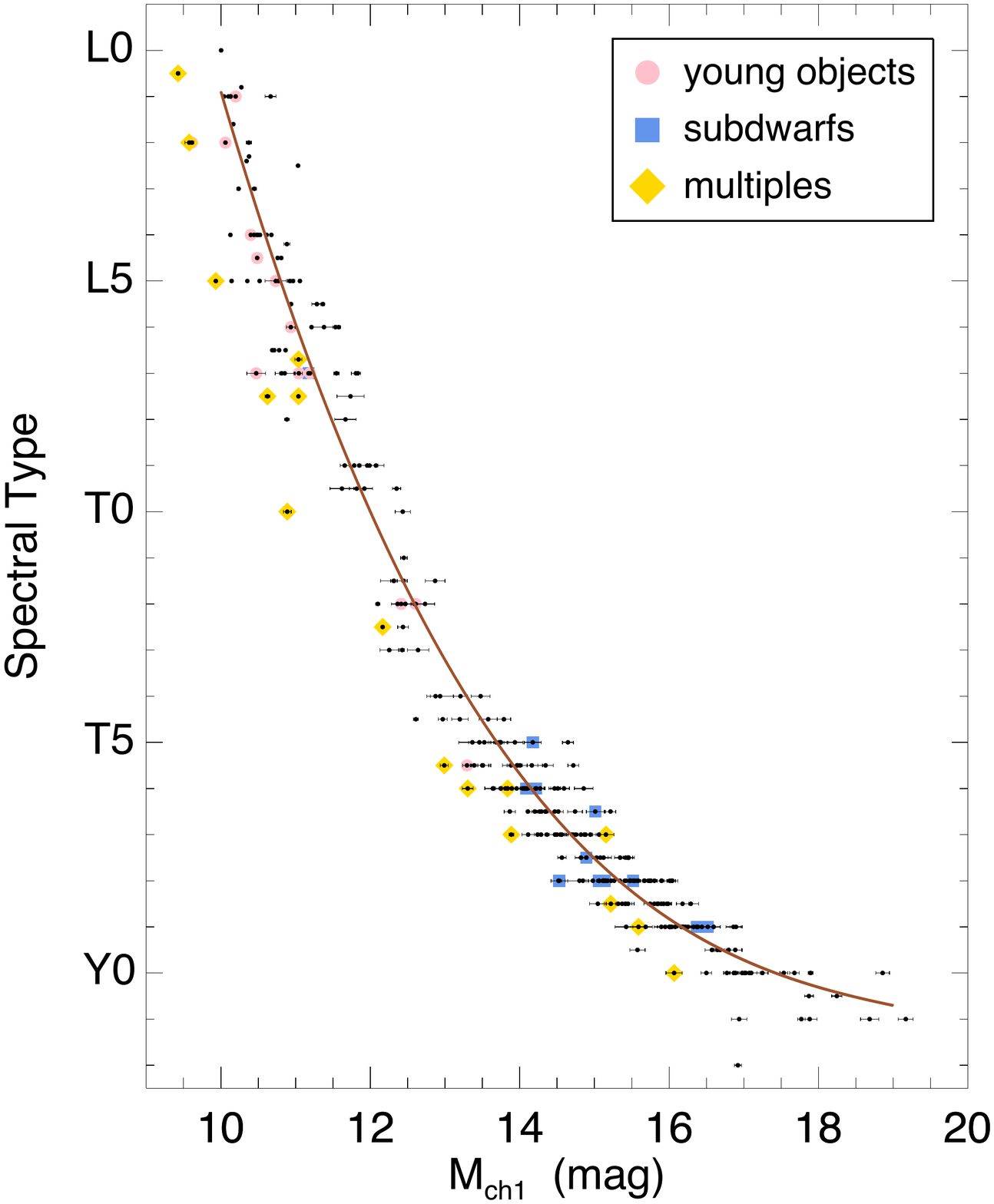}{0.25\textwidth}{(c)}
          \fig{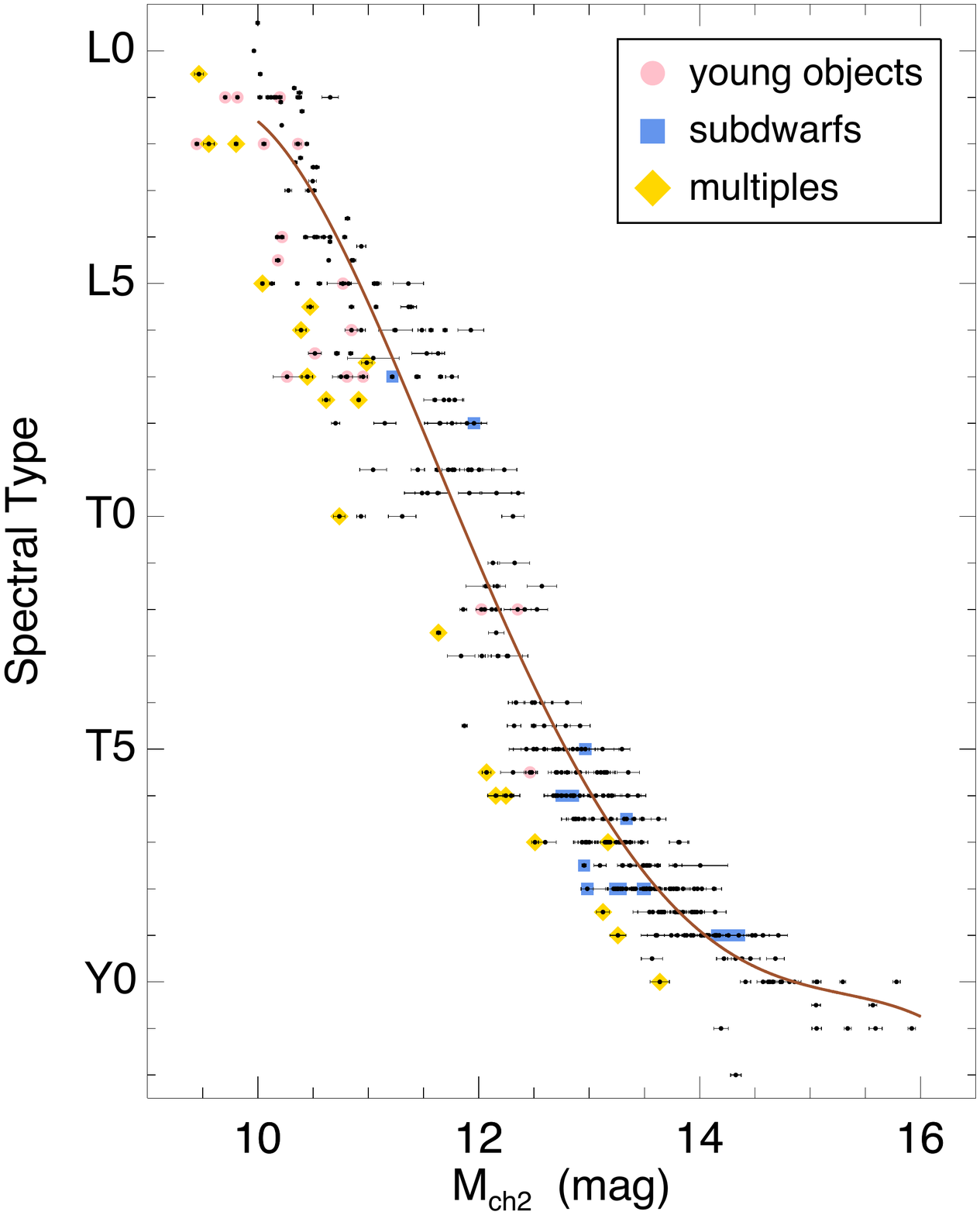}{0.25\textwidth}{(d)}}
\gridline{\fig{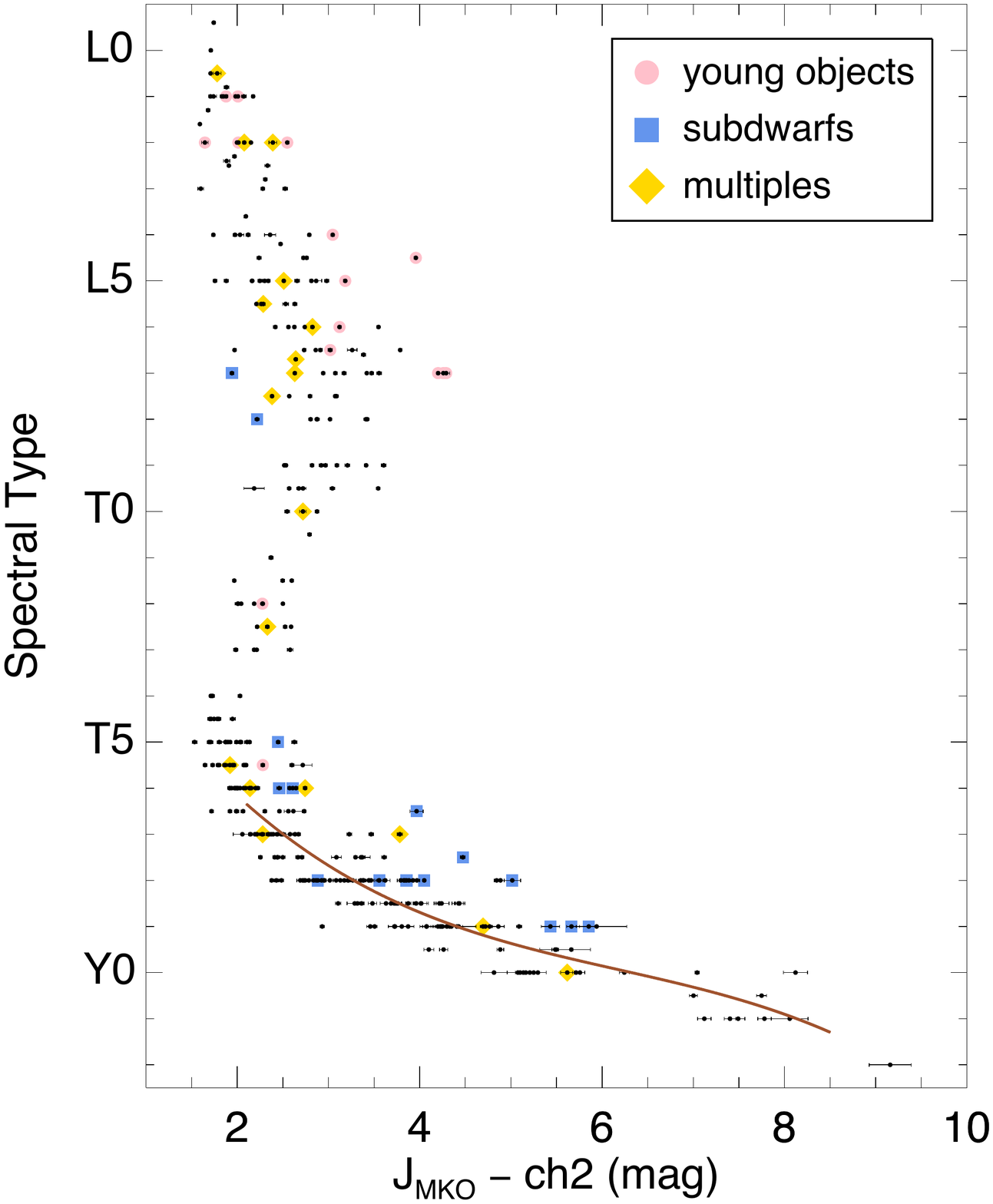}{0.25\textwidth}{(e)}
          \fig{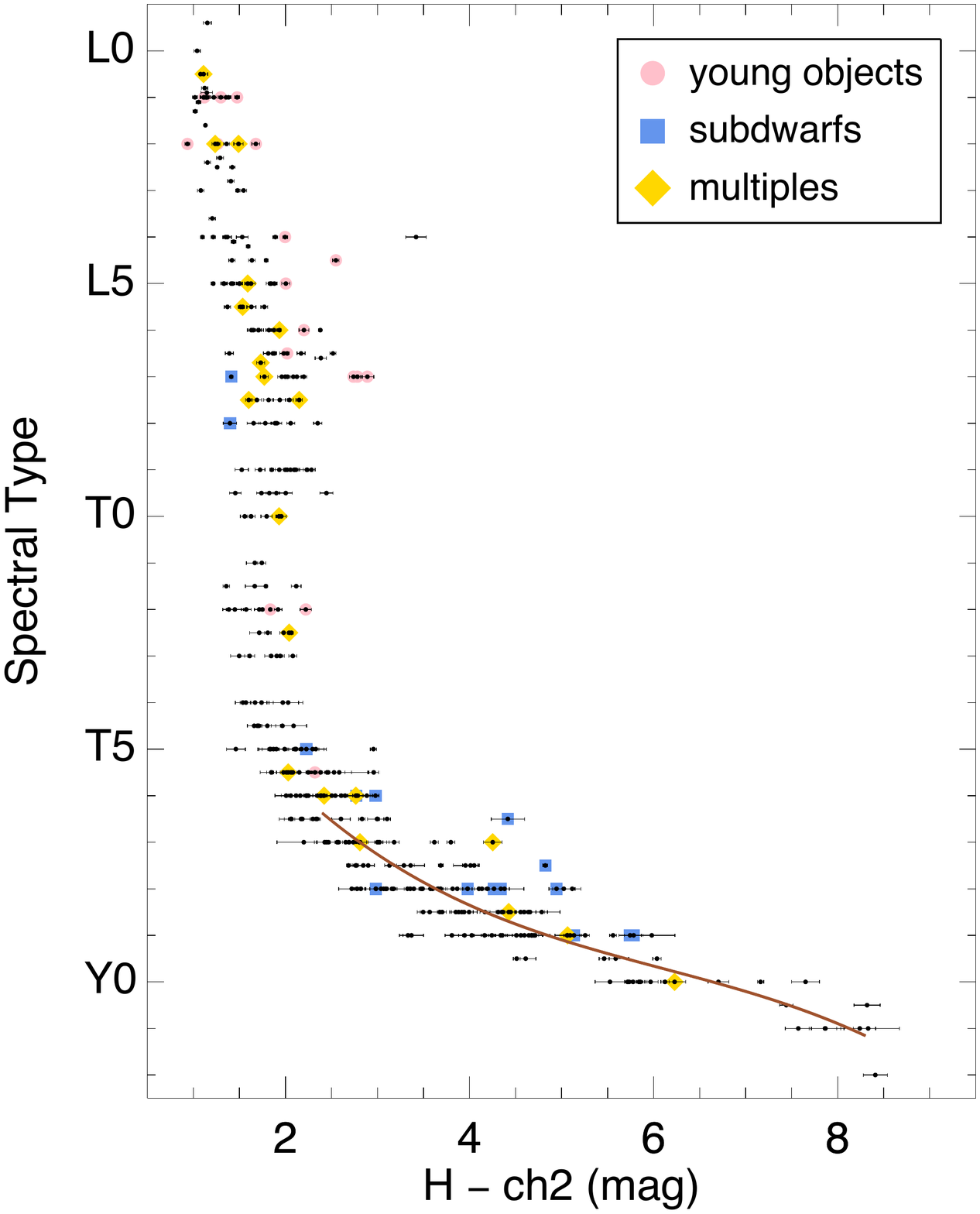}{0.25\textwidth}{(f)}
          \fig{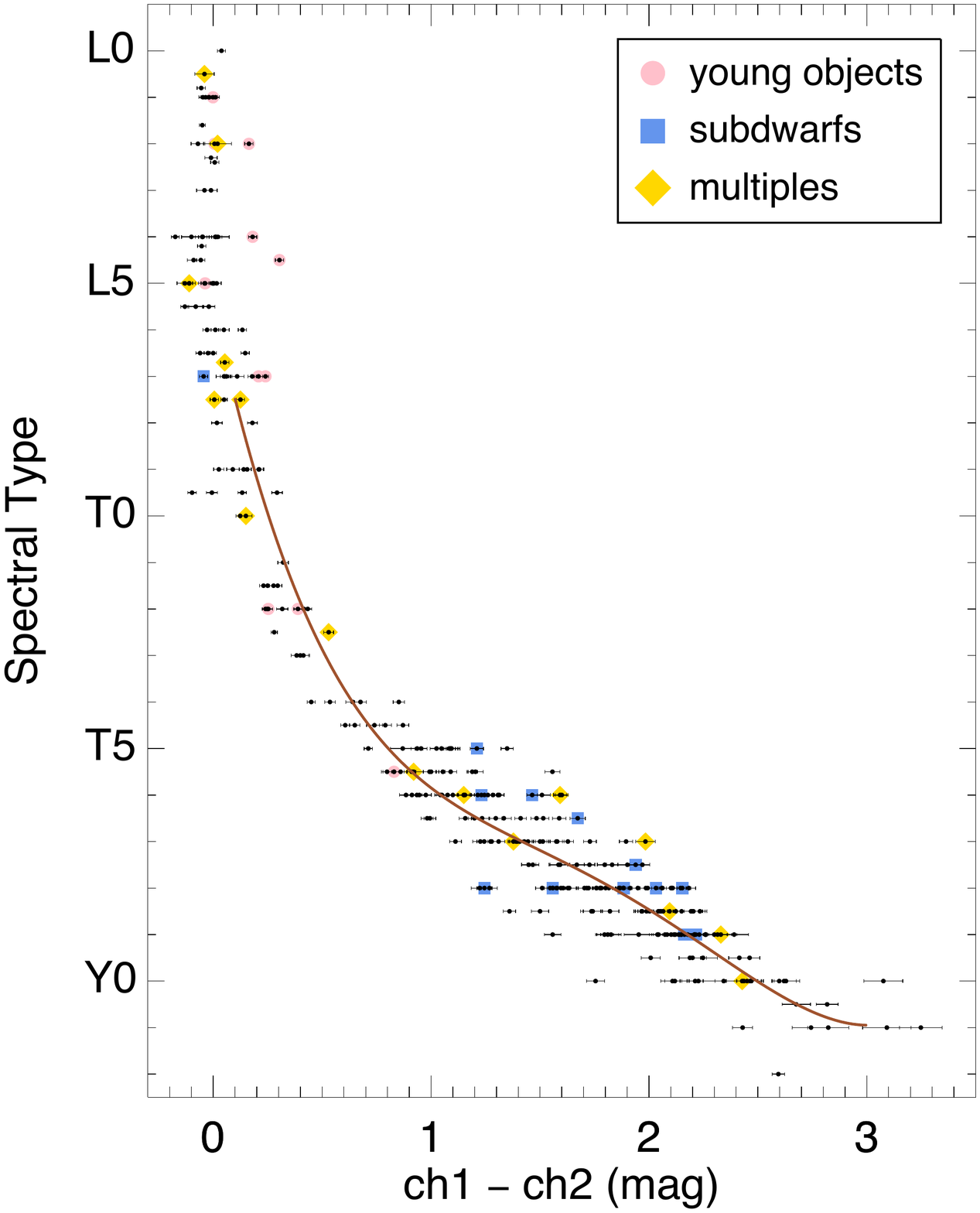}{0.25\textwidth}{(g)}
          \fig{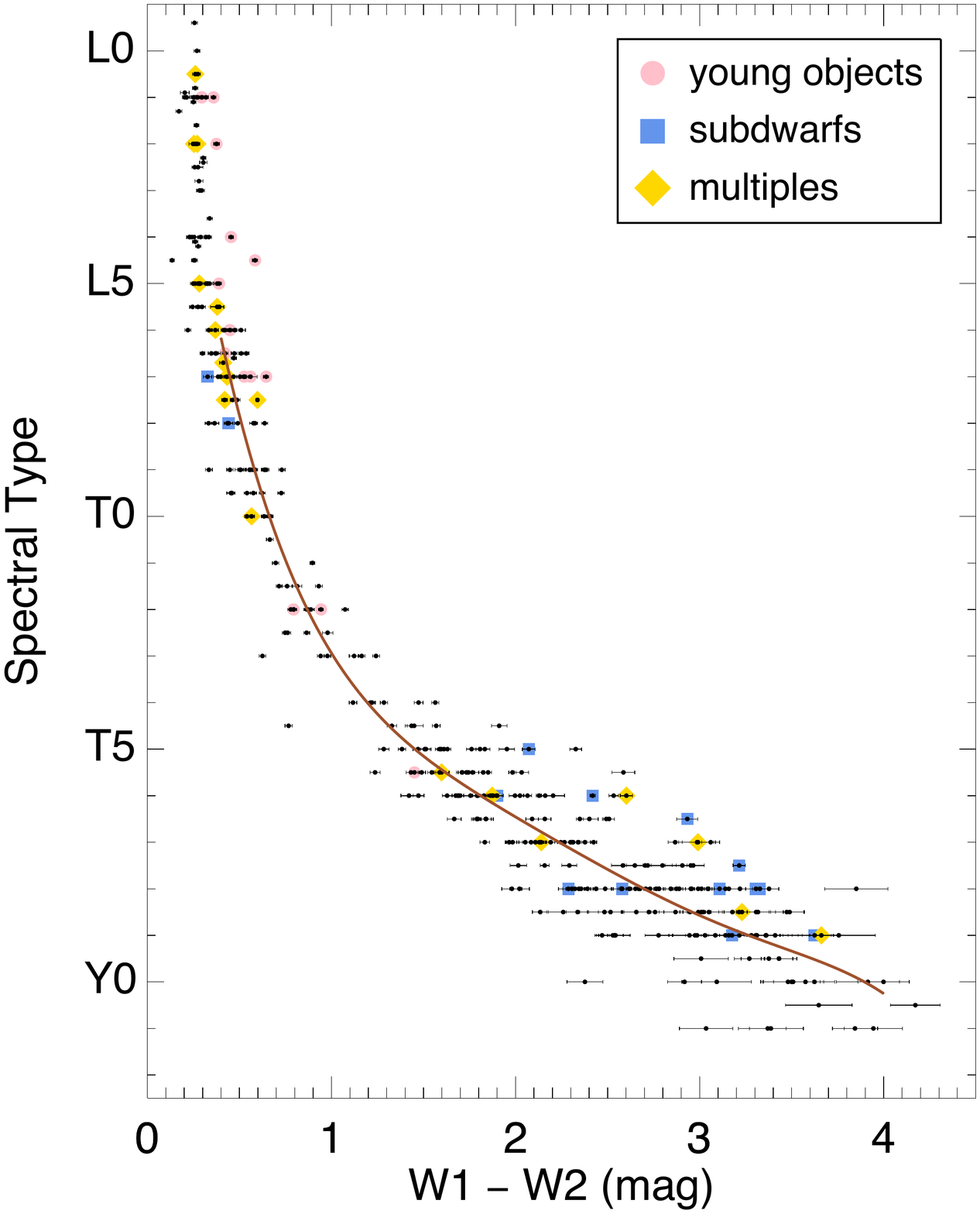}{0.25\textwidth}{(h)}}
\caption{Plots identical to those of Figure~\ref{figure:xaxis_spectype}, except that the $x$ and $y$ axes have been reversed. Polynomial fits to provide a translation from absolute magnitude or color into spectral type are shown in brown and described in Table~\ref{table:equations}. These fits exclude known young objects (pink circles, section~\ref{section:known_youngs}), subdwarfs (blue squares, section~\ref{section:known_sds}), and multiple systems (yellow diamonds, section~\ref{section:known_multiples}).\label{figure:yaxis_spectype}}
\end{figure*}

In Figure~\ref{figure:yaxis_spectype}(a-h), we show these same plots as above, but with the axes flipped. This is to provide researchers with fits to convert absolute magnitudes or colors to a spectral type. As is illustrated in the plots, it is not always possible to provide simple polynomial fits over the entire range of absolute magnitude or color because of degeneracies. For example, a color of $J_{\rm MKO} - {\rm ch2} = 3.0$ mag corresponds to either a mid/late-L dwarf or a mid/late-T dwarf. Users are urged to check the notes in Table~\ref{table:equations} to check the ranges over which these fits are valid.

\begin{figure*}
\figurenum{16}
\gridline{\fig{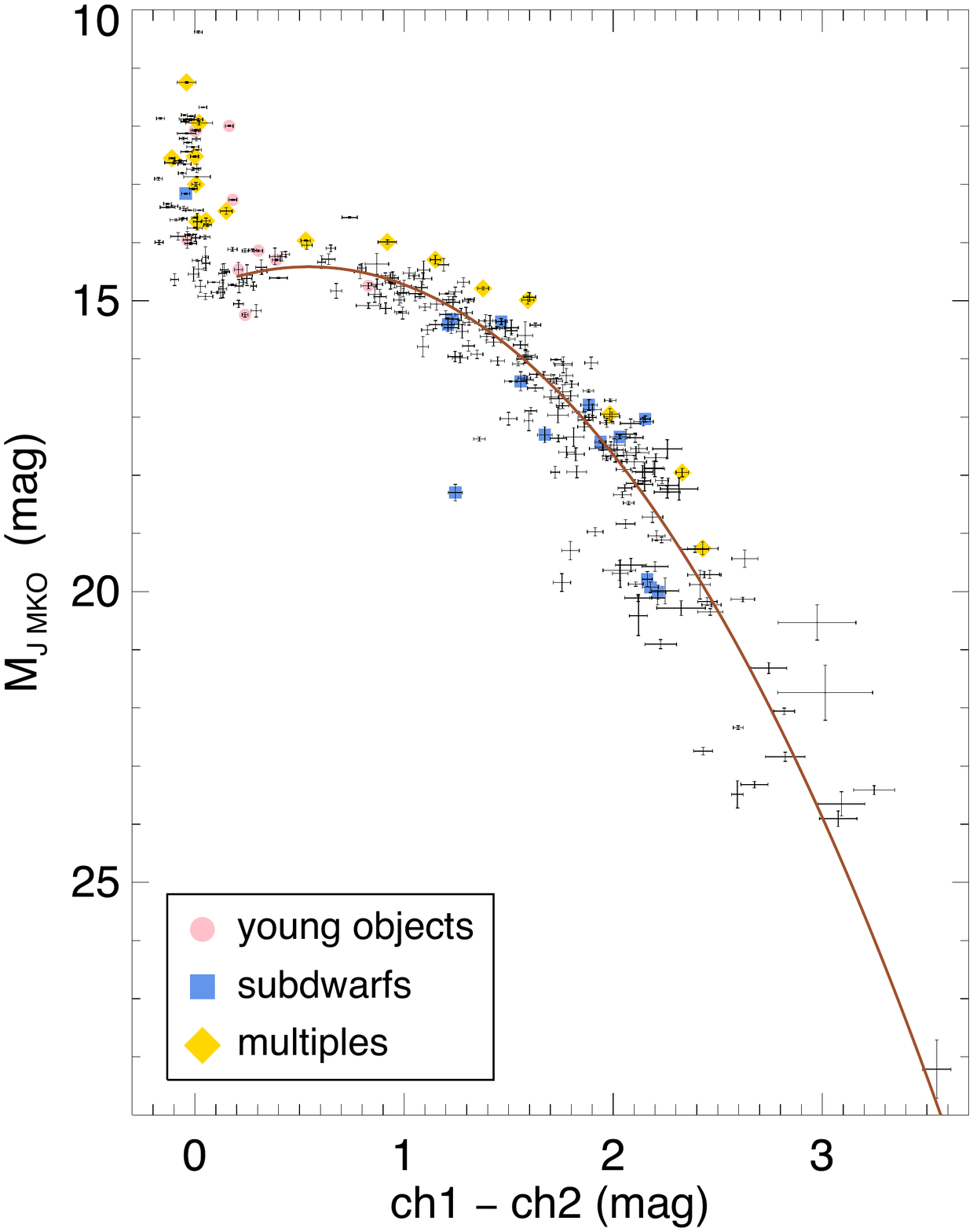}{0.33\textwidth}{(a)}
          \fig{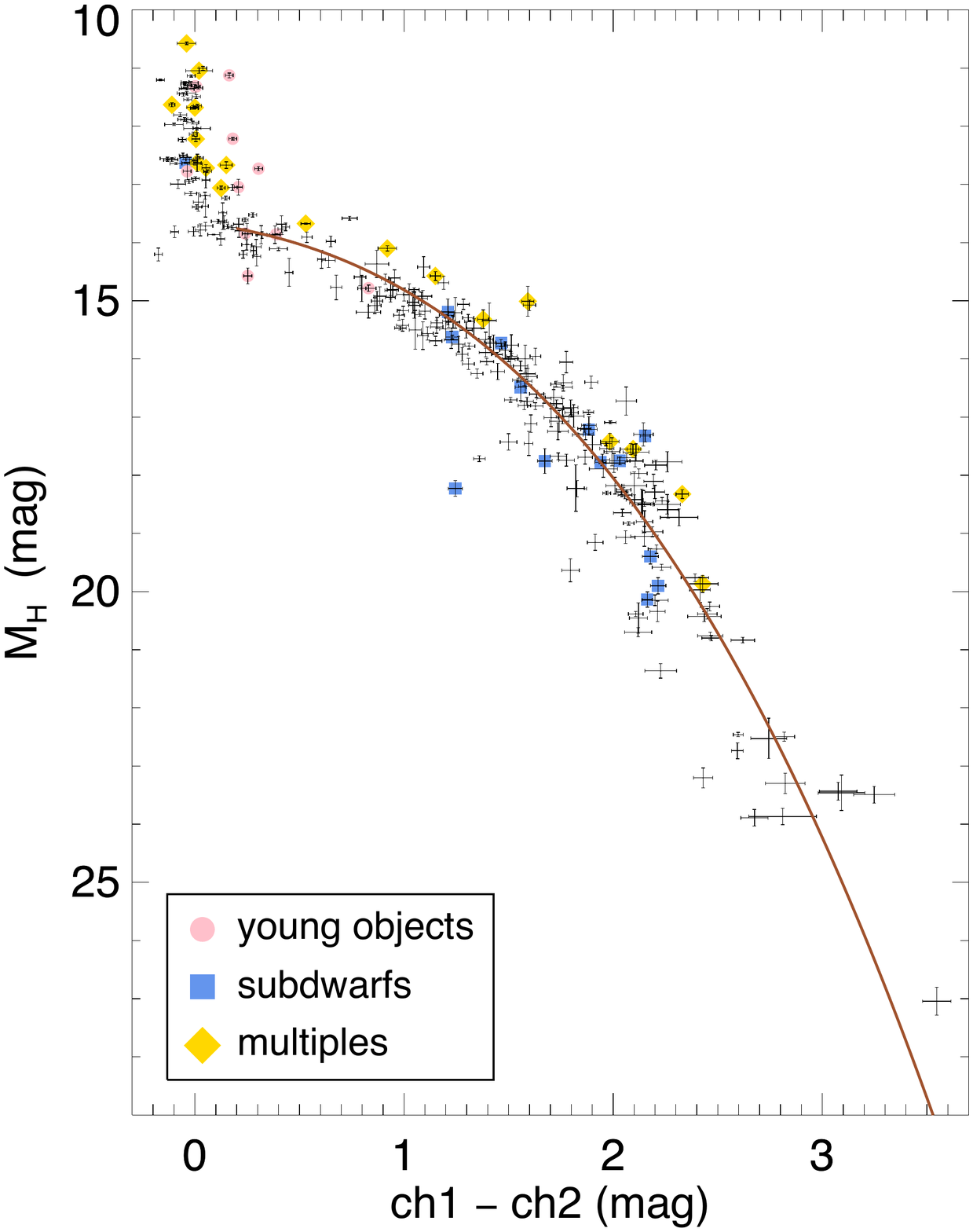}{0.33\textwidth}{(b)}
          \fig{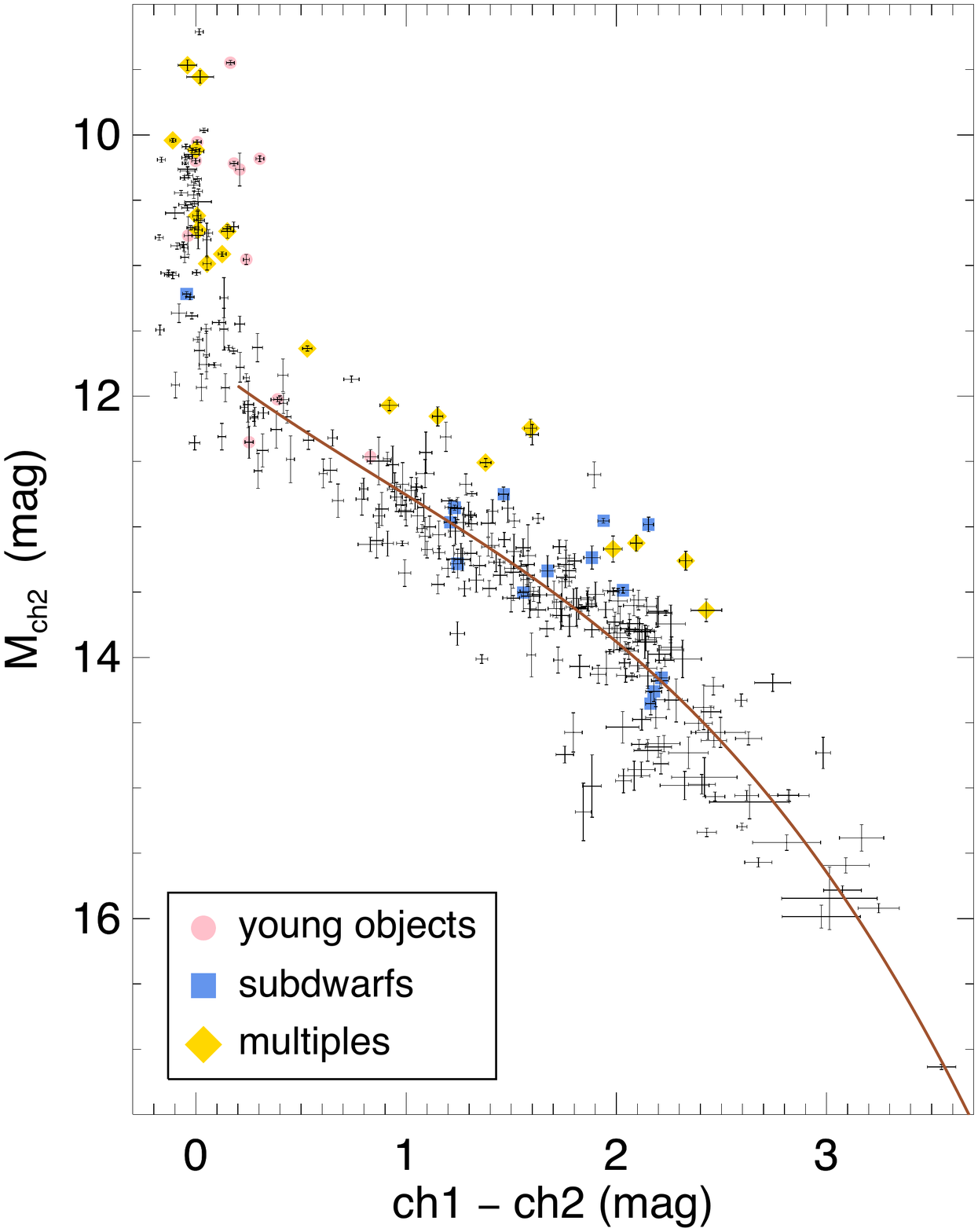}{0.33\textwidth}{(c)}}
\gridline{\fig{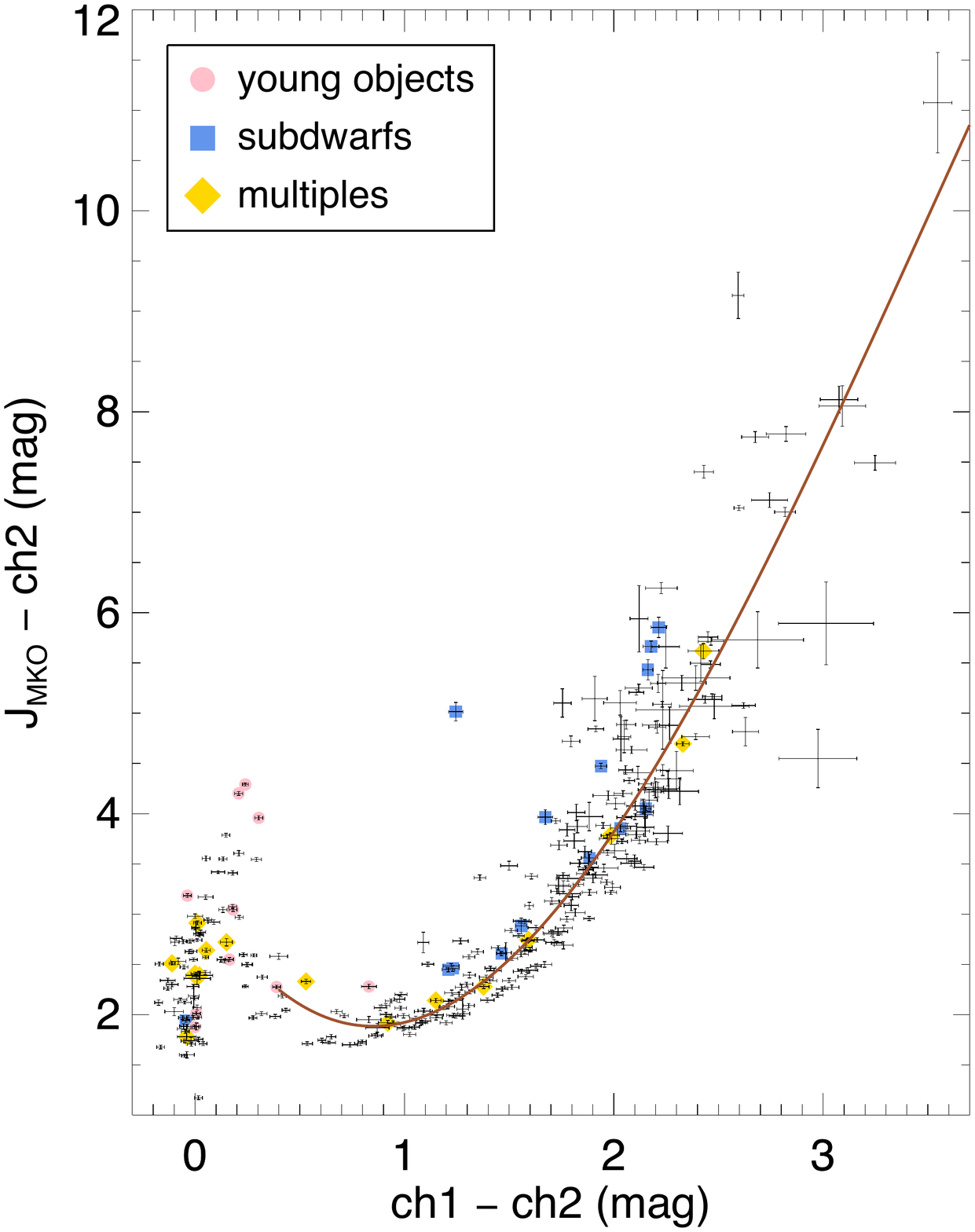}{0.33\textwidth}{(d)}
          \fig{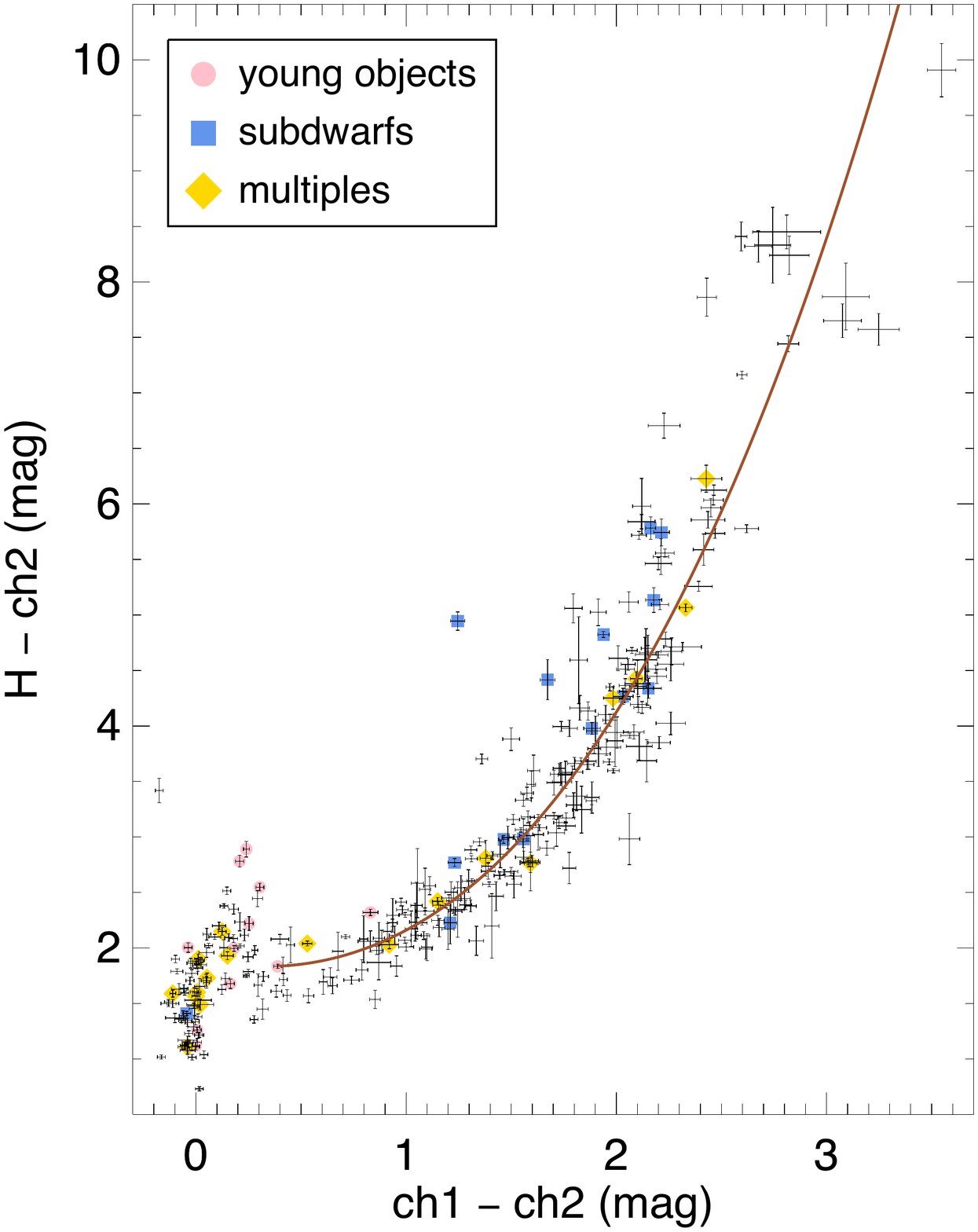}{0.33\textwidth}{(e)}
          \fig{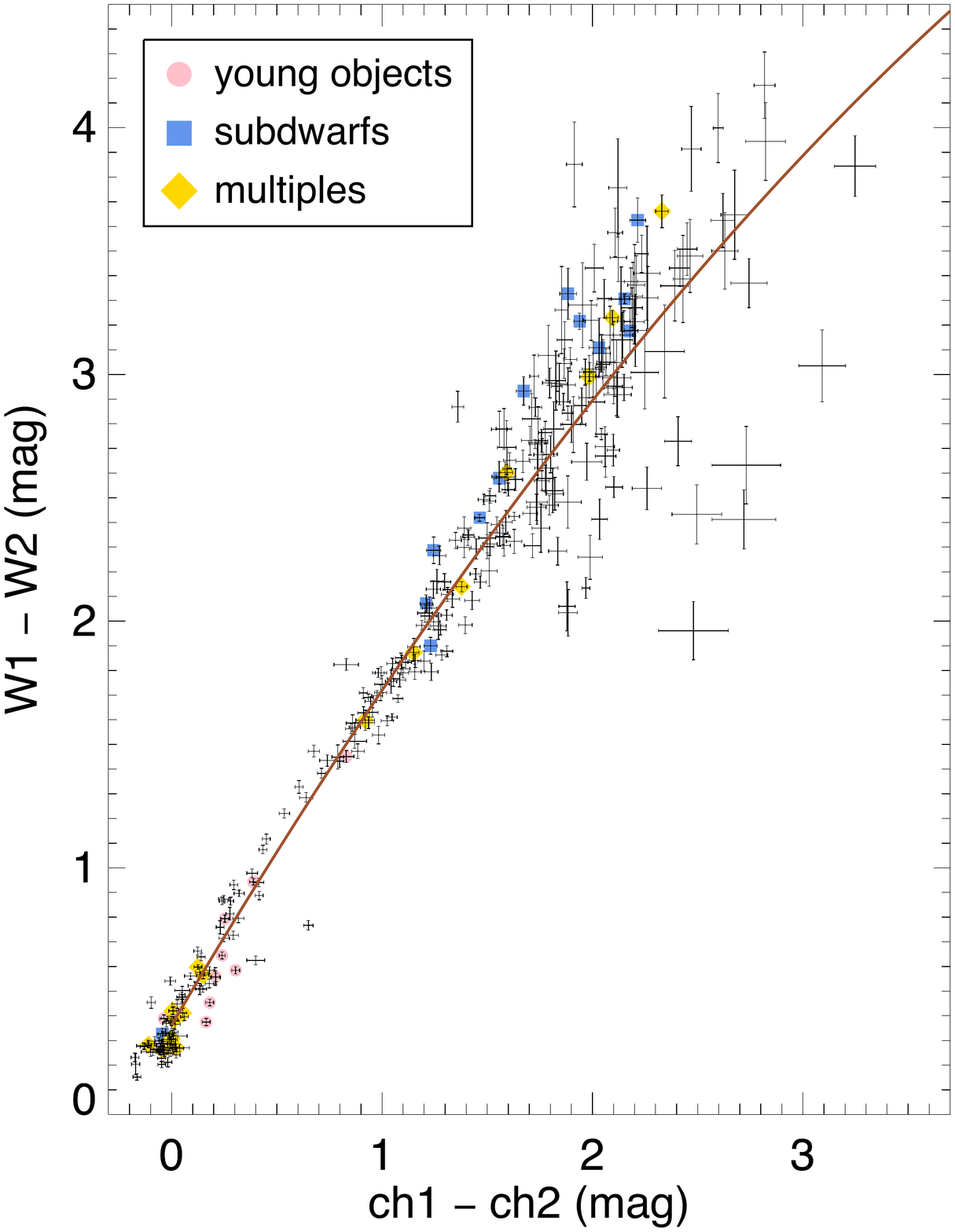}{0.33\textwidth}{(f)}}
\caption{Plots of various absolute magnitudes (a-c) and colors (d-f) as a function of ${\rm ch1} - {\rm ch2}$ color. Only members of the 20-pc census are shown, and plots a-c show only the subset of 20-pc objects having parallaxes measured to better than 12.5\%. Polynomial fits that exclude known young objects (pink circles, section~\ref{section:known_youngs}), subdwarfs (blue squares, section~\ref{section:known_sds}), and multiple systems (yellow diamonds, section~\ref{section:known_multiples}) are shown in brown and described in Table~\ref{table:equations}. Fits include only those points with ${\rm ch1} - {\rm ch2} > 0.2$ mag for panels a-e.
\label{figure:xaxis_ch1ch2}}
\end{figure*}

In Figure~\ref{figure:xaxis_ch1ch2}(a-f), we illustrate trends of absolute magnitudes and colors as a function of ${\rm ch1} - {\rm ch2}$ color. In the plots of absolute magnitude, multiples are seen as overluminous, as expected, and only the most metal poor T subdwarf, WISE 2005+5424 ([Fe/H] = $-0.64{\pm}0.17$) is well removed from the trend in $M_{J \rm MKO}$ and $M_H$. On the color plots, the T subdwarfs are redder in $J_{\rm MKO} - {\rm ch2}$, $H - {\rm ch2}$, and ${\rm W1} - {\rm W2}$ at a fixed value of ${\rm ch1} - {\rm ch2}$.

\begin{figure*}
\figurenum{17}
\gridline{\fig{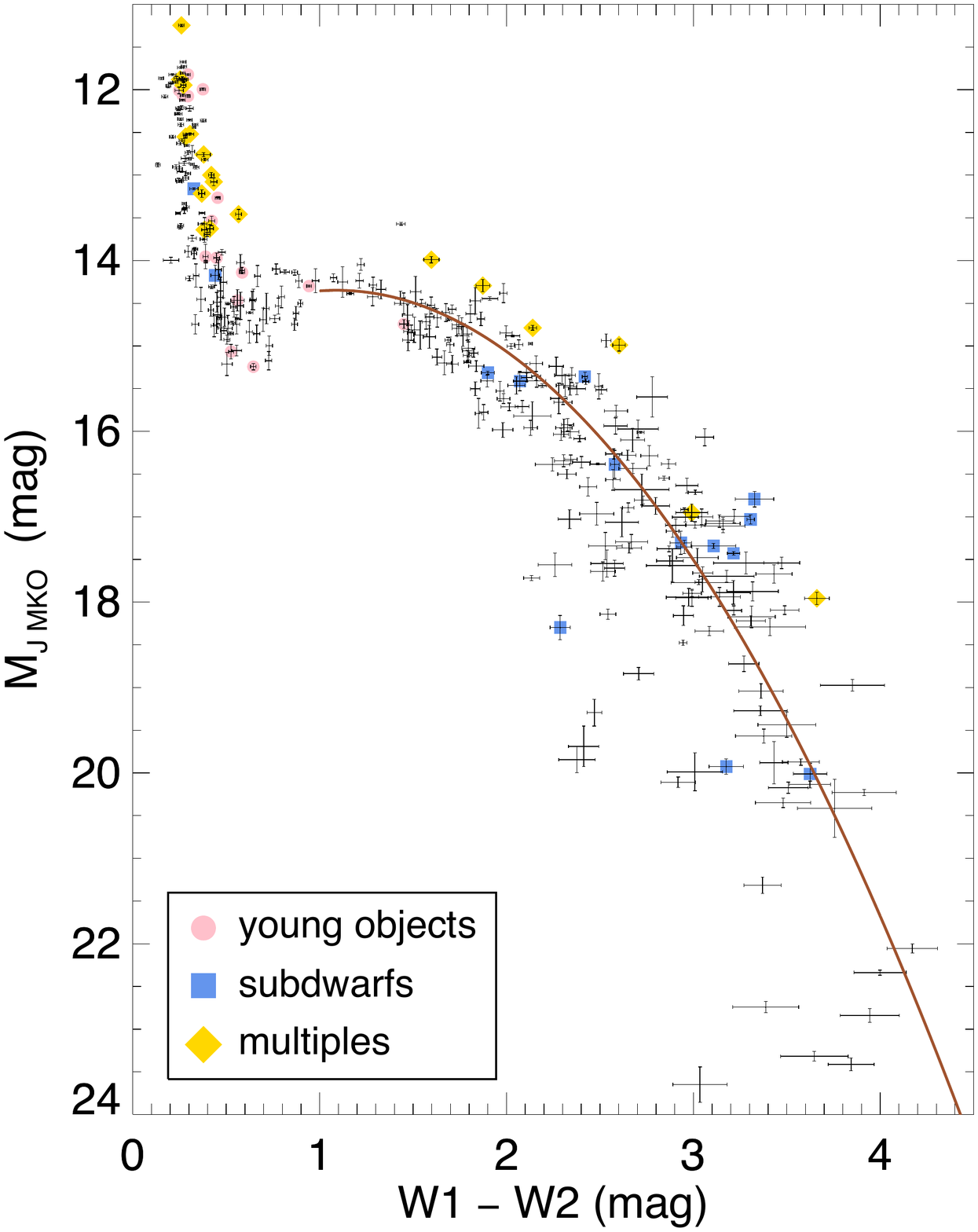}{0.33\textwidth}{(a)}
          \fig{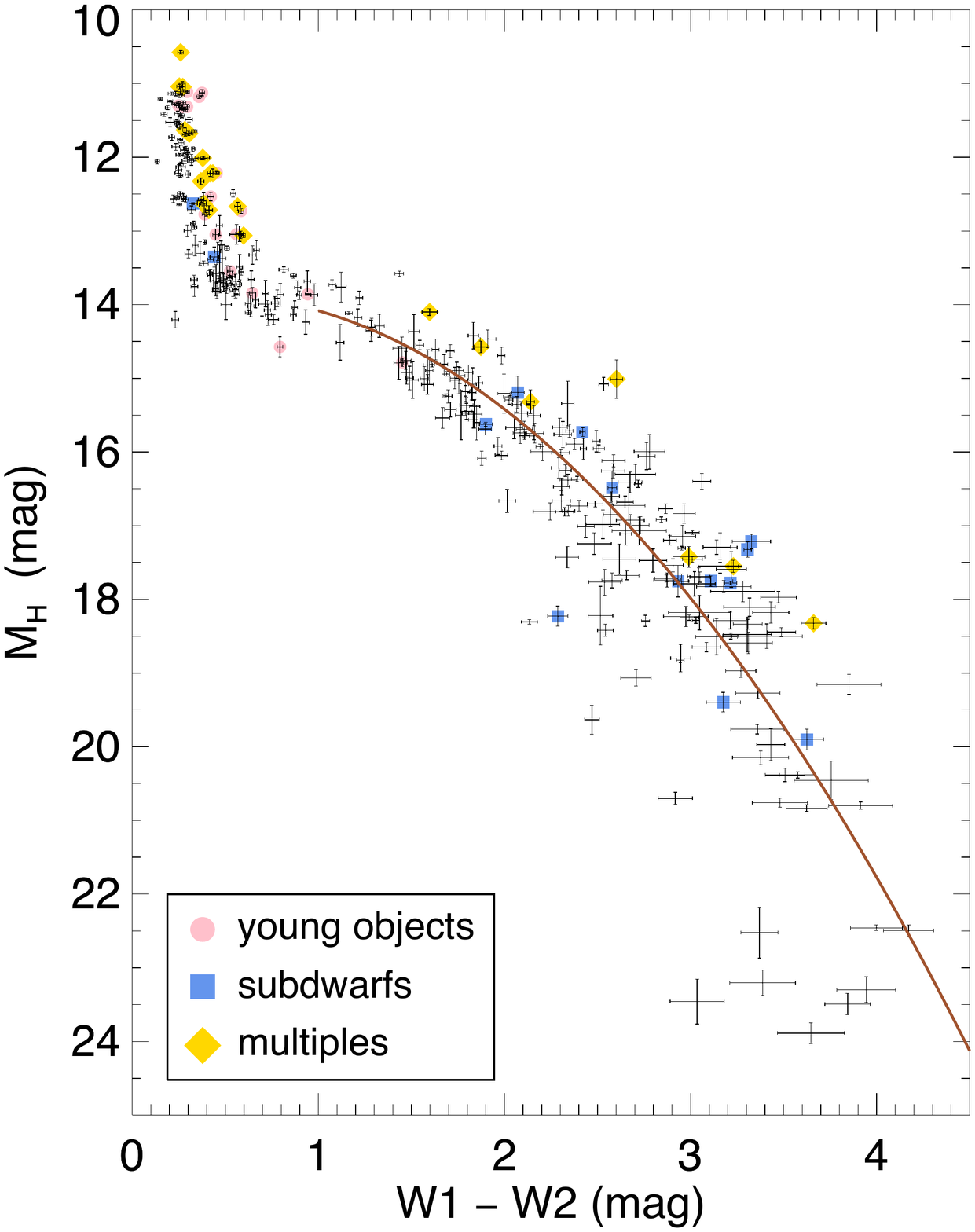}{0.33\textwidth}{(b)}
          \fig{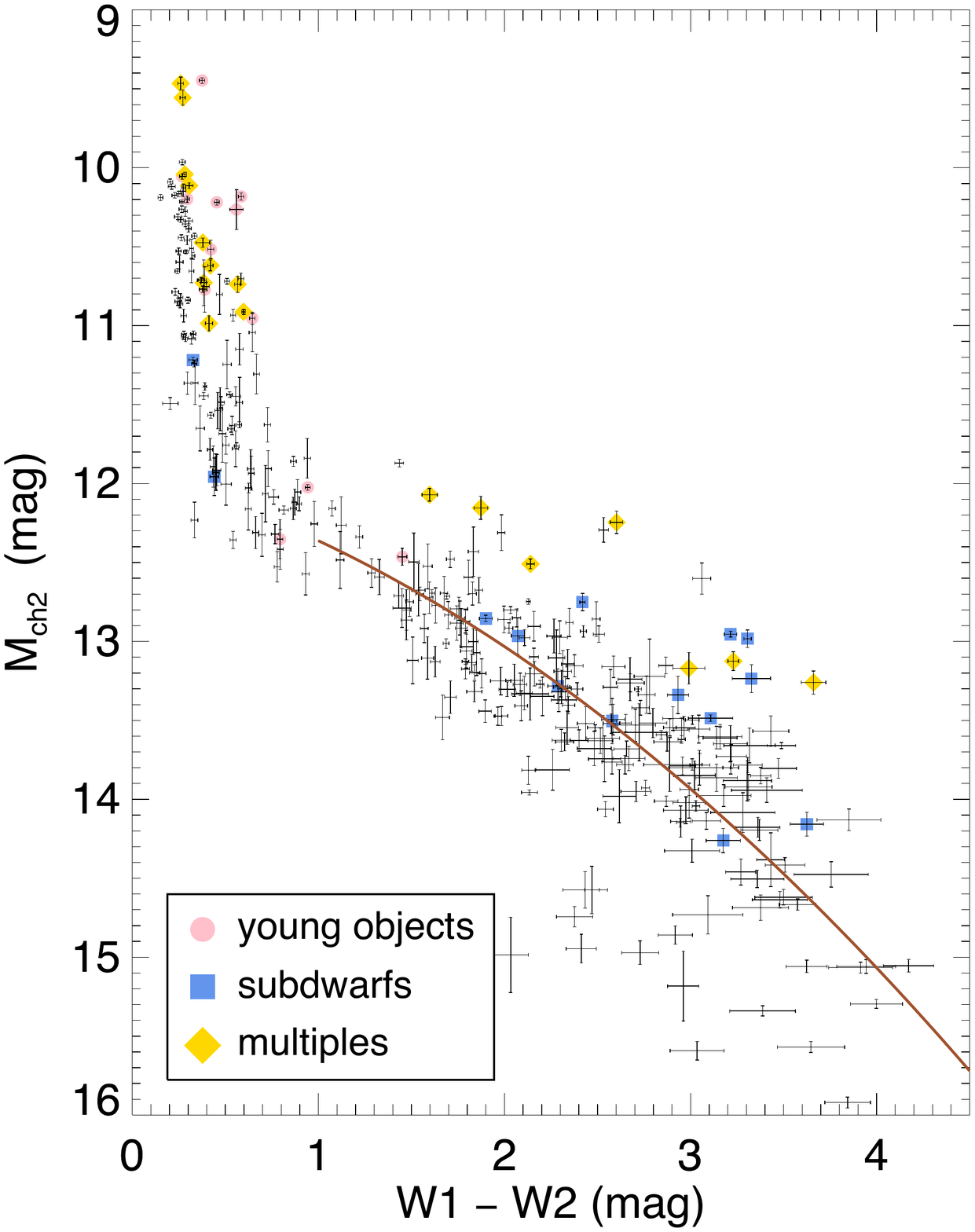}{0.33\textwidth}{(c)}}
\gridline{\fig{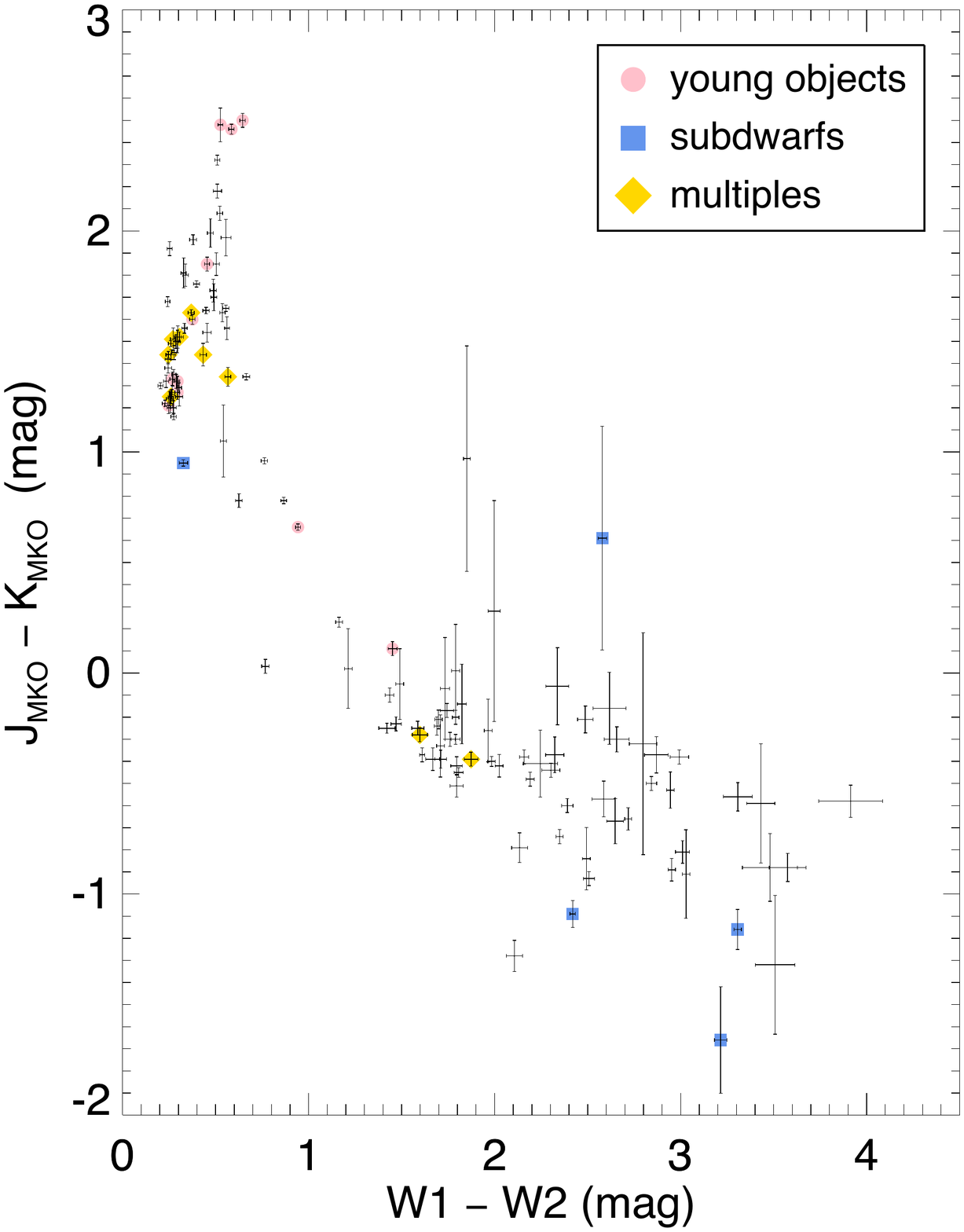}{0.33\textwidth}{(d)}
          \fig{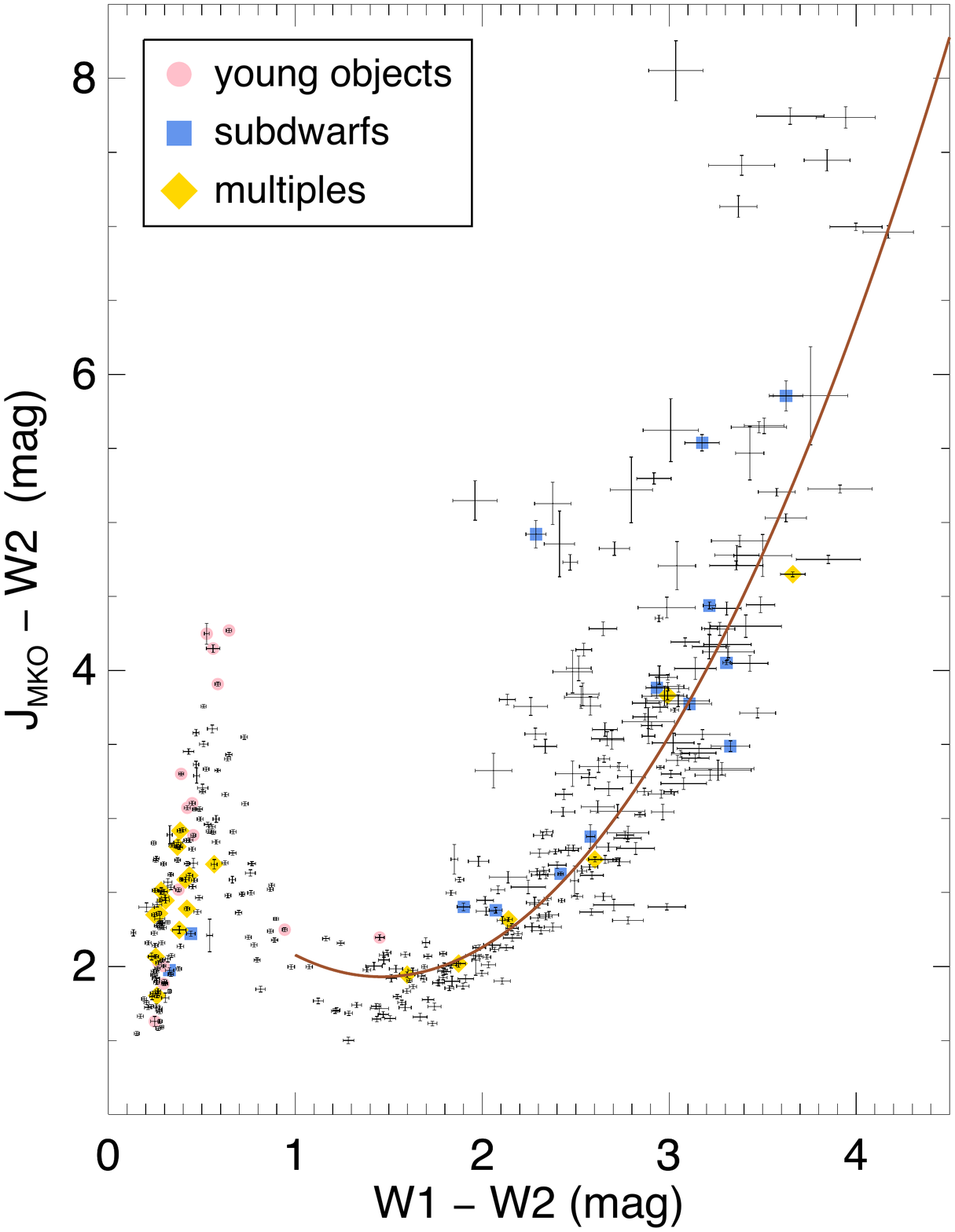}{0.33\textwidth}{(e)}
          \fig{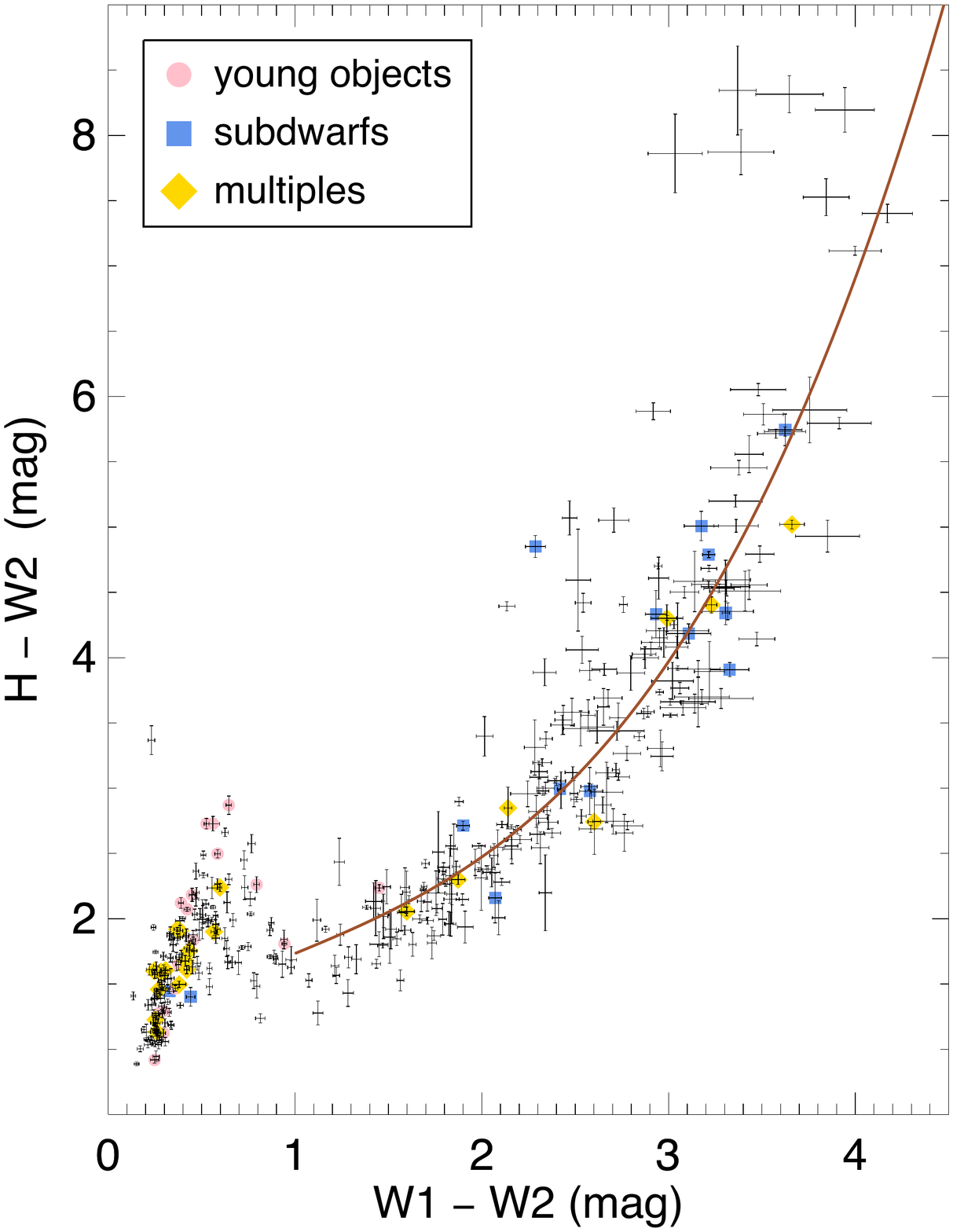}{0.33\textwidth}{(f)}}
\caption{Plots of various absolute magnitudes (a-c) and colors (d-f) as a function of ${\rm W1} - {\rm W2}$ color. Only members of the 20-pc census are shown, and plots a-c show only the subset of 20-pc objects having parallaxes measured to better than 12.5\%. Polynomial fits that exclude known young objects (pink circles, section~\ref{section:known_youngs}), subdwarfs (blue squares, section~\ref{section:known_sds}), and multiple systems (yellow diamonds, section~\ref{section:known_multiples}) are shown in brown and described in Table~\ref{table:equations}. In panels a-c, the fits include only those points with ${\rm W1} - {\rm W2} > 1.0$ mag, and in panels e-f the fits include only those points with ${\rm W1} - {\rm W2} > 0.8$ mag
\label{figure:xaxis_W1W2}}
\end{figure*}

Plots of absolute magnitude and color as a function of ${\rm W1} - {\rm W2}$ color are shown in Figure~\ref{figure:xaxis_W1W2}(a-f). The same trends as those mentioned above in ${\rm ch1} - {\rm ch2}$ color are seen. 

\begin{figure*}
\figurenum{18}
\gridline{\fig{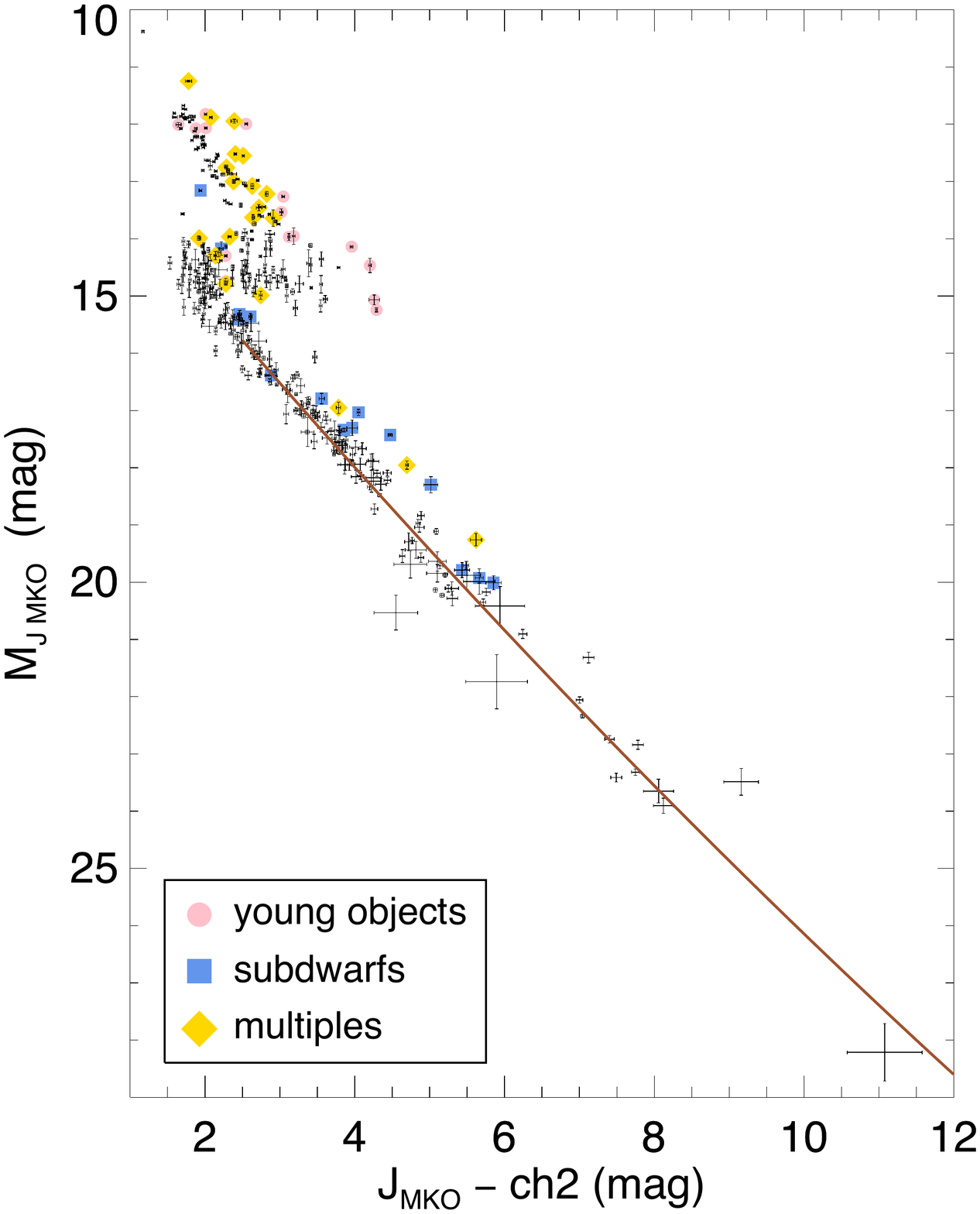}{0.33\textwidth}{(a)}
          \fig{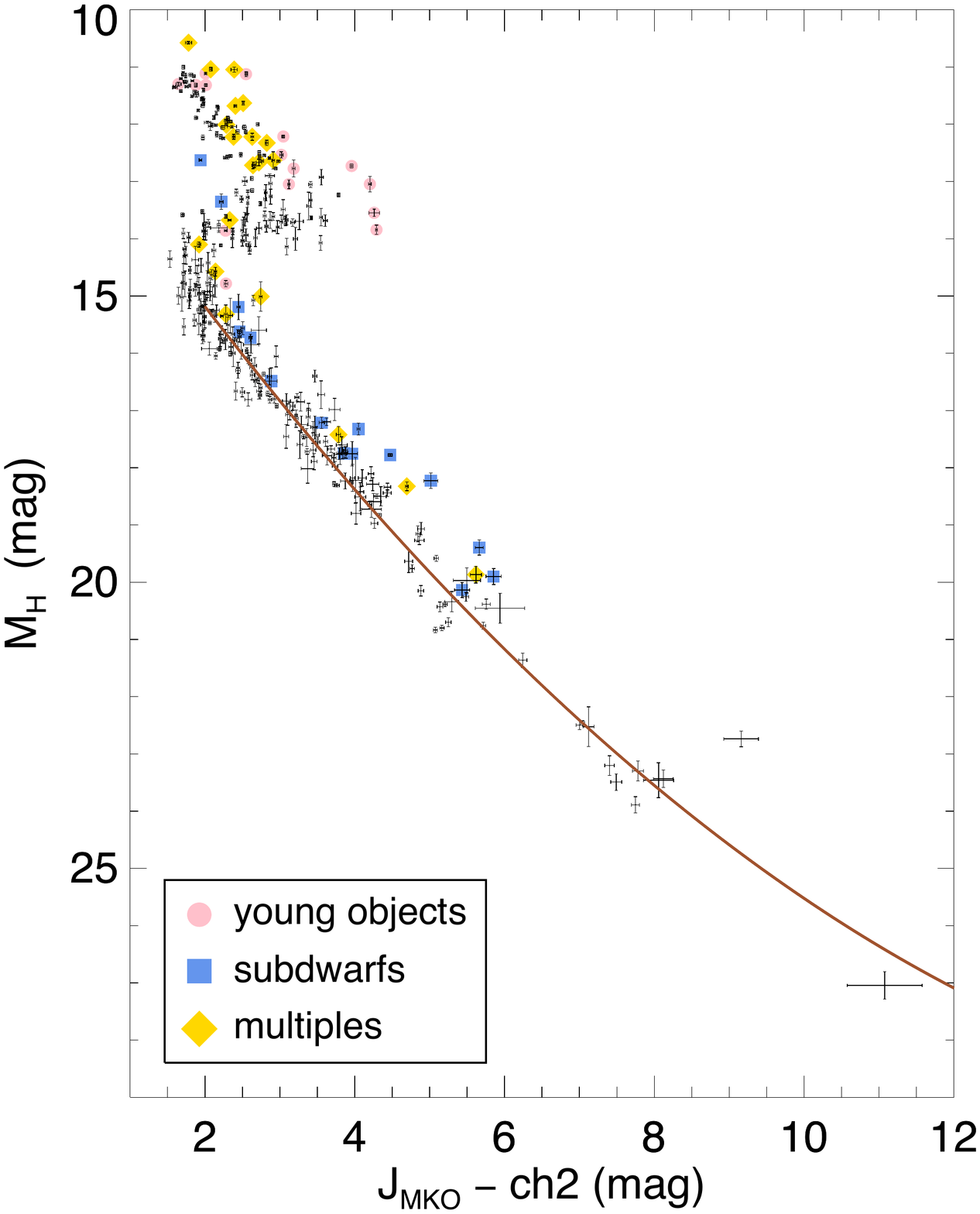}{0.33\textwidth}{(b)}
          \fig{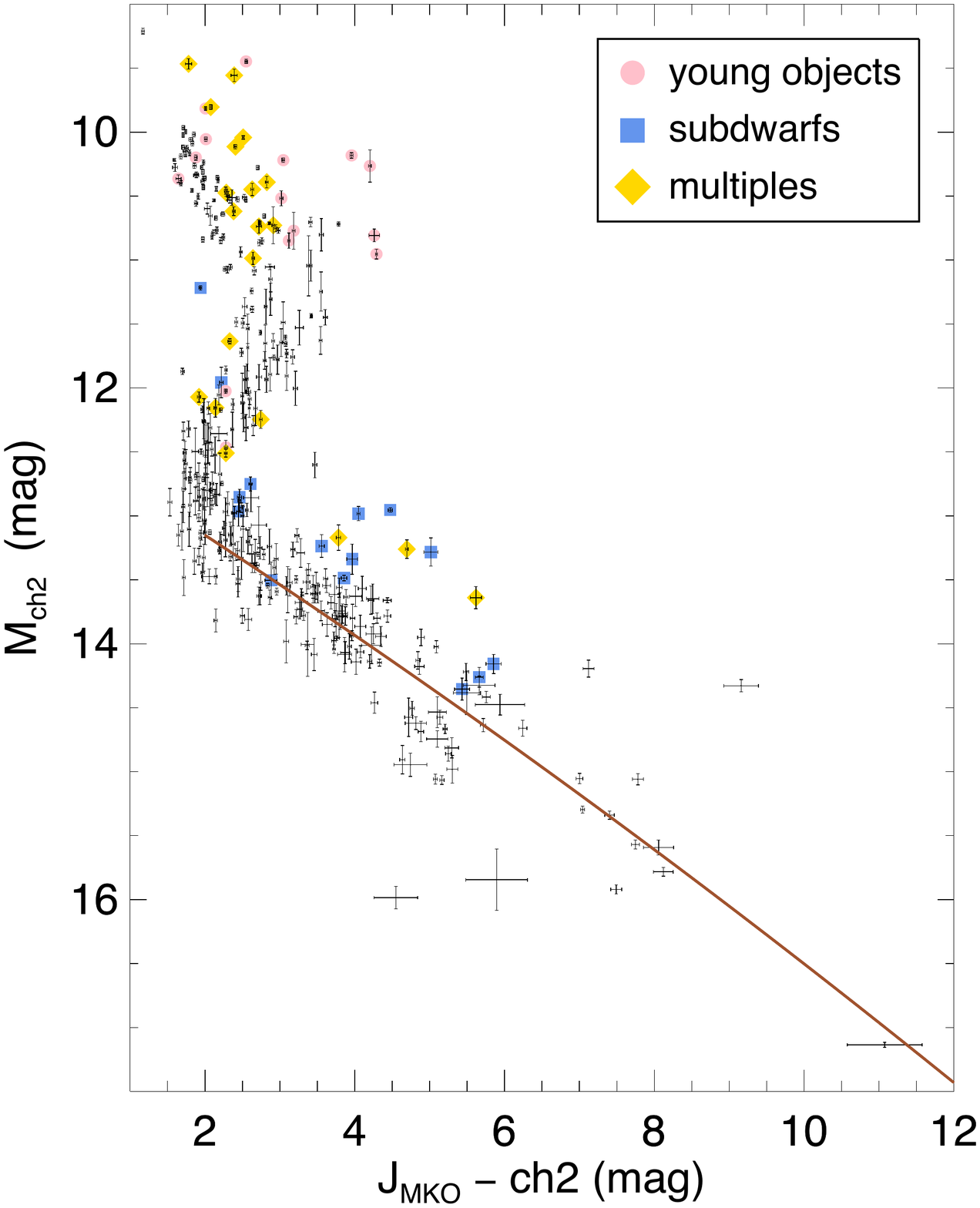}{0.33\textwidth}{(c)}}
\gridline{\fig{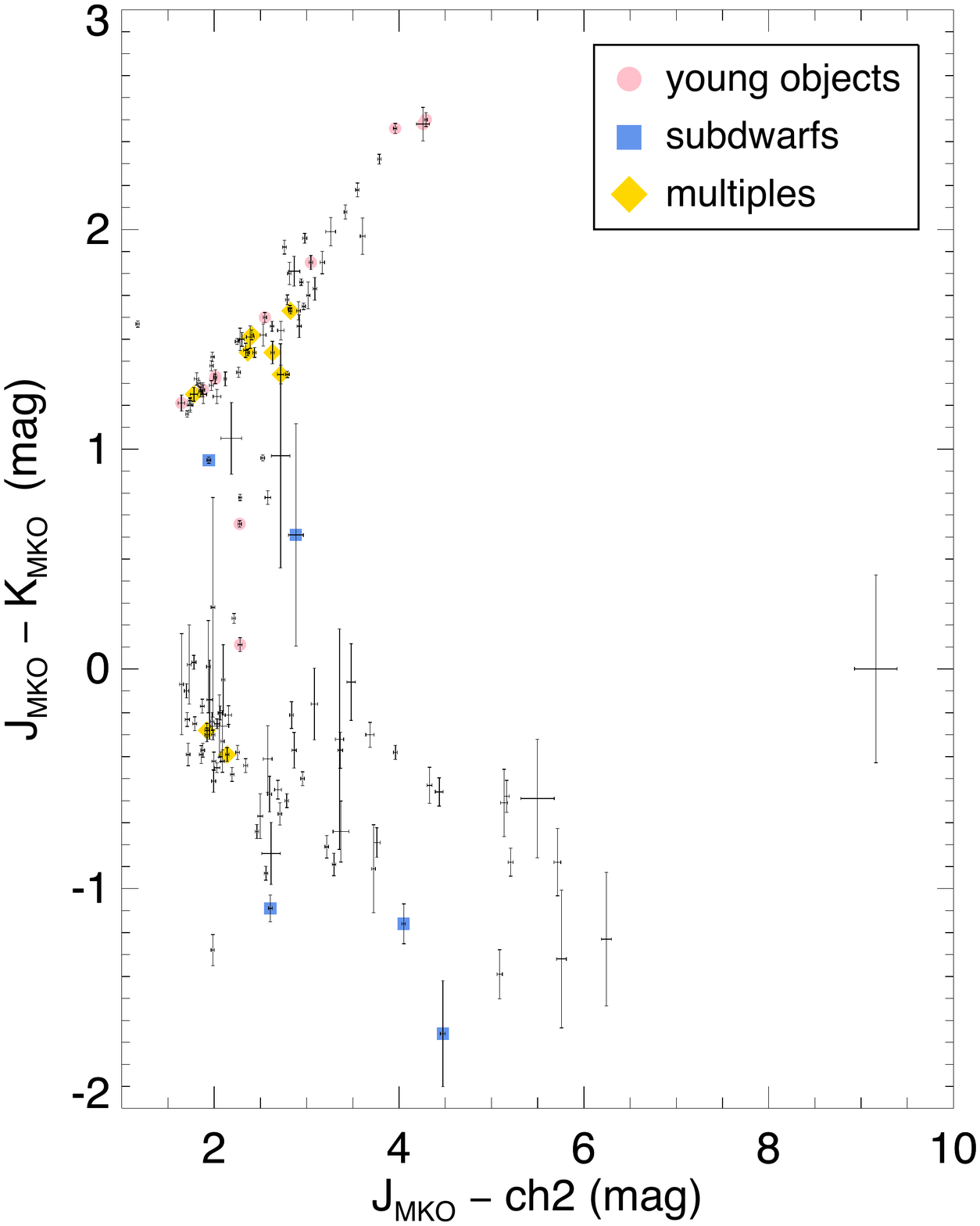}{0.33\textwidth}{(d)}
          \fig{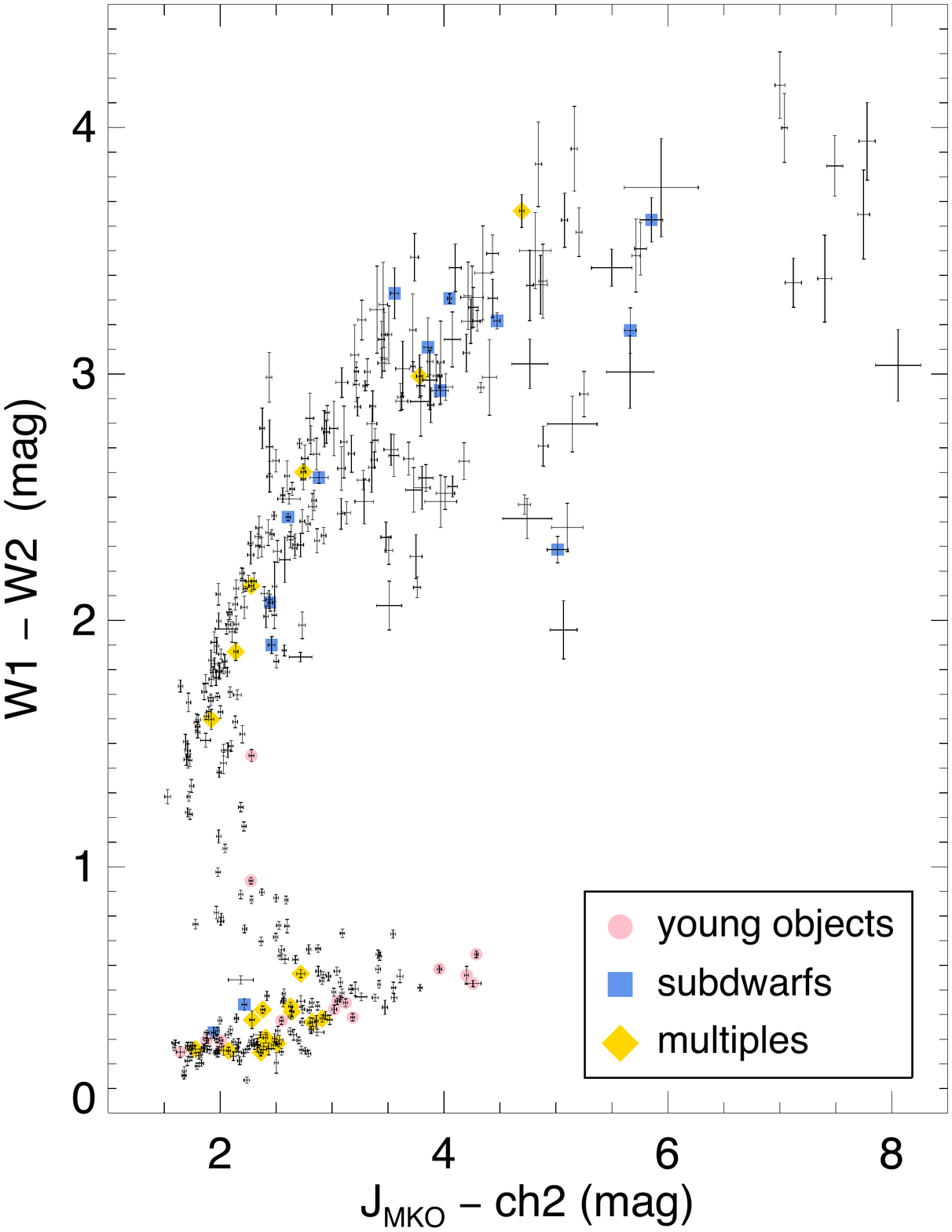}{0.33\textwidth}{(e)}}
\caption{Plots of various absolute magnitudes (a-c) and colors (d-e) as a function of $J_{\rm MKO} - {\rm ch2}$ color. Only members of the 20-pc census are shown, and plots a-c show only the subset of 20-pc objects having parallaxes measured to better than 12.5\%. All five panels are supplemented with W2 magnitudes when ch2 is not available, as described in section~\ref{section:plot_analysis}. Polynomial fits that exclude known young objects (pink circles, section~\ref{section:known_youngs}), subdwarfs (blue squares, section~\ref{section:known_sds}), and multiple systems (yellow diamonds, section~\ref{section:known_multiples}) are shown in brown and described in Table~\ref{table:equations}. These fits are restricted to points with $M_{J \rm MKO} \ge 16.0$ mag in panel a, $M_H \ge 15.0$ mag in panel b, and $M_{\rm ch2} \ge 13.0$ mag in panel c.
\label{figure:xaxis_Jch2}}
\end{figure*}

\begin{figure*}
\figurenum{19}
\gridline{\fig{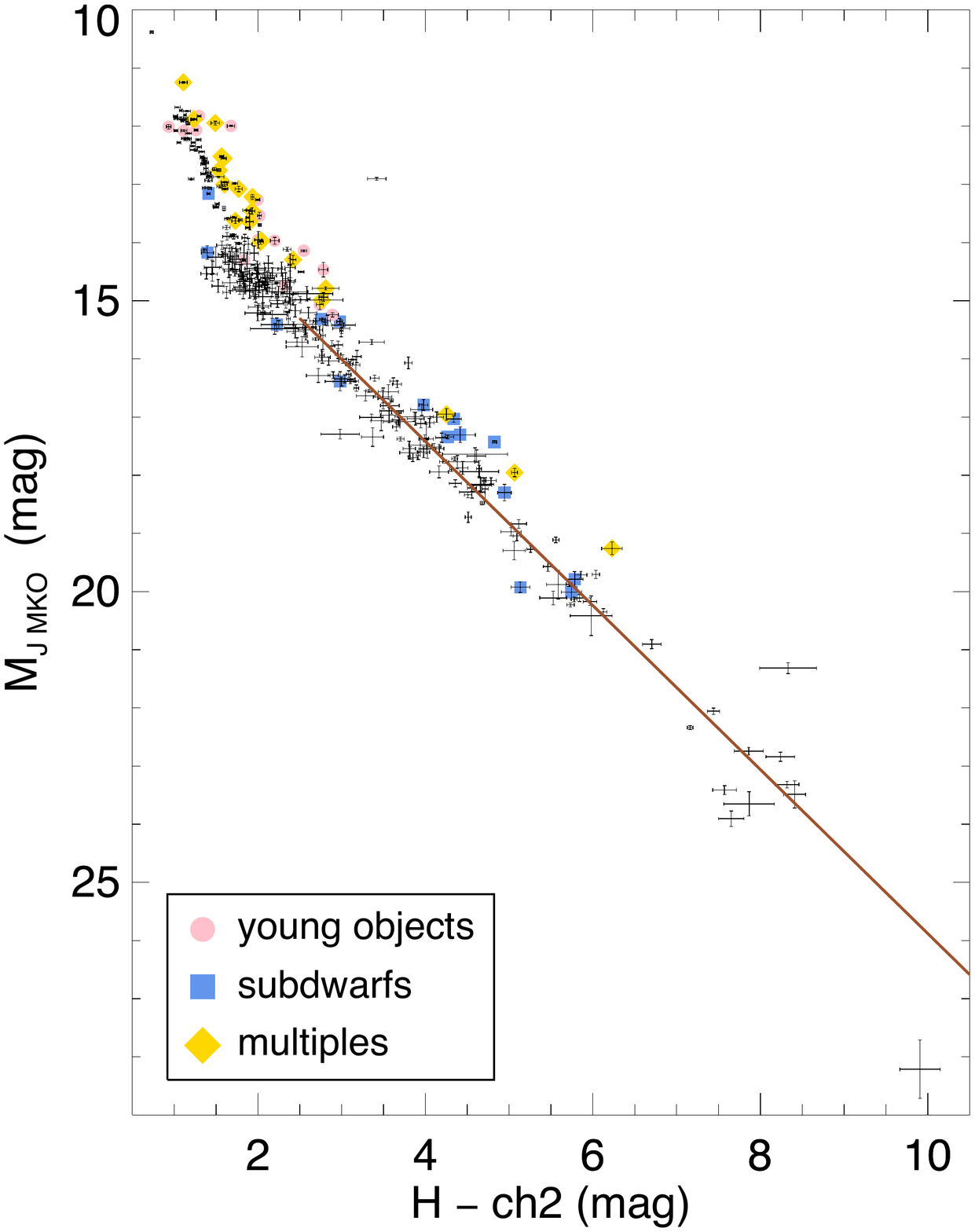}{0.33\textwidth}{(a)}
          \fig{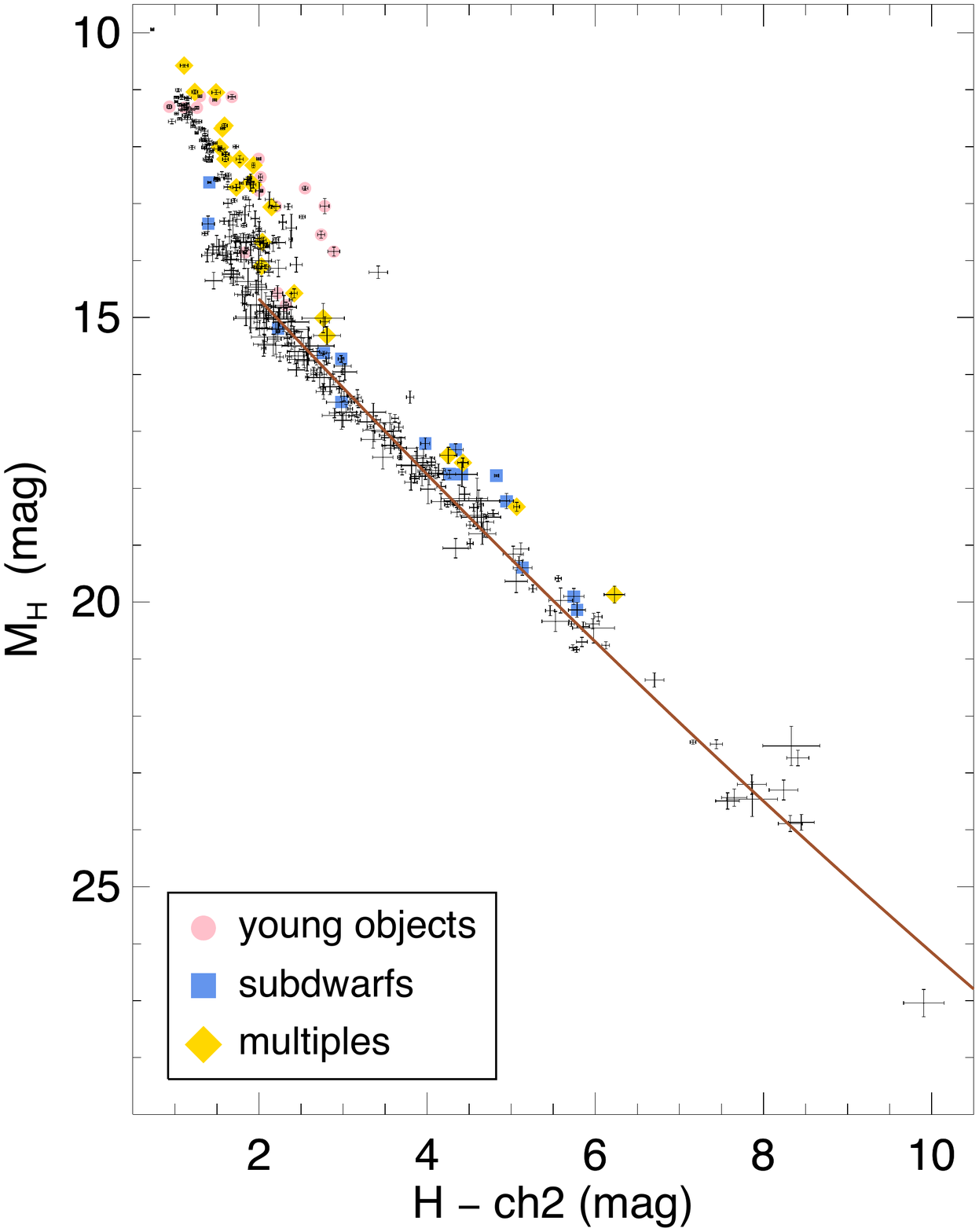}{0.33\textwidth}{(b)}
          \fig{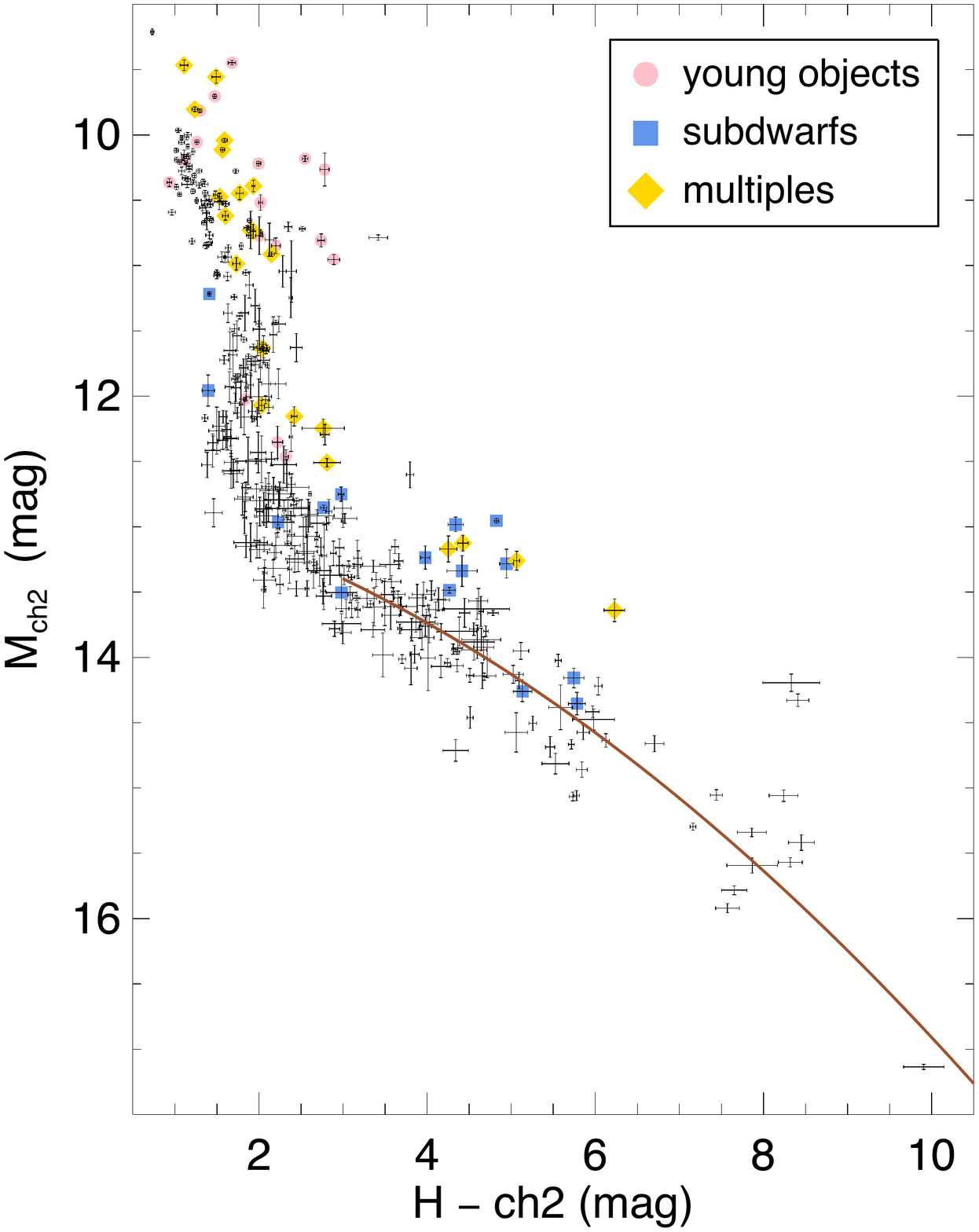}{0.33\textwidth}{(c)}}
\gridline{\fig{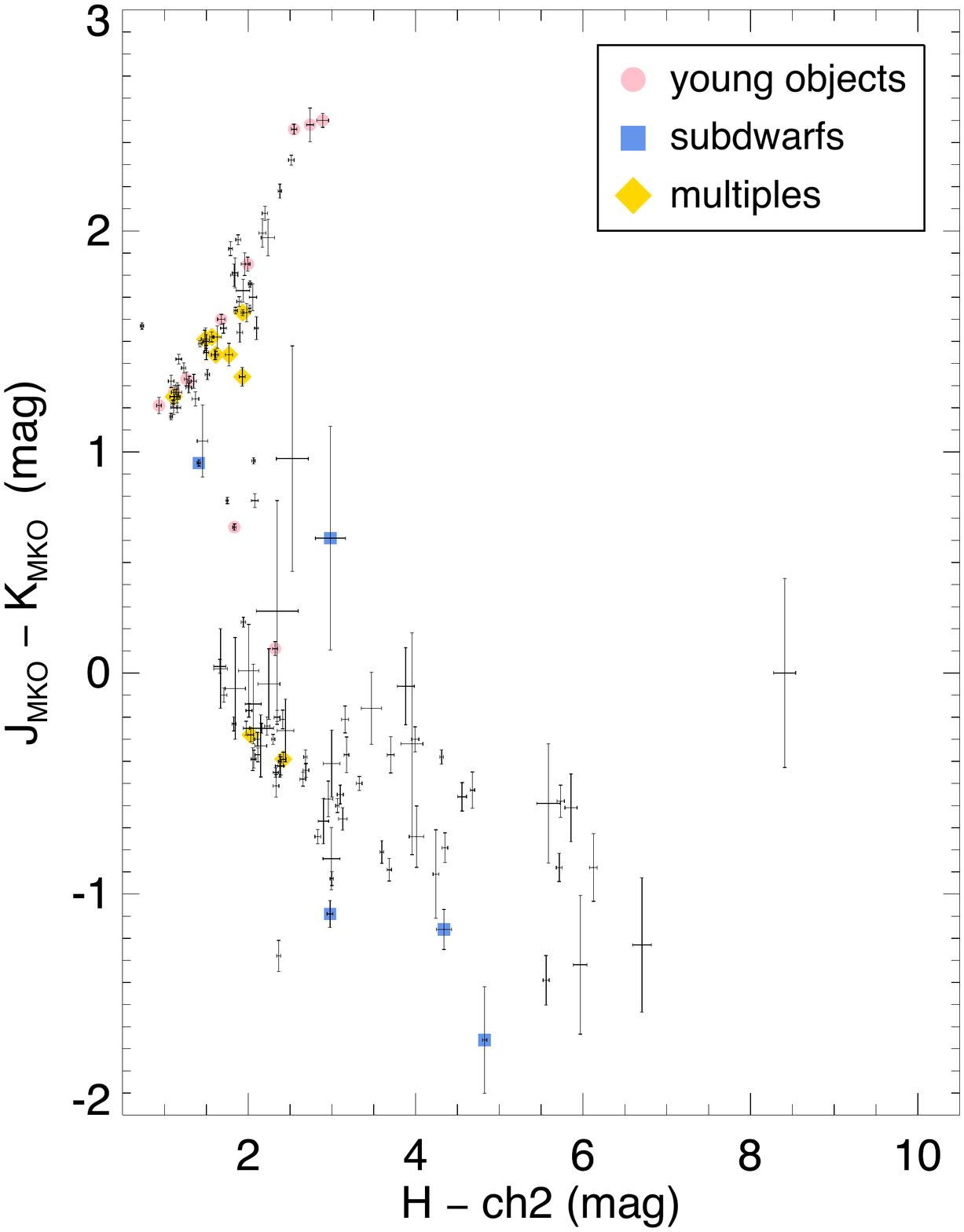}{0.33\textwidth}{(d)}
          \fig{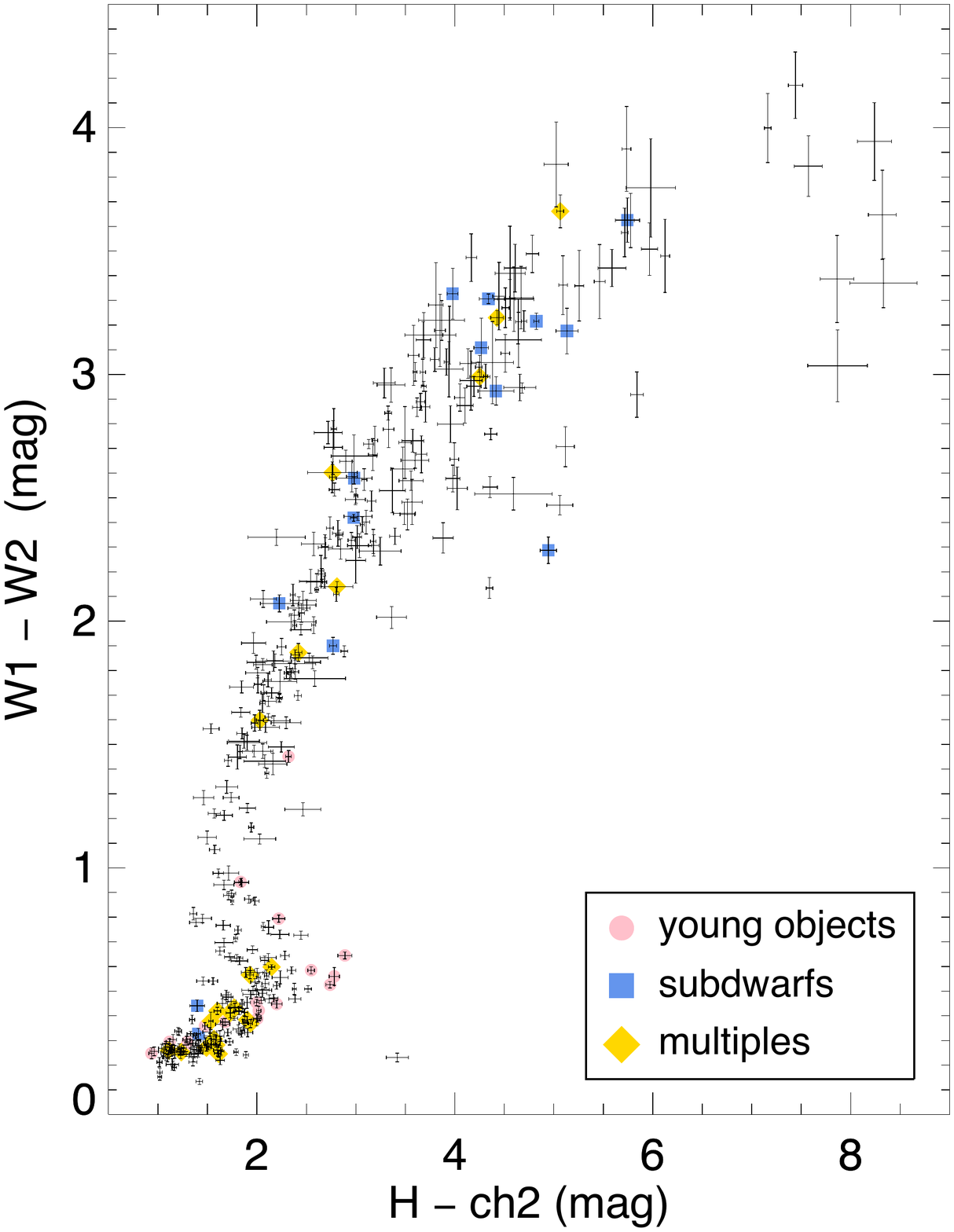}{0.33\textwidth}{(e)}}
\caption{Plots of various absolute magnitudes (a-c) and colors (d-e) as a function of $H - {\rm ch2}$ color. Only members of the 20-pc census are shown, and plots a-c show only the subset of 20-pc objects having parallaxes measured to better than 12.5\%. All five panels are supplemented with W2 magnitudes when ch2 is not available, as described in section~\ref{section:plot_analysis}. Polynomial fits that exclude known young objects (pink circles, section~\ref{section:known_youngs}), subdwarfs (blue squares, section~\ref{section:known_sds}), and multiple systems (yellow diamonds, section~\ref{section:known_multiples}) are shown in brown and described in Table~\ref{table:equations}. These fits are restricted to points with $M_{J \rm MKO} \ge 16.0$ mag in panel a, $M_H \ge 15.0$ mag in panel b, and $M_{\rm ch2} \ge 13.0$ mag in panel c.
\label{figure:xaxis_Hch2}}
\end{figure*}

Plots of absolute magnitude and color as a function of $J_{\rm MKO} - {\rm ch2}$ and $H - {\rm ch2}$ color are shown in Figure~\ref{figure:xaxis_Jch2}(a-e) and Figure~\ref{figure:xaxis_Hch2}(a-e). At a given absolute magnitude in $M_{J \rm MKO}$, $M_H$, and $M_{\rm ch2}$, young L dwarfs are shown to be redder than field objects, as are T subdwarfs, although L subdwarfs appear bluer. On the color-color plots, the reddest of the young L dwarfs are the reddest objects of all in $J_{\rm MKO} - K_{\rm MKO}$; at their ${\rm W1} - {\rm W2}$ colors, they are also the reddest objects in $J_{\rm MKO} - {\rm ch2}$ and $H - {\rm ch2}$.

Having established the locations of unusual objects on these diagrams, we examine the evidence for other, previously unrecognized (or, in some cases, previously suspected) young dwarfs, subdwarfs, and multiples in the 20-pc census. These are discussed in the next three subsections.

\subsection{Potential Young Objects}

No newly recognized young object candidates were identified from these diagrams.

\subsection{Potential Subdwarfs\label{sec:suspected_subdwarfs}}

A number of objects, not discussed in section~\ref{section:known_sds} above, appear to fall along the subdwarf locus in Figures~\ref{figure:xaxis_spectype} through \ref{figure:xaxis_Hch2}. These are addressed below.

\begin{itemize}

    \item WISE 0316+4307: This T8 dwarf falls along the locus of subdwarfs in the color-type plots shown in Figures~\ref{figure:xaxis_spectype}e and f. It also appears as a color outlier on the color-color plot \ref{figure:xaxis_W1W2}f. \cite{mace2013} acquired separate $J$- and $H$-band spectra of the object and did not note any peculiarities, although a spectrum across the full $JHK$ wavelength range could elucidate whether the telltale $K$-band suppression seen in T subdwarfs is confirmed.
    
    \item WISE 0359$-$5401: This Y0 dwarf falls along the locus of subdwarfs in Figure~\ref{figure:xaxis_ch1ch2}d. No Y dwarfs have yet been classified as subdwarfs, but 
    \cite{leggett2017} found this object indeed falls in the part of the $J - {\rm ch2}$ vs.\ ${\rm ch1} - {\rm ch2}$ diagram where substellar models predict low-metallicity objects to fall. We consider this to be a normal Y dwarf in subsequent analysis, pending the empirical spectroscopic identification of other Y subdwarfs.
    
    \item WISE 0430+4633: This T8 dwarf falls along the locus of subdwarfs in the color-type plots of  Figures~\ref{figure:xaxis_spectype}e and f. It is also a color outlier on the color-type plot of Figure~\ref{figure:xaxis_spectype}h and the color-color plot of Figure~\ref{figure:xaxis_ch1ch2}f. The spectral classification of this object is based on only a $J$-band spectrum by \cite{mace2013}. As with WISE 0316+4307 above, a spectrum across the full $JHK$ wavelength range is needed to confirm whether a subdwarf classification is warranted.
    
    \item UGPS 0521+3640: This T8.5 dwarf falls along the subdwarf locus in the absolute magnitude-color plot of Figure~\ref{figure:xaxis_ch1ch2}b. It is also an outlier on the color-color plot of Figure~\ref{figure:xaxis_ch1ch2}f. However, this source's photometry may be confused by the halo of a much brighter star. The near-infrared spectrum by \cite{burningham2011} shows no peculiarities, so we think it is only the poor photometry that is causing this object to appear as an outlier.

    \item WISE 0751$-$7634: This T9 dwarf falls along the subdwarf locus in the absolute magnitude-color plots of Figures~\ref{figure:xaxis_ch1ch2}a,b and \ref{figure:xaxis_W1W2}a,b, as well as in the color-color plot of Figure~\ref{figure:xaxis_W1W2}e. It is also an outlier on the color-color plot of Figure~\ref{figure:xaxis_W1W2}f. The near-infrared spectrum shown by \cite{kirkpatrick2011} has low S/N in the $K$-band and may show the flux suppression typical of T subdwarfs, but an improved spectrum is needed to verify this. \cite{leggett2017} notes that this object falls within the locus on the $J - {\rm ch2}$ vs.\ ${\rm ch1} - {\rm ch2}$ diagram where substellar models predict low-metallicity objects to fall. We await improved spectroscopic data before classifying this object as a subdwarf.
    
    \item WISE 1112$-$3857: This T9 dwarf falls along the subdwarf locus in the color-type plots of Figures~\ref{figure:xaxis_spectype}e,f, and the color-color plot of Figure~\ref{figure:xaxis_ch1ch2}d. The near-infrared spectrum presented in \cite{tinney2018} does not extend to the $K$-band but appears to show excess flux on the blueward side of $Y$-band, as seen in other T subdwarfs (see section~\ref{section:known_sds}). A more complete spectrum at higher S/N is needed to confirm the subdwarf hypothesis.
    
    \item WISE 1141$-$3326: This is a Y0 dwarf that falls along the subdwarf locus in the absolute magnitude-color plots of Figures~\ref{figure:xaxis_ch1ch2}a and \ref{figure:xaxis_W1W2}a, and the color-color plots of  Figure~\ref{figure:xaxis_ch1ch2}d and \ref{figure:xaxis_W1W2}e. As noted in \cite{kirkpatrick2019}, however, these anomalies can likely be attributed to photometric contamination at earlier epochs when the source was passing in front of a background galaxy.

    \item WISE 1818$-$4701: A spectrum of this object has not yet been acquired, but it is believed to be a late-T dwarf. It falls along the subdwarf locus in the absolute magnitude-color plot of Figure~\ref{figure:xaxis_W1W2}a and color-color plot of Figure~\ref{figure:xaxis_W1W2}e. A spectrum is required to confirm or refute the subdwarf hypothesis.
    
    \item GJ 836.7B (2144+1446): This T3 dwarf, also known as HN Peg B, appears along the subdwarf sequence in the color-color plot of Figure~\ref{figure:xaxis_W1W2}f and is an outlier on the color-type plot of Figure~\ref{figure:xaxis_spectype}h and the color-color plot of Figure~\ref{figure:xaxis_ch1ch2}f. \cite{luhman2007} cite an age of $\sim$300 Myr for the system, and \cite{valenti2005} find that the primary has [M/H] $\approx -0.01$. Since this object is obviously not a subdwarf, we suspect that the CatWISE2020 photometry may be corrupted due to the proximity of the bright primary itself. The AllWISE and CatWISE2020 photometry (Table~\ref{table:monster_table}) differ in both W1 and W2 by $>5{\sigma}$, indicating that the the automated measurements are likely poor. Further evidence that the ${\rm W1} - {\rm W2}$ color may be suspect is the fact that similar plots with ${\rm ch1} - {\rm ch2}$ color (Figures~\ref{figure:xaxis_spectype}g and \ref{figure:xaxis_ch1ch2}e) show this source falling along the locus of normal field dwarfs.
    
    \item GJ 1263B (2146$-$0010): This T8.5 dwarf, also known as Wolf 940B, lies along the subdwarf locus in Figures~\ref{figure:xaxis_W1W2}a,b. \cite{burningham2009} find that the primary has an age of $\sim$3.5 Gyr and metallicity of [Fe/H] $= -0.06{\pm}0.20$, so the B component cannot be a subdwarf. As with GJ 836.7B above, the AllWISE and CatWISE2020 photometry (Table~\ref{table:monster_table}) differ in both W1 and W2, in this case by $>10{\sigma}$ and $>6{\sigma}$, respectively. Further evidence that the ${\rm W1} - {\rm W2}$ color may be suspect is the fact that similar plots with ${\rm ch1} - {\rm ch2}$ color (Figures~\ref{figure:xaxis_ch1ch2}a,b) show this source to fall along the normal locus. We suspect that the bright primary has corrupted the {\it WISE} photometry of the secondary.
    
\end{itemize}

\subsection{Potential Multiples\label{sec:suspected_multiples}}

Several L, T and Y dwarfs within the 20-pc census have been previously published as suspected multiples and either remain unconfirmed or have subsequently been discounted. Several others are newly addressed here as suspected binary systems. Suspected companions are denoted by brackets ("[B]" or "[C]") around the suffix both in the text below and in Table~\ref{20pc_sample}.

\begin{itemize}

    \item WISE 0309$-$5016A[B]: This T7 dwarf is an outlier on the absolute magnitude-type plot of Figure~\ref{figure:xaxis_spectype}d and on the absolute magnitude-color plots of Figure~\ref{figure:xaxis_ch1ch2}b,c; \ref{figure:xaxis_W1W2}a,b,c; \ref{figure:xaxis_Jch2}a,b,c; and \ref{figure:xaxis_Hch2}a,b,c. The consistent overluminosity of this object across colors and bands strongly points to its being an unresolved double with components of near-equal magnitude. As we did in \cite{kirkpatrick2019}, we consider it to be a two-body system in subsequent analysis.
    
    \item WISE 0350$-$5658: This Y1 dwarf falls well above the mean trend in Figure~\ref{figure:xaxis_ch1ch2}b. Oddities in absolute magnitude-type plots were also noted in \cite{kirkpatrick2019}. Few Y1 dwarfs are presently known, so it is unclear the extent to which this is just cosmic scatter for normal dwarfs of this spectral type. We consider this object to be single.
    
    \item WISE 0535$-$7500: This $\ge$Y1: dwarf falls well above the mean trend on the absolute magnitude-type plot of Figure~\ref{figure:xaxis_spectype}d and on the absolute magnitude-color plots of Figures~\ref{figure:xaxis_ch1ch2}c; \ref{figure:xaxis_Jch2}a,c; and \ref{figure:xaxis_Hch2}a,b,c. This overluminosity was also noted by \cite{tinney2014}, \cite{leggett2017}, and \cite{kirkpatrick2019}. \cite{opitz2016} used adaptive-optics imaging to rule any equal-magnitude companion at a separation greater than $\sim$1.9 AU. As with WISE 0350$-$5658 above, it is unclear the extent to which this may just be cosmic scatter for normal dwarfs of this spectral type, since few are known. We consider this object to be single.
    
    \item WISE 0546$-$0959: This T5 dwarf falls above the mean locus on the $M_H$ vs.\ ${\rm ch1} - {\rm ch2}$ diagram of Figure~\ref{figure:xaxis_ch1ch2}b and the $M_H$ vs.\ ${\rm W1} - {\rm W2}$ diagram of Figure~\ref{figure:xaxis_W1W2}b. Because it appears overluminous only in $H$-band, we consider this object to be single.

    \item 2MASS 0559$-$1404: This mid-T dwarf falls well above the mean locus on all of the plots based on absolute magnitude in Figures~\ref{figure:xaxis_spectype}, \ref{figure:xaxis_ch1ch2}, and \ref{figure:xaxis_W1W2}. It is also an outlier on the $M_{J \rm MKO}$ vs.\ $J_{\rm MKO} - {\rm ch2}$ plot of Figure~\ref{figure:xaxis_Jch2}a. Two hypotheses have been proposed to explain the overluminosity, which was first noted by \cite{dahn2002}: (1) \cite{burgasser2001} suggested that the object was an equal-magnitude binary. (2) \cite{burgasser2003c} later proposed that the quick dissipation of clouds near the L-to-T dwarf transition could be responsible for the overluminosity, which is largest at $J$-band. However, both of these hypotheses have encountered problems in the intervening years. The cloud disruption theory was largely invoked to explain the $J$-band overluminosity (\citealt{tsuji2003}), but as our figures show, this overluminosity is present across all bands from $J$ through W2. The binary theory has yet to be confirmed, either. High-resolution {\it HST} imaging by \cite{burgasser2003c} showed no indication of a hidden companion down to a separation of 0$\farcs$09. Using radial velocity measurements covering a 4.4-yr period, \cite{zapatero2007} found no velocity variations (to $1\sigma = 0.5$ km s$^{-1}$). Other radial velocity measurements by \cite{prato2015} were able to rule out a companion with a period of a day or less, but these authors stress that there is still orbital parameter space between their sampled region and the 0$\farcs$09 (0.9 AU) limit by the {\it HST} imaging mentioned above. Given the inability of observers to confirm the binary hypothesis for this object, we will assume the object is a single dwarf in subsequent analysis.

    \item PSO 0652+4127: \cite{best2013} label this object as a possible binary based on the fact that some near-infrared spectral indices better match a L8+T2.5 composite that the single T0 type. Their single-object photometric distance suggests the object falls at 14.2$\pm$1.2 pc, whereas the binary hypothesis suggests 20.1$\pm$2.4 pc. Our {\it Spitzer} parallax gives a distance of 17.4$\pm$1.0 pc, which is intermediate between the two estimates. In the absence of data confirming a companion, we consider this object to be single.

    \item SDSS 0758+3247: This early T dwarf was discovered by \cite{knapp2004}. It was identified by \cite{burgasser2010} as a weak candidate for unresolved binarity due to its near-infrared spectral morphology. However, as stated in that paper, the single object spectral fit outperformed that of the best binary fit. Nonetheless, the spectral type listed in the SIMBAD database shows this as a composite type. \cite{bardalez2015} list this system as a "visual spectral binary" but surmise that it is comprised of two components with types of T2.2$\pm$0.0 and T2.3$\pm$0.0 despite the fact that it is not possible to detect a binary comprised of identical components using low-resolution spectral morphology alone. Our plot of M$_H$ vs.\ near-infrared spectral type, for example, shows no overluminosity of this object compared to other early-T dwarfs, ruling out the equal-magnitude binary hypothesis. We thus consider this object to be a single brown dwarf.
    
    \item SDSS 0857+5708: This L8 dwarf falls above the mean trend on the plots of $M_{\rm ch1}$ and $M_{\rm ch2}$ vs.\ spectral type in Figures~\ref{figure:xaxis_spectype}c,d. Given that there is no evidence of overluminosity in other diagrams and that there is no indication in the literature of binarity, we consider this to be a single object.
    
    \item WISE 0920+4538: Given that this L9 dwarf is labeled only as a weak binary candidate in \cite{mace2013} and that some of its peculiarities may be attributed to spectroscopic variations (\citealt{best2013}), we consider this to be a single object. 
    
    \item 2MASS 0939$-$2448A[B]: This T8 dwarf has been considered an unresolved, equal-magnitude binary for many years based on its overluminosity, as discussed in \cite{kirkpatrick2019}. In section~\ref{section:known_sds}, we noted that the spectrum shows signs of low-metallicity as well. Thus, we consider this to be a T subdwarf binary.

    \item PSO 0956$-$1447: \cite{best2015} list this late-L dwarf as a marginal spectral binary candidate. In the absence of any confirming high-resolution imaging, we consider this to be a single object. 
    
    \item SDSS 1048+0111: This early- to mid-L dwarf falls above the mean locus on the plots of absolute magnitude vs.\ spectral type in Figures~\ref{figure:xaxis_spectype}a,b. \cite{reid2006} did not find any evidence of binarity in high-resolution {\it HST} imaging. Furthermore, we note that our perceived overluminosity vanishes if we plot against the optical spectral type of L1 instead of the near-infrared type of L4 (Table~\ref{20pc_sample}). We consider this to be a single object.

    \item 2MASS 1231+0847: This T5.5 dwarf is overluminous for its ${\rm ch1} - {\rm ch2}$ and ${\rm W1} - {\rm W2}$ color on Figures~\ref{figure:xaxis_ch1ch2}a,b,c and \ref{figure:xaxis_W1W2}a,b,c. The object was observed with high-resolution imaging on {\it HST} by \cite{aberasturi2014}, who found no companion with a separation $> 0{\farcs}3$ down to $\Delta{J} \approx 2.5$ mag (their Figure 7). As discussed in \cite{kirkpatrick2019}, \cite{burgasser2004} proposed that this object's broad \ion{K}{1} lines might indicate a higher gravity that is the consequence of lower metallicity. Given the uncertain cause of this object's peculiarities, we will consider it to be a single dwarf of normal metallicity in subsequent analysis.
    
    \item WISE 1318$-$1758: This T8 dwarf is overluminous on the $M_H$ vs.\ ${\rm ch1} - {\rm ch2}$ plot of Figure~\ref{figure:xaxis_ch1ch2}b and the $M_H$ vs.\ $J_{\rm MKO} - {\rm ch2}$ plot of Figure~\ref{figure:xaxis_Jch2}b. Because the object does not appear overluminous on other plots, we consider it to be single.
    
    \item WISE 1322$-$2340: This late-T dwarf is overluminous only on the $M_H$ vs.\ ${\rm ch1} - {\rm ch2}$ plot of Figure~\ref{figure:xaxis_ch1ch2}b although \cite{kirkpatrick2019} noted it was an outlier in $H - {\rm ch2}$ color as well. However, the object does not distinguish itself on other plots, and \cite{gelino2011} ruled out any companion with a separation $> 0{\farcs}2$ down to $\Delta{H} \approx 4.0$ mag. We consider this object to be single.

    \item ULAS 1416+1348: In \cite{kirkpatrick2019}, we considered this (sd)T7.5 to be an unresolved double based on its overluminosity with respect to normal late-T dwarfs and with respect to the few sdT dwarfs identified in that paper. However, it now appears that overluminosity with respect to normal T dwarfs of the same color or spectral type is a trait shared with a wider variety of low-metallicity T dwarfs. We therefore now consider this to be a single object. 

    \item WISE 1627+3255A[B]: This mid-T dwarf is overluminous on the absolute magnitude-color plots of Figures~\ref{figure:xaxis_ch1ch2}a,b,c and \ref{figure:xaxis_W1W2}a,b,c. Although \cite{gelino2011}
    found no companion down to ${\Delta}H \approx 5$ mag at separations $> 0{\farcs}2$, we consider this object to nonetheless be a tight unresolved binary, just as \cite{kirkpatrick2019} concluded.

    \item DENIS 1705$-$0516: \cite{kendall2004} discovered this early-L dwarf. \cite{reid2006}, using {\it HST}/NICMOS imaging in 2005 Jun, found a faint source separated by 1$\farcs$36 and consistent with either a distant (1-2 kpc), unrelated mid-M dwarf or a physically related early-T dwarf. Our analysis of more recent imaging by {\it HST}/WFC3 (Program 13724; PI: T.\ Henry) as well as $J$ and $K_S$ imaging by VHS show that the putative companion is a stationary background source, the motion of the early-L dwarf having increased the separation between the two objects to 2$\farcs$9 arcsec by 2015 Mar. We consider this L dwarf to be a single object.
    
    \item WISE 1804+3117: This late-T dwarf is overluminous only on the $M_{\rm ch1}$ vs.\ spectral type diagram of Figure~\ref{figure:xaxis_spectype}c. This object has both an uncertain type of T9.5: and falls close to the Y dwarf regime where the identification of binarity has proven to be problematic. Therefore, as \cite{kirkpatrick2019} also concluded, we will consider this object to be single in our subsequent analysis.
    
    \item Gaia 1831$-$0732: This object does not yet have a measured spectral type, but if a classification of L0 is verified, it is overluminous relative to other L0 dwarfs on the absolute magnitude vs.\ type plots of Figure~\ref{figure:xaxis_spectype}a,c,d. It is also overluminous on the absolute magnitude vs.\ color plots of Figure~\ref{figure:xaxis_ch1ch2}a,b,c, but this overluminosity would vanish if the object were actually a late-M dwarf. The fact that it is an outlier on the color-color plot of Figure~\ref{figure:xaxis_ch1ch2}e strongly suggests that it is, indeed, an M dwarf. Given the evidence that this object is earlier than L0, we exclude it from subsequent analyses.
    
    \item Gl 758B (1923+3313): This late-T dwarf companion was discovered using Subaru/HiCIAO by \cite{thalmann2009}, who also reported a possible third member of the system. Using the same instrument, \cite{janson2011} confirmed that this purported Gl 758"C" was a background star based on data with a $\sim$1.5-yr baseline.
    
    \item 2MASS 2126+7617A[B]: This object appears overluminous on Figure~\ref{figure:xaxis_spectype}b. \cite{kirkpatrick2010} note that this object has peculiar spectra in both the optical and near-infrared, and the spectral types are discrepant between the two -- L7 in the optical, and T0 pec in the near-infrared. These authors also found that a spectral binary comprised of an L7 dwarf and a T3.5 dwarf accounts for the main peculiarities in the near-infrared spectrum. Given that this is a strong case for a spectral binary, we tentatively include the B component in our subsequent analysis.
    
    \item 2MASS 2139+0220: This early-T dwarf was identified as a possible unresolved binary based on its near-infrared spectral morphology by \cite{burgasser2010}. Individual components of types L8.5 and T3.5 were suggested, although it was noted that the synthetic composite type still failed to reproduce important features in the observed spectrum. This object is now noted for its extreme variability (26\% at $J$-band), leading  \cite{radigan2012} to conclude that the object's variations were caused either by multi-layered clouds or a cloud layer with holes. \cite{bardalez2015} conjectures that some candidate spectral binaries may instead be single objects whose photospheres are comprised of multi-component cloud layers of differing temperatures. We consider 2MASS 2139+0220 to be a single object.
    
\end{itemize}

\subsection{Other Outliers}

\begin{itemize}

    \item SDSS 0000+2554: This T4.5 dwarf is an outlier on the ${\rm W1} - {\rm W2}$ vs.\ spectral type plot of Figure~\ref{figure:xaxis_spectype}h, the ${\rm W1} - {\rm W2}$ vs.\ ${\rm ch1} - {\rm ch2}$ plot of Figure~\ref{figure:xaxis_ch1ch2}f, and the $J_{\rm MKO} - K_{\rm MKO}$ vs.\ ${\rm W1} - {\rm W2}$ plot of Figure~\ref{figure:xaxis_W1W2}d. Examination of the {\it WISE} images shows this object to be buried within the halo of the bright star Z Pegasi, which must be corrupting the {\it WISE} colors.

    \item WISE 0715$-$1145: This object appears as a color outlier on at least nine of the previous plots (Figures~\ref{figure:xaxis_spectype}b,f; \ref{figure:xaxis_ch1ch2}e; \ref{figure:xaxis_W1W2}b,f; \ref{figure:xaxis_Hch2}a,b,c,e) but does not fall in the locus of known young objects, subdwarfs, or unresolved multiples. It is an L4 pec (blue) dwarf whose near-infrared spectrum is much bluer than the standard L4 dwarf but lacks indications of low-metallicity (\citealt{kirkpatrick2014}), and it is one of just six blue L dwarfs known in the 20-pc census -- the others being SIPS J0921$-$2104, 2MASS 1300+1912, 2MASS 1721+3344, VVV 1726$-$2738, WISE 2141$-$5118. Only three of these others (2MASS 1300+1912, 2MASS 1721+3344, VVV 1726$-$2738) appear as outliers on the previous plots, and these distinguish themselves only in Figure~\ref{figure:xaxis_spectype}, which is based on spectral type. WISE 0715$-$1145 therefore appears to be the most extreme color outlier of the 20-pc blue L dwarfs. \cite{faherty2009} noted that the general population of blue L dwarfs, despite not showing obvious signs of low metallicity, nonetheless have kinematics consistent with an old age.
    
    \item WISE 1828+2650: This Y dwarf is overluminous on Figures~\ref{figure:xaxis_spectype}a,b,c,d; \ref{figure:xaxis_Jch2}b,c; and \ref{figure:xaxis_Hch2}b,c. It also falls along the subdwarf locus in Figure~\ref{figure:xaxis_ch1ch2}d. This object was discussed in section 8.2.47 of \cite{kirkpatrick2019}. Compared to all other Y dwarfs with near-infrared spectra, this object has a unique spectrum that does not compare well with the known suite of theoretical models (Cushing et al., in prep.).
    
\end{itemize}

\section{Temperatures and Space Densities\label{section:temps_densities}}

\subsection{Assigning Each Object to a $T_{\rm eff}$ Bin\label{subsection:assigning_temps}}

Finding the functional form of the mass function from our 20-pc census is not a straightforward exercise because mass is not an observable quantity. Moreover, since most of the objects in our L, T, and Y dwarf census are brown dwarfs, they continue to cool as they age, and as a result there is no direct mapping from spectral type to mass unless the age of the object is known. Only a small number of the objects within the census have age estimates -- i.e., confirmed members of young moving groups and companions to higher mass stars whose ages are known through other means.

Because the bulk of our objects have no age estimates, we rely instead on simulating empirical distributions using various assumed forms of the mass function, an assumed star formation rate over the interval of interest, and theoretical models to evolve each object to the current epoch. This work is described in detail in sections 9.1 and 9.2 of \cite{kirkpatrick2019}. The evolutionary models allow us to transform the predictions into distributions of either effective temperature or bolometric luminosity. Both of these quantities have their own limitations, however. Effective temperature is not a directly observable quantity and requires either forward modeling (comparison to atmospheric models), inverse modeling ("retrieval" analysis), or calculation via the Stefan-Boltzmann Law. Measuring effective temperature via the Stefan-Boltzmann equation would require only a measurement of the bolometric luminosity and an assumption about the object's radius which, fortunately for most of these old brown dwarfs, can be assumed to be ${\sim}1R_{\rm Jup}$ due to their electron degeneracy. However, if bolometric luminosities were already measured, we could forgo temperature determinations entirely and simply compare our observed luminosity distributions to the simulations. At present, however, we have insufficient data with which to compute accurate bolometric luminosities for most of these objects, although more complete spectral coverage over the bulk of these objects' spectral energy distribution will soon be obtainable using the {\it Spectro-Photometer for the History of the Universe, Epoch of Reionization and Ices Explorer} ({\it SPHEREx}; \citealt{dore2016,dore2018}), supplemented at longer wavelengths with data from {\it WISE} and the {\it James Webb Space Telescope} ({\it JWST}; \citealt{gardner2006}).

For now, we are left to convert our sample into a distribution of effective temperature. \cite{filippazzo2015} calculated bolometric luminosities for a large number of late-M, L and T dwarfs, and used those to compute effective temperatures once a radius was deduced from model calculations. (These radii were very close to ${\sim}1 R_{\rm Jup}$ as expected, since most of these objects are old brown dwarfs that have contracted to their final equilibrium radius.) Those authors then plotted various observable parameters against the resulting effective temperature measurements and found that the relation with the smallest scatter was $T_{\rm eff}$ $vs.\ M_H$. For objects in our 20-pc sample that are thought to be old field objects, we can therefore use $M_H$ to transform into $T_{\rm eff}$. However, a few objects do not have $H$-band measurements, and for those we can use the measured spectral type (or its estimate) as the arbiter of effective temperature.

The relations presented in \cite{filippazzo2015} predate the release of {\it Gaia} DR2 and do not extend into the Y dwarf regime. Therefore, we have updated the data presented in that paper to include new {\it Gaia} parallaxes and improved parallaxes from {\it Spitzer}, and have also updated $H$-band values where more accurate photometry is now available from VHS or other follow-up surveys. Those results are given in Table~\ref{table:teff}. We have extended this list into the Y dwarf regime by including objects from Table 10 of \cite{kirkpatrick2019} whose effective temperatures were calculated from published values computed using forward and inverse modeling techniques.

\startlongtable
\begin{deluxetable*}{lrcccc}
\tabletypesize{\footnotesize}
\tablecaption{Late-M, L, T, and Y Dwarfs with $T_{\rm eff}$ Measurements\label{table:teff}}
\tablehead{
\colhead{Name\tablenotemark{a}} & 
\colhead{SpT\tablenotemark{b}} &
\colhead{$\varpi_{abs}$} &
\colhead{$T_{\rm eff}$} &
\colhead{$H$} &
\colhead{Ref\tablenotemark{c}} \\
\colhead{} & 
\colhead{} &
\colhead{(mas)} &
\colhead{(K)} &
\colhead{(mag)} &
\colhead{} \\
\colhead{(1)} &                          
\colhead{(2)} &  
\colhead{(3)} &  
\colhead{(4)} &
\colhead{(5)} &
\colhead{(6)} \\
}
\startdata
 SDSS 0000+2554             & 14.5&  70.8$\pm$1.9     & 1227$\pm$95& 14.731$\pm$0.074&  TTFT\\
 2MASS J00034227-2822410    & -2.5&  24.351$\pm$0.201 & 2871$\pm$76& 12.376$\pm$0.028&  FGFF\\
 BRI B0021-0214             & -0.5&  79.965$\pm$0.221 & 2390$\pm$80& 11.084$\pm$0.022&  FGFF\\
 2MASS 0034+0523            & 16.5& 118.8$\pm$2.7     &  899$\pm$82& 15.58$\pm$0.01  &  TTFT\\
 ULAS 0034-0052             & 18.5&  68.7$\pm$1.4     &  583$\pm$75& 18.49$\pm$0.04  &  TTKT\\
 2MASS 0036+1821            &  4.0& 114.417$\pm$0.209 & 1869$\pm$64& 11.59$\pm$0.03  &  TTFT\\
 Gl 27B (0039+2115)         & 18.0&  89.789$\pm$0.058 &  793$\pm$35& 16.72$\pm$0.03  &  TTFT\\
 2MASS 0050-3322            & 17.0&  94.6$\pm$2.4     &  836$\pm$71& 16.04$\pm$0.10  &  TTFT\\
\enddata
\tablecomments{(This table is available in its entirety in a machine-readable form in the online journal. A portion is shown here for guidance regarding its form and content.)}
\tablenotetext{a}{For objects also listed in Table~\ref{table:monster_table}, the abbreviated name is given; full designations can be found in Table~\ref{table:monster_table} itself. For all other objects, the full name is presented.}
\tablenotetext{b}{This is the (near-infrared) spectral type encoded as follows: M5 = -5.0, L0 = 0.0, L5 = 5.0, T0 = 10.0, T5 = 15.0, Y0 = 20.0, etc.}
\tablenotetext{c}{This is a four-character code that gives the reference for the spectral type, parallax, effective temperature, and $H$-band magnitude, respectively:
    C = \citealt{gelino2011},
    D = \citealt{dupuy2015},
    F = \citealt{filippazzo2015},
    G = \citealt{gaia2018},
    J = \citealt{faherty2012},
    K = \citealt{kirkpatrick2019}, 
    L = \citealt{liu2012},
    T = Table~\ref{table:monster_table} in this paper,
    W = \citealt{weinberger2016},
    X = \citealt{faherty2009}.}
\end{deluxetable*}

\begin{figure}
\figurenum{20}
\gridline{\fig{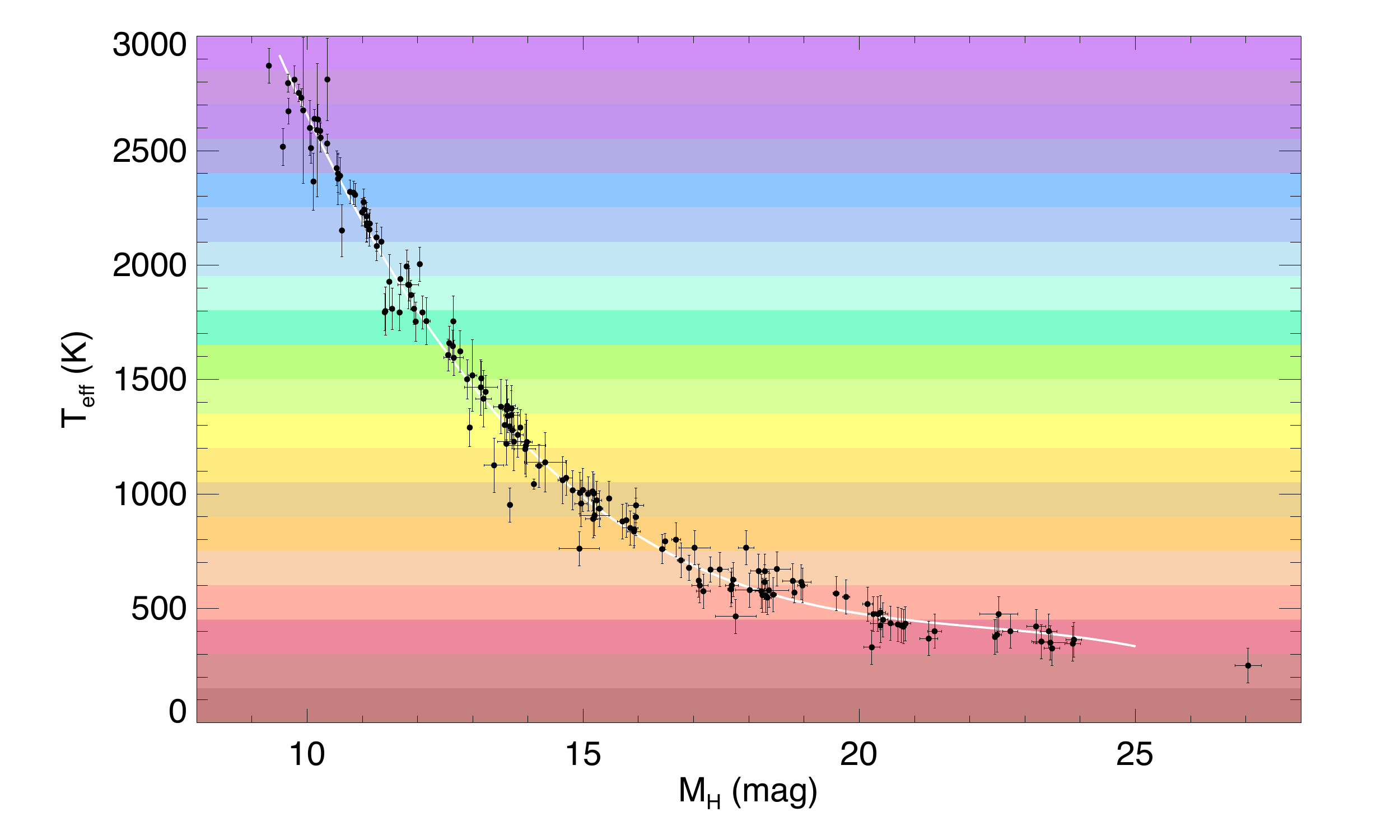}{0.5\textwidth}{(a)}}
\gridline{\fig{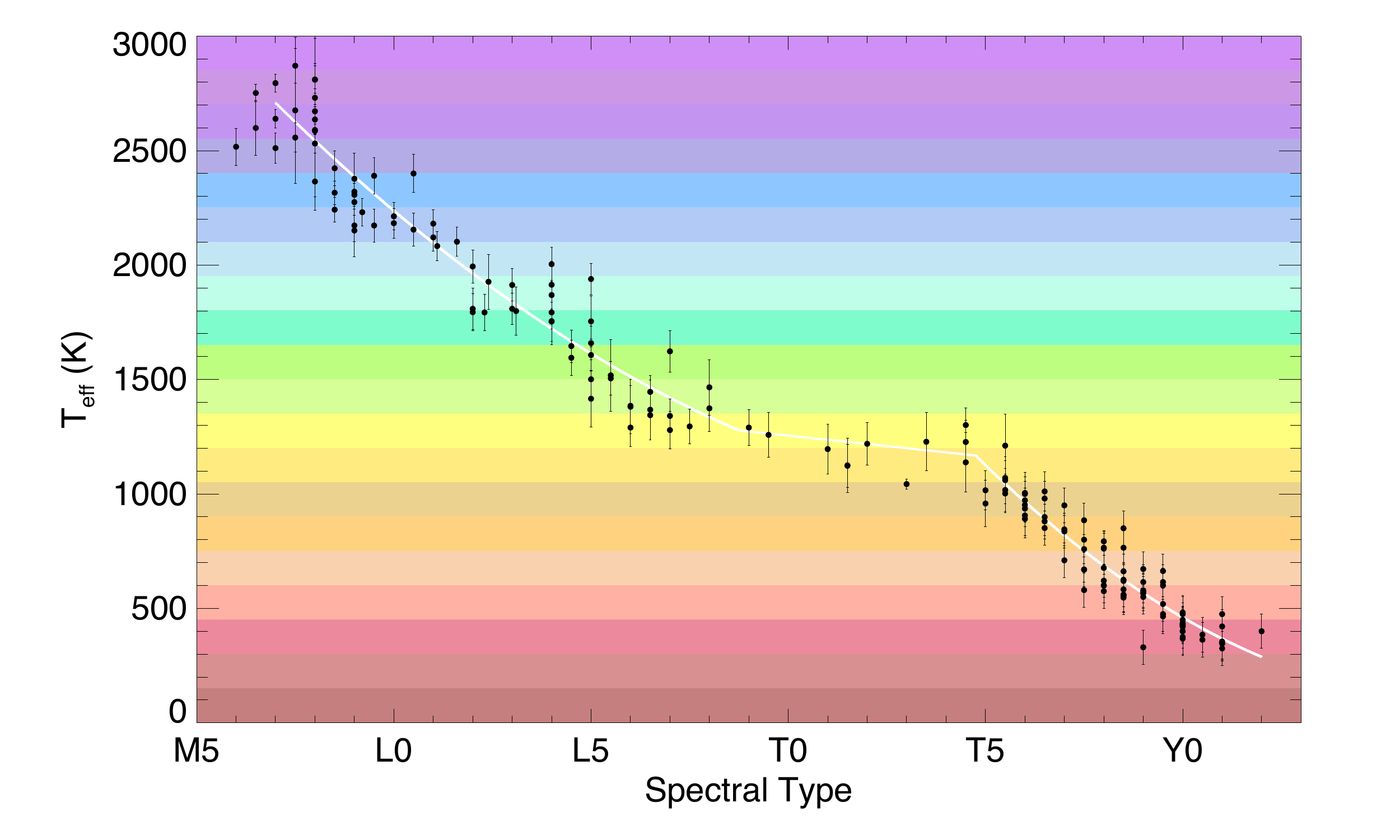}{0.5\textwidth}{(b)}}
\caption{Plots showing the trend of absolute $H$-band magnitude with effective temperature (a) and spectral type with effective temperature (b), using the data (black points) from Table~\ref{table:teff}. Functional fits to the trends, shown by the white curves, can be found in Table~\ref{table:equations}. The colored bands on each plot depict each of the 150K-wide temperature bins into which the data will be sorted in the following section.
\label{figure:yaxis_teff}}
\end{figure}

These results are plotted in Figure~\ref{figure:yaxis_teff} and the fitted relations given in Table~\ref{table:equations}. The plot in panel $a$ shows that from early-L through mid-T ($10.5 < M_H < 15$ mag), each 150K bin in $T_{\rm eff}$ corresponds to a fairly narrow range of $M_H$. However, at spectral types later than mid-T ($M_H > 15$ mag), each 150K temperature bin encompasses a larger and larger range of $M_H$ values. In panel $b$ we see the well-known result that objects in the L/T transition between types of late-L to mid-T span a very narrow range in $T_{\rm eff}$. Outside of this spectral type range, there is a monotonic trend of decreasing temperature with later spectral type.

For the 525 individual objects in the 20-pc census, we have assigned values of $T_{\rm eff}$ as follows; these values can be found in column 10 of Table~\ref{20pc_sample}. For old field dwarfs of normal gravity, we take the measured values of $T_{\rm eff}$ from \cite{filippazzo2015} if the object has a computed value there. Otherwise, we assign a $T_{\rm eff}$ value using the relation in Figure~\ref{figure:yaxis_teff}a using the object's measured $M_H$ if an $H$-band magnitude exists and the parallax is known to better than 12.5\%. If these conditions are not met, we use the spectral type contained in the $SpAd$ column of Table~\ref{table:monster_table} along with the relation shown in Figure~\ref{figure:yaxis_teff}b. The only exception is WISE 0855$-$0714, which is assigned a 250K value, as was done in \cite{kirkpatrick2019}.

For low-gravity (young) objects, we take the $T_{\rm eff}$ value computed by \cite{faherty2016} if the object has a value there; otherwise, we take the value from \cite{filippazzo2015}. For other objects noted as young in column 11 of Table~\ref{table:monster_table} but lacking measured values, we assign temperatures using an updated version (Faherty, priv.\ comm.) of the optical spectral type to $T_{\rm eff}$ relation of \cite{faherty2016}. When no optical type is available, we use the near-infrared type as a proxy.

For low-metallicity (subdwarf) objects, we take $T_{\rm eff}$ measurements directly from \cite{filippazzo2015}, when available. However, no relation between absolute magnitude (or spectral type) and temperature exists for these subdwarfs. Three mild, and presumably single, subdwarfs in our sample have measurements in \cite{filippazzo2015}: 2MASS 0729$-$3954 (752$\pm$69K), 2MASS 0937+2931 (881$\pm$74K), and ULAS 1416+1348 (656$\pm$54K). The field relation would suggest values of 749K, 858K, and 610K for these same three objects, respectively, showing that values from the field relation are consistent with the actual measurements. In fact, the most extreme subdwarf in the 20-pc sample, WISE 2005+5424, has a model fit temperature of 600-900K (\citealt{mace2013b}), which is also roughly consistent with the field estimate of 574K. Thus, as was done for the old field objects above, we assign temperatures to the other subdwarfs using the field relations of Figure~\ref{figure:yaxis_teff}.

\subsection{Space Densities vs.\ $T_{\rm eff}$ and Spectral Type}

To aid in comparison to our mass function simulations, we present our final space densities as a function of temperature. Specifically, these are shown as histograms binned in 150K-wide increments of $T_{\rm eff}$. To ease other empirical comparisons, we also present space densities as a function of spectral type, binned via integral subtypes.

Before computing these space densities, we must first determine whether the data contributing to each of these bins is complete to our target distance of 20 pc. For this, we use the $V/V_{max}$ test advocated by \cite{schmidt1968}. The basis of this test is as follows. Consider a proposed completeness limit of $d_{max}$. For each object $i$ at distance $d_i$ within that distance, the test computes the ratio of the volume interior to that object's position, $V_i = (4/3){\pi}{d_i}^3$, to the total volume being considered, $V_{max} = (4/3){\pi}{d_{max}}^3$. The average of these ratios, $\langle{V}/{V_{max}}\rangle = (1/n) \times \sum_{i=0}^{n}(V_i/V_{max})$, should be $\sim$0.5 for a complete, isotropically distributed sample. Values that fall significantly below 0.5 indicate that there is incompleteness in the outer parts of the volume being considered. In other words, if the outer half-volume has significantly less than half of all objects within the total volume, the sample is likely incomplete to that distance.

\begin{figure*}
\figurenum{21}
\includegraphics[scale=0.85,angle=0]{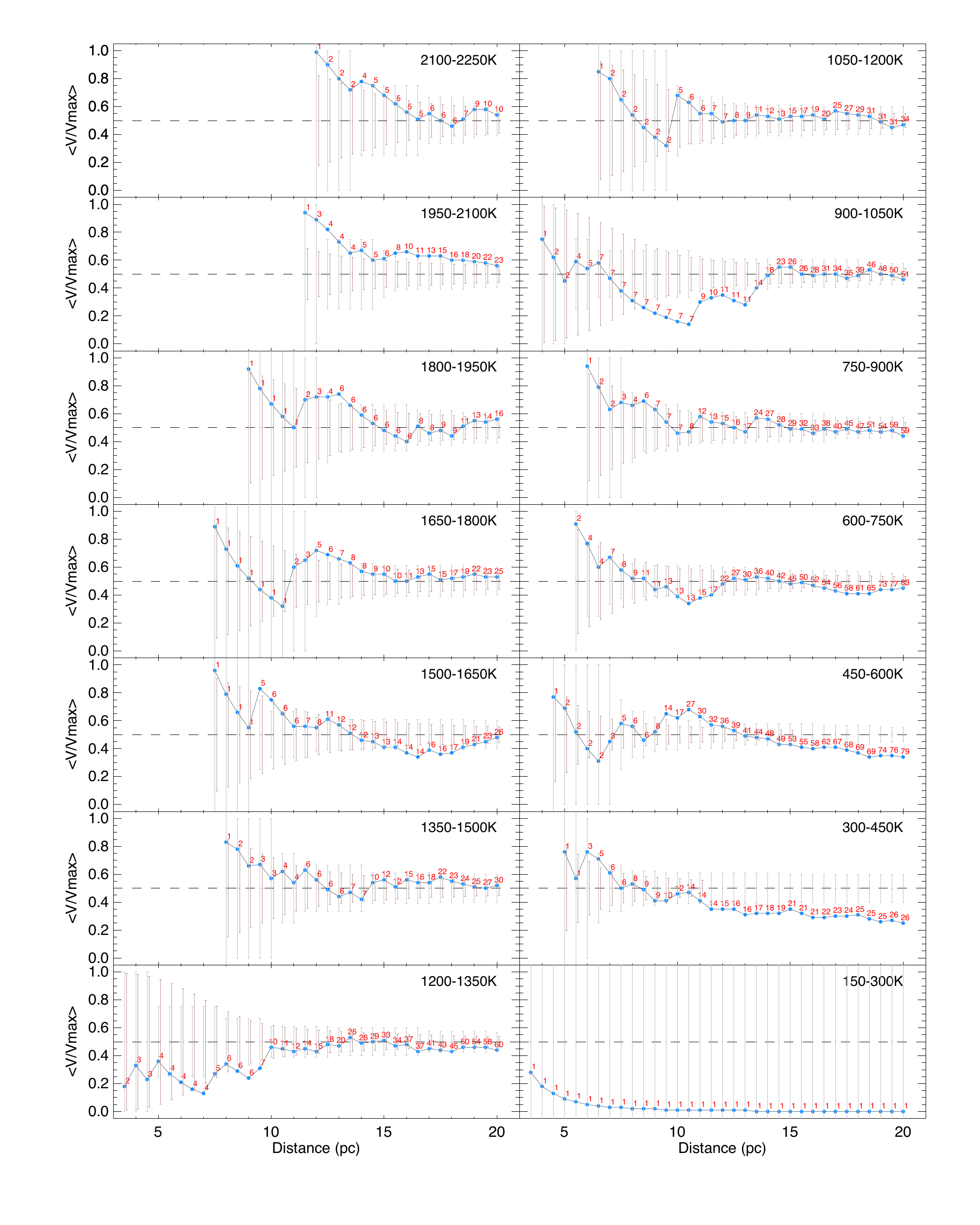}
\caption{The average $V/V_{\rm max}$ value in 0.5-pc intervals for fourteen 150-K bins encompassing our 20-pc L, T, and Y dwarf census. Blue dots represent our empirical sample. Red labels mark the number of objects in the computation at each 0.5-pc interval. The black dashed line shows the $\langle{V}/{V_{max}}\rangle = 0.5$ level indicating a complete sample. The grey error bars show the approximate 1$\sigma$ range around $\langle{V}/{V_{max}}\rangle$ = 0.5 that a complete sample of the size indicated by the red number would exhibit, given random statistics. The brown error bars, offset by +0.05 pc from the grey error bars for clarity, show the 1$\sigma$ variation around 0.5 obtained by 10,000 Monte Carlo simulations having the number of objects and completeness limit listed in Table~\ref{table:space_densities}. 
\label{figure:vmax_teff}}
\end{figure*}

\begin{figure*}
\figurenum{22}
\includegraphics[scale=0.85,angle=0]{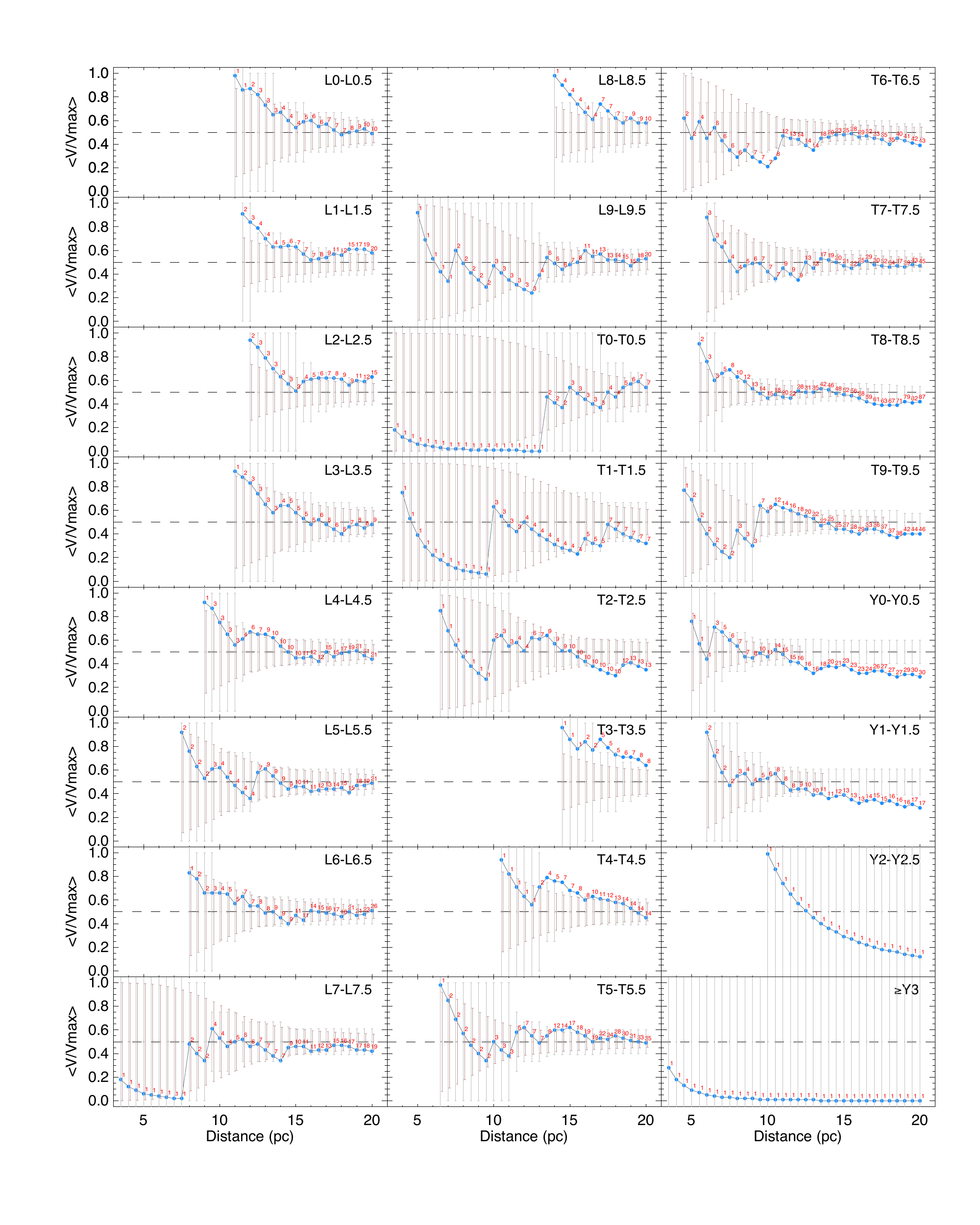}
\caption{The average $V/V_{\rm max}$ value in 0.5-pc intervals for twenty-four integral spectral type bins encompassing our 20-pc L, T, and Y dwarf census. See the caption to Figure~\ref{figure:vmax_teff} for more details. 
\label{figure:vmax_SpTy}}
\end{figure*}

We compute $\langle{V}/{V_{max}}\rangle$ at half-parsec steps within each bin. The computation starts with the first half-parsec step falling just beyond the distance of the closest object in the bin and continuing out to $d = 20$ pc. These computations are graphically illustrated in Figure~\ref{figure:vmax_teff} for each bin in $T_{\rm eff}$ and in Figure~\ref{figure:vmax_SpTy} for each bin in spectral type. Practically, though, what does "significantly below 0.5" mean for $\langle{V}/{V_{max}}\rangle$? \cite{kirkpatrick2019} proposed two ways to address this. First, a Poisson formalism was developed that establishes a 68\% likelihood threshold (the equivalent of 1$\sigma$ for a continuous distribution) that the $\langle{V}/{V_{max}}\rangle$ is significantly different from 0.5, given the number of objects in the sample. These thresholds are shown as the light grey error bounds in Figure~\ref{figure:vmax_teff} and \ref{figure:vmax_SpTy}. Second, a run of 10,000 Monte Carlo simulations for a sample size of $n$ objects was used to identify the range of $\langle{V}/{V_{max}}\rangle$ around 0.5 that contains 68\% of all simulated outcomes. Here, $n$ is the number of objects in the most distant bin for which the Poisson formalism determined the sample to be complete. These simulated likelihoods are shown by the brown error bounds in the figures.

Using these methods, we find that our sample is likely complete\footnote{As explained later in this section, the 2100-2250K bin is complete only for L dwarfs, but some late-M dwarfs are also expected to populate this temperature range. Hence, the space density derived for this bin should be considered a lower limit.} to 20 pc for all bins between 600 and 2250K in $T_{\rm eff}$. For cooler bins the completeness limit drops to 15 pc for 450-600K and to 11 pc for 300-450K. (The coolest bin with data, 150-300K, has only one object in it, WISE 0855$-$0714, so the completeness cannot be computed.) We note, however, that the 300-450K bin is likely complete over only a fraction of its 150K interval because the coldest assigned $T_{\rm eff}$ for any object in this bin is 367K. We further note that two sources within the 525-object L, T, and Y dwarf 20-pc census -- G 239-25B (1442+6603) and LSPM J1735+2634B -- have assigned $T_{\rm eff}$ values (Table~\ref{20pc_sample}) that are hotter than the hottest temperature bin considered here. Finally, our measured space density in the 2100-2250K bin should be considered as a lower limit, too, because if we were to have included late-M dwarfs in our 20-pc census some fraction of them would have populated this bin. These results  are shown in the first three columns of the upper portion of Table~\ref{table:space_densities}. 

Bins of integral spectral subtype, which generally have poorer statistics can, by extension, be assumed complete out to 20 pc for types warmer than 600K, which is roughly late-T (Figure~\ref{figure:yaxis_teff}b). A close look at Figure~\ref{figure:vmax_SpTy} shows that the census appears to be complete for spectral types from L0 through T7.5. The completeness limit drops to $\sim$17 pc for types T8-T9.5 and to $\sim$13 pc for types Y0-Y1.5. Later types than this have only one representative per bin -- WISE 1828+2650 at Y2 and WISE 0855$-$0714 at a type presumably later than that -- so completeness limits cannot yet be determined. Results are shown in the first three columns of the lower portion of Table~\ref{table:space_densities}. 

\begin{deluxetable}{ccccccc}
\tablecaption{Space Densities for Early-L through Early-Y Dwarfs\label{table:space_densities}}
\tablehead{
\colhead{T$_{\rm eff}$ or} & 
\colhead{Complete-} &
\colhead{Raw} &
\colhead{Adjusted} &
\colhead{Corr.} &
\colhead{Adopted} \\
\colhead{SpT} & 
\colhead{ness Limit} &
\colhead{No.\ of} &
\colhead{No.\ of} &
\colhead{Factor} &
\colhead{Space Density\tablenotemark{b}} \\
\colhead{Bin\tablenotemark{a}} & 
\colhead{(pc)} &
\colhead{Objects} &
\colhead{Objects} &
\colhead{} &
\colhead{($\times$10$^{-3}$ pc$^{-3}$)} \\
\colhead{} &                          
\colhead{$d_{max}$} &
\colhead{$raw$} &
\colhead{$adj$} &
\colhead{$corr$} &
\colhead{$dens$} \\
\colhead{(1)} &                          
\colhead{(2)} &
\colhead{(3)} &
\colhead{(4)} &
\colhead{(5)} &
\colhead{(6)} \\
}
\startdata
2100-2250K& 20.0\tablenotemark{c}& 10& 10.9$\pm$2.5& 1.05& $>$0.31\\ %$\pm$0.13
1950-2100K& 20.0& 23& 19.3$\pm$3.2& 1.05& 0.72$\pm$0.18\\
1800-1950K& 20.0& 16& 21.2$\pm$3.6& 1.05& 0.50$\pm$0.17\\
1650-1800K& 20.0& 25& 24.0$\pm$3.8& 1.05& 0.78$\pm$0.20\\
1500-1650K& 20.0& 26& 24.7$\pm$3.9& 1.05& 0.81$\pm$0.20\\
1350-1500K& 20.0& 30& 32.2$\pm$4.5& 1.05& 0.94$\pm$0.22\\
1200-1350K& 20.0& 60& 50.9$\pm$5.2& 1.09& 1.95$\pm$0.30\\
1050-1200K& 20.0& 34& 44.0$\pm$5.2& 1.09& 1.11$\pm$0.25\\
 900-1050K& 20.0& 51& 48.6$\pm$5.2& 1.13& 1.72$\pm$0.30\\
  750-900K& 20.0& 59& 58.4$\pm$5.8& 1.13& 1.99$\pm$0.32\\
  600-750K& 20.0& 83& 76.0$\pm$6.3& 1.13& 2.80$\pm$0.37\\
  450-600K& 15.0& 53& 44.9$\pm$4.9& 1.13& 4.24$\pm$0.70\\
  300-450K& 11.0& 14& 16.7$\pm$3.0& 1.13& $>$2.84\\ %$\pm$0.97
  150-300K& \nodata&  1&  \nodata& \nodata& \nodata\\
\hline
L0-L0.5& 20.0& 10&  8.1$\pm$0.8& 1.05& 0.31$\pm$0.10 \\
L1-L1.5& 20.0& 20& 21.7$\pm$0.9& 1.05& 0.63$\pm$0.14 \\
L2-L2.5& 20.0& 15& 13.7$\pm$1.0& 1.05& 0.47$\pm$0.13 \\
L3-L3.5& 20.0&  9&  9.6$\pm$1.1& 1.05& 0.28$\pm$0.10 \\
L4-L4.5& 20.0& 21& 20.5$\pm$0.9& 1.05& 0.66$\pm$0.15 \\
L5-L5.5& 20.0& 21& 21.8$\pm$0.9& 1.05& 0.66$\pm$0.15 \\
L6-L6.5& 20.0& 26& 22.6$\pm$1.4& 1.05& 0.81$\pm$0.17 \\
L7-L7.5& 20.0& 19& 21.6$\pm$1.5& 1.05& 0.60$\pm$0.14 \\
L8-L8.5& 20.0& 10& 11.2$\pm$0.8& 1.05& 0.31$\pm$0.10 \\
L9-L9.5& 20.0& 20& 19.5$\pm$0.5& 1.05& 0.63$\pm$0.14 \\
T0-T0.5& 20.0&  7&  7.0$\pm$0.7& 1.13& 0.24$\pm$0.09 \\
T1-T1.5& 20.0&  7&  7.5$\pm$0.5& 1.13& 0.24$\pm$0.09 \\
T2-T2.5& 20.0& 13& 13.0$\pm$0.1& 1.13& 0.44$\pm$0.12 \\
T3-T3.5& 20.0&  8&  7.0$\pm$0.7& 1.13& 0.27$\pm$0.10 \\
T4-T4.5& 20.0& 14& 14.5$\pm$0.9& 1.13& 0.47$\pm$0.13 \\
T5-T5.5& 20.0& 35& 34.5$\pm$0.9& 1.13& 1.18$\pm$0.20 \\
T6-T6.5& 20.0& 43& 43.5$\pm$1.0& 1.13& 1.45$\pm$0.22 \\
T7-T7.5& 20.0& 45& 43.5$\pm$1.2& 1.13& 1.52$\pm$0.23 \\
T8-T8.5& 16.5& 59& 58.0$\pm$1.5& 1.13& 3.54$\pm$0.47 \\
T9-T9.5& 17.5& 37& 37.0$\pm$1.8& 1.13& 1.86$\pm$0.32 \\
Y0-Y0.5& 12.0& 16& 17.0$\pm$0.7& 1.13& 2.50$\pm$0.63 \\
Y1-Y1.5& 13.5& 11& 10.0$\pm$1.0& 1.13& 1.21$\pm$0.36 \\
Y2-Y2.5& \nodata&  1& \nodata& \nodata& \nodata\\
$\ge$Y3& \nodata&  1& \nodata& \nodata& \nodata\\
\enddata
\tablenotetext{a}{The $SpAd$ spectral type from Table~\ref{table:monster_table}, which defaults to near-infrared types, is used here.}
\tablenotetext{b}{This value is computed via the equations $$dens = \Bigl(raw\Bigr)\Bigl(corr\Bigr)\bigg/\Bigl(\frac{4}{3}{\pi}{d_{max}}^3\Bigr)$$
and $$\sigma_{dens} = \sqrt{\bigl({\sigma_{raw}}^2 + {\sigma_{adj}}^2\bigr)}\Bigl(corr\Bigr)\bigg/\Bigl(\frac{4}{3}{\pi}{d_{max}}^3\Bigr)$$ where $\sigma_{raw} = \sqrt{raw}$.}
\tablenotetext{c}{This bin is complete only for its L dwarf complement. Since late-M dwarfs are also expected to populate this bin, the derived space density is considered to be a lower limit.}
\end{deluxetable}

\begin{figure*}
\figurenum{23}
\includegraphics[scale=0.24,angle=0]{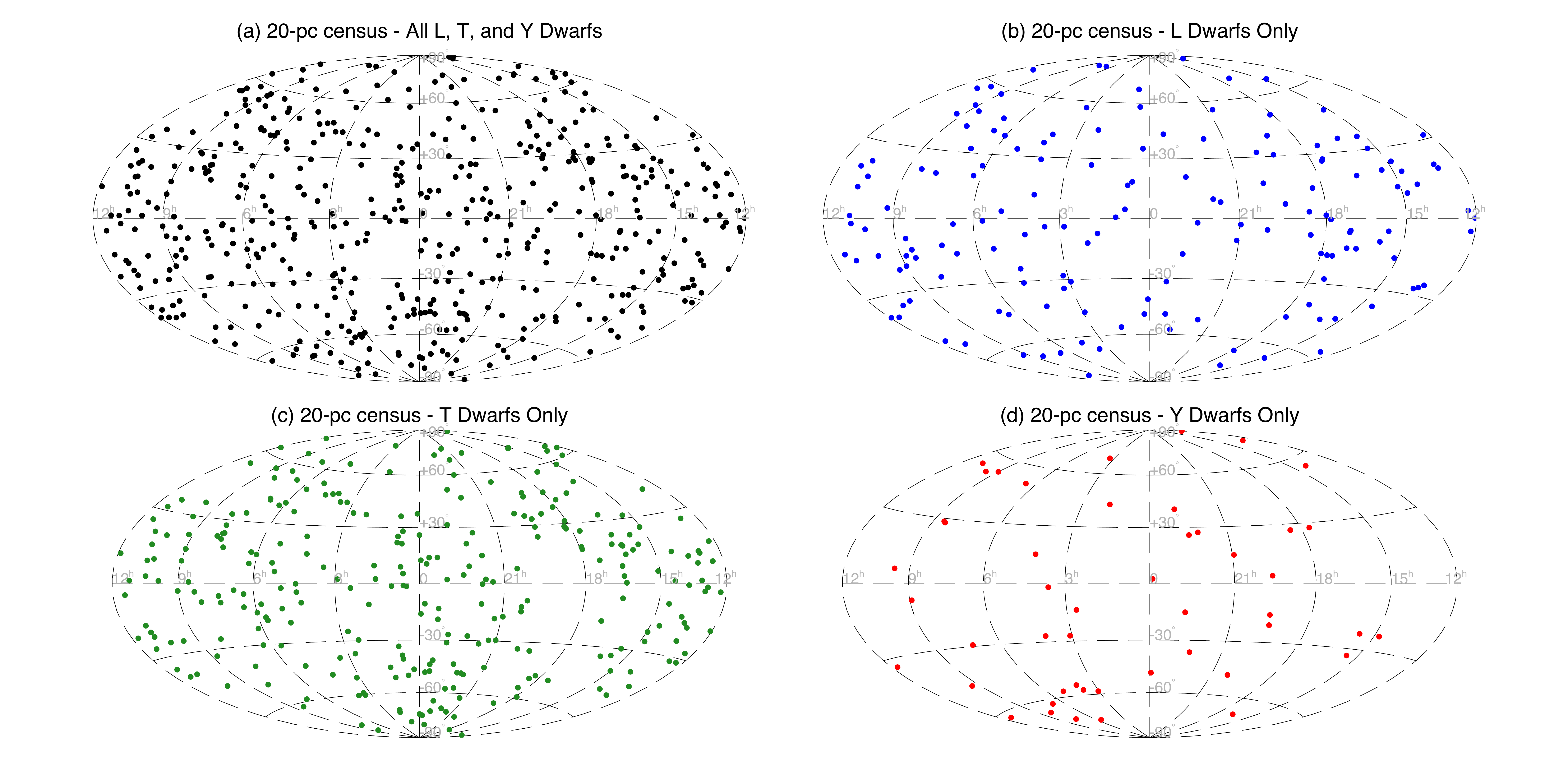}
\caption{Plots of the 20-pc L, T, and Y dwarf sample in equatorial coordinates. The four panels display the sample in its entirety (black), only the L dwarfs (blue), only the T dwarfs (green), and only the Y dwarfs (red).
\label{figure:sky_plot_equatorial}}
\end{figure*}

\begin{figure*}
\figurenum{24}
\includegraphics[scale=0.24,angle=0]{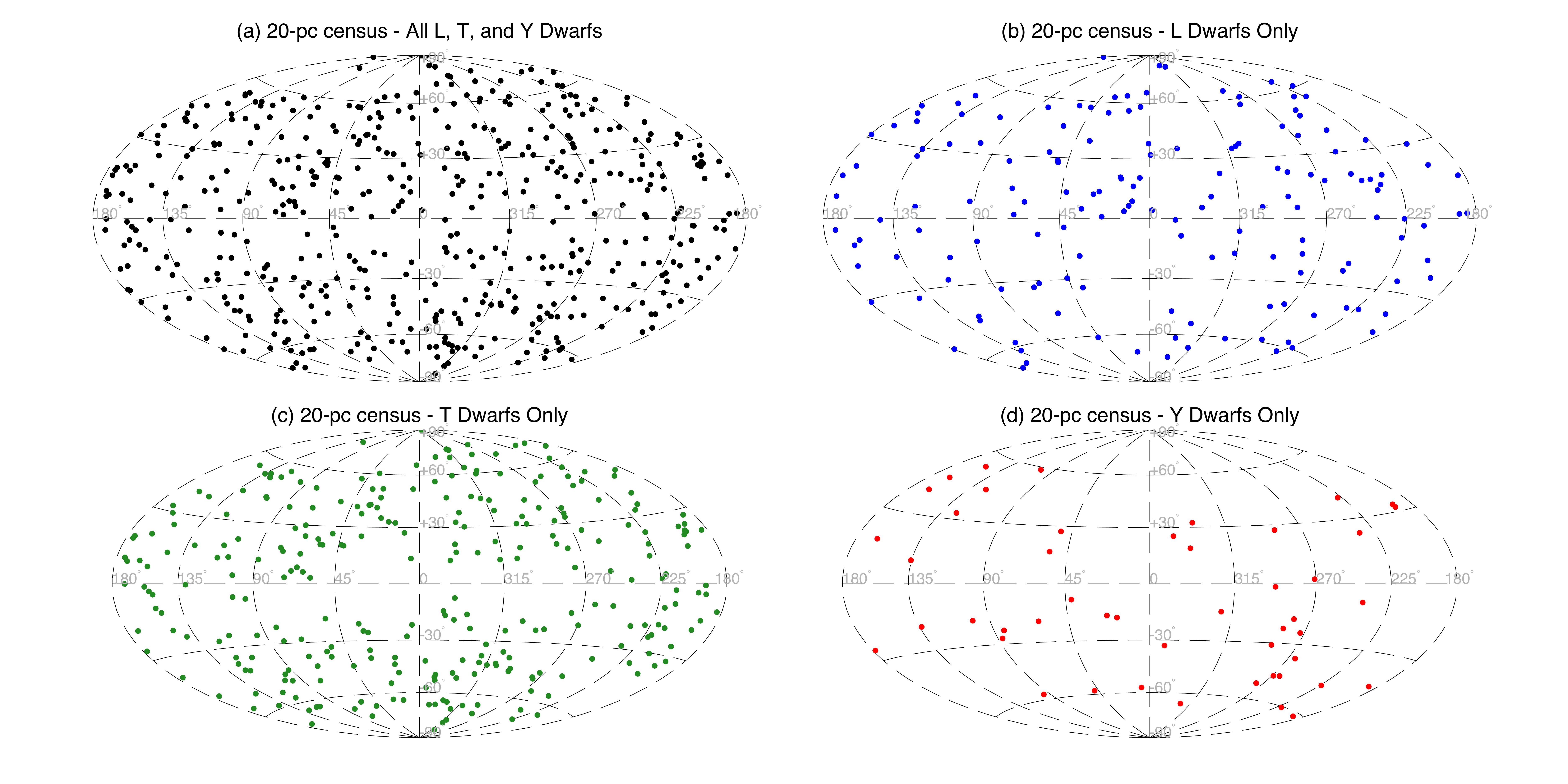}
\caption{Plots of the 20-pc L, T, and Y dwarf sample in Galactic coordinates. See the caption to Figure~\ref{figure:sky_plot_equatorial} for more details.
\label{figure:sky_plot_galactic}}
\end{figure*}

The bins in our $T_{\rm eff}$ and spectral type histograms are fixed, but our confidence in placing an object in a particular bin is directly related to the uncertainties in these quantities. For example, some of our objects have errors on $T_{\rm eff}$ that are comparable to our 150K bin size, and the errors on some of our spectral types are also comparable to the integral spectral type bin size used. The lack of precision in these values is our biggest uncertainty in fixing the space densities in each bin. To address what the size of this uncertainty should be, we have run 10,000 Monte Carlo simulations for both the $T_{\rm eff}$ and  spectral type distributions. For $T_{\rm eff}$ we have taken the error bars listed in Table~\ref{20pc_sample}, which were taken either from literature values (see Table~\ref{table:teff}) or assigned via the root-mean-square scatter from whichever relation in Table~\ref{table:equations} was used for the $T_{\rm eff}$ estimate. For spectral type, we have assigned the standard 0.5-subclass uncertainty to all types except those with uncertainties already specified explicitly or for those with brackets or colons, for which we have assigned 1.0-subclass uncertainties. For each simulation, we take the $T_{\rm eff}$ or spectral type uncertainty, and multiply it by a random value generated from a normal distribution having a mean of 0 and a standard deviation of 1. We add this uncertainty onto the measured value, and then rebin. We then compute the means and standard deviations across all 10,000 simulations and report these in column 4 of Table~\ref{table:space_densities}. 

These simulations have a drawback, however, because the $T_{\rm eff}$ bins at either end of our 150-2250K range are incomplete. Firstly, the 1950-2100K bin will contain objects that scatter into the 2100-2250K bin, but this loss in the cooler bin will not be mitigated by a concomitant gain from the warmer bin because the object count in that latter bin is incomplete. Secondly, over the 300-750K range, we encounter differing completeness limits in distance across the three bins that span this range as well as having an incompleteness in temperature in the 300-450K bin. For example, objects that scatter from the 600-750K bin into the 450-600K bin will be lost if they have a distance larger than the completeness limit of that colder bin. Objects scattering in the other direction will not be similarly lost. The same is true of objects scattering between the 450-600K bin and the 300-450K bin. Given these biases, we adopt a methodology whereby we use the raw number counts in each bin to set the space density, but we use the uncertainties from the simulations to set a conservative limit on their 1$\sigma$ errors. 

Although most of our bins pass the $\langle{V}/{V_{max}}\rangle$ completeness test to 20 pc, this does not address whether there are inhomogeneities in the all-sky distribution. \cite{kirkpatrick2019} found an inhomogeneity in the T and Y dwarf counts toward the Galactic Plane, in which source confusion limits our ability to select objects in the faintest, coldest bins. We re-investigate this here. Plots of our all-sky distributions broken down by broad spectral class are shown in Figures~\ref{figure:sky_plot_equatorial} and \ref{figure:sky_plot_galactic}. The plot of T dwarfs appears to show a thinner area of coverage around and just south of the Galactic Plane in  Figure~\ref{figure:sky_plot_galactic}c. 

\begin{figure}
\figurenum{25}
\includegraphics[scale=0.55,angle=0]{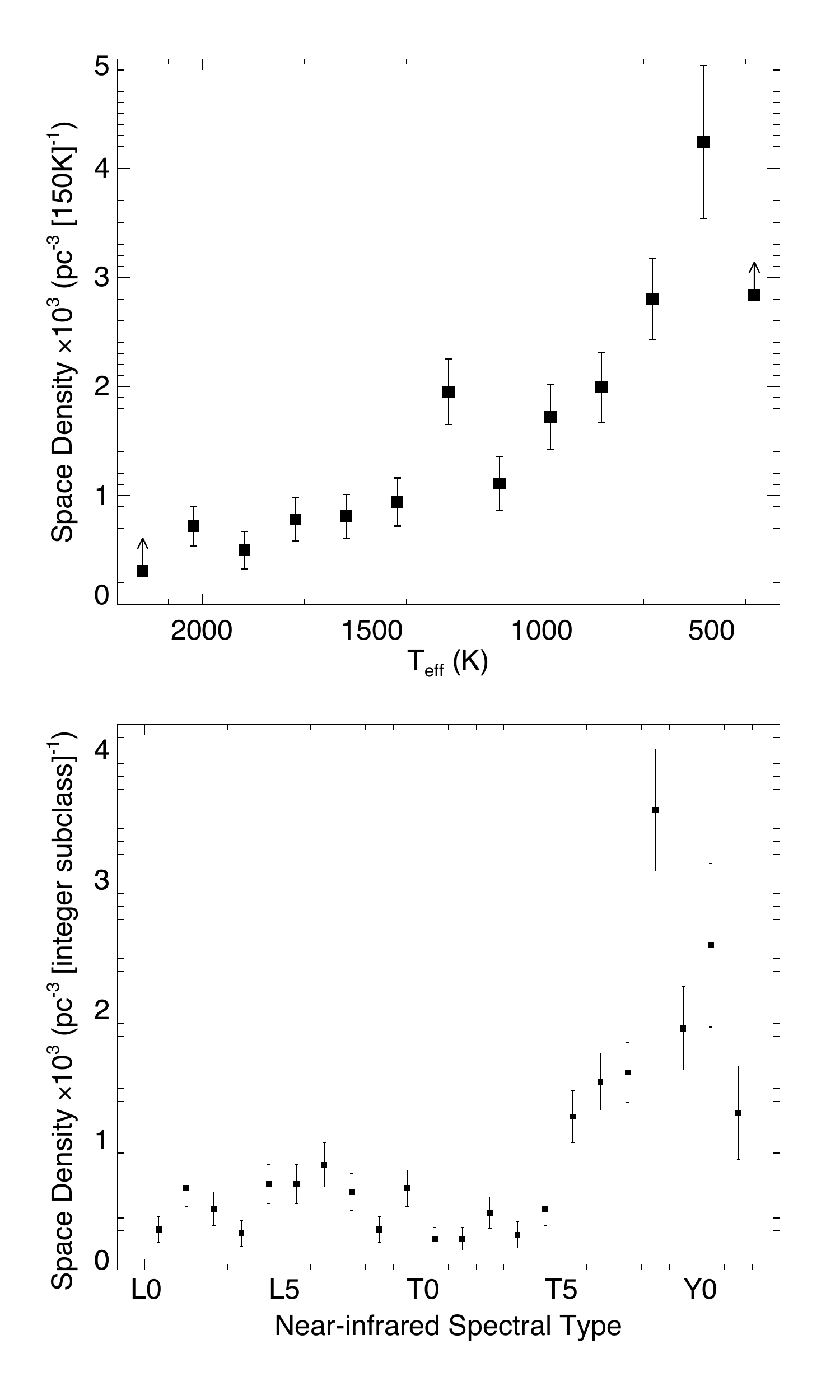}
\caption{Our measured space densities from Table~\ref{table:space_densities}. (Top) Space densities as a function of effective temperature. (Bottom) Space densities as a function of near-infrared spectral type.
\label{figure:space_densities_Teff_SpType}}
\end{figure}

We address this further by dividing objects in our 20-pc census into two sectors, one for the objects having an absolute Galactic latitude ($|glat|$)  $< 14{\fdg}48$ (the "Plane" sector) and the second for objects having $|glat| \ge 14{\fdg}48$. This cut on $glat$ was selected so that the first sector covers one quarter of the sky and the second covers the other three quarters. For each temperature and spectral type bin, we can therefore determine if the numbers in the Plane sector, when tripled, appear to be significantly lower than those found in the second sector. Using the complete samples as defined in Table~\ref{table:space_densities}, we find 27 Y0-Y1.5 dwarfs. Of these, 23 lie outside of the Plane sector, meaning that we would expect $23/3 \approx 8$ similar objects to lie in the Plane sector itself. However, only 4 are found there, for a shortfall of 4, or 15\% of the total sample. Using the same methodology and combining spectral bins to increase the statistical significance of each binned population, we find shortfalls of 13\% for T8-T9.5 (96 objects total), 10\% for T6-T7.5 (88 objects total), 14\% for T4-T5.5 (49 objects total), 12\% for T0-T3.5 (35 objects total), 5\% for L6-L9.5 (75 objects total), and 5\% for L0-L5.5 (96 objects total). We thus apply an adjustment factor of 1.05 across the L dwarf densities and 1.13 across the T and Y dwarf densities. We apply these same factors to the $T_{\rm eff}$-based densities, and use an average adjustment factor of 1.09 to the 1050-1350K bins that cross the L/T transition. These factors are listed in the fourth column of Table~\ref{table:space_densities}. To compute the space densities, we used the formulae given in the footnotes of Table~\ref{table:space_densities}. These final values are given in column 6 and are represented graphically in Figure~\ref{figure:space_densities_Teff_SpType}. 

We can compare these results to other recent determinations in the literature. At early-L types, \cite{bardalez2019} find space densities of [0.75$\pm$0.13, 1.02$\pm$0.16, 0.78$\pm$0.14, 0.58$\pm$0.12, 0.88$\pm$0.15, 1.44$\pm$0.19] ${\times}10^{-3}{\rm pc}^{-3}$ per integral spectral type bins of [L0-L0.5, L1,L1.5, L2-L2.5, L3-L3.5, L4-L4.5, L5-L5.5]. Our space density determinations across each of these bins differ by an average of 2.1$\sigma$, and the \cite{bardalez2019} results are consistently a factor of $\sim$1.9 higher. However, Bardalez Gagliuffi (priv.\ comm.) find that their published densities included a pessimistic set of assumptions in their completeness calculation. Our Table~\ref{table:space_densities} values compare favorably to the $T_{\rm eff}$-binned values of \cite{kirkpatrick2019}, the biggest deviations being a 1.2$\sigma$ variation (difference factor of 0.84 between \citealt{kirkpatrick2019} and this paper) in the 750-900K bin and a 1.4$\sigma$ variation in the opposite direction (difference factor of 1.27) in the adjacent 600-750K bin. 

\section{Determining the Mass Function\label{determining_the_MF}}

In \cite{kirkpatrick2019} we developed a formalism for translating various forms of the mass function into the observational domain, since mass is not an observable quantity for most objects within the 20-pc census. There are several steps in doing this, which we summarize below.

First, we considered a variety of functional forms of the mass function that have been proposed in the literature. These include power laws ($dN/dM \propto M^{-\alpha}$) with $\alpha$ values ranging from $-$1.0 to 1.5, the log-normal distribution ($dN/dM \propto e^{-(ln(M) - \mu)^2/2\sigma^2}$) with values of the mean ($\mu$) and standard deviation ($\sigma$) taken from \cite{chabrier2001}, \cite{chabrier2003a}, and \cite{chabrier2003b}, and a bi-partite power law favored by \cite{kroupa2013}. These forms determine the distribution of masses produced.

Second, a stellar birthrate that has remained constant in time over the past 10 Gyr was assumed. \cite{burgasser2004b} found that the stellar luminosity function for T dwarfs is largely invariant to the birthrate assumed, although the L dwarf regime can still bear an imprint from recent events if star formation is more episodic. \cite{allen2005} explored this further and found that changes in the luminosity function produced by the underlying mass function were much larger than those produced by variations in the birthrate.

Third, because most of the objects in our simulations are brown dwarfs, the observable quantity we use for the empirical determinations ($T_{\rm eff}$) changes with time as the brown dwarf ages and cools. Hence, we tie each simulated object to an evolutionary path applicable to its mass, so that we can determine its current $T_{\rm eff}$. Two sets of evolutionary models were employed for this, resulting in two different sets of simulated $T_{\rm eff}$ distributions. The first were the solar-metallicity COND models from \cite{baraffe2003} that, because they neglect dust opacity, are most applicable to mid-M dwarfs and mid- to late-T dwarfs believed to be free of photospheric clouds. These model grids are sampled at five different ages (0.1, 0.5, 1, 5, and 10 Gyr) and sample the temperature range $125K \lesssim T_{\rm eff} \lesssim 2800K$, which corresponds to masses around $0.01 M_\odot < M < 0.10M_\odot$. The second set of models were the hybrid suite of solar-metallicity models from \cite{saumon2008} that assume cloud-free atmospheres only in the late-M and late-T zones but account for cloud growth and subsequent clearing in and around the transition from L dwarfs to T dwarfs. The evolutionary model grids are sampled at twenty-six different ages in the 3 Myr $<$ age $<$ 10 Gyr range and cover the range $300K \lesssim T_{\rm eff} \lesssim 2400K$, which corresponds to the mass range $0.002M_\odot < M < 0.085M_\odot$.

Fourth, we used the inverse transform sampling method to turn the various forms of the mass function into space densities binned in $T_{\rm eff}$. The process is as follows. Each normalized mass function can be used as a probability density function, which gives the likelihood of drawing at random an object of a certain mass from within that distribution. In a practical sense, this random drawing is done by integrating under the probability density function to produce a cumulative distribution function, reversing the dependent and independent variables, and re-solving for the dependent variable, thus creating the inverse cumulative distribution function which then provides a mapping from the a random seed to an actual mass. The seed is produced via a random sampling of a uniform distribution over the range zero to one.

Fifth, we performed the simulations by creating $3{\times}10^6$ random seeds, each of which was assigned an age according to its order of selection. These ages were distributed uniformly over the subset of 0-10 Gyr interval over which each evolutionary model is valid. The seed was then passed through the inverse cumulative distribution function to assign its mass, then the assigned age and mass were passed through the evolutionary models to get the current $T_{\rm eff}$. Because the evolutionary models are sampled only on a sparse grid, bilinear interpolation between neighboring points was used to assign the temperature.

Finally, simulations were produced for each of the twelve assumed functional forms of the mass function, each of which was run through the two different evolutionary model grids. Furthermore, each simulation was run with three different values of a cutoff mass (10$M_{Jup}$, 5$M_{Jup}$, or 1$M_{Jup}$,), which is the lowest mass product that can be created. This resulted in a grid of seventy-two simulated $T_{\rm eff}$ distributions.

\subsection{Mass Function Fits}

Here, we have compared our measured space densities to these seventy-two simulations. To determine the simulation that fits best, we have used the IDL routine {\tt mpfit} (\citealt{markwardt2009}) to perform a weighted least-squares fit between the data and the simulations, where the only adjustable parameter is the scaling between the arbitrary number counts in the models and our measured space densities. For the calculation, we use only the eleven values in the upper portion of Table~\ref{table:space_densities} that cover the range 450-2100K, as the other values are lower limits only. The best fit to each model produces a reduced $\chi^2$ value.

\begin{figure*}
\figurenum{26}
\includegraphics[scale=0.85,angle=0]{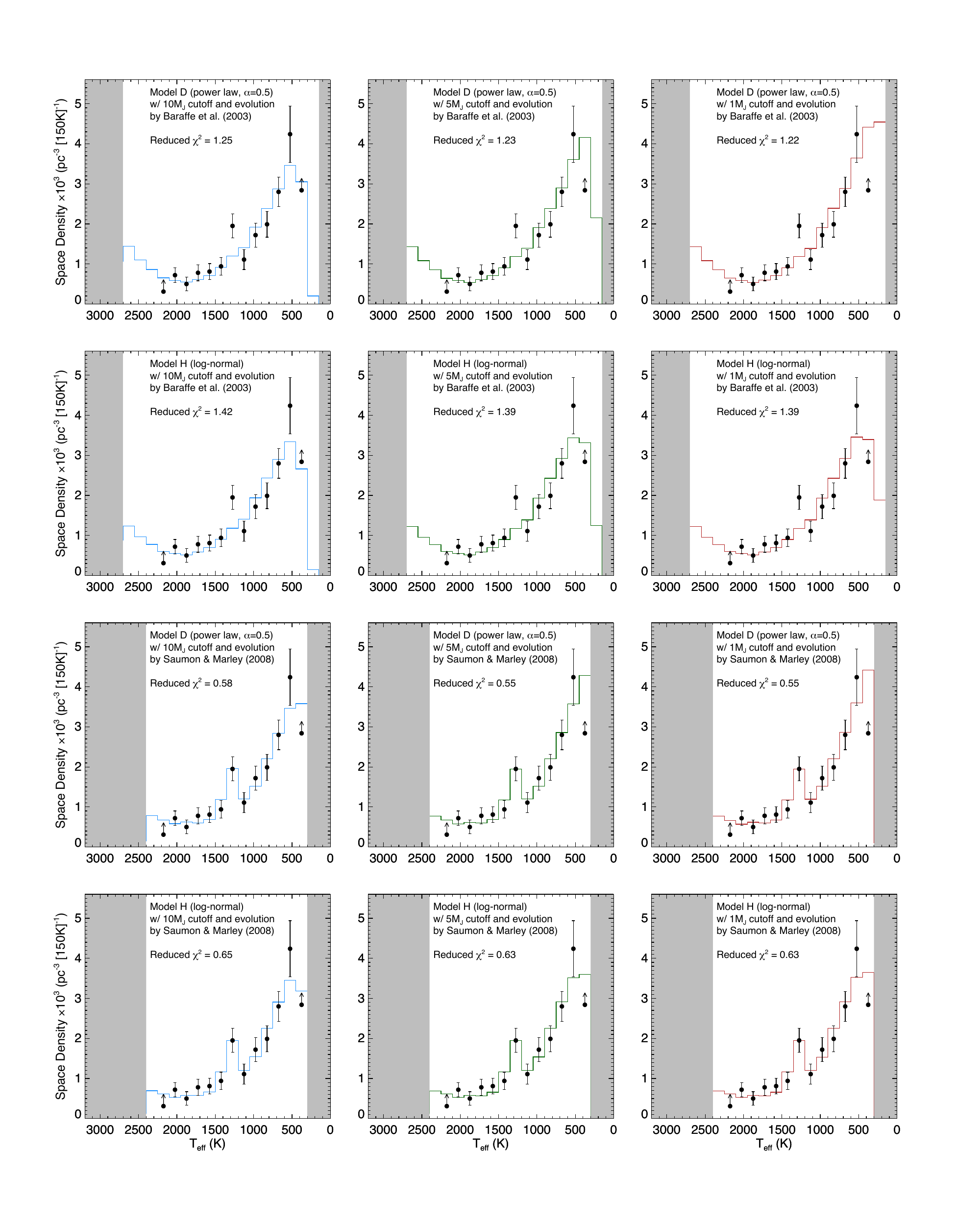}
\caption{The best fits between the simulations and our measured space densities. Of the simulations that use the evolutionary tracks of \cite{baraffe2003}, the two with the smallest reduced $\chi^2$ values are shown in the top two rows. Of the simulations that use the evolutionary tracks of \cite{saumon2008}, the two that provide the best fits are show in the two bottom rows. "Model D" refers to the power law with $\alpha = 0.5$, and "Model H" refers to the single-object log-normal form of \cite{chabrier2001}. See \cite{kirkpatrick2019} for additional information on these simulations. Each row shows the same model with a different low-mass cutoff: 10$M_{Jup}$ (blue) in the left panel, 5$M_{Jup}$ (dark green) in the middle panel, and 1$M_{Jup}$ (red) in the  right panel. Our measured space densities and their uncertainties are shown in black. Grey zones denote areas not covered by the simulations.
\label{figure:mf_fits_full_grid}}
\end{figure*}

Figure~\ref{figure:mf_fits_full_grid} shows the fits for which this value is minimized. These best fits are identical to the best fits found by \cite{kirkpatrick2019}, and involved the single power law and log-normal forms. For each evolutionary model, the power law form is slightly favored over the log-normal based on the best-fit $\chi^2$ minimization values. In contrast to the results of \cite{kirkpatrick2019}, we now find that the evolutionary code of \cite{saumon2008} is highly favored over that of \cite{baraffe2003}, and the reason for this is the inclusion in this paper of space density measurements over the cloudy-to-clear transition that the \cite{saumon2008} models were designed to address. Specifically, the space density spike in the 1200-1350K bin of Figure~\ref{figure:mf_fits_full_grid} is well produced by simulations incorporating the \cite{saumon2008} models, and this bin is the one covering spectral types from $\sim$L8 to $\sim$T3 (the yellow zone in Figure~\ref{figure:yaxis_teff}b) over which cloud building and subsequent break-up have been hypothesized. These models not only predict the position of the spike but also correctly predict its magnitude. Furthermore, they also predict the magnitude of the drop-off and recovery at cooler types once clouds have cleared and cooling once again proceeds as normal.

\begin{figure}
\figurenum{27}
\includegraphics[scale=0.50,angle=0]{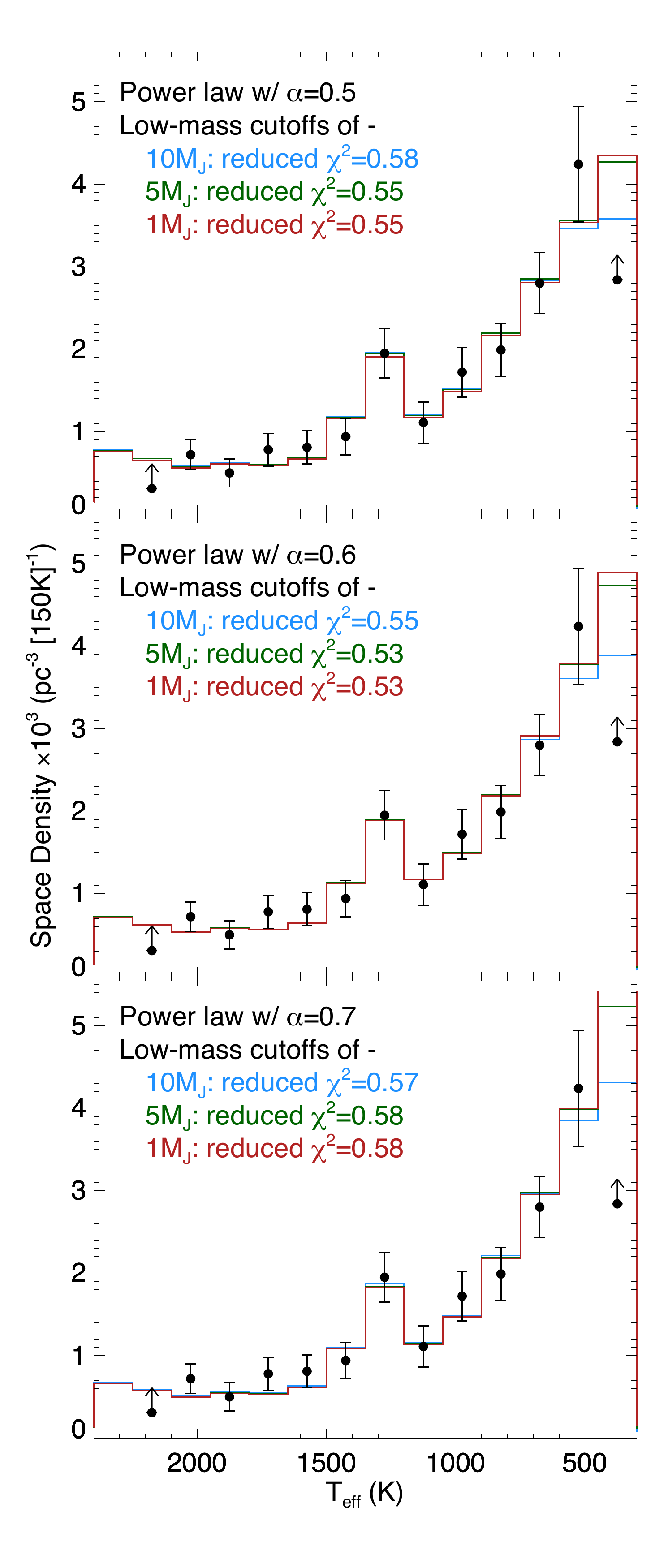}
\caption{Fits of power laws with $\alpha = 0.5$ (top panel), $\alpha = 0.6$ (middle panel), and $\alpha = 0.7$ (bottom panel) to our observational data (black points). These predicted $T_{\rm eff}$ distributions have been passed through the evolutionary models of \cite{saumon2008}. Each panel shows simulations for three low-mass cutoffs: 10 $M_{Jup}$ (blue), 5 $M_{Jup}$ (green), and 1 $M_{Jup}$ (red). The minimum reduced $\chi^2$ values are found for the $\alpha=0.6$ model. 
\label{figure:mf_fits_three_best}}
\end{figure}

The best fits across the coarse grid of 72 models are those with the single power law of $\alpha = 0.5$. Figure~\ref{figure:mf_fits_three_best} illustrates a few supplemental simulations to show that the minimum $\chi^2$ value across a finer grid of models is actually reached at $\alpha = 0.6$, which was the same conclusion found by \cite{kirkpatrick2019}. There is however, no significant difference between the $\chi^2$ values of the $\alpha = 0.5$, 0.6, and 0.7 models. Obtaining a more accurate space density in the 450-600K bin is critical to pinning down the true value of $\alpha$.

As a closer look at Figure~\ref{figure:mf_fits_three_best} reveals, the preferred value of $\alpha$ rests largely with the steepness of the curve over the 1200-450K region, and most of the power falls in that region's final bin (450-600K), for which the space density is the highest. If we use the densities implied by our temperature randomizations (column 4 of Table~\ref{table:space_densities}), we find a best fit of $\alpha = 0.4$, although, as discussed earlier, the density for that bin is likely biased low. This leads us to conclude that our measurements of the space density support a value of $\alpha = 0.6{\pm}0.1$.

\subsection{The Low-mass Cutoff}

Whereas the 450-600K bin is critical in determining the value of the power law's exponent, the next cooler bins are critical in determining the cutoff mass. The best fits to our observed space densities currently do not have a strong dependence on the low-mass cutoff. As the plots in Figure~\ref{figure:mf_fits_three_best} show, this is because the lower limit to the density in the 300-450K bin is consistent with all three values of the cutoff mass (10, 5, and 1 $M_{Jup}$ ). An increase of just 40\% in the value of this lower limit would enable us to confidently claim a cutoff mass below 10$M_{Jup}$. (In \citealt{kirkpatrick2019} we had claimed to push the cutoff mass below {\bf 5}$M_{Jup}$, but this was based on a number of objects in the 300-450K bin that was half as large as the sample we are now using.) This bin is comprised mostly of Y0.5 to Y2 dwarfs (Figure~\ref{figure:yaxis_teff}b), which are challenging objects to uncover given their faint absolute magnitudes ($M_J \approx M_H > 23$ mag, $M_{W2} = M_{\rm ch2} > 15$ mag; Figure~\ref{figure:xaxis_spectype}).

Even more critical to defining the low-mass cutoff is the next cooler bin, 150-300K, which presently has only one known object in it, WISE 0855$-$0714. Finding more representative objects in this bin would even more readily determine the cutoff mass, as the top row of Figure~\ref{figure:mf_fits_full_grid} shows. For the $\alpha =0.5$ model, the space density values in this bin vary wildly -- from ${\sim}0.2{\times}10^{-3} {\rm pc}^{-3}$ for a 10$M_{Jup}$ cutoff, to ${\sim}2.2{\times}10^{-3} {\rm pc}^{-3}$ for a 5$M_{Jup}$ cutoff, to ${\sim}4.5{\times}10^{-3} {\rm pc}^{-3}$ for a 1$M_{Jup}$ cutoff. Finding objects in this bin is an even more challenging proposition, as WISE 0855$-$0714 itself has absolute magnitudes of $M_J \approx 28$ mag, $M_H \approx 27$ mag, and $M_{W2} = M_{\rm ch2} \approx 17$ mag.

Nonetheless, we can use objects of known mass within the 20-pc census to help further refine the cutoff value. Most notably, a number of census members are known to belong to young moving groups and associations (section~\ref{section:known_youngs}), and these objects will have hotter temperatures and earlier spectral types than older counterparts in the field of the same mass. Hence, finding an object of exceedingly low mass is a far less daunting challenge if is it younger and brighter. Young members of the 20-pc census are listed along with their assigned $T_{\rm eff}$ values and published masses in Table~\ref{table:masses}.

\begin{figure*}
\figurenum{28}
\gridline{\fig{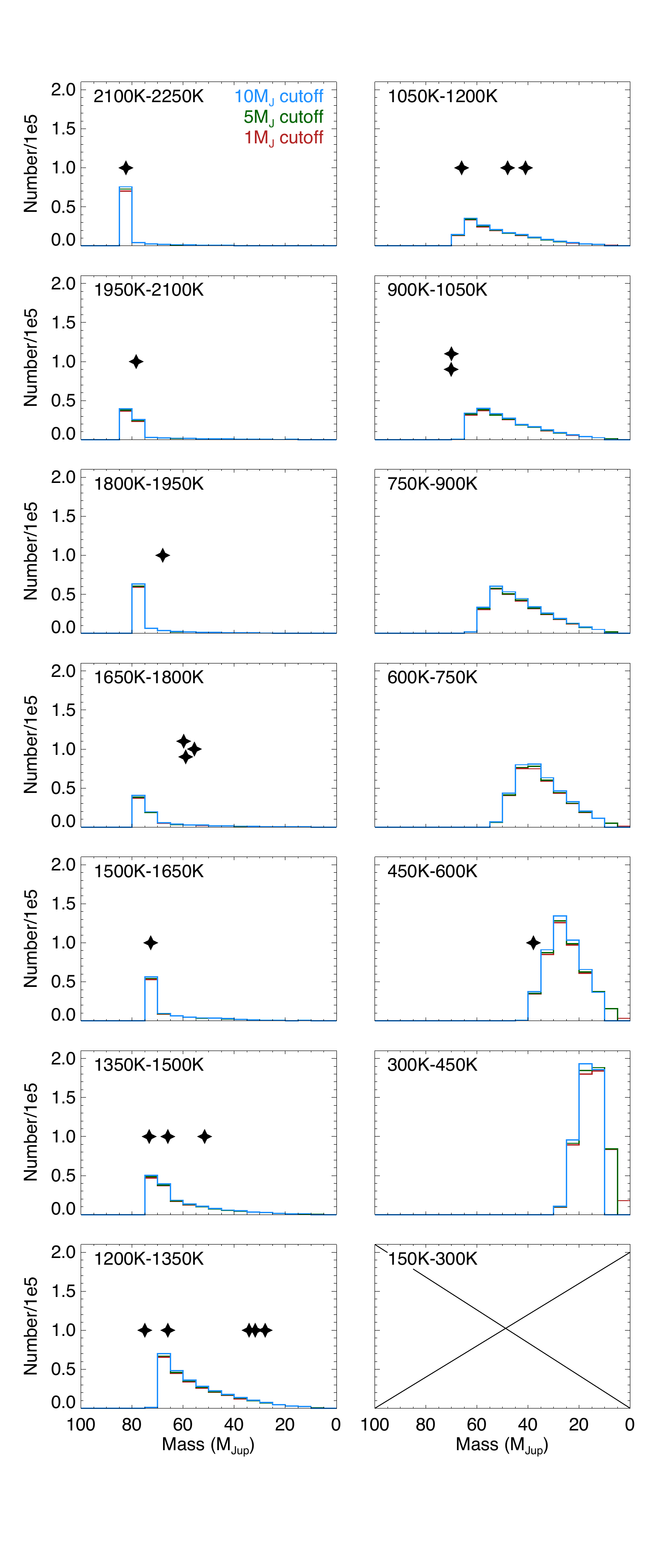}{0.45\textwidth}{(a)}
          \fig{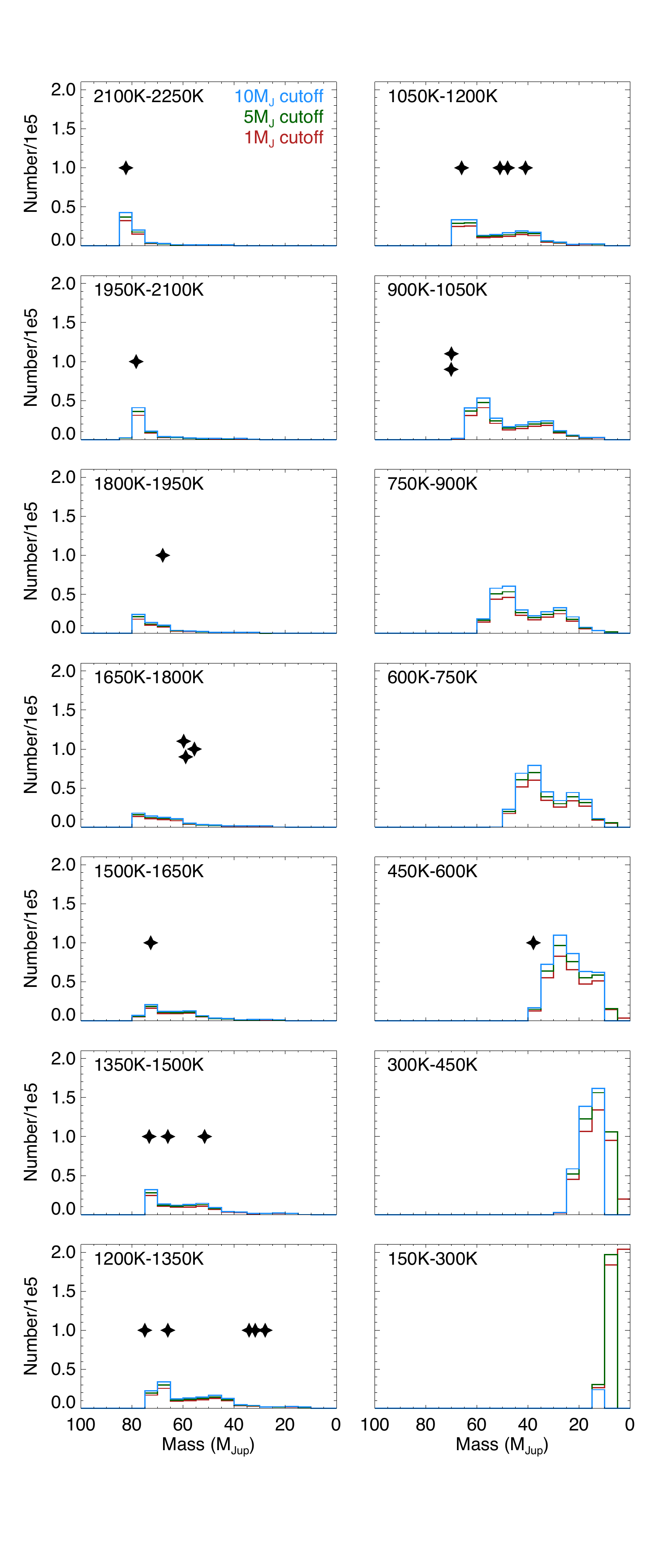}{0.45\textwidth}{(b)}}
\caption{Simulated mass distributions for each of the 150K $T_{\rm eff}$ bins. (a) The single power law of $\alpha = 0.5$ coupled with the \cite{saumon2008} evolutionary tracks. (b) The same, but coupled with the \cite{baraffe2003} evolutionary tracks. Because the \cite{saumon2008} models do not extend below 300K, the bin at lower right in panel (a) is empty. For ease of comparison, the same $x$ and $y$ scaling is used for all subpanels. Objects from Table~\ref{table:masses} that have dynamically measured masses (filled black stars) are plotted in their $T_{\rm eff}$ bins at the $x$ location corresponding to their mass; their $y$ positions are arbitrary.
\label{figure:mf_mass_distributions}}
\end{figure*}

Before exploring these masses, though, we note that such determinations are direct comparisons to evolutionary models and thus fail to provide an independent check of the theory. Are the masses coming from the evolutionary models trustworthy? To answer this, we have also listed in Table~\ref{table:masses} those multiple systems within the 20-pc census whose masses have been measured dynamically. These objects are identified with their corresponding $T_{\rm eff}$ bin and indicated in Figure~\ref{figure:mf_mass_distributions}. This figure shows, for both the \cite{saumon2008} and \cite{baraffe2003} evolutionary tracks, the expected mass distributions from our simulations for each of our 150K bins. The simulations show a tight distribution of masses for the hotter bins, but the range of masses quickly expands for the colder bins. In the \cite{saumon2008} models, a wide range of masses is expected to inhabit each of the temperature bins from 750K to 1500K. At colder temperatures, though, the mass range reduces dramatically, with the 300-450K bin containing only objects with masses below $\sim30 M_{Jup}$. (Using the \citealt{baraffe2003} models, which explore even colder temperatures, we find that the mass range shrinks to $< 15 M_{Jup}$ for the 150-300K bin.)

For the warm bins with the narrowest mass distributions (2100-2250K and 1950-2100K), the two objects in Table~\ref{table:masses} with dynamical measures have masses in accordance with the model predictions. Good agreement is seen at cooler bins as well. The only objects with measures that may be discrepant with expectations are the four objects in the 1650-1950K range (Gl 584B and C, DENIS 2252+1730A, 2MASS 0700+3157A) in panel (a), the highest mass object in the 1200-1350K bin (Gl 845B) along with the two objects in the 900-1050K bin (Gl 229B and Gl 845C) of both panels, and the three lowest mass objects (SDSS 0423$-$0414B and WISE 1049$-$5319AB) in the 1200-1350K bin of panel (b). These latter three objects can be explained as the inability of the older \cite{baraffe2003} models to account for clouds in this range, since these objects do not appear unusual when compared to the expectations from \cite{saumon2008}.

The other objects deserve closer scrutiny:

\begin{itemize}
    \item Gl 564BC: This pair has  masses lower than 85\% of objects in the 1650-1800K bin. Objects of this mass, according to our simulations, would have a relatively young age of $\sim$580$\pm$67 Myr. \cite{potter2002} note that the primary in this system, the G2 dwarf Gl 564A, is chromospherically active, a fast rotator, and an object of high lithium abundance, which places its age at $<$800 Myr. After a more careful analysis, \cite{dupuy2009} adopt an age for the primary of $790^{+220}_{-150}$ Myr, which accords with the young age expected by our simulations. 
    
    \item DENIS 2252$-$1730A: The is the third other object in the 1650-1800K bin. It has a dynamical mass intermediate between Gl 564B and Gl 564C and would thus be expected from our simulations to have a similarly young age. However, there does not appear to be independent verification of a young age in the literature, such as a measurement of lithium absorption in the A component (\citealt{dupuy2017}). 
    
    \item 2MASS 0700+3157A: This object falls in the 1800-1950K bin. Our simulations find that it has a mass lower than 85\% of objects in its temperature bin, implying another relatively young age of 755$\pm$101 Myr. There is no independent assessment of age for this object, although \cite{dupuy2017} also note the model-implied young age for the primary. As stated in that work, \cite{thorstensen2003} report no lithium in the joint spectrum of the AB pair, which would likely mean only that the age is $>$200 Myr. 
    
    \item Gl 845BC: The masses of both components are surprisingly high for their respective temperature bins. In our simulations that use the \cite{saumon2008} evolutionary models, we find $\sim$250,000 objects in our 3-million-object simulation that fall in the 1200-1350K bin inhabited by Gl 845B but none of these simulated objects has a mass as high as Gl 845B. Likewise, of our $\sim$190,000 simulated objects in the 900-1050K bin, none has a mass as high as Gl 845C. This system is not believed to be exceptionally old, either (see \citealt{dieterich2018}), which might partly explain the ultra-high masses. Switching to the \cite{baraffe2003} evolutionary code instead gives a similar result. The published mass measurements for this system are completely at odds with theoretical expectations. 
    
    \item Gl 229B: This object has an ultra-high mass for its effective temperature. Its measured mass is almost identical to that of Gl 845C, so the arguments for Gl 845C above also apply to Gl 229B. \cite{brandt2020} note that an exceptionally old age for the Gl 229 system is disfavored, making Gl 229B another T dwarf whose mass measurement is at odds with expectations.
\end{itemize}

In summary, then, the masses expected from our simulations are consistent with the measured dynamical masses in Table~\ref{table:masses} for most objects for which direct comparisons can be done. The exceptions are Gl 229B and Gl 845BC, which remain puzzles.

The consistency between most of the measurements and the expected values at higher masses gives us a cautious confidence -- but not independent confirmation -- in trusting model-implied values at lower masses. Of the 20-pc moving group members listed in Table~\ref{table:masses}, the ones of lowest mass are between 10 to 12 $M_{Jup}$. So, within the 20-pc census, we are not able to push the cutoff mass below 10 $M_{Jup}$ through either a critical analysis of the entire L, T, and Y sample or through an analysis of the subset with moving group membership. Despite this limitation, we can look at the young moving group members in a larger sample volume, which strongly hint at a low-mass cutoff substantially below 10 $M_{Jup}$. As discussed in section~\ref{section:known_youngs}, PSO J318.5338$-$22.8603, 2MASSW J1207334$-$393254b, and 2MASS J11193254$-$1137466AB are believed to have masses in the 4-7 $M_{Jup}$ range, and other objects identified in Table~\ref{table:young_objects} could possibly lower the limit within the 20-pc census itself.

\begin{deluxetable*}{lccccc}
\tabletypesize{\footnotesize}
\tablecaption{Masses for L, T, and Y Members of the 20-pc Census\label{table:masses}}
\tablehead{
\colhead{Object} &                          
\colhead{Sp.} &
\colhead{T$_{\rm eff}$} &
\colhead{Mass} &
\colhead{Method} &
\colhead{Mass} \\
\colhead{} &                          
\colhead{Type} &
\colhead{(K)} &
\colhead{(M$_{Jup}$)} &
\colhead{} &
\colhead{Ref.} \\
\colhead{(1)} &                          
\colhead{(2)} &  
\colhead{(3)} &  
\colhead{(4)} &
\colhead{(5)} &
\colhead{(6)} \\
}
\startdata
2MASS 0045+1634      & L2$\gamma$  & 2059$\pm$45& $24.98{\pm}4.62$    &  MovGp&  F\\
WISE 0047+6803       & L6-8$\gamma$& 1230$\pm$27& $11.84{\pm}2.63$    &  MovGp&  F\\
SIMP 0136+0933       & T2          & 1051$\pm$198& $12.7{\pm}1.0$     &  MovGp&  G\\
2MASS 0355+1133      & L3-6$\gamma$& 1478$\pm$58& $21.62{\pm}6.14$    &  MovGp&  F\\
%2MASS 0421-6306     & L5$\gamma$  & 1388$\pm$197& {\bf TBD}          &  MovGp&  ?\\
SDSS 0423-0414A      & L6.5:       & 1465$\pm$134& $51.6^{+1.5}_{-2.5}$& dynam&  D\\
SDSS 0423-0414B      & T2          & 1218$\pm$79& $31.8^{+1.5}_{-1.6}$&  dynam&  D\\
AB Dor Cb(0528-6526) & \nodata     & \nodata    & $14{\pm}1$          &  MovGp&  C\\
Gl 229B(0610-2152)   & T7 pec      &  927$\pm$77& $70{\pm}5$          &  dynam&  A\\
2MASS 0700+3157A     & L3:         & 1838$\pm$134& $68.0{\pm}1.6$     &  dynam&  D\\
2MASS 0700+3157B[C]  & L6.5:       & 1465$\pm$134& $73.3^{+2.9}_{-3.0}$& dynam&  D\\
WISE 0720-0846B      & [T5.5]      & 1183$\pm$88 & $66{\pm}4$          & dynam&  T\\
2MASS 0746+2000A     & L0          & 2237$\pm$134& $82.4^{+1.4}_{-1.5}$& dynam&  D\\
2MASS 0746+2000B     & L1.5        & 2029$\pm$134& $78.4{\pm}1.4$     &  dynam&  D\\
WISE 1049-5319A      & L7.5        & 1334$\pm$58 & $34.2^{+1.3}_{-1.1}$& dynam&  V\\
WISE 1049-5319B      & T0.5:       & 1261$\pm$55 & $27.9^{+1.1}_{-1.0}$& dynam&  V\\
%2MASS 1108+6830     & L1$\gamma$  & 1951$\pm$197& {\bf TBD}          &  MovGp&  ?\\
SDSS 1110+0116       & T5.5        &  926$\pm$18& 10-12               &  MovGp&  I\\
LHS 2397aB(1121-1313)& [L7.5]      & 1282$\pm$88& $66{\pm}4$          &  dynam&  D\\
2MASS 1324+6358      & T2: pec     & 1051$\pm$197& 11-12              &  MovGp&  H\\
DENIS 1425-3650      & L4$\gamma$  & 1535$\pm$53& $22.52{\pm}6.07$    &  MovGp&  F\\
Gl 564B(1450+2354)   & L4          & 1722$\pm$134& $59.8^{+2.0}_{-2.1}$& dynam&  D\\
Gl 564C(1450+2354)   & L4          & 1722$\pm$134& $55.6^{+2.0}_{-1.9}$& dynam&  D\\
2MASS 1534-2952A     & T4.5        & 1172$\pm$79& $51{\pm}5$          &  dynam&  D\\
2MASS 1534-2952B     & T5          & 1125$\pm$79& $48{\pm}5$          &  dynam&  D\\
LSPM 1735+2634B      & L0:         & 2274$\pm$88& $87{\pm}3$          &  dynam&  D\\
%WISE 1741-4642      & L6-8$\gamma$& 1145$\pm$197& {\bf TBD}          &  MovGp&  ?\\
Gl 758B (1923+3313)  & T7:         &  581$\pm$88& $37.9^{+1.4}_{-1.5}$&  dynam&  B\\
Gl 779B (2004+1704)  & L4.5$\pm$1.5& 1533$\pm$88& $72.7{\pm}0.8$      &  dynam&  B\\
Gl 802B (2043+5520)  & [L5-L7]     & 1483$\pm$88& $66{\pm}5$          &  dynam&  M\\
Gl 845B (2204-5646)  & T1          & 1236$\pm$79& $75.0{\pm}0.8$      &  dynam&  S\\
Gl 845C (2204-5646)  & T6          &  965$\pm$79& $70.1{\pm}0.7$      &  dynam&  S\\
2MASS 2244+2043      & L6-8$\gamma$& 1184$\pm$10& $10.46{\pm}1.49$    &  MovGp&  F\\
DENIS 2252-1730A     & [L4:]       & 1722$\pm$134& $59{\pm}5$         &  dynam&  D\\
DENIS 2252-1730B     & [T3.5]      & 1190$\pm$79& $41{\pm}4$          &  dynam&  D\\
\enddata
\tablecomments{Legend for method:
{\it MovGp} = mass comes from evolutionary models combined with the known age of the moving group or young association with which this object is a member;
{\it dynam} = mass is measured dynamically.}
\tablecomments{Reference code for mass determination: 
A = \citealt{brandt2020},
B = \citealt{brandt2019},
C = \citealt{climent2019},
D = \citealt{dupuy2017},
F = \citealt{faherty2016},
G = \citealt{gagne2017},
H = \citealt{gagne2018},
I = \citealt{gagne2015},
M = \citealt{ireland2008},
S = \citealt{dieterich2018},
T = \citealt{dupuy2019},
V = \citealt{garcia2017}.
}
\end{deluxetable*}

\subsection{The Age Distribution}

\begin{figure*}
\figurenum{29}
\gridline{\fig{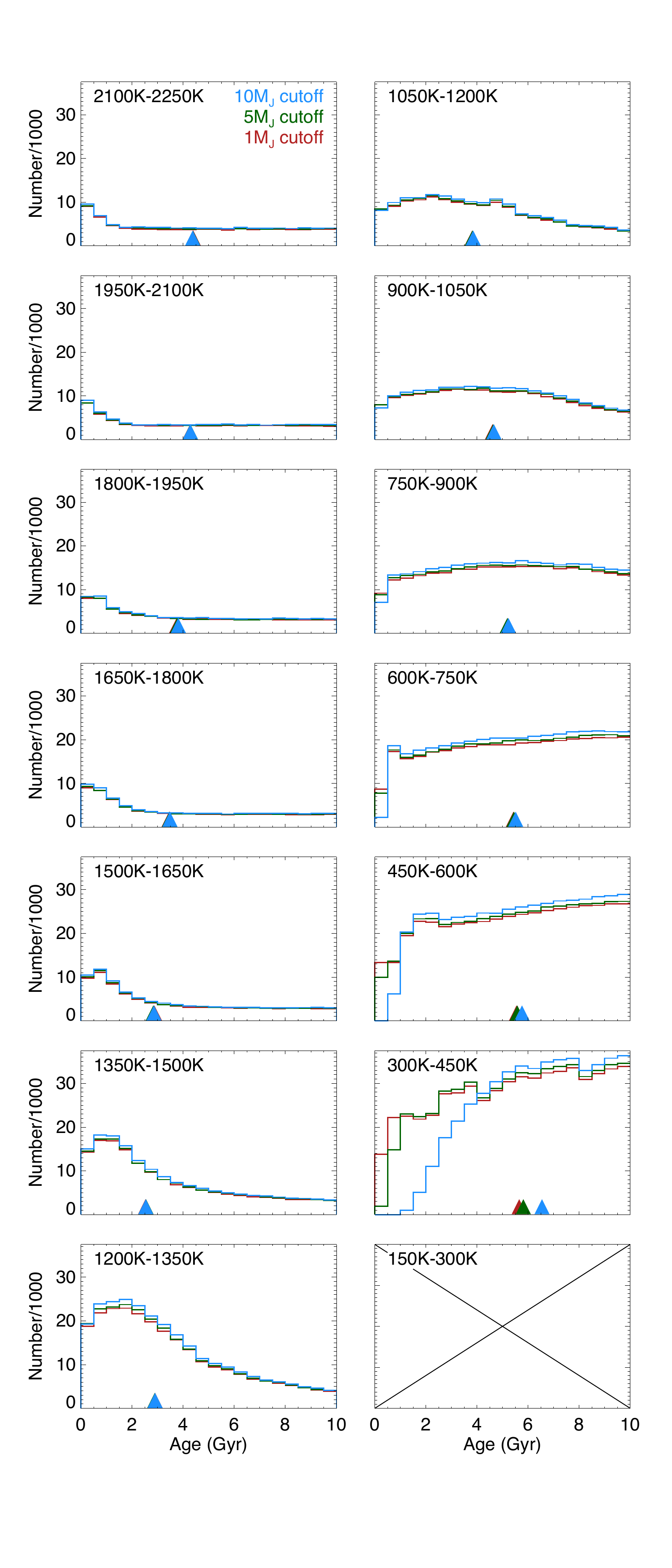}{0.45\textwidth}{(a)}
          \fig{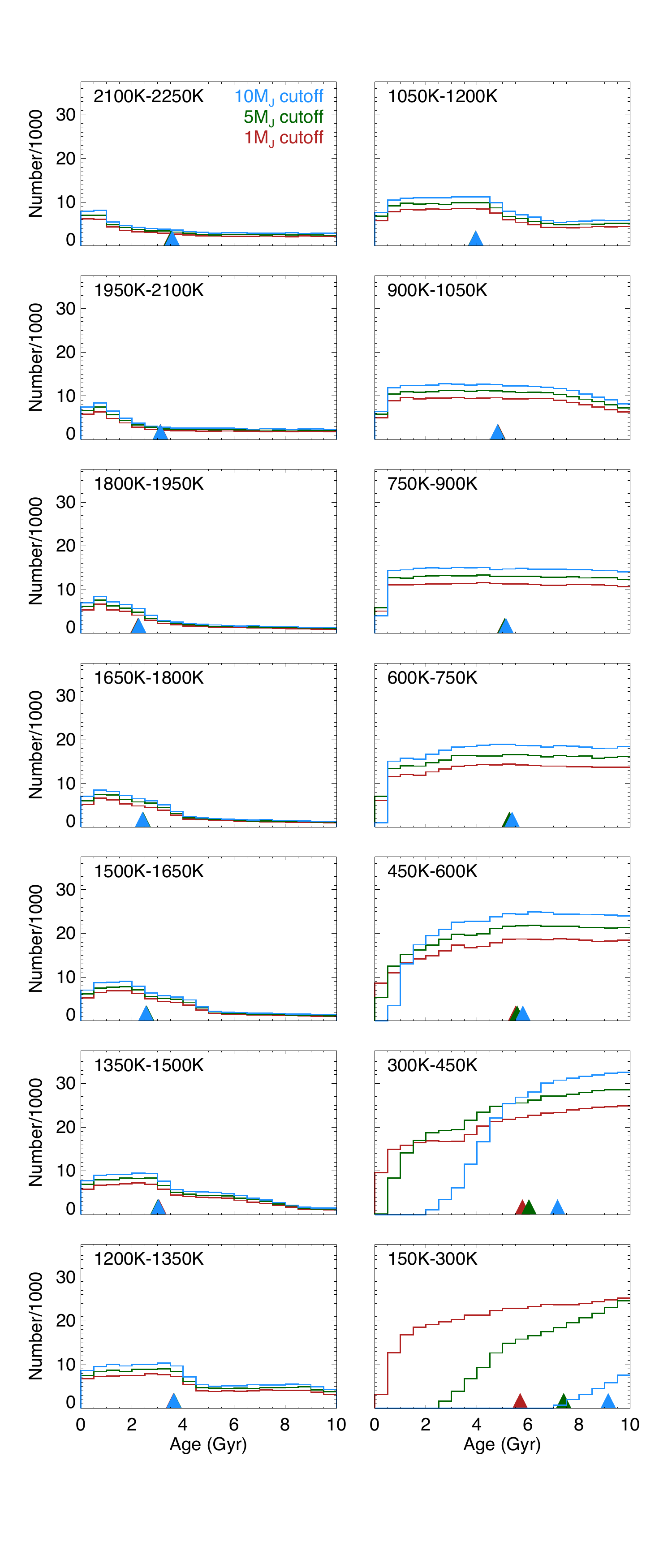}{0.45\textwidth}{(b)}}
\caption{Simulated age distributions for each of the 150K $T_{\rm eff}$ bins. (a) The single power law of $\alpha = 0.5$ coupled with the \cite{saumon2008} evolutionary tracks. (b) The same, but coupled with the \cite{baraffe2003} evolutionary tracks. Because the \cite{saumon2008} models do not extend below 300K, the bin at lower right in panel (a) is empty. For ease of comparison, the same $x$ and $y$ scaling is used for all subpanels. The colored triangles along the bottom edge of each subpanel show the median age for cutoff masses of 10$M_{Jup}$ (blue), 5$M_{Jup}$ (green), and 1$M_{Jup}$ (red); these triangles overlap in all but the coldest bins.
\label{figure:mf_age_distributions}}
\end{figure*}

We can also compare the expected age distributions with our limited knowledge of the ages for objects in the census. Figure~\ref{figure:mf_age_distributions} shows plots analogous to the mass distributions shown in Figure~\ref{figure:mf_mass_distributions}. For the \cite{saumon2008} evolutionary tracks in the 900-2250K regime, the age distributions cover the entire range of 0-10 Gyr ages but with a skew toward young ages. The age distribution then flattens across the 600-900K range, although the youngest ages ($<$0.5 Gyr) start to disappear. A skew toward old ages appears below 600K, with the skew becoming more severe with higher cutoff mass. The \cite{baraffe2003} evolutionary tracks show that this skew toward old ages is exacerbated in the coldest bin (150-300K). Here, a 10 $M_{Jup}$ cutoff mass would imply no objects with ages $<$7 Gyr, whereas a 1 $M_{Jup}$ cutoff would give a much more uniform age distribution, albeit with few objects having ages below 1 Gyr.

\begin{figure}
\figurenum{30}
\includegraphics[scale=0.56,angle=0]{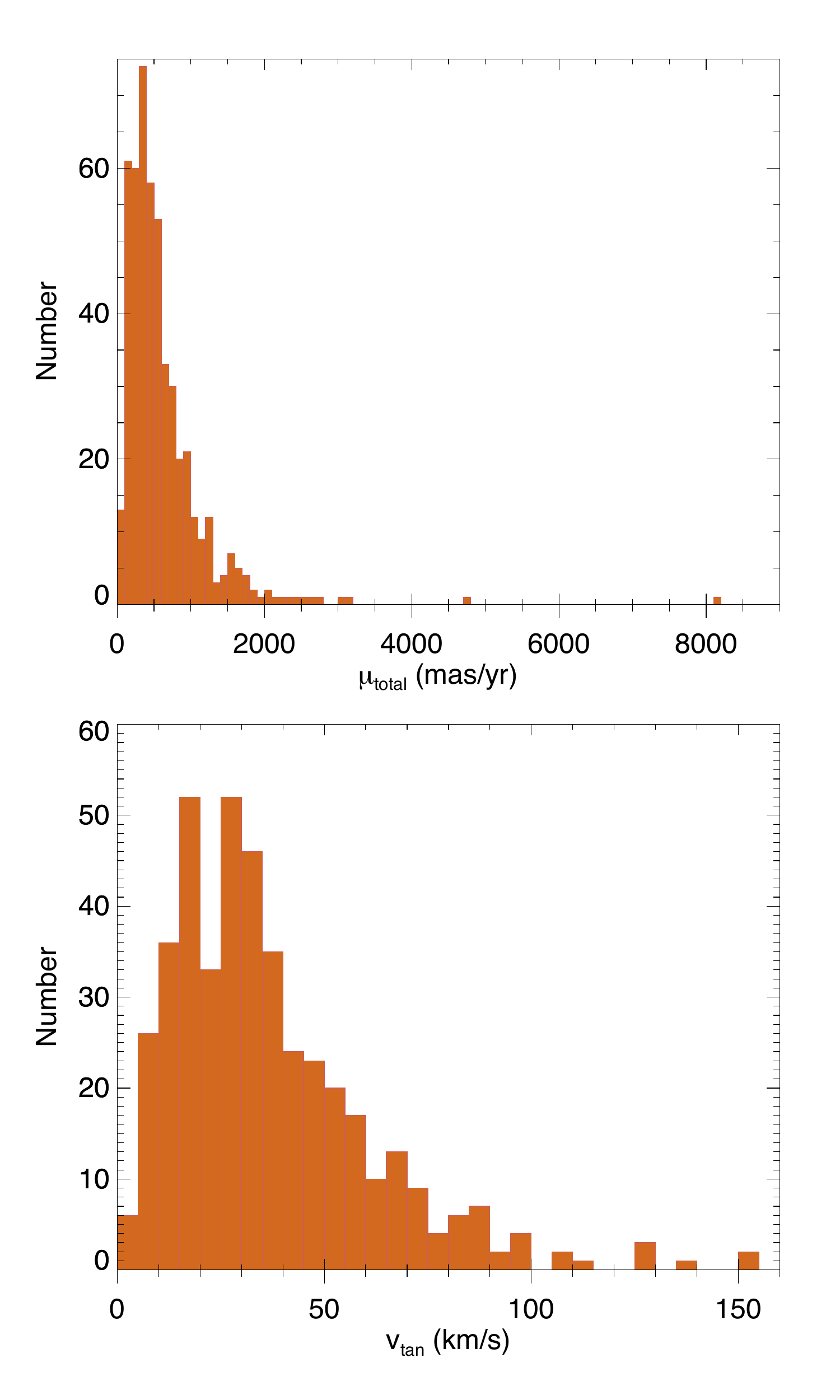}
\caption{Histograms of the total proper motion and $v_{tan}$ for the L, T, and Y dwarfs in the 20-pc census. In the upper diagram, the total motion is shown for all systems in the census. In the lower diagram, the tangential velocity is shown only for those systems having parallax measures with uncertainties below 12.5\%. The median $v_{tan}$ value for objects in the lower panel is 30.8 km s$^{-1}$.
\label{figure:hist_mutot_vtan}}
\end{figure}

Most of the objects in the 20-pc L, T, and Y dwarf census lack age information, but we can examine this using tangential velocities as proxies of dynamical heating. Figure~\ref{figure:hist_mutot_vtan} shows the census' total proper motion and tangential velocity distributions. A total of 2\% of the objects -- nine in total -- have $v_{tan} > 100$ km s$^{-1}$. These objects are 2MASS 0251$-$0352 (112 km s$^{-1}$), 2MASS 0645-6646 (139 km s$^{-1}$), WISE 0833+0052 (106 km s$^{-1}$), 2MASS 1126$-$5003 (127 km s$^{-1}$), 2MASS 1231+0847 (106 km s$^{-1}$), DENIS 1253$-$5709 (128 km s$^{-1}$), 2MASS 1721+3344 (151 km s$^{-1}$), WISE 2005+5424 (129 km s$^{-1}$), and Gl 802B (154 km s$^{-1}$). Three of these are subdwarfs discussed in section~\ref{section:known_sds}, one is a possible subdwarf discussed in section~\ref{sec:suspected_subdwarfs}, two are blue/peculiar L dwarfs, and one is a companion to a mid-M binary believed to be $\sim$10 Gyr old (\citealt{ireland2008}).

For the entire 20-pc census, we can check whether the expected inflation of the velocities at older ages is seen in our empirical data. To accomplish this, we compare the median ages expected from our simulations to the median $v_{tan}$ values from our actual measurements. In Figure~\ref{figure:mf_age_distributions} we illustrate the median age at each 150K bin for our $\alpha = 0.5$ power law simulation. We also plot the measured tangential velocity against effective temperature in Figure~\ref{figure:teff_vs_vtan}, along with the median tangential velocity value in each of the 150K bins. In  Figure~\ref{figure:mf_age_distributions}, we see that the median age shifts to younger values from 2250K down to 1500K and reaches a minimum in the 1350-1500K bin before reversing course and trending to increasingly older values for increasingly cooler bins. Our measured $v_{tan}$ values in Figure~\ref{figure:teff_vs_vtan} show only a little variation across the 500-2250K regime but increase substantially in the 300-450K bin.

Although the agreement is qualitatively the same -- in the sense that the colder, older objects have higher velocities indicative of dynamical heating -- the coldest portion of our sample may be biased toward higher velocities anyway. Objects in the coldest bins are Y dwarfs that are uncovered almost exclusively with {\it WISE} data and should have very red colors of W1$-$W2 $>$ 4 mag. However, given their intrinsic faintness, they are usually not detected at W1, leading to W1$-$W2 color limits only. As the W2 mags themselves grow fainter, this color limit becomes less useful, and thus a detection of proper motion is the best way to discern W2-only Y dwarfs from background chaff. This reliance on a proper motion signature -- which at faint magnitudes is itself only reliable if the motion is large -- leads to a kinematic bias. Thus, the larger median velocity in the 300-450K bin may be a consequence of relying more heavily on motion as a selection criterion.

\subsection{Where are the WISE 0855$-$0714 Analogs?\label{section:vtan}}

In the next fainter bin, 150-300K, WISE 0855$-$0714 is the only object recognized despite concentrated efforts to find other examples by both the Backyard Worlds and CatWISE teams. (With additional follow-up, WISE 0830+2837 from \citealt{bardalez2020} may prove to be the second known member of this $T_{\rm eff}$ bin.) As Figure~\ref{figure:mf_age_distributions}b demonstrates, objects in this bin should be heavily skewed old unless the low-mass cutoff is substantially less than 1$M_{Jup}$. Such a heavy skew to old ages also implies that such objects will be on average more metal poor than the Sun. 

It is possible that analogs to WISE 0855$-$0714 have already been cataloged in the thousands of faint motion candidates already identified by the Backyard Worlds and CatWISE teams but remain unrecognized? After all, many of the objects have W1$-$W2 color limits only and were never imaged by {\it Spitzer} to provide more diagnostic ch1$-$ch2 colors. The answer is almost certainly "no," for the following reason. One of the criteria used to prioritize follow-up observations is the reduced proper motion, $H_{W2} = W2 + 5\log{\mu_{tot}} + 5$, which is a crude measure of the object's intrinsic faintness based on its apparent magnitude and the size of its transverse motion. If any of the motion candidates lacking solid color had distinguished themselves with an exceptionally faint $H_{W2}$ value -- WISE 0855$-$0714 has $H_{W2} = 23.4$ mag (Figure 1 of \citealt{bardalez2020}) -- it would certainly have been noticed. WISE 0830+2837 from \cite{bardalez2020}, with $H_{W2} = 22.6$ mag, is the nearest contender now known.

Four possible scenarios to explain our lack of success in finding additional objects in the 150-300K bin are (1) they are exceedingly rare, (2) their intrinsic faintness places them too close to the W2 detection limit of {\it WISE} for motion searches to identify them confidently, (3) their motions are so high that coadds cannot be used to push the {\it WISE} detection limits deeper, and (4) their colors and magnitudes differ significantly from expectations. We discuss each of these scenarios below:

(1) The coldest objects are rare: Our result that the mass function is best fit with a power law of $\alpha = 0.6$ and that the cutoff mass is likely at or below 5$M_{Jup}$ would imply a distribution of objects in the 150-300K bin like that shown in the green curve in the lower right panel of Figure~\ref{figure:mf_age_distributions}b. This implies a space density of at least $2{\times}10^{-3}pc^{-3}$, which makes objects in this bin as common as T6 or T7 dwarfs. It is thus hard to reconcile these results with the hypothesis that such cold objects are extremely rare. Furthermore, it would be an unbelievable stroke of luck\footnote{It is already an oddity that our G star has, as its four closest neighbors, systems that harbor 1 G dwarf, 1 K dwarf, 2 M dwarfs, 1 L dwarf, 1 T dwarf, and 1 Y dwarf, since a random draw of the overall mass function would be heavily weighted toward M dwarfs plus a random K or T dwarf but weighted against rarer G or L dwarfs. See \citealt{kirkpatrick2012} for the full-sky 8-pc sample.} that our Sun falls a mere 2.3 pc from such an extremely rare, cold object, as it does with WISE 0855$-$0714.  So we reject rarity as a possible cause.

(2) {\it WISE} is too shallow: History has shown us that all-sky surveys can lead to curious results when researchers push those surveys near their limits. The bottom of the main sequence in the 1980s appeared to fall at late-M (\citealt{probst1983}; \citealt{reid1987}) based on the dominant discovery engine of its time, the Palomar Observatory Sky Survey (\citealt{minkowski1963}; \citealt{reid1991}). We now know, of course, that the reason for this is the low space density of early-L dwarfs (see Figure~\ref{figure:mf_fits_three_best}) and the fact that the POSS-I and POSS-II $B$ and $R$ plates failed to survey enough volume to detect all but the nearest L dwarf examples. The L/T pair WISE 1049$-$5319 is present on the southern UK Schmidt photographic plates but was not selected as a motion source (\citealt{luhman2013}); we find that Willem Luyten, despite having cataloged over 58,000 proper motion stars using photographic data (\citealt{luyten1979}), failed to catalog any of the 20-pc L dwarfs in Table~\ref{20pc_sample}. In the case of {\it WISE}, \cite{wright2014} have used the relatively bright W2 magnitude of WISE 0855$-$0714 (W2 = 13.82 mag), its distance (2.3 pc), and the fact that it lies $\sim$2 magnitudes above the limit of the AllWISE Catalog to argue that there should be another 4 to 35 similar objects already detected in AllWISE itself. The CatWISE and CatWISE2020 Catalogs (see below) have increased the sensitivity to lower motions at fainter magnitudes, thus making the identification of these detected objects even easier. Hence, it is unlikely that the survey that found WISE 0855$-$0714 is too shallow to find other analogs.

(3) High motions confound deeper searches: The data sets using the longest time baseline of {\it WISE} data are CatWISE Preliminary (\citealt{eisenhardt2020}) and CatWISE2020 (\citealt{marocco2020b}). Most points on the celestial sphere are visited by {\it WISE} during a several-day window every six months. Both the CatWISE Preliminary and CatWISE2020 processing leveraged these repeats to measure proper motions of all sources. Full-depth coadditions, which took all of the available data to create a single, deep image, were used for source detection. Those source detections were then characterized through the stack of epochal coadds (from each six-month window) to measure photometry and astrometry for each source. Sources with significant proper motions could then be selected from the resulting source tables. Sources that fail to move a significant portion of a full-depth coadd's W2 FWHM (${\sim}6\arcsec$; \citealt{meisner2019}) benefit from the coaddition, as their S/N increases by roughly the square root of the number of epochs. However, sources with higher motions do not see this benefit; a very high motion source will appear as a tracklet of separate sources in the full-depth coadd, and each separate apparition contains the background noise component from all epochs but the source signal from only one. Therefore, faint, high-motion sources can be lost in this process. If many of the coldest brown dwarfs are older kinematically, as Figure~\ref{figure:mf_age_distributions}a and b suggest, their concomitant high proper motions may quash their identification by the CatWISE pipeline.

\begin{figure}
\figurenum{31}
\includegraphics[scale=0.34,angle=0]{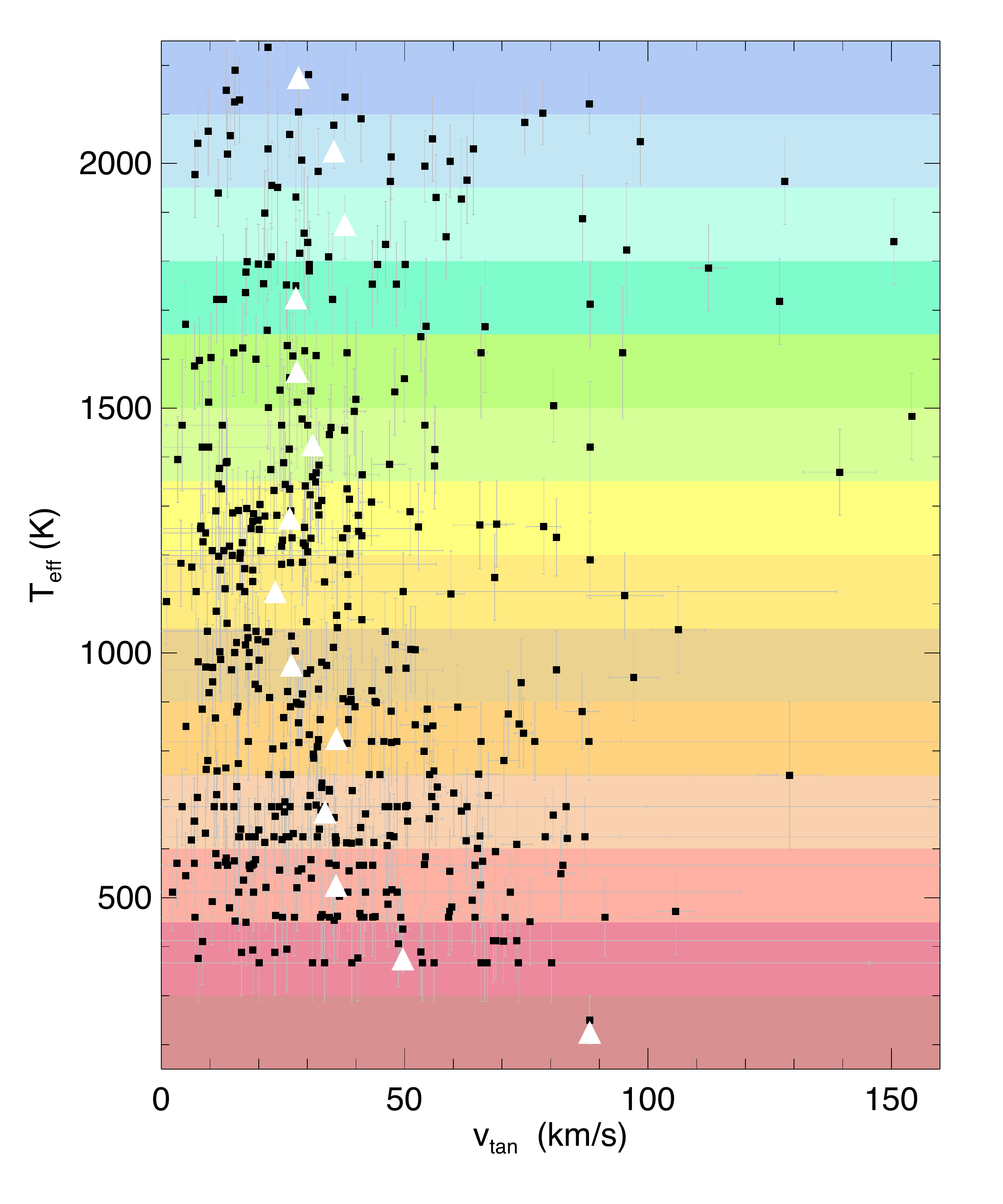}
\caption{Tangential velocities plotted against effective temperature for L, T and Y dwarfs in the 20-pc census. Only those objects having parallax measurements with uncertainties $<$12.5\% are shown. Individual objects are shown as black squares and the median $v_{tan}$ values in each 150K bin are shown as white triangles.
\label{figure:teff_vs_vtan}}
\end{figure}

(4) Cold objects have unexpected colors or magnitudes: The analysis from \cite{wright2014} inherently assumed that WISE 0855$-$0714 is a representative member of the Y dwarfs populating the 150-300K bin. What if WISE 0855$-$0714 is atypical? It has $v_{tan} = 88.0$ km s$^{-1}$, which, although in the highest 4\% of all $v_{tan}$ values in Figure~\ref{figure:teff_vs_vtan}, is not exceptional. If the majority of objects in the 150-300K bin are much older and have higher kinematics, then their high motions may suggest that point (3) above is a contributing cause. In addition, however, their older ages would also suggest a somewhat lower metallicity in general. If we look at the 20-pc T subdwarfs (section~\ref{section:known_sds}) that have metallicity measurements, we find that values as low as $[M/H] = -0.3$ dex produce noticeable changes in the spectra of mid- to late-T dwarfs. Values of $[M/H] = -0.6$ dex begin to move objects into unfamiliar loci on color-magnitude diagrams. Inasmuch as molecular absorption strengths dictate the overall spectral energy distribution of Y dwarfs (Figure 15 of \citealt{dore2016}), slight changes in metallicity could affect the relative importance of these bands and dramatically alter Y dwarf spectra and colors. Recent discoveries at early-T from \cite{schneider2020} and \cite{meisner2021} underscore the point that warmer brown dwarfs with presumably lower metallicity ([Fe/H] $\le -1$ dex) exist; their spectra are vastly different, at least in the near-infrared, from those of solar-metallicity T dwarfs. These may be harbingers of the photometric and spectroscopic bizarreness we can expect from the majority of later Y dwarfs, even if these Y dwarfs in general have less extreme metallicities.

In summary, other nearby objects with temperatures comparable to WISE 0855$-$0714 must exist, based on evidence from the mass function shape and knowledge of its low-mass cutoff. However, the expected higher motions and lower metallicities of objects in this 150-300K bin, may make them a challenge to identify, especially when coupled with their intrinsic faintness. 

\section{Conclusions\label{conclusions}}

Our results, which use the final trigonometric parallaxes we have measured using {\it Spitzer}, confirm the result of \cite{kirkpatrick2019} that the 20-pc brown dwarf portion of the mass function, which is based here on 525 L through Y dwarfs, can be best described as a power law with an exponent of $\alpha = 0.6{\pm}0.1$. We have not yet, however, extended this analysis to higher masses to investigate how the mass function behaves over the entire mass range within 20 pc. Earlier analyses have indicated that the higher mass portion can be described as a two-part power law (\citealt{kroupa2013}) or log-normal form (\citealt{chabrier2003b}). New data, particularly data from {\it Gaia} DR2 and subsequent releases can be used to refine our knowledge of the A through M dwarfs (and white dwarfs) with the 20-pc census as well as providing important astrometric information to help identify companions to those stars. Developing a database containing all knowledge of our stellar and substellar neighbors within this volume will enable us to explore the individual-object mass function with unprecedented detail.

Our results have also shown that the cutoff mass for star formation, is constrained to be lower than ${\sim}10M_{Jup}$ and that analysis of young moving group members over a wider sample likely constrains this value to ${\sim}5M_{Jup}$. Obtaining a more solid value for the cutoff mass requires volume-complete subsets of a substantial number of Y dwarfs colder than 450K, and particularly below $\sim$350K, a regime in which we have only one confirmed Y dwarf. Although {\it WISE} has provided a trove of Y dwarf discoveries, probing a substantial volume colder than $\sim$350K may require other resources. One such resource currently being planned is the {\it Near Earth Object Surveyor} (formerly called {\it NEOCam}) that is due to launch in 2025. As discussed in \cite{kirkpatrick2019-decadal}, {\it NEO Surveyor} will cover 64\% of the celestial sphere in two bands, NC1 and NC2, that cover wavelengths of 4.0-5.2 $\mu$m and 6.0-10.0 $\mu$m. Portions of the sky will be repeatedly scanned during their 75-day visibility windows then scanned again roughly 215 days later when the next visibility window opens. The mission, although planned for five years, has a design lifetime of twelve years.

The absolute NC1 fluxes of 350K Y dwarf and a 250K Y dwarf are 103 $\mu$Jy and 26 $\mu$Jy, respectively. The use of image differencing for high-motion objects in {\it NEO Surveyor} data will theoretically allow us to achieve single-epoch S/N=5 sensitivities of $\sim$4 $\mu$Jy at NC1, thereby greatly increasing the distances to which we can detect these coldest brown dwarfs. However, {\it NEO Surveyor} is run through NASA's Planetary Defense Coordination Office, so no funding is being provided for the additional processing needed for astrophysical studies. For a relatively small investment, NASA Astrophysics could realize the full potential of {\it NEO Surveyor} data for stellar astrophysical research, of which cold brown dwarf discovery would be a major beneficiary. 

\acknowledgments
This publication makes use of data products from {\it WISE}, which is a joint project of the University of California, Los Angeles, and the Jet Propulsion Laboratory (JPL), California Institute of Technology (Caltech), funded by the National Aeronautics and Space Administration (NASA). Work in this paper is based on observations made with the {\it Spitzer Space Telescope}, which is operated by JPL/Caltech, under a contract with NASA. Support for this work was provided to Davy Kirkpatrick by NASA through a Cycle 14 award issued by JPL/Caltech. Some of the data presented here were obtained at the W.\ M.\ Keck observatory, which is operated as a scientific partnership among Caltech, the University of California, and NASA. The Observatory was made possible by the generous financial support of the W.\ M.\ Keck Foundation. The authors wish to recognize and acknowledge the very significant cultural role and reverence that the summit of Maunakea has always had within the indigenous Hawaiian community. We are most fortunate to have the opportunity to conduct observations from this mountain. Results here are partly based on observations obtained at the Hale Telescope, Palomar Observatory, as part of a continuing collaboration between Caltech, NASA/JPL, Yale University, and the National Astronomical Observatories of China. We would like to thank SURF students Tea Freedman-Susskind, Emily Zhang, Yerong Xu, and Feiyang Liu for help with the spectroscopic observation of WISE 2126+2530 from Palomar. 

This work has made use of data from the European Space Agency (ESA) mission {\it Gaia} (\url{https://www.cosmos.esa.int/gaia}), processed by the {\it Gaia} Data Processing and Analysis Consortium (DPAC, \url{https://www.cosmos.esa.int/web/gaia/dpac/consortium}). Funding for the DPAC has been provided by national institutions, in particular the institutions participating in the {\it Gaia} Multilateral Agreement. This research has made use of IRSA, which is operated by JPL/Caltech, under contract with NASA.  This research has also made use of the SIMBAD database, operated at CDS, Strasbourg, France. Federico Marocco acknowledges support from grant \#80NSSC20K0452 under the NASA Astrophysics Data Analysis Program. Alfred Cayago gratefully acknowledges financial support through the Fellowships and Internships in Extremely Large Data Sets (FIELDS) Program, a National Aeronautics and Space Administration (NASA) science/technology/engineering/math (STEM) grant administered by the University of California, Riverside. Emily Martin is supported by an NSF Astronomy and Astrophysics Postdoctoral Fellowship under award AST-1801978. Eileen Gonzales acknowledges support from an LSSTC Data Science Fellowship. Christopher Theissen acknowledges support for this work through NASA Hubble Fellowship grant HST-HF2-51447.001-A awarded by the Space Telescope Science Institute, which is operated by the Association of Universities for Research in Astronomy, Inc., for NASA, under contract NAS5-26555. The Backyard Worlds: Planet 9 team thanks Zooniverse volunteers who have participated in the project. Backyard Worlds research was supported by NASA grant 2017-ADAP17-0067 and by the NSF under grants AST-2007068, AST-2009177, and AST-2009136. CatWISE is led by JPL/Caltech, with funding from NASA's Astrophysics Data Analysis Program. This research was partly carried out at JPL/Caltech, under contract with NASA. We thank the referee for a quick report despite difficulties imposed by the current pandemic.

\facilities{Spitzer(IRAC), WISE, Gaia, IRSA, CTIO:2MASS, FLWO:2MASS, Blanco(NEWFIRM, ARCoIRIS), Mt.Bigelow(2MASS), Gemini:South(FLAMINGOS-2), Magellan:Baade(PANIC, FIRE), PAIRITEL, Hale(WIRC, DBSP), SOAR(OSIRIS), LDT(NIHTS), Keck:II(NIRES), IRTF(SpeX), HST(WFC3)}

\software{IDL (https://www.harrisgeospatial.com/Software-Technology/IDL),
          MOPEX/APEX (http://irsa.ipac.caltech.edu),
          mpfit (\citealt{markwardt2009}),
          WiseView (\citealt{caselden2018}).}

\appendix
\restartappendixnumbering

\section{Spectral Types, Astrometry, and Photometry for Systems}

For systems in Tables~\ref{spitzer_results_highq}, \ref{spitzer_results_lowq}, \ref{spitzer_results_poorq}, \ref{ancillary_spitzer_photometry}, \ref{spectroscopic_followup}, \ref{table:distance_estimates}, and \ref{20pc_sample}, we have collected spectroscopic, astrometric, and photometric data from both this paper and the literature. These data are listed in Table~\ref{table:monster_table}. The various sections of the table are described in detail below. Close binaries are generally entered as a single entry with joint photometry unless there are components of the multiple with spectral types earlier than L0. For a full accounting of individual L, T, and Y components within the 20-pc census, refer to Table~\ref{20pc_sample}.

\subsection{Origin and Name}

Column $T$ indicates the table(s) from which the source originates. Objects in the 20-pc census (Table~\ref{20pc_sample}) are indicated by "T". Users are encouraged to use this column rather than the parallax column if they wish to select the same set of objects that we included in our 20-pc census. Objects that are not listed in our 20-pc census (Table~\ref{20pc_sample}) but were nonetheless part of our {\it Spitzer} parallax program (Tables~\ref{spitzer_results_highq}-\ref{spitzer_results_poorq}) are indicated by "P". Objects that are not from any of these tables but were part of our photometric or spectroscopic follow-up campaigns (Tables~\ref{ancillary_spitzer_photometry} and \ref{spectroscopic_followup}) are indicated by "F". Objects considered for the 20-pc census but ultimately not included (Table~\ref{table:distance_estimates}) are indicated by "C".

Column $Short Name$ gives the abbreviated prefix and suffix of the full source name. This prefix is generally the survey of origin, and the abbreviated suffix is the sexagesimal RA and Dec of the source in the form $hhmm{\pm}ddmm$. As examples, CWISEP J193518.59$-$154620.3 is denoted as CWISE 1935$-$1546, and PSO J149.0341$-$14.7857 is denoted as PSO 0956$-$1447. Exceptions are made for objects with common names like Gl 570D and LHS 2397aB, whose full names are used instead. 

\subsection{Spectral Types}

Columns $SpO$ and $SpIR$ list the optical and near-infrared spectral types, respectively, if known. These are converted to a decimal scale, and any qualifying criteria such as "pec", "$\beta$", and "sd" are dropped. The convention for the decimal scale is L0 = 0.0, T0 = 10.0, and Y0 = 20.0. As examples, an object with a spectral type of sdT8 is given as 18.0, and one with a type of L7: VL-G is given as 7.0. The two objects listed in Table~\ref{20pc_sample} with types of "extremely red" in \cite{mace2013} are given in this table as 9.5. Column $SpAd$ is the adopted spectral type, which is the same as $SpIR$ if that value is not null; otherwise, it is the same as $SpO$. If both of those quantities are null, a spectral type estimate is given. A few objects, however, have null values for $SpAd$, and these are objects believed to be background interlopers and not brown dwarfs.

The source of the spectral type is given in column $OI$. An explanation of the double-letter code for this column can be found in the table comments.

\subsection{Astrometric Data}

Columns $\varpi_{abs}$, $\mu_\alpha$, and $\mu_\delta$ list the best measured trigonometric parallax and proper motion values in RA and Dec. The "best" astrometry is simply that data set with the smallest quoted uncertainty in the parallax or, for objects lacking a parallax measurement, the data set with the smallest quoted uncertainty in the total proper motion. All parallaxes are given on the absolute reference grid; data from \cite{tinney2003} and \cite{tinney2014}, along with USNO data from \cite{kirkpatrick2019}, were converted from relative to absolute as described in section 8 of \cite{kirkpatrick2019}. The values listed for proper motion are a mixture of relative and absolute measurements. Readers are encouraged to cite the source of those values if this distinction is important for their research.

The source of the astrometry is given in column $AS$. An explanation of the single-letter code for this column can be found in the table comments.

\subsection{$JHK$ Photometry}

Column $J_{MKO}$ lists $J$-band photometry on the MKO system, $J_{2MASS}$ lists $J$-band photometry one the 2MASS system, $H$ lists $H$-band photometry on either the MKO or 2MASS system, $K_{MKO}$ lists $K$-band photometry on the MKO system, and $K_{S(2MASS)}$ lists $K_S$-band photometry on the 2MASS system.  See section~\ref{section: JHK} for details. Photometric values listed without corresponding errors are magnitude limits.

The source of the photometry is given in column $PhotS$. An explanation of the five-letter code for this column can be found in the table comments.

\subsection{CatWISE2020 Data}

Columns $RA\_C2$, $Dec\_C2$, $pmra\_C2$, $pmdec\_C2$, $W1mag\_C2$, $W2mag\_C2$, and $par\_C2$ contain astrometric information from the CatWISE2020 Catalog and Reject Table (\citealt{marocco2020b}). The first two columns are the J2000 equinox RA and Dec positions from the moving-object solution at epoch MJD 57170.0, the next two columns are the measured proper motion and their uncertainties in RA and Dec, the next two columns are the moving-object PSF-fit photometry in {\it WISE} bands W1 and W2, and the final column is a crude measurement of the object's parallax (called {\fontfamily{pcr}\selectfont par\_pm} in the documentation).

The source of the CatWISE2020 data is given in column $C2S$. Upper-case "C2" refers to the Catalog and lower-case "c2" refers to the Reject Table.

\subsection{AllWISE Data}

Columns $W1mag$, $W2mag$, and $W3mag$ provide stationary-object PSF-fit measurements (primarily from AllWISE) in {\it WISE} bands W1, W2, and W3. These are provided for two reasons. First, CatWISE2020 does not provide any W3 data, since this band was not available for the post-cryogenic phases of the {\it WISE} and {\it NEOWISE} missions. Second, the short, six-month time baseline of AllWISE means that this stationary-object photometry should be robust for all sources except those of exceptionally large motion, and thus the W1 and W2 photometry can be compared to the moving-object photometry from CatWISE2020 to provide another photometric check.

The source of the stationary-object photometry is given in column $WS$. In most cases, this is the AllWISE Source Catalog or Reject Table. Some sources, however, were not detected until {\fontfamily{pcr}\selectfont crowdsource} (\citealt{schlafly2018}) was used on the unWISE images underlying the CatWISE2020 processing. In this case, the stationary-object W1 and W2 photometry from CatWISE2020 is listed instead.

\subsection{{\it Spitzer} Data}

Columns $ch1mag$ and $ch2mag$ provide the {\it Spitzer} channel 1 (3.6 $\mu$m) and channel 2 (4.5$\mu$m) photometry. The source of this photometry is given in column $SS$, the single-character code for which is described in the table comments.

\subsection{Note and Full Designation}

Column $Note$ lists a one-letter code indicating whether the object is an unresolved multiple (M); a young, low-gravity object (Y), or an old, subdwarf (S). Column $FullName$ gives the full discovery designation of the system.

\movetabledown=65mm
\begin{rotatetable}
\begin{deluxetable*}{clrrccccccccccccccc}
\tabletypesize{\scriptsize}
%\tablenum{1}
\tablecaption{Amassed Spectroscopic, Astrometric, and Photometric Data for Objects Listed in Tables~\ref{spitzer_results_highq}, \ref{spitzer_results_lowq}, \ref{spitzer_results_poorq}, \ref{ancillary_spitzer_photometry}, \ref{spectroscopic_followup}, \ref{table:distance_estimates}, and \ref{20pc_sample}\label{table:monster_table}}
\tablehead{
\colhead{T} &
\colhead{Name} &
\colhead{SpO} &
\colhead{SpIR} &
\colhead{SpAd} &
\colhead{OI} &
\colhead{$\varpi_{abs}$} &
\colhead{$\mu_\alpha$} &
\colhead{$\mu_\delta$} &
\colhead{AS} &
\colhead{$J_{MKO}$} &
\colhead{$J_{2MASS}$} &
\colhead{$H$} &
\colhead{$K_{MKO}$} &
\colhead{$K_{S(2MASS)}$} &
\colhead{PhotS} &
\colhead{RA\_C2} &
\colhead{Dec\_C2} &
\colhead{...} \\
\colhead{} &
\colhead{} &
\colhead{} &
\colhead{} &
\colhead{} &
\colhead{} &
\colhead{(mas)} &
\colhead{(mas/yr)} &
\colhead{(mas/yr)} &
\colhead{} &
\colhead{(mag)} &
\colhead{(mag)} &
\colhead{(mag)} &
\colhead{(mag)} &
\colhead{(mag)} &
\colhead{} &
\colhead{(deg)} &
\colhead{(deg)} &
\colhead{} \\
\colhead{(1)} &
\colhead{(2)} &
\colhead{(3)} &
\colhead{(4)} &
\colhead{(5)} &
\colhead{(6)} &
\colhead{(7)} &
\colhead{(8)} &
\colhead{(9)} &
\colhead{(10)} &
\colhead{(11)} &
\colhead{(12)} &
\colhead{(13)} &
\colhead{(14)} &
\colhead{(15)} &
\colhead{(16)} &
\colhead{(17)} &
\colhead{(18)} &
\colhead{(19)} 
}
\startdata
T& SDSS  0000+2554&     15.0&  14.5& 14.5& TT&   70.8   $\pm$ 1.9   &   $-$19.1 $\pm$  1.5 &   126.7 $\pm$  1.3 &   D&  14.85$\pm$0.01&  15.06$\pm$0.04&  14.73$\pm$0.07&  14.82$\pm$0.03&  14.84$\pm$0.12& U22D2&   0.0563043&    25.9054854\\ 
T& GJ 1001BC      &      5.0&   5.0&  5.0& kT&   82.0946$\pm$ 0.3768&   671.09$\pm$  0.35& $-$1498.16$\pm$  0.51&   G&  12.98$\pm$0.01&  13.11$\pm$0.02&  12.06$\pm$0.03&  \nodata       &  11.39$\pm$0.01& V22-V&   1.1491771&   $-$40.7415963\\ 
T& WISE  0005+3737&  \nodata&  19.0& 19.0& -T&  126.9   $\pm$ 2.1   &   997.3 $\pm$  1.0 &  $-$271.6 $\pm$  1.0 &   T&  17.58$\pm$0.04&  \nodata       &  17.98$\pm$0.02&  \nodata       &  16.28$\pm$0.31& U-k-2&   1.3250452&    37.6219054\\      
T& 2MASS 0014$-$4844&      2.5&   2.5& 2.5& TT&   50.1064$\pm$ 0.3898&   870.72$\pm$  0.27&   281.46$\pm$  0.43&   G&  13.91$\pm$0.01&  14.05$\pm$0.04&  13.26$\pm$0.01&  \nodata       &  12.78$\pm$0.01& V2V-V&   3.7386552&   $-$48.7367024\\ 
T& WISE  0015$-$4615&  \nodata&  18.0& 18.0& -T&   75.2   $\pm$ 2.4   &   413.4 $\pm$  1.1 &  $-$687.8 $\pm$  1.0 &   T&  17.67$\pm$0.02&  \nodata       &  17.91$\pm$0.07&  \nodata       &  \nodata       & V-V--&   3.7755685&   $-$46.2558784\\ 
T& 2MASS 0015+3516&      2.0&   1.0& 1.0& TT&   58.6085$\pm$ 0.3664&    55.17$\pm$  0.45&  $-$257.09$\pm$  0.28&   G&  13.71$\pm$0.01&  13.88$\pm$0.03&  12.89$\pm$0.04&  \nodata       &  12.26$\pm$0.02& UKK-2&   3.9368016&    35.2663391\\ 
\enddata
%\end{deluxetable*}
\onecolumngrid
\tablecomments{Only a portion of the table's rows and columns is shown here. The full table is available from the journal in machine-readable format.}
\tablecomments{References for OI, where the reference for the optical (O) spectral type is given as the first character and that for the near-infrared (I) spectral type is given as the second character:
(a) \citealt{albert2011},
(A) \citealt{thompson2013},  
(b) \citealt{burningham2010},  
(B) \citealt{burgasser2010b},  
(c) \citealt{cushing2011},
(C) \citealt{cushing2018}, 
(d) \citealt{kirkpatrick2012},
(D) \citealt{kirkpatrick2000},
(e) \citealt{martin2018},  
(E) \citealt{reid2001a},  
(f) \citealt{kirkpatrick2010}, 
(F) \citealt{faherty2014},
(g) \citealt{burgasser2006},
(G) \citealt{bardalez2014}, 
(h) \citealt{hawley2002}, 
(H) \citealt{dhital2011}, 
(i) \citealt{chiu2006},
(I) \citealt{koen2017}, 
(J) \citealt{kirkpatrick1999}, 
(j) \citealt{kirkpatrick2008},  
(k) \citealt{kirkpatrick2001}, 
(K) \citealt{kirkpatrick2011}, 
(l) \citealt{kendall2007},
(L) \citealt{kendall2003},  
(m) \citealt{artigau2011},  
(M) \citealt{mace2013}, 
(n) \citealt{scholz2003},  
(N) \citealt{king2010},  
(p) \citealt{potter2002},
(P) \citealt{pineda2016}, 
(q) \citealt{gizis2002},
(Q) \citealt{cruz2007},
(r) \citealt{reid2008},
(R) \citealt{reid2006}, 
(s) \citealt{schneider2014},   
(S) \citealt{schneider2015},  
(t) \citealt{tinney2018},
(T) See Tables~\ref{spectroscopic_followup}-\ref{table:young_objects} in this paper for references,
(u) \citealt{burningham2013},
(U) \citealt{burgasser2007}, 
(v) \citealt{schneider2017},
(V) \citealt{kirkpatrick2016},
(w) \citealt{kendall2004}, 
(W) \citealt{best2013},   
(X) \citealt{burgasser2003-opt_spec},  
(x) \citealt{thorstensen2003}, 
(y) \citealt{deacon2014},    
(Y) \citealt{reyle2014},  
(z) \citealt{burgasser2010},    
(Z) \citealt{fan2000}.}
\tablecomments{References for AS, the source of the astrometric data:
(A) \citealt{dahn2002}, 
(b) \citealt{burgasser2008}, 
(B) \citealt{bartlett2017}, 
(c) CatWISE2020 Catalog,
(C) \citealt{tinney2014},
(d) \citealt{dahn2017},
(D) \citealt{dupuy2012},
(E) \citealt{dupuy2019},
(F) \citealt{faherty2012},
(G) Gaia DR2 - quoted astrometry is for the actual source listed,
(g) Gaia DR2 - quoted astrometry is that of the brighter primary in the system,
(J) \citealt{kirkpatrick2011},
(H) Hipparcos - \citealt{vanleeuwen2007}, 
(K) \citealt{kirkpatrick2019} for NTT and UKIRT parallaxes,
(k) \citealt{kirkpatrick2019} for USNO parallaxes, 
(l) \citealt{leggett2012},
(L) \citealt{liu2016},
(m) \citealt{manjavacas2013},
(M) \citealt{marocco2010}, 
(r) Smart, priv.\ comm.,
(R) \citealt{smart2018}, 
(S) \citealt{casewell2008}, 
(s) \citealt{smart2013},
(t) \citealt{tinney2003},
(T) This paper,
(V) \citealt{vrba2004},
(W) \citealt{best2020},
(z) \citealt{dupuy2020},  
(Z) \citealt{lazorenko2018}.}
\tablecomments{References for PhotS, the source of the  $J, H, K$ photometry: 
(2) 2MASS \citealt{skrutskie2006},
(a) \citealt{meisner2020a},
(A) \citealt{meisner2020b},
(b) \citealt{bardalez2020}: Note that the {\it HST} F125W magnitude limit for WISE 0830+2837 is used as its value for $J_{MKO}$,
(B) Bigelow/2MASS,
(c) \citealt{boccaletti2003},
(C) CTIO-4m/NEWFIRM,
(d) \citealt{kirkpatrick2012},
(D) Database of Ultracool Parallaxes as of 2020 April: \citealt{dupuy2012}, \citealt{dupuy2013}, and \citealt{liu2016}, 
(e) \citealt{martin2018},
(E) \citealt{mcelwain2006}, 
(f) \citealt{faherty2012},  
(F) \citealt{freed2003}, 
(g) \citealt{mamajek2018},
(G) Gemini-South/FLAMINGOS2,
(h) \citealt{pinfield2014b}, 
(H) \citealt{pinfield2014a},
(i) \citealt{ireland2008},
(I) \citealt{dupuy2019},
(j) \citealt{janson2011},
(J) \citealt{faherty2014b},
(k) \citealt{kirkpatrick2019}, 
(K) \citealt{kirkpatrick2011},
(m) \citealt{mace2013},
(M) Magellan/PANIC,
(p) PAIRITEL,
(P) Palomar/WIRC,
(q) \citealt{dhital2011},
(Q) \citealt{deacon2017},
(r) \citealt{deacon2012b},
(s) \citealt{schneider2015},
(S) SOAR/OSIRIS,
(t) \citealt{tinney2014},
(T) \citealt{thompson2013},
(u) ULAS, UGPS, or UGCS,
(U) UHS,
(v) VVV,
(V) VHS,
(w) \citealt{wright2013},
(W) \citealt{best2020}.}
\tablecomments{References for C2S, the source of the CatWISE2020 data:
(C2) CatWISE2020 Catalog,
(c2) CatWISE2020 Reject Table.}
\tablecomments{References for WS, the source of the {\it WISE} photometry:
(AW) AllWISE Source Catalog,
(aw) AllWISE Reject Table,
(C2) CatWISE2020 Catalog,
(c2) CatWISE2020 Reject Table.}
\tablecomments{References for SS, the source of the {\it Spitzer} photometry:
(0) This paper,
(f) \citealt{marocco2020},
(F) \citealt{filippazzo2015},
(K) \citealt{kirkpatrick2019}, 
(L) \citealt{leggett2007}, 
(M) \citealt{meisner2020a},
(m) \citealt{meisner2020b},
(P) \citealt{patten2006}, 
(S) \citealt{metchev2015}.}
\end{deluxetable*}
\end{rotatetable}

\end{document}